%% file: Diplomarbeit.tex
\documentclass[10pt,a4paper,fleqn,twoside]{report}
\usepackage{amssymb}
\usepackage{amsmath}
\usepackage{graphics}
\usepackage{epsfig}
\usepackage{feynmf}
\usepackage{slashed}
\usepackage[german,english]{babel}
\renewcommand{\vec}[1]{{\bf #1}}
\def\bra{\langle}
\def\ket{\rangle}
\newcommand{\eqb}{\begin{equation}}
\newcommand{\eqe}{\end{equation}}
\newcommand{\dmb}{\begin{displaymath}}
\newcommand{\dme}{\end{displaymath}}
\newcommand{\pd}{\partial}

\newcommand{\eab}{\begin{eqnarray}}
\newcommand{\eae}{\end{eqnarray}}
\newcommand{\ra}{\right\rangle}
\newcommand{\la}{\left\langle}
\newcommand{\e}{\mbox{e}}
\newcommand{\be}{\begin{equation}}
\newcommand{\ee}{\end{equation}}

\begin{document}
\rm
\pagenumbering{roman}
\begin{titlepage}
\centering
\renewcommand{\baselinestretch}{1.5}
\sc
\huge Faculty of Physics and Astronomy \\
\LARGE University of Heidelberg \\
\vfill
\rm
\large
{\bf
Diploma thesis \\
in Physics \\
submitted by \\
Jochen Keller \\
born in Speyer, Germany. \\
2008
}
\end{titlepage}

\thispagestyle{empty}
\cleardoublepage

\begin{titlepage}
\centering
\renewcommand{\baselinestretch}{1.5}
\vspace*{3cm}
\sc
\huge
Gauge-invariant two-point correlator of energy density in deconfining \\
$SU(2)$ Yang-Mills thermodynamics\\
\vfill
\rm
\large
{\bf
This diploma thesis has been carried out by \\
Jochen Keller \\
at the \\
Institut f\"ur Theoretische Physik \\
under the supervision of \\
Priv.-Doz. Dr. Ralf Hofmann
}
\end{titlepage}
\thispagestyle{empty}
\cleardoublepage

\include{abstract}

\setcounter{tocdepth}{3}
\tableofcontents
\cleardoublepage

\include{introduction}

\include{YMTD}

\include{energydensity}

\include{stringtension}

\include{summary}

\appendix

\include{FTFT}

\include{details1}

\include{details2}

\include{details3}

\include{details4}

\cleardoublepage
\chapter*{Danksagung}
\selectlanguage{german}

An erster Stelle m\"ochte ich mich bei meinem Betreuer \textit{Ralf Hofmann} bedanken.
Die Zusammenarbeit w\"ahrend des letzten Jahres, vor allem das Verfassen eines gemeinsamen Artikels, war eine sehr lehrreiche Erfahrung f\"ur mich. Obwohl der etwas komplizierten Situation, die sich aufgrund der Tatsache ergab, dass wir seit M\"arz nicht mehr zusammen im Heidelberger Institut t\"atig waren, war der Kontakt doch stets gegeben und man f\"uhlte sich nie allein gelassen. Auch ich kann mich meinen Vorg\"angern nur anschlie\ss\,en, dass das Interesse welches \textit{Ralf Hofmann} seinen Studenten entgegenbringt, mit Sicherheit bemerkenswert ist.

Desweiteren bedanke ich mich bei \textit{Jan Martin Pawlowski} f\"ur die \"Ubernahme der Zweitkorrektur dieser Diplomarbeit.

Auch bedanken m\"ochte ich mich bei meinem Kollegen. Zun\"achst bei \textit{Markus Schwarz} in Karlsruhe, f\"ur das Bereitstellen der Ergebnisse seiner Diplomarbeit, die in meine Arbeit eingingen. Desweiteren danke ich auch meinen Kollegen im Heidelberger Institutskeller, \textit{Michal Szopa} und \textit{Andreas von Manteuffel}, welche stets hilfreich zur Seite standen, als ich mit Mathematica, \LaTeX, etc. etwas \"uberfordert war.

Ein besonderer Dank geht an meine Eltern \textit{Manfred} und \textit{Ilse Keller} f\"ur die Unterst\"utzung w\"ahrend des Studiums.

\newpage
\thispagestyle{empty}
\
\newpage
\thispagestyle{empty}
\subsection*{Erkl\"arung:}

Ich versichere, dass ich diese Arbeit selbstst\"andig verfasst und keine anderen
als die angegebenen Quellen und Hilfsmittel benutzt habe.\\[2cm]
Heidelberg, den ......................\hfill.....................................\\[-0.5cm]
\begin{flushright}
\small Unterschrift
\end{flushright}


\end{document}

%% file: abstract.tex
\begin{abstract}
The thesis is considering aspects of $SU(2)$ Yang-Mills thermodynamics in its deconfining high-temperature phase.

We calculate the two-point correlation function of the energy density of the photon in a thermalized gas, at first in the conventional $U(1)$ gauge theory, followed by a calculation, where the photon is identified with the massless gauge mode in deconfining $SU(2)$ Yang-Mills thermodynamics. Apart from the fact, that this calculation is interesting from a technical point of view, we can consider several aspects of phenomenological relevance.
Since we interpret the two-point correlator of energy density as a measure for the energy transfer, and thus for the electromagnetic interaction of microscopic objects, such as atoms immersed into a photon gas, we are able to give an explanation for the unexpected stability of cold, innergalactic clouds consisting of atomic hydrogen.

Subsequently, we evaluate the spatial string tension in deconfining $SU(2)$ Yang-Mills thermodynamics, which can be regarded as measure for the magnetic flux through the area enclosed by the associated Wilson loop. On the level of on-shell polarization effects for the massless mode we observe a perimeter-law, and we speculate that the lattice-obtained area-law is induced by off-shell contributions to the polarization tensor. Moreover, we discuss an interesting two-loop result for the pressure which seems to be associated with the presence of screened magnetic monopoles being responsible for an area-law.
\vspace{35mm}
\selectlanguage{german}
\begin{center}
{\bf Zusammenfassung}
\end{center}
\vspace{2mm}
In dieser Diplomarbeit werden Aspekte der $SU(2)$ Yang-Mills Thermodynamik in ihrer dekonfinierten Hochtemperatur-Phase betrachtet.

Wir berechnen die Zwei-Punkt Korrelationsfunktion der Energiedichte des Photons in einem thermalisiertem Gas, zun\"achst in der konventionellen $U(1)$-Eichtheorie. Danach betrachten wir den Fall, dass das Photon mit dem masselosen Eichfeld in einer dekonfinierten $SU(2)$ Yang-Mills Theorie identifiziert wird. Neben der Tatsache, dass diese Rechnung vom technischen Standpunkt aus von Interesse ist, k\"onnen wir mehrere Aspekte ph\"anomenologischer Natur betrachen.
Sofern wir den Zwei-Punkt Korrelator der Energiedichte als ein Ma\ss\,  f\"ur den Energietransfer, und damit auch f\"ur die elektromagnetische Wechselwirkung mikroskopischer Objekte, wie in ein Photongas eingebettete Atome, ansehen, sind wir in der Lage, eine Erkl\"arung f\"ur die unerwartete Stabilit\"at kalter, innergalaktischer Wolken, bestehend aus atomarem Wasserstoff, zu geben.

Anschlie\ss end berechnen wir die r\"aumliche Stringspannung in dekonfinierter $SU(2)$ Yang-Mills Thermodynamik. Diese Gr\"osse kann als Ma\ss\,  f\"ur den magnetischen Flu\ss\, durch  die vom entsprechenden Wegner-Wilson Loop umschlossene Fl\"ache interpretiert werden. Auf dem Niveau von on-shell Polarisationseffekten f\"ur die masselose Mode beobachten wir ein Perimeter-Gesetz, und wir vermuten, dass das aus Gittersimulationen best\"atigte Fl\"achen-Gesetz auf off-shell-Beitr\"age des Polarisationstensors zur\"uckzuf\"uhren ist. Weiterhin diskutieren wir ein interessantes Zwei-Schleifen Resultat f\"ur den Druck, das mit der Pr\"asenz abgeschirmter magnetischer Monopole in Zusammenhang zu stehen scheint, welche wiederum verantwortlich f\"ur ein Fl\"achen-Gesetz w\"aren.
\end{abstract}

\selectlanguage{english}
\cleardoublepage

%% file: introduction.tex
\chapter{Introduction}
\pagenumbering{arabic}
The importance of gauge theories in theoretical physics is evident, since they provide a deep, mathematical framework for the formulation of quantum field theories. Fundamental interactions are described by gauge theories. Interactions, as they occur in the standard model of particle physics (SM), between fermionic fields are mediated via the exchange of bosonic gauge fields. The interaction is completely determined through gauge symmetry. In particular, non-abelian gauge groups proved to be an important step in describing Nature on the level of elementary particles.

In the SM gauge theories are considered in the framework of perturbation
theory. Many results derived on this basis are verified up to enormous
precision. For examples we refer to the anomalous magnetic moment of the
electron in Quantum Electrodynamics (QED) or the asymptotic freedom in
Quantum Chromodynamics (QCD). Nevertheless, it is not the case that
perturbative methods leave no open questions. At first we have to
mention that not even the most elementary processes of QED are
mathematically described in a sound way. Perturbative solutions are
asymptotic series, the philosophy is an expansion in the small gauge
coupling. Thus the perturbative concepts are only viable for small
couplings. In case of a strongly coupled system, the philosophy of
perturbation theory collapses. For instance, the dependence of the gauge
coupling on the energy scale is characterized by the
$\beta$-function. In QCD, for increasing energy, the coupling decreases
logarithmically and thus the calculation of the $\beta$-function in
terms of a perturbation series seems to be well justified for large
momentum transfer, see also \cite{Hofmann/Giacosa Landaupole}.
This phenomenon is called asymptotic freedom and was discovered in 1973
by Gross, Wilczek and Politzer \footnote{We recall that already in 1969 I. Khriplovich \cite{Khriplovich} and in 1972 G. 't Hooft \cite{'t Hooft asymptotic freedom} had performed according calculations.}. As for the thermalized situation
perturbation
theory leads to infrared instabilities, resulting from weakly screened soft magnetic modes \cite{Linde80/Polyakov75}. This manifests itself in a gluon plasma at high temperature, and is in contradiction to naive expectation, since one would expect perturbative methods to yield reasonable results, because the gauge coupling approaches zero for temperature $T$ exceeding the Yang-Mills scale many times over. However, a perturbative expansion of the thermodynamical pressure in a Yang-Mills gas merely is applicable up to order $g^6\ln g$.

Beyond this order we have to regard the fact, that in pure non-abelian gauge theories exist infrared stabilizing, topologically nontrivial field configurations in the euclidean formulation, while in perturbation theory the expansion is about the trivial vacuum.

In addition, there are a number of experimental and observational results in elementary particle physics and cosmology, where the SM either yields no explanation or even leads to contradiction with the experiments and observation. For instance, there is the fact that the SM does not provide for any theory about the phenomenon of dark matter and dark energy, which corresponds to about 95\% of the total energy density content of the Universe. Furthermore the direct experimental detection of the Higgs particle is still open. This leaves room for speculations concerning the electroweak symmetry breaking of the SM.
Another unexplained observation is the missing power in the CMB temperature-temperature correlation at large angular scales \cite{Kogut, Hofmann/Szopa}. Finally we refer to the existence of a large, cold, old, dilute, innergalactic cloud (GSH130-03-69) in-between spiral arms of the Milky Way \cite{Brunt/Knee01, Dickey01}. The astrophysical origin of this and similar structures is mysterious, its high content of atomic hydrogen in view of its unexpectedly large age clashes with the SM, see Sec.\,\ref{MilkyWay}.

In order to overcome these problems an investigation of (non-abelian) gauge theories involving nonperturbative methods seems to be essential. In \cite{Hofmann05} an analytic and nonperturbative approach to $SU(2)$ and $SU(3)$ Yang-Mills thermodynamics was developed. The construction of this effective theory is close in spirit to the Ginzburg-Landau-Abrikosov theory, a macroscopic, phenomenonological description of superconductivity in metals, where the microscopic processes leading to a nontrivial ground state are summarized by scalar fields and pure-gauge configurations. In the deconfining phase of $SU(2)$/$SU(3)$ Yang-Mills thermodynamics \cite{Herbst2005} the thermal ground state is determined by an adjoint scalar field $\phi$ and a pure-gauge field configuration $a_\mu^{bg}$, solving the Yang-Mills equation in the presence of the background $\phi$. This macroscopic field $\phi$ originates from spatial coarse-graining over microscopic processes. The field $\phi$ is compose
 d of noninteracting calorons (and anticalorons) of topological charge one and trivial holonomy. This composite adjoint scalar field governs the dynamics of the spatially averaged, BPS saturated, topological nontrivial and noninteracting part of the thermal ground state, it is quantum mechanically and statistically inert, and thus it indeed acts as a background for the topological trivial sector of the theory, which is responsible for caloron interactions. The sector of trivial topology is minimally coupled to the coarse-grained caloron configurations and manifested as the already mentioned pure-gauge configuration $a_\mu^{bg}$. The macroscopic adjoint scalar acts as a Higgs field, and it breaks the $SU(2)$ gauge symmetry down to $U(1)$. Thereby, on tree-level, two out of three gauge bosons acquire a temperature-dependent mass, while the third gauge mode remains massless. Another remarkable property of the effective theory is its finiteness in the ultraviolet and infrared. An
  infrared cutoff results from caloron induced gauge boson masses, an ultraviolet cutoff is provided by the compositeness scale $|\phi|$ which constrains momenta in loop calculations in the only physical gauge: unitary-Coulomb. Thus no renormalization procedure to absorb divergencies is necessary. The evolution of the effective gauge coupling $e$ is derived from the requirement of thermodynamical self-consistency. That is, Legendre transformations following from the fundamental partition function, need to be kept intact after spatial coarse-graining. In the deconfining phase of $SU(2)$ Yang-Mills thermodynamics the effective coupling has a plateau value $e(T)=\sqrt{8}\pi$, and diverges logarithmically at the critical temperature. This is the phase boundary to the preconfining (magnetic) phase. In this thesis we consider the high-temperature deconfining (electric) phase of $SU(2)$ Yang-Mills thermodynamics only.

The thesis is organized as follows: Chapter two gives an introduction into the concepts of Yang-Mills theories. Nonperturbative, topological field configurations in euclidean spacetime are presented, especially instanton, or at finite temperature, caloron configurations. The next section contains a review of the developments in nonperturbative, analytic $SU(2)$ Yang-Mills thermodynamics. The generation of the nonperturbative ground state and the construction of the effective theory is presented, followed by a brief overview of the other two phases of the theory. Chapter three contains the calculation of the two-point correlation function of energy density, at first in the conventional $U(1)$ electrodynamics, followed by a calculation in deconfining $SU(2)$ Yang-Mills thermodynamics. We compare our results for both cases and observe a sizeable suppression of the $SU(2)$ correlator at low temperatures and large distances. Since we interpret the two-point correlation of energy d
 ensity as a measure for the energy transfer and thus for the interaction between single atoms, immersed in a photon gas, we give a possible explanation concerning the unexpected stability of cold clouds consisting of atomic hydrogen located in-between spiral arms of our galaxy. Chapter four is dedicated to a quantity called spatial string tension. This object is defined as the logarithm of the spatial Wilson loop and is a measure for the dual magnetic flux through the area enclosed by the loop. In lattice gauge theory the high-temperature properties of the string tension were calculated. On the level of on-shell 'photon' polarization effects the generally accepted lattice results are not reproduced, but we are able to offer an explanation why our approach is insufficient at this level. Finally, in Chapter five we give a summary of the thesis and an outlook on further research. In the appendix, we recall some general concepts of quantum field theory at finite temperature and 
 state the Feynman rules in the real-time formulation. In addition, we give some technical details concerning various calculations in this work. 

%% file: YMTD.tex
\chapter{Yang-Mills Thermodynamics}

\section{Lie groups and Lie algebras}

The following section is dedicated to the mathematical requirements for the formulation
formulation of non-abelian gauge theories. We give a short introduction into the theory
of Lie groups and Lie algebras. More material concerning the theory of Lie groups can be found in \cite{Frankel, Nakahara, Georgi}.

\underline{\textit{Lie group:}}
A Lie group is a differentiable manifold $G$ endowed with an associative,
differentiable, invertible map
\begin{equation}
G\times G\rightarrow G \hspace{20mm} (g_1,g_2)\mapsto g_1\cdot g_2
\quad \forall g_1,g_2\in G,
\end{equation}
such that $G$ obeys the group properties. The existence of a neutral element $e$ and an inverse element $g^{-1}$ is provided.

As examples for Lie groups we regard the
group of regular invertible linear transformations of a vector space $V$, $Gl(V)$. Of particular interest in theoretical physics are the Lie groups $Gl(N,\mathbb{R})$ and $Gl(N,\mathbb{C})$ and some of their subgroups. It is a crucial fact that every subgroup of a Lie group is a Lie group itself. If $V$ is given as the
$N$-dimensional euclidean space $\mathbb{R}^N$, the general linear group $Gl(N,\mathbb{R})$ consists of all real $N\times N$-matrices with $\det g\neq 0$. $Gl(N,\mathbb{R})$ is a manifold of dimension $N^2$.

If $V$ is given as the complex vector space $\mathbb{C}^N$, the Lie group
$Gl(N,\mathbb{C)}$ is the set of all nonsingular $N\times N$-matrices with complex entries. As subgroup of physical relevance we should mention the $N^2-1$-dimensional special unitary group $SU(N)$, consisting of all unitary matrices $M\in U(N)$ with unit determinant.

The center of a group G is the subgroup containing all those elements x in G, that commute with all other elements, that is $xg=gx$ for all g in G.

\underline{\textit{Lie algebra:}}
A Lie algebra is a vector space together with a bilinear operator $[\cdot,\cdot]$
\begin{equation}
V\times V\rightarrow V \hspace{20mm} (x,y)\mapsto [x,y] \quad\forall x,y\in V\,,
\end{equation}
called the Lie bracket, which is anticommutative,
\begin{equation}
[x,y]=-[y,x] \quad\forall x,y\in V\,,
\end{equation}
and satisfies the Jacobi identity
\begin{equation}
[x,[y,z]]+[z,[x,y]]+[y,[z,x]]=0 \quad\forall x,y,z\in V\,.
\end{equation}
If we constrain ourselves to Matrix groups, the Lie bracket is identical with the commutator
of two matrices. For a Lie group $Gl(V)$ the associated Lie algebra is denoted as $\mathfrak{gl}(V)$.

As an example for a Lie algebra we point out that the tangent space at an arbitrary element g of a Lie group $G$ always possesses the structure of a Lie algebra. We assume that the group elements $g\in G$ are parametrized by a set of continuous parameters $\alpha^{a}\,(a=1,2,\dots,N)$ in a way that $\,\alpha=0\,$ corresponds to the identity, i.e. $g(\alpha^{a}=0)=e$. The set of parameters represent a differentiable manifold.
Since the group elements depend analytically on the parameters $\alpha^{a}$, any element $g\in G$ (at least in the vicinity of the identity element) can be Taylor expanded about the unit element $e$ of the group. It holds
\begin{equation}
g(\alpha)=e+i\alpha^{a}t^{a}+\mathcal{O}(\alpha^2),\qquad\mbox{where}\quad
t^{a}=\left.\left(\frac{\partial g(\alpha)}{\partial\alpha^{a}}\right)\right|_{\alpha=0}\,.
\end{equation}
The objects $t^{a}$ are referred to as infinitesimal generators of the Lie group.
The generators of a Lie group always establish a Lie algebra, where the dimension of the vector space V (and with this in mind the dimension of the Lie algebra) corresponds to the number of generators $K$. In order to illustrate
this fact we consider the concrete case of a group manifold, parametrized by coordinates $x^1,x^2,\dots,x^N$. A basis for the tangent space is spanned by the directional derivatives $\partial_i=\frac{\partial}{\partial x^{i}}$. A tangent vector $v^{i}\partial_i$ acts on a differentiable scalar function $f$ defined on $G$ as
\begin{equation}
\vec{v}(f)=v^{i}\frac{\partial f(x^1,\dots,x^N)}{\partial x^{i}}\,.
\end{equation}
Of fundamental importance is the tangent space $T_eG$ at the unit element of $G$ and the Lie bracket of two tangent vector fields $x$ and $y$, acting on a differentiable scalar function $f$ on $G$, defined as
\begin{equation}
[x,y]=x(y(f))-y(x(f))\,,
\end{equation}
building up a Lie algebra.

Since the generators constitute a basis, the Lie bracket of two generators is proportional to another basis vector
\begin{equation}
\label{generator}
[t^{a},t^{b}]=i f^{abc}t^{c},
\end{equation}
where the real and antisymmetric factors $f^{abc}$ are called structure constants of the algebra. For the Lie algebra $\mathfrak{su}(2)$ a possible basis is given by the Pauli matrices $\sigma^{1,2,3}$ multiplied with the imaginary unit $i$, these three skew hermitian matrices are generators of $\mathfrak{su}(2)$.
The commutator relation (\ref{generator}) is in this particular case
\begin{equation}
\left[\frac{\sigma^{a}}{2},\frac{\sigma^{b}}{2}\right]=
i\varepsilon^{abc}\frac{\sigma^{c}}{2}\,,
\end{equation}
where $\varepsilon^{abc}$ denotes the Levi-Civita tensor. Analogously $\mathfrak{su}(3)$
is spanned by the set of Gell-Mann matrices.

A Lie algebra is connected to its associated Lie group via the exponential map, defined as the power series $\exp(A)=\sum_{n=0}^\infty\frac{A^n}{n!}$. If the matrix $A$ is an element of the algebra, then $\exp(A)$ is an element of the group. Thus two conditions on the matrices are implied: $A$ ought to be anti-hermitian $(A^\dagger=-A)$ and traceless
$(\textrm{tr} A=0)$.

The rank of a Lie algebra is then defined as the dimension of the Cartan subalgebra, which is given as the maximal algebra of commuting generators.

\underline{\textit{Linear representation:}}
A linear representation of a Lie group G on a vector space V is a homomorphism
which connects every element $x\in G$ with a linear transformation $R(x)$ acting on $V$. The dimension of the representation $R(x)$ is equal to the dimension of the vector space
$V$. If a basis of $V$ is assigned, $R(x)$ can be expressed as a matrix.

A representation is called (ir)reducible if the representation can(not) be transformed
into the blockdiagonal form
\begin{equation}
R(x)=\left( \begin{array}{cc}
             R^1(x) & A(x)   \\
             0      & R^2(x) \\
            \end{array}
\right)\,.
\end{equation}
If A(x) vanishes for any $x\in G$, $R(x)$ is called fully reducible.

A linear representation of the Lie algebra is formed by matrices which satisfy the commutation relation (\ref{generator}) of the Lie algebra. Since the elements of a matrix group are linear transformations, the group elements themselves generate the fundamental representation of the group.

The $r$ generators of a Lie group provide a
representation of dimension $r$.
A  basis of this adjoint representation is then provided through the structure
constants, $R_{ab}t_{c}=i f_{abc}$. The structure constants obey the Jacobi
identity and the commutation relation as required.

\section{SU(N) Yang-Mills theory: Foundations and topological aspects}

In this section we consider the formulation of non-abelian gauge
theories: The Lagrangian density and the equation of motion for the gauge fields is derived. We give a review of solitonic field configurations solving the Yang-Mills equation. As a first example for an object of topological origin we motivate the emergence of magnetic monopoles in thermalized, pure Yang-Mills theory according to the construction by 't Hooft and Polyakov. Subsequently, the main focus lies on
a solution to the equation of motion in euclidean field theory, which is stabilized by topology, called instanton. This is followed by a presentation of its finite temperature generalization, the caloron.
An introduction into foundations of non-abelian gauge theories and topological aspects of quantum field theory can for instance be found in \cite{Peskin, WeinbergII, Ryder, 't Hooft course}. The topic of instanton physics is covered in \cite{Shifman, Coleman} on a fundamental level. The generalization of instantons to finite temperature field theory is reviewed in \cite{Gross/Pisarski/Yaffe81, Shuryak98}.
In this thesis we use the convention that Greek indices correspond to four-dimensional Lorentz indices, while the Latin indices (from the middle of the alphabet) denote spatial indices. Lie algebra indices are denoted with Latin indices from the beginning of the alphabet. The Einstein summation convention is implied if not stated otherwise. In Minkowskian spacetime we have to distinguish between upper and lower indices. Only contractions between covariant and contravariant indices are possible. In Euclidean spacetime, a distinction between upper and lower indices is not necessary.

\subsection{The Yang-Mills Equation}

In the following section we contemplate a four-dimensional Minkowskian spacetime.
In 1954 Yang and Mills \cite{Yang/Mills54} constructed a Lagrangian density, which is invariant under non-abelian gauge transformations. This Lagrangian describes a nontrivial interacting field theory. While in the original work the gauge symmetry was taken to be $SU(2)$ of isospin rotations, soon a generalization to arbitrary non-abelian gauge groups took place. The perturbative analysis of these theories was a merit of Feynman, Faddeev, Popov, DeWitt, and many other authors in the 60's and 70's, when Feynman rules for their perturbative quantization were developed. The tremendous physical relevance of non-abelian gauge theories as a mathematical framework for the description of fundamental interactions is generally acknowledged.

We designate an $SU(N)$ Yang-Mills theory as being pure, if only gauge fields and no matter fields are involved. We demand from the Lagrangian of an $SU(N)$ Yang-Mills theory gauge invariance under any local $SU(N)$ transformation $U(x)$. In order to construct gauge-invariant kinetic quantities from the gauge fields in the Lagrangian, all partial
derivatives, acting on non-gauge fields in a non-pure Yang-Mills theory, are replaced by covariant derivatives, defined as
\begin{equation}
D_\mu=\partial_\mu-igA_\mu\,,
\end{equation}
where $A_\mu$ is a gauge field and $g$ denotes the gauge coupling of the fundamental
theory. This gauge field is an element of the Lie algebra, and therefore an expansion
exists in terms of the generators $\lambda^{a}$ of the algebra
\begin{equation}
A_\mu=A_\mu^{a}\frac{\lambda^{a}}{2}\,.
\end{equation}
The generators are normalized as
\begin{equation}
\label{Norm generator}
\textrm{tr} \lambda^{a}\lambda^{b}=\frac{1}{2}\delta^{ab}\,.
\end{equation}
For determining the effect of a covariant derivative on a non-gauge field $\phi$, we have to consider the representation of the field. If $\phi$ transforms under the fundamental representation
we obtain
\begin{equation}
D_\mu\phi=\partial_\mu\phi-igA_\mu^{a}\frac{\lambda^{a}}{2}\phi\,,
\end{equation}
which is different from the case when $\phi$ is in adjoint representation. Then we have
\begin{equation}
(D_\mu\phi)=\partial_\mu\phi-ig[A_\mu,\phi]\,.
\end{equation}
Since we require that the covariant derivative is constructed in a manner, so that the derivative of the field obeys the same transformation law as the field itself, the gauge field transforms under a gauge transformation
$U(x)\in SU(N)$ according to
\begin{equation}
A_\mu(x)\rightarrow U(x)A_\mu(x)U^\dagger(x)+\frac{i}{g}U(x)\partial_\mu(x)U^\dagger(x)\,.
\end{equation}
In the language of differential geometry the gauge field can be regarded as a connection on an analytic manifold, the gauge group. The field strength tensor is associated with the curvature on the manifold and is defined through the Lie bracket of covariant derivatives:
\begin{equation}
F_{\mu\nu}=\frac{i}{g}[D_\mu,D_\nu]
=\partial_\mu A_\nu-\partial_\nu A_\mu-ig[A_\mu,A_\nu]\,,
\end{equation}
and in component notation
\begin{equation}
F_{\mu\nu}^{a}=\partial_\mu A_\nu^{a}-\partial_\nu A_\mu^{a}+gf^{abc}A_\mu^{b}A_\nu^{c}\,.
\end{equation}
In non-abelian gauge theories the field strength tensor is not gauge-invariant, but transforms homogeneously
\begin{equation}
F_{\mu\nu}(x)\rightarrow U(x)F_{\mu\nu}(x)U^\dagger(x)\,.
\end{equation}
Therefore, the quantity $\textrm{tr}\,F_{\mu\nu}F^{\mu\nu}$ is a gauge invariant Lorentz scalar, and the Lagrangian density is given as
\begin{equation}
\label{YMLagrangian}
\mathcal{L}=-\frac{1}{2}\textrm{tr}\,F_{\mu\nu}F^{\mu\nu}
=-\frac{1}{4} F_{\mu\nu}^{a}F^{\mu\nu a}\,,
\end{equation}
where the normalization of the generators (\ref{Norm generator}) was used.
In the case of Abelian gauge groups the commutator vanishes and therefore no self-interaction of the gauge bosons occurs. In contrast, for non-abelian gauge groups self-interactions governed by three- and four-boson-vertices are present, due to the fact that the Lagrangian contains additional cubic and quartic terms in the gauge fields. Although a mass term for gauge fields is for reasons of gauge invariance not admissible, the gauge bosons can acquire a mass by dynamical symmetry breaking via the
Higgs mechanism, if the accordingly charged scalar sector is added to the Lagrangian in Eq.\,(\ref{YMLagrangian}).
\footnote{To construct a theory of gauge fields interacting with fermions we have to add the Dirac Lagrangian to the gauge-field Lagrangian (\ref{YMLagrangian}):
\begin{equation}
\mathcal{L}=-\frac{1}{4} F_{\mu\nu}^{a}F^{\mu\nu a}+i\bar{\psi}\slashed{D}\psi-m\bar{\psi}\psi\,.
\end{equation}
The Euler-Lagrange equations are obtained by varying the resulting action and we obtain
\begin{equation}
(i\slashed{D}-m)\psi=0\,,
\end{equation}
as the Dirac equation for the fermion field, and
\begin{equation}
(D^\mu F_{\mu\nu})^a=-gj_\nu^{a}\,,
\end{equation}
as equation of motion for the gauge field, where $j_\nu^{a}=\bar{\psi}\gamma_\nu t^{a}\psi$ is the current density of the fermion field.}

The Euler-Lagrange equations are derived from minimizing the action associated with the Yang-Mills Lagrangian (\ref{YMLagrangian}). We get
\begin{equation}
\label{Yang-Mills equation}
D_\mu F^{\mu\nu}=0\,,
\end{equation}
or in component notation
\begin{equation}
\label{Yang-Mills equation II}
\partial_\mu F^{\mu\nu a}+gf^{abc}A^{\mu b}F^{\mu\nu c}=(D_\mu F^{\mu\nu})^{a}=0\,.
\end{equation}
According to Eq.\,(\ref{Yang-Mills equation II}), the fact that a pure $SU(N)$ Yang-Mills theory is interacting is reflected in the
nonlinearity of the field strength tensor and the term proportional to $g$.

\subsection{The 't Hooft-Polyakov monopole}
An abelian $U(1)$-gauge theory of electromagnetism containing magnetic monopoles in addition to electric charges is more symmetric than the standard Maxwell theory.
Despite that, there is no necessity for the existence of magnetic monopoles. However,
if the gauge theory of electromagnetism would be described by an underlying non-abelian gauge group (like $SU(2)$) subject to dynamical symmetry breaking, the field
equations would yield solutions with magnetic monopole constituents. The existence of magnetic monopoles at finite temperature is enabled by the nontrivial homotopy group $\pi_2(SU(2)/U(1))=\mathbb{Z}$ \cite{Nahm80/84, Lee/Lu98, Kraan/vanBaal98}.
The theoretical possibility of magnetic monopoles in gauge theories was elucidated the first time when 't Hooft \cite{'tHooft74} and Polyakov \cite{Polyakov74} considered an $SU(2)$-Yang-Mills
theory with an isovector adjoint Higgs field $\phi^{a}$. The Lagrangian density is then given as
\begin{equation}
\label{'tHooftPolyakov}
\mathcal{L}=-\frac{1}{4}F_{\mu\nu}^{a}F^{\mu\nu a}
+\frac{1}{2}(D_\mu\phi^{a})(D^\mu\phi^{a})-\frac{m^2}{2}\phi^{a}\phi^{a}
-\lambda(\phi^{a}\phi^{a})^2\,.
\end{equation}
If the parameter $m^2$ is chosen negative, the vacuum expectation value of the
Higgs field minimizes the action at $|\phi|^2=\frac{-m^2}{4\lambda}=:F$.

In 1974 't Hooft derived non-singular solutions with finite energy to the field equations resulting from the Lagrangian (\ref{'tHooftPolyakov}) by considering static configurations. Setting $A_0^{a}=0$, the $SU(2)$ gauge field and the Higgs field has the following asymptotic behavior $(|\vec{x}|\rightarrow\infty)$:
\begin{eqnarray}
\label{hedgehog}
A_i^{a}&=&-\varepsilon_{i}^{ab}\frac{x^b}{e|\vec{x}|^2}\\
\phi^{a}&=&F\frac{x^{a}}{|\vec{x}|}\,.
\end{eqnarray}
It is worth mentioning that in these configurations space and isospace indices are mixed. For a field configuration of the form (\ref{hedgehog}) the name 'hedgehog' was introduced by Polyakov.

The $SU(2)$ gauge symmetry is broken due to the nonvanishing vacuum expectation value of the Higgs field. The field strength is determined by the $SU(2)$-gauge invariant
't Hooft-tensor
\begin{equation}
\label{'tHooft tensor}
\mathcal{F}_{\mu\nu}=\frac{\phi^{a}}{|\phi|}F_{\mu\nu}^{a}
-\frac{1}{e|\phi|^2}\varepsilon^{abc}\phi^{a}(D_\mu\phi^{b})(D_\nu\phi^{c})\,,
\end{equation}
where $|\phi|\equiv\sqrt{\phi^{a}\phi^{a}}$.
Assuming that the gauge field and the scalar field point into the three direction of
the Lie algebra, we insert
\begin{equation}
A_\mu^3=\frac{\phi^{a}}{|\phi|}A_\mu^{a},\qquad\phi^3=F\neq0\,.
\end{equation}
Fixing the scalar field $\phi$ in isospace in such a way, the 't Hooft tensor reduces to the usual abelian $U(1)$ electromagnetic field tensor for the gauge field $A_\mu^3$. Inserting the hedgehog-Ansatz (\ref{hedgehog}) into Eq.\,(\ref{'tHooft tensor}) leads to an asymptotic static field strength
\begin{equation}
F_{ij}=-\frac{1}{e|\vec{x}|^3}\varepsilon_{ij}^{a} x^{a}\,.
\end{equation}
This corresponds to a radial magnetic field
\begin{equation}
B_a=\frac{x_a}{e|\vec{x}|^3}\,.
\end{equation}
The magnetic flux follows from
\begin{equation}
\Phi=\oint\limits_{S^2}d\sigma_a B_a=\frac{4\pi}{e}\,,
\end{equation}
and the magnetic charge as
\begin{equation}
q_m=\frac{\Phi}{4\pi}=e^{-1}
\end{equation}
In the BPS-limit ($m\rightarrow 0,\lambda\rightarrow 0$) the monopole mass is evaluated as $M_m=\frac{4\pi}{e^2}M_W$, where $M_W$ denotes the vector boson mass, given as $M_W=e|\phi|$.
We define the current
\begin{equation}
K^\mu=\partial_\nu \tilde{F}^{\mu\nu}
=\frac{1}{2}\varepsilon^{\mu\nu\kappa\lambda}\partial_\nu F_{\kappa\lambda}
=-\frac{1}{2e}\varepsilon^{\mu\nu\kappa\lambda}\varepsilon^{abc}
\partial_\nu\hat{\phi}^{a}\partial_\kappa\hat{\phi}^{b}\partial_\lambda\hat{\phi}^{b}\,,
\end{equation}
where $\hat{\phi}^{a}\equiv \frac{\phi^{a}}{|\vec{\phi}|}$.

The magnetic charge can then be written as the integral over a sphere at infinity:
\begin{equation}
q_m=\frac{1}{4\pi}\int \!d^3x\, K^0
=-\frac{1}{8\pi e}\oint\limits_{S^2}d\sigma^{i}\,\varepsilon^{ijk}\varepsilon^{abc}
\hat{\phi}^{a}\partial_i\hat{\phi}^{b}\partial_j\hat{\phi}^{b}\,.
\end{equation}
This equation leads to two important statements. The magnetic charge depends only on the asymptotic behavior of the fields and the magnetic current merely has contributions from the Higgs field. The current $K_\mu$ is identically conserved, its divergence vanishes. Notice that this current conservation is a consequence of topology, and does not follow from a continuous symmetry of the Lagrangian.

\subsection{Instanton solution}

In order to construct a four-dimensional well localized solution to the gauge field equation in the case of a pure $SU(2)$ Yang-Mills theory, we have to consider this theory in euclidean spacetime. Thus we perform a
Wick-rotation in the zero-coordinate, $t\rightarrow -i\tau$, where $\tau\in\mathbb{R}$.
The Yang-Mills action is then given as
\footnote{In the remainder of this section we absorb the gauge coupling into the gauge field. This is a common convention for considering nonperturbative aspects.}
\begin{equation}
S=\frac{1}{4g^2}\int d^4x\,\textrm{tr} F_{\mu\nu}F_{\mu\nu}
=\frac{1}{4g^2}\int d^4x\,\textrm{tr}(\vec{E}^2+\vec{B}^2)\,,
\end{equation}
with the color electric and magnetic fields $\vec{E^{a}}$ and $\vec{B^{a}}$.
A solution to the equation of motion is determined by detecting the minima of the action.
The existence of the configuration is guaranteed by $\pi_3(SU(2))=\mathbb{Z}$.
The $SU(N)$ Yang-Mills vacuum is degenerate, that is there exist apart from the vacuum $A_\mu=0$ infinitely many configurations of the kind
\begin{equation}
A_\mu=iU(x)\partial_\mu U^\dagger(x)\,,
\end{equation}
which differ from the trivial vacuum only by an $SU(N)$ gauge transformation $U(x)$, and thereby have zero field strength and energy momentum tensor. These so-called pure gauges fall into distinct topological classes, which are characterized by an integer winding
number
\begin{equation}
\label{Pontryagin index}
n_W=\frac{1}{24\pi^2}\int d^3x\,\varepsilon^{ijk}
\,\textrm{tr}(U^\dagger\partial_i U)(U^\dagger\partial_j U)(U^\dagger\partial_k U)\,.
\end{equation}
There exists no continuous transformation between two pure gauges with different winding number, so a transition can only be realized via a tunneling process with probability amplitude $\e^{-S}$. Such a tunneling solution is called instanton.

We define the dual field strength tensor as
\begin{equation}
\tilde{F}_{\mu\nu}=\frac{1}{2}\varepsilon_{\mu\nu\kappa\lambda}F_{\mu\nu}\,.
\end{equation}
In the dual field strength the electric fields are replaced by the magnetic fields, and vice versa. Now we are able to write the Yang-Mills equation in the form of a Bogomolnyi decomposition \cite{Bogomolnyi78}
\begin{eqnarray}
\label{Bogomolnyi}
S&=&\frac{1}{4g^2}\int d^4x\,\textrm{tr} F_{\mu\nu}F_{\mu\nu}\nonumber\\
&=&\frac{1}{4g^2}\int d^4x\,\textrm{tr} \left(\pm F_{\mu\nu}\tilde{F}_{\mu\nu}
+\frac{1}{2}(F_{\mu\nu}\mp\tilde{F}_{\mu\nu})(F_{\mu\nu}\mp\tilde{F}_{\mu\nu})\right)\,.
\end{eqnarray}
The first summand in the Yang-Mills action (\ref{Bogomolnyi}) is the topological charge
also denoted as Pontryagin index
\begin{equation}
Q=\frac{1}{16\pi^2}\int d^4x\,\textrm{tr}\,F_{\mu\nu}\tilde{F}_{\mu\nu}
=\frac{1}{32\pi^2}\int d^4x\,\textrm{tr}\,F_{\mu\nu}^{a}\tilde{F}_{\mu\nu}^{a}\,.
\end{equation}
The integrand is called Pontryagin density and is equal to the total divergence of the
Chern-Simons current $K_\mu$. It holds that\footnote{Consider for instance the case of $SU(2)$ as gauge group \cite{Ryder, 't Hooft course}. The Chern-Simons current can be written as
\begin{equation}
K_\mu=\frac{1}{16\pi^2}\varepsilon_{\mu\nu\kappa\lambda}
\left(A_\nu^{a}\partial_\kappa A_\lambda^{a}+\frac{1}{3}\varepsilon^{abc}
A_\nu^{a}A_\kappa^{b}A_\lambda^{c}\right),
\end{equation}
where $\varepsilon^{abc}$ denotes the Levi-Civita tensor. We obtain this form of $K_\mu$ by employing the commutator relation $\left[\frac{\tau^a}{2},\frac{\tau^b}{2}\right]=i\varepsilon^{abc}\frac{\tau^c}{2}$ and the normalization condition of the generators of the Lie-algebra $\textrm{tr}\,\frac{\tau^a}{2}\frac{\tau^b}{2}=\frac{1}{2}\delta^{ab}$, where $\tau^{a,b,c}$ denote the Pauli matrices. Thus $K_\mu$ can be written as trace
\begin{equation}
K_\mu=\frac{1}{4\pi^2}\varepsilon_{\mu\nu\kappa\lambda}
\,\textrm{tr}\,\left(\frac{1}{2}A_\nu\partial_\kappa A_\lambda-\frac{i}{3}A_\nu A_\kappa A_\lambda \right)\,.
\end{equation}
With respect to the cyclicity of the trace the divergence of this quantity is then given as
\begin{equation}
\partial_\mu K_\mu=K_\mu=\frac{1}{4\pi^2}\varepsilon_{\mu\nu\kappa\lambda}
\,\textrm{tr}\,\left(\frac{1}{2}(\partial\mu A_\nu)(\partial_\kappa A_\lambda)-i (\partial_\mu A_\nu) A_\kappa A_\lambda \right)\,.
\end{equation}
Now we calculate $\textrm{tr}\,F_{\mu\nu}\tilde{F}_{\mu\nu}
=\frac{1}{2}\varepsilon_{\mu\nu\kappa\lambda}F_{\mu\nu}F_{\kappa\lambda}$\,, 
with $F_{\mu\nu}=\partial_\mu A_\nu-\partial_\nu A_\mu-i[A_\mu,A_\nu]$\,.
We obtain
\begin{equation}
\textrm{tr}\,F_{\mu\nu}\tilde{F}_{\mu\nu}=2\varepsilon_{\mu\nu\kappa\lambda}\,\textrm{tr}\,(\partial\mu A_\nu)(\partial_\kappa A_\lambda)-2i\varepsilon_{\mu\nu\kappa\lambda}\,\textrm{tr}\,(\partial_\mu A_\nu) A_\kappa A_\lambda-2i\varepsilon_{\mu\nu\kappa\lambda}\,\textrm{tr}\,A_\mu A_\nu (\partial_\kappa A_\lambda)-2\varepsilon_{\mu\nu\kappa\lambda}\,\textrm{tr}\,\,A_\mu A_\nu A_\kappa A_\lambda\,.
\end{equation}
Due to the cyclicity property of the trace the second and the third terms are equal and the last term vanishes for reasons of symmetry. Thus in the case of $SU(2)$
\begin{equation}
\partial_\mu K_\mu=\frac{1}{32\pi^2}F_{\mu\nu}^{a}\tilde{F}_{\mu\nu}^{a}
\end{equation}
is proven.}
\begin{equation}
\partial_\mu K_\mu=\frac{1}{32\pi^2}F_{\mu\nu}^{a}\tilde{F}_{\mu\nu}^{a},
\end{equation}
with
\begin{equation}
K_\mu=\frac{1}{16\pi^2}\varepsilon_{\mu\alpha\beta\gamma}
\left(A_\alpha^{a}\partial_\beta A_\gamma^{a}+\frac{1}{3}f^{abc}
A_\alpha^{a}A_\beta^{b}A_\gamma^{c}\right).
\end{equation}
The winding number defined in (\ref{Pontryagin index}) is connected to the charge of the Chern-Simons current by virtue of
\begin{equation}
n_W=\int d^3x K^0=\frac{1}{16\pi^2}\int d^3x\,
\varepsilon_{ijk}
\left(A_i^{a}\partial_j A_k^{a}+\frac{1}{3}f^{abc}A_i^{a}A_j^{b}A_k^{c}\right).
\end{equation}
Notice that the Chern-Simons current is not a gauge invariant quantity, in contrast to the Pontryagin density. Provided that the integrand is nonsingular and
rapidly decreasing in the limit of spatial infinity, Gauss' theorem yields again the
topological charge
\begin{eqnarray}
\label{topological charge}
Q&=&\int d^4x\,\partial_\mu K_\mu=\nonumber\\
&=&\int d\tau\frac{d}{d\tau}\int d^3x\,K_0+\int d\tau\int d\sigma_i\,K_i\nonumber\\
&=&\int\limits_{\tau\rightarrow\infty} d^3x\,K_0
+\int\limits_{\tau\rightarrow -\infty} d^3x\,K_0\nonumber\\
&=&n_W(\tau=\infty)-n_W(\tau=-\infty)\,.
\end{eqnarray}
We recognize that a tunneling field configuration from one vacuum to another topologically different is associated with a Pontryagin index $Q\neq0$.
Thus the second term in the Bogomolnyi decomposition (\ref{Bogomolnyi}) is a square, and the euclidean action is bounded from below (Bogomolnyi bound). The minimum in a given topological sector is reached for an selfdual or a anti-selfdual field configuration
\begin{equation}
\label{selfduality equation}
F_{\mu\nu}^{a}=\pm\tilde{F}_{\mu\nu}^{a}\,.
\end{equation}
A selfdual field corresponds to an instanton, an anti-selfdual field to an antiinstanton. The minimum of the action is then given as $S=\frac{8\pi^2}{g^2}|Q|$. We mention that the nonperturbative character of instantons is reflected in the essential zero at $g=0$ in the transition amplitude, thus perturbative expansions completely disregard instanton solutions.

The selfduality-equation (\ref{selfduality equation}) constitutes a differential equation of first order, in contrast to the second order like Euler-Lagrange equation (\ref{Yang-Mills equation}). The Bianchi identity,
\begin{equation}
D_\mu F_{\kappa\lambda}+D_\lambda F_{\mu\kappa}+D_\kappa F_{\mu\lambda}=0,
\end{equation}
provides that (anti-)selfdual field configurations satisfy the equations of motion automatically\footnote{\begin{eqnarray}
D_\mu F_{\mu\nu}^{a}&=&\pm D_\mu \tilde{F}_{\mu\nu}^{a}\nonumber\\
&=&\pm\frac{1}{2}\varepsilon_{\mu\nu\kappa\lambda}D_\mu F_{\mu\nu}^{a}\nonumber\\
&=&\frac{1}{6}\varepsilon_{\mu\nu\kappa\lambda}
\left(D_\mu F_{\kappa\lambda}^{a}+D_\kappa F_{\lambda\mu}^{a}
+D_\lambda F_{\mu\kappa}^{a}\right)=0\nonumber\\
\nonumber
\end{eqnarray}.}.
(Anti-)selfdual configuration have a vanishing energy-momentum tensor. This is also called BPS-saturation\cite{Prasad/Sommerfield75}.

In order to construct an explicit solution to (\ref{selfduality equation}) we have to demand that the field strength vanishes at infinity. This condition is realized, if the gauge field approaches a pure gauge solution on the boundary of euclidean spacetime.

A gauge field which satisfies the mentioned requirements, should show an asymptotic behavior like
\begin{equation}
\label{instanton ansatz}
A_\mu^{a}\rightarrow \eta_{\mu\nu}^{a}\frac{x_\nu}{x^2}\qquad(|x|\rightarrow\infty),
\end{equation}
where $|x|$ denotes $\sqrt{x_\mu x_\mu}$ and the definition of the 't Hooft symbols is
given as
\begin{eqnarray}
\eta_{\mu\nu}^{a}&=&\varepsilon_{a\mu\nu}
+\delta_{a\mu}\delta_{\nu 4}-\delta_{a\nu}\delta_{\mu 4}\,,\\
\bar{\eta}_{\mu\nu}^{a}&=&\varepsilon_{a\mu\nu}
-\delta_{a\mu}\delta_{\nu 4}+\delta_{a\nu}\delta_{\mu 4}\,.\\
\nonumber
\end{eqnarray}
Eq.\,(\ref{instanton ansatz}) is generalized to the case of finite $|x|$ as
\begin{equation}
A_\mu^{a}(x)=2\eta_{a\mu\nu}x_\nu\frac{f(x^2)}{x^2}\,,
\end{equation}
where $f$ fulfills the boundary conditions
\begin{equation}
\label{boundary conditions}
\lim_{x^2\rightarrow\infty}f(x^2)\rightarrow 1\hspace{25mm}
\lim_{x^2\rightarrow 0}f(x^2)\rightarrow const\cdot x^2.
\end{equation}
The request for selfduality determines a differential equation for $f$, which subject to Eq.\,(\ref{boundary conditions}) is solved by
\begin{equation}
f=\frac{x^2}{x^2+\rho^2}.
\end{equation}
The constant of integration $\rho$ is called the instanton radius. The translational invariance ensures that the center of the instanton can be shifted to an arbitrary
point $z^\mu$.

Summarizing we obtain the Belavin-Polyakov-Schwarz-Tyupkin instanton \cite{Belavin/Polyakov/Schwartz/Tyupkin75}
\begin{equation}
\label{BPST instanton}
A_\mu^{a}(x)=2\eta_{a\mu\nu}\frac{(x-z)_\nu}{(x-z)^2+\rho^2}
\end{equation}
with the corresponding field strength
\begin{equation}
F_{\mu\nu}^{a}(x)=-4\eta_{a\mu\nu}\frac{\rho^2}{\left[(x-z)^2+\rho^2\right]^2}\,.
\end{equation}
The antiinstanton is obtained by substituting $\eta_{a\mu\nu}\rightarrow\bar{\eta}_{a\mu\nu}$. This field configuration is well localized in spacetime, since the field strength decreases like $x^{-4}$.

The topological charge of the instanton is derived via evaluation of Eq.\,(\ref{topological charge}) with the asymptotic Chern-Simons current
\begin{equation}
\lim_{|x|\rightarrow\infty}K_\mu=\frac{1}{2\pi^2}\frac{x_\mu}{x^4}\,.
\end{equation}
The BPST-instanton carries the topological charge $Q=1$ (respectively $Q=-1$ for the BPST-antiinstanton).

The BPST-instanton (\ref{BPST instanton}) has eight free parameters (also denoted as collective coordinates): four $z^\mu$ due to the translation of the instanton center,
the instanton radius, associated with dilatations and three coordinates in color space. The action of the BPST-instanton is independent of these parameters, so each of them gives rise to a zero mode, i.e. they are moduli of the instanton solution.

The BPST-instanton so far was considered in regular gauge. This implies that the configuration is well behaved except at infinity where a singularity emerges. However,
this singularity can be shifted from infinity to the center of the instanton by application of a suitable gauge transformation. In this singular gauge the gauge field
and the field strength transform to
\begin{eqnarray}
A_\mu^{a}(x)&=&
2\bar{\eta}_{a\mu\nu}(x-x_0)_\nu\frac{\rho^2}{(x-x_0)^2\left[(x-z)^2+\rho^2\right]}\nonumber\\
&=&-\bar{\eta}_{a\mu\nu}\partial_\nu\ln\left[1+\frac{\rho^2}{(x-x_0)^2}\right]\\
F_{\mu\nu}^{a}(x)&=&
-8\left[\frac{(x-x_0)_\mu(x-x_0)_\rho}{(x-x_0)^2}-\delta_{\mu\rho}\right]
\bar{\eta}_{a\nu\rho}\frac{\rho^2}{\left[(x-x_0)^2+\rho^2\right]^2}
-[\mu\leftrightarrow\nu].\,\\
\nonumber
\end{eqnarray}
In the gauge potential there now occurs a singularity at the instanton center, but nevertheless the field strength remains finite. The fact that in singular gauge the action density as well as the topological charge $Q$ is localized at the instanton center enables us to construct field configurations with topological charge higher than unity, so-called multi-instanton solutions. This construction can be realized by generalization of the singular gauge BPST-instanton to the 't Hooft-ansatz for the gauge field
\begin{equation}
\label{'tHooft-ansatz}
A_\mu^{a}(x)=-\bar{\eta}_{a\mu\nu}\partial_\nu\ln\Pi(x)\,,
\end{equation}
where $\Pi(x)$ denotes a pre-potential. The required selfduality of the field strength $F_{\mu\nu}^{a}$ yields a Laplace-like equation for the pre-potential
\begin{equation}
\label{pre-potential}
\frac{\partial_\nu\partial_\mu\Pi(x)}{\Pi(x)}=0\,.
\end{equation}
A general solution with topological charge $Q=K$ evaluates to
\begin{equation}
\label{prepotential}
\Pi(x)=1+\sum\limits_{i=1}^K\frac{\rho_i}{(x-z_i)^2}\,,
\end{equation}
i.e. it describes instantons with centers at $z_i$ and radii $\rho_i$. (The multi-instanton solution with $Q=K$ is constructed by substituting $\bar{\eta}_{a\mu\nu}$ through $\eta_{a\mu\nu}$ in the 't Hooft-ansatz (\ref{'tHooft-ansatz}). A multi-instanton with Pontryagin index $Q=K$ can be considered as a superposition of $K$ single instantons, thus $8K$ parameters generate the moduli space. But we should mention that the pre-potential (\ref{prepotential}) does not represent the most general solution, since all $K$ instantons have the same orientation in the color space. The most general construction of a multi-instanton was obtained in 1978 by Atiyah, Drinfeld, Hitchin and Manin \cite{Atiyah/Drinfeld/Hitchin/Manin78}.

\subsection{Caloron solution}

Quantum field theory at finite temperature (\ref{FTFT}) can be formulated in euclidean signature. Thereby temperature $T$ is identified with the euclidean time $\tau$ compactified on a circle with perimeter $\beta=T^{-1}$. The fields have to show $\beta$-periodicity, at least up to a gauge transformation
\begin{equation}
A_\mu(\beta,\vec{x})=U(0,\vec{x})A_\mu(0,\vec{x})U^{\dagger}(0,\vec{x})
+\frac{i}{g}U(0,\vec{x})\partial_\mu(0,\vec{x})U^{\dagger}(0,\vec{x})\,.
\end{equation}
The intention is then to construct an instanton solution which obeys this condition. The $\beta$-periodic field configuration is called caloron, the $\beta$-periodic anti-selfdual solution anticaloron \cite{Harrington/Shepherd78}.

The initial point concerning this construction is the 't Hooft-ansatz (\ref{'tHooft-ansatz}) with a pre-potential as defined in (\ref{pre-potential}). We obtain the caloron centered at $(0,0)$ associated pre-potential by evaluating the infinite sum over instanton pre-potentials centered at $(n\beta,0),n\in\mathbb{Z}$, and equal radii $\rho_i$. In 1978 Harrington and Shepherd calculated this quantity
\begin{equation}
\Pi(x)=1+\sum\limits_{-\infty}^\infty\frac{\rho^2}{(\tau-n\beta,\vec{x})^2}
=1+\frac{\pi\rho^2}{\beta r}\frac{\sinh\left(\frac{2\pi}{\beta}r\right)}
{\cosh\left(\frac{2\pi}{\beta}r\right)-\cos\left(\frac{2\pi}{\beta}\tau\right)}\,,
\end{equation}
where $r=|\vec{x}|$. Integrating the associated Chern-Simons current over a three-sphere (with center at the singularity), we obtain the topological charge of the caloron (anticaloron) as plus one (minus one). The action of the Harrington-Shepherd caloron calculates as equal to the instanton action $\frac{8\pi^2}{g^2}$: An on the classical level temperature-independent value.

In analogy to the construction of multi-instanton configurations in singular gauge a caloron of topological charge higher then unity is built as a superposition of the pre-potentials of the invoked single calorons. Such a multi-caloron solution possesses among the radius more dimensionful moduli. However, they do not yield any contribution to the thermodynamics, as we explain below.

Calorons are categorized in field configurations with trivial and nontrivial holonomy. The essential quantity is the value of the Polyakov loop at spatial infinity, which is a topological invariant. The Polyakov loop is defined as a time-like Wegner-Wilson loop at finite temperature
\begin{equation}
P(x)=\mathcal{P}\e^{ig\int\limits_0^\beta d\tau A_4(\tau,\vec{x})}\,,
\end{equation}
where $\mathcal{P}$ is the path-ordering operator. Calorons with a Polyakov loop lying in the center of the Lie group are said to have trivial holonomy, otherwise they have nontrivial holonomy. In particular, for the interesting case of $SU(2)$ calorons with $P(|\vec{x}|\rightarrow\infty)=\pm1$ corresponds to trivial holonomy. For the Harrington-Shepherd caloron we obtain for the Polyakov loop $P=1$ (in the limit $|x\rightarrow\infty|$).

In the early 80's Nahm \cite{Nahm80/84} pointed out the existence of caloron solutions with nontrivial holonomy, in 1998 an explicit construction was accomplished indepedently by Lee and Lu \cite{Lee/Lu98} and Kraan and van Baal \cite{Kraan/vanBaal98}.
Lee and Lu parametrized
\begin{equation}
A_4^C(\tau,|\vec{x}|\rightarrow\infty)=-i\frac{u}{2}\lambda_3\,,
\end{equation}
where $\lambda_3$ is the skew-Hermitian generator and $u$ denotes the holonomy, which is constrained to $0\leq u\leq \frac{2\pi}{\beta}.$ The solution of the ADHMN-equations \cite{Nahm80/84, Atiyah/Drinfeld/Hitchin/Manin78} yield a selfdual field configuration consisting of BPS-magnetic (anti-)monopole constituents. This is the nontrivial holonomy caloron. The masses of the constituents are given as
\begin{equation}
m_1=4\pi u\,,\qquad m_2=4\pi\left(\frac{2\pi}{\beta}-u\right)\,.
\end{equation}
The one-loop effective action $S_{\tiny\mbox{eff}}$ (or the one-loop quantum weight) of a trivial-holonomy caloron was computed by Gross, Pisarski and Yaffe \cite{Gross/Pisarski/Yaffe81}, while the calculation in the nontrivial-holonomy case was performed by Diakonov, Gromov and Slizovskiy in 2004 \cite{Diakonov/Gromov/Slizovskiy04}. Summarizing those results, we can conclude that the kind of interaction between BPS monopoles and antimonopoles in the caloron is depending on its holonomy. Small holonomy leads to an attractive potential, large holonomy to a repulsive potential. Calorons with nontrivial holonomy are no stable objects, since they either decay into a into a monopole-antimonopole pair or they annihilate each other and, then subsequently the caloron collapses back onto trivial holonomy. In contrast to the emergence of small holonomy calorons, the genesis of a large holonomy caloron $\left(u\sim\frac{\pi}{\beta}\right)$ is strongly Boltzmann-suppressed subject to $\e
 ^{-\beta(m_1+m_2)}\sim\e^{-8\pi^2}$. As a consequence the ground-state physics is essentially governed by small holonomy calorons.

\section{The deconfining phase of SU(2) Yang-Mills thermodynamics}
\label{Deconfining YMTD}

In this section we give an outline of \cite{Hofmann05} where fundamental concepts of SU(2)-Yang-Mills thermodynamics are developed. An investigation of the microscopic dynamics of individual calorons is a sophisticated task, and for the derivation of macroscopic quantities as pressure and energy density, not convenient if not impossible. But as long as we are focussing on the thermodynamics a spatial average via coarse-graining over the physics resulting from the topological nontrivial sector is a feasible method, which yields exact results.

\subsection{Construction of the composite adjoint Higgs field}

The spatial average over calorons leads us to the concept of a macroscopic thermal ground state, which is characterized by a macroscopic field $\phi$ and a pure-gauge configuration $a_\mu^{gs}$. The adjoint scalar field $\phi$ is generated by calorons with topological charge one and trivial holonomy. We require for $\phi$ the following properties:
\begin{enumerate}
\item {Due to spatial isotropy $\phi$ must be a Lorentz scalar.}
\item {In a pure Yang-Mills theory all local fields (as elements of the Lie algebra) transform under the adjoint representation of the gauge group. So $\phi$ as composite of this fields transform under the adjoint representation as well.}
\item{Since $\phi$ describes a homogeneous ground state of a thermal system its modulus is independent of space and time. This modulus is governed by a dynamically generated Yang-Mills scale $\Lambda$ and temperature.}
\item{The color orientation of $\phi$, also referred to $\phi$'s phase, is $\tau$-dependent. Since $\phi$ is composed of calorons the phase of $\phi$ is periodic in euclidean time. The classical caloron action $S=\frac{8\pi^2}{g^2}$ is independent on temperature, so $\phi$'s phase is not explicitly time independent.}
\end{enumerate}
Thus the field can be written as a product of modulus and phase
\begin{equation}
\phi^{a}=|\phi|(\Lambda,\beta)\frac{\phi^{a}}{|\phi|}\left(\frac{\tau}{\beta}\right)\,.
\end{equation}
In the following we sketch the evaluation of $\phi$'s modulus and phase and present the results.

\underline{\textit{$\phi$'s phase:}}
In \cite{Hofmann05, Hofmann/Herbst04} it is shown that the phase of the field can be evaluated as follows:
\begin{equation}
\label{phi's phase}
\frac{\phi^{a}}{|\phi|}\left(\frac{\tau}{\beta}\right)\sim
\textrm{tr}\int d^3x\int d\rho\,\lambda^{a} F_{\mu\nu}(\tau,0)\left\{(\tau,0),(\tau,\vec{x})\right\}
F_{\mu\nu}(\tau,\vec{x})\left\{(\tau,\vec{x}),(\tau,0)\right\},
\end{equation}
Here the field strength belongs to a Harrington-Shepherd caloron with caloron scale parameter $\rho$, and
\begin{eqnarray}
|\phi|^2&\equiv&\frac{1}{2}\textrm{tr}\phi^2\,,\\
\left\{(\tau,0),(\tau,|\vec{x}|)\right\}&=&
\mathcal{P}\exp\left[i\int\limits_{(\tau,0)}^{(\tau,\vec{x})}dz_\mu A_\mu(z)\right]\,,\\
\left\{(\tau,|\vec{x}|),(\tau,0)\right\}&=&
\mathcal{P}\exp\left[i\int\limits_{(\tau,\vec{x})}^{(\tau,0)}dz_\mu A_\mu(z)\right]\,.\\
\nonumber
\end{eqnarray}
The Wilson lines are calculated along straight lines. The $\sim$-sign declares that $\frac{\phi^{a}}{|\phi|}$ as well as the trace in equation (\ref{phi's phase}) solve the same homogeneous evolution equation in $\tau$
\begin{equation}
\label{eom phase}
\mathcal{D}\left[\frac{\phi}{|\phi|}\right]=0\,,
\end{equation}
where $\mathcal{D}$ represents a linear second-order differential equation and turns out to be uniquely determined as
\begin{equation}
\mathcal{D}=\partial_\tau^2+\left(\frac{2\pi}{\beta}\right)^2\,.
\end{equation}
Eq.\,(\ref{phi's phase}) is to evaluate on both caloron and anticaloron field configuration and the results have to be added up. Indeed, it transforms like an adjoint scalar. We underline that the construction of $\phi$'s phase is well justified and reasonable, since every generalization leads to explicit temperature dependencies or additional scales.
\begin{enumerate}
\item {Due to the selfduality of calorons and anticalorons every local definition of $\phi$'s phase yields zero.}
\item {Inclusion of higher $n$-point functions $(n\geq 3)$ would force us to introduce additional factors $\beta^{n-2}$ in order to get a dimensionless contribution. However this would lead to an explicit $\beta$-dependence of the action on the classical level.}
\item {Inclusion of nontrivial-holonomy calorons is not admissible, because of the divergence of the one-loop effective action in the thermodynamical limit  \cite{Gross/Pisarski/Yaffe81}.}
\item {A shift of the initial point of the Wilson line from $(\tau,0)$ to an arbitrary point $(\tau,\vec{z})$ would introduce a finite mass scale $|\vec{z}|^{-1}$ which is not existent on the classical level.}
\item {The calculation of the Wilson lines needs to be performed along straight lines, because spatial curvature would correspond to an additional mass-scale, which would have to be compensated by a factor of $\beta$. A non-flat measure of $\rho$ has the same effect.}
\end{enumerate}
Another condition to the macroscopic field is a consequence of BPS-saturation. Since $\phi$ is constructed from non-interacting BPS-saturated calorons, the energy-momentum tensor vanishes identically for the composite object. Thus $\phi$ solves a first-order differential equation, the BPS-equation
\begin{equation}
\label{BPS equation}
\partial_\tau\phi=V^{(1/2)}\,,
\end{equation}
where $V^{(1/2)}$ is a 'square-root' of a the potential $V(\phi)=\textrm{tr}(V^{(1/2)})^\dagger V^{(1/2)}$. This BPS-equation is determined from the evolution equation for the phase up to global gauge rotations and a choice of winding sense. In order to evaluate the phase we have to find a traceless, hermitian configuration which solves the second-order equation of motion (\ref{eom phase}) for the phase as well as the first-order BPS-equation (\ref{BPS equation}). In \cite{Hofmann05, Hofmann/Herbst04} a solution in the $(1,2)$-plane of the Lie algebra is derived as
\begin{eqnarray}
\label{phase winding gauge}
\frac{\phi}{|\phi|}&=&C_1\lambda_1\cos\left(\frac{2\pi}{\beta}(\tau-\tau_0)\right)
+C_2\lambda_2\sin\left(\frac{2\pi}{\beta}(\tau-\tau_0)\right)\\
&=&C\lambda_1\exp\left(\mp\frac{2\pi i}{\beta}\lambda_3(\tau-\tau_0)\right)\,,
\nonumber
\end{eqnarray}
with undetermined integration constants. Since the phase is $\tau$-dependent, we refer to this gauge as winding gauge. This solution corresponds to circular polarisation in the $x_1$-$x_2$ plane of color space. $\phi$ winds along an $S^1$ on the group manifold $S^3$ of $SU(2)$. Notice that a global gauge ambiguity associated with the choice of the polarization plane of this oscillation remains.

\underline{\textit{$\phi$'s modulus:}}
Initially, we assume the existence of an externally given, constant Yang-Mills scale $\Lambda$. $\Lambda$ is a renormalization group invariant. This is only realized by an effect resulting from trivial-topology fluctuations called dimensional transmutation.
If we require that the phase Eq.\,(\ref{phase winding gauge}) is reproduced, $V^{(1/2)}$ ought to be a linear function of $\phi$. The temperature-dependence of $\phi$ cannot be explicit, since the average over the caloron-anticaloron system is $\beta$-independent. But $V(\phi)$ may depend on temperature through the periodicity of $\phi$. Furthermore, the potential should be an analytic function of $\phi$. The conditions lead us to a BPS-equation for $\phi$
\begin{equation}
\partial_\tau\phi=\pm i\Lambda^3\lambda_3\frac{\phi}{|\phi|^2}\,.
\end{equation}
Inserting the expression for the phase, we obtain for $\phi$'s modulus  \cite{Hofmann05, Hofmann/Herbst04}
\begin{equation}
|\phi|(\beta,\Lambda)=\sqrt{\frac{\beta\Lambda^3}{2\pi}}=
\sqrt{\frac{\Lambda^3}{2\pi T}}=\frac{\Lambda}{\sqrt{\lambda}}\,,
\end{equation}
where $\lambda$ denotes the dimensionless temperature $\lambda=2\pi T\Lambda^{-1}$. Thus the field $\phi$ vanishes proportional to $T^{-\frac{1}{2}}$ with increasing temperature. An important fact is that the modulus $|\phi|$ corresponds to the maximal resolving power allowed in the effective theory.

Since $\phi$'s phase and modulus are determined we are able to calculate the potential $V(\phi)$ from the BPS-equation Eq.\,(\ref{BPS equation}) as
\begin{equation}
V(\phi)=\Lambda^6 \textrm{tr}\phi^{-2}=4\pi T\Lambda^3\,.
\end{equation}
The euclidean Lagrangian governing the dynamics of the Higgs field $\phi$ is then given as
\begin{equation}
\mathcal{L}_\phi=\textrm{tr}\left((\partial_\tau\phi)^2+\Lambda^6\phi^{-2}\right)\,.
\end{equation}
We consider now the statistical and quantum mechanical inertness of $\phi$ \cite{Hofmann05}. A field is called inert, if the ratio of the squared masses of fluctuations $\delta\phi$, defined as $\frac{\partial^2V}{\partial_{|\phi|}^2}$ an the squared compositeness scale $|\phi|$, or respectively the temperature, are both larger than unity. We obtain
\begin{equation}
\frac{\partial_{|\phi|}^2V}{\partial_{|\phi|}^2}=12\lambda^3\qquad\mbox{and}\qquad
\frac{\partial_{|\phi|}^2V}{T^2}=48\pi^2\,.
\end{equation}
In the deconfining phase of $SU(2)$ Yang-Mills thermodynamics it holds $\lambda>\lambda_c=13.87$, and the criterions for inertness are satisfied. We conclude that quantum fluctuations of the field $\phi$ do not exist, since their masses would exceed the compositeness scale many times over.

\subsection{The effective theory}

At first we recall the validity of a unique decomposition for every $SU(2)$ gauge field of the fundamental theory. Each gauge field $A_\mu$ decays into a topologically nontrivial, BPS-saturated part $A_\mu^{top}$ represented by calorons and a topologically trivial remainder $a_\mu$. Until now, we have confined ourselves to the topologically nontrivial sector, described by the adjoint scalar field $\phi$ after spatial average. In order to include caloron interactions we have to consider the topological trivial (i.e. $Q=0$) sector. The topologically trivial fluctuations are responsible for a change of short-lived holonomy of calorons and for interactions between magnetic monopole constituents of nontrivial holonomy. In this situation the inert scalar field $\phi$ acts as a background after coarse-graining.

The influence of the configurations $a_\mu$ is taken into account via minimal coupling
\begin{equation}
\partial_\mu\phi\rightarrow D_\mu\phi=\partial_\mu\phi-ie[a_\mu,\phi]\,,
\end{equation}
where $e$ denotes the gauge coupling in the effective theory (which is not equal to the coupling $g$ of the fundamental theory).

We are interested in an effective theory, after the application of spatial coarse-graining. Integration of topological defects into an inert field $\phi$ and the renormalizability of Yang-Mills theory \cite{'t Hooft/Veltman71/72} assures form-invariance of the effective Lagrangian including the topologically trivial sector under the applied spatial coarse-graining. The resulting gauge-invariant, effective Yang-Mills action determining the dynamics of $Q=0$-fluctuations $a_\mu$ and the background field $\phi$ in the deconfining phase is then given as
\begin{equation}
S[a_\mu]=\int\limits_0^\beta d\tau\int d^3x \,\textrm{tr} \left(\frac{1}{2}G_{\mu\nu}G_{\mu\nu}+(D_\mu\phi)^2+\frac{\Lambda^6}{\phi^2}\right)\,,
\end{equation}
where $G_{\mu\nu}=G_{\mu\nu}^{a}\frac{\lambda^{a}}{2}
=\partial_\mu a_\nu-\partial_\nu a_\mu-ie[a_\mu,a_\nu]$. The Euler-Lagrange equation concerning the field $a_\mu$ follows as
\begin{equation}
D_\mu G_{\mu\nu}=ie[\phi,D_\nu\phi]\,.
\end{equation}
For reasons of rotational invariance this equation of motion requires a pure-gauge solution $a_\mu^{gs}$, implying $D_\nu\phi=0$ \cite{Hofmann05}. It holds that
\begin{equation}
a_\mu^{gs}=\frac{i}{e}(\partial_\mu\Omega)\Omega^\dagger\,,\qquad
\mbox{with}\qquad\Omega=e^{\pm i\frac{2\pi}{\beta}\tau\frac{\lambda_3}{2}}\,.
\end{equation}
At this point, the ground-state is completely determined by $a^{gs}$ and $\phi$. The energy-momentum tensor associated with this ground-state is then
\begin{equation}
\Theta_{\mu\nu}=4\pi\Lambda^3T\delta_{\mu\nu}\,,
\end{equation}
implying a ground-state equation connecting pressure and energy density:
\begin{equation}
P^{gs}=-4\pi\Lambda^3T=-\rho^{gs}\,.
\end{equation}
Since the ground state contains not only contributions of the BPS-saturated fields, but as well from topological configurations, $P^{gs}$ and $\rho^{gs}$ attain finite values after spatial-coarse graining. Thereby the Yang-Mills scale becomes gravitationally measurable.

In the following we discuss propagating fluctuations $\delta a_\mu$ in the effective theory. It is convenient to transform from winding gauge to unitary gauge via the transformation
\begin{equation}
U=\e^{-i\frac{\pi}{4}\lambda_2}\Omega\,.
\end{equation}
In unitary gauge the pure-gauge configuration vanishes, while the adjoint scalar field is rotated into the three-direction of the algebra:
\begin{equation}
a^{gs}=0\,,\qquad\phi=|\phi|\lambda_3\,.
\end{equation}
This gauge transformation switches the value of the Polyakov loop from $\mathcal{P}[a_\mu^{gs}=\delta_{\mu4}\frac{\pi}{e\beta}\lambda_3]=-1$ to $\mathcal{P}[a_\mu^{gs}=0]=1$. Therefore we can conclude that the theory possesses two equivalent ground states. The fact that the ground state is degenerate with respect to a global electric $Z_2(SU2)$ center symmetry designates the electric phase to be deconfining.
$\phi$ can be beheld as an adjoint Higgs field, it breaks the $SU(2)$ gauge symmetry dynamically down to the Abelian subgroup $U(1)$. Due to the adjoint Higgs mechanism two out of the three gauge modes $\delta a_\mu^{(1,2)}$ acquire a temperature-dependent mass
\begin{equation}
m=2e(T)|\phi|=2e(T)\sqrt{\frac{\Lambda^3}{2\pi T}}
=2e(T)\frac{\Lambda}{\sqrt{\lambda}}\,,
\end{equation}
while the third gauge mode remains massless. Here $e(T)$ is again the temperature-dependent effective gauge coupling, which will be investigated in detail below. The massive bosons are referred to tree-level heavy (TLH) and are denoted as $V^+$ and $V^-$, the massless boson as tree-level massless (TLM) and denoted as $\gamma$\,.

Finally, we review the emergence of compositeness constraints on the maximal off-shellness of quantum fluctuations. The existence of a maximal resolution $|\phi|$ resulting from spatial coarse-graining introduces limits for the propagating momenta. Momenta exceeding the compositeness scale $|\phi|$ would lead to a higher resolution and are thus forbidden. We state the constraints:
\begin{enumerate}
\item {The maximal off-shellness of a gauge mode is restricted to
  \begin{equation}
  \label{compositeness1}
  |p^2-m^2|\leq|\phi|^2\,,
  \end{equation}
  where p and m denote four-momentum and mass of the quantum fluctuation.}
\item {The four-momentum flux in a four-vertex is constraint subject to
  \begin{equation}
  \label{compositeness2}
  |(p+k)^2|\leq|\phi|^2\,.
  \end{equation}
  In this case two of the four legs have to form a loop, otherwise a differentiation between s-, t- and u-channel scattering is mandatory.}
\end{enumerate}
Since the phase space for fluctuations is tremendously limited, the massive excitations $V^\pm$ are very weakly interacting thermal quasi-particles. That implies the negligibility of higher-loop contributions, thus (at large temperatures) the order of magnitude from loop-order to loop-order is about $10^{-3}-10^{-4}$.
It is worth mentioning that the generated masses yield infrared cutoffs, and the arising of infrared divergencies in loop expansions is avoided automatically.
Furthermore, as a side-effect the compositeness constraints avoid the emergence of ultraviolet divergencies, so no standard renormalization procedure is necessary in the effective theory.

The evolution equation for the effective gauge coupling  $e(T)$ is derived from the demand for thermodynamical self-consistency. Thermodynamical quantities are connected via Legendre transformations in the fundamental theory. Self-consistency is provided, if a reformulation of the theory into an effective Lagrangian does not alter the Legendre transformation interrelating the thermodynamical quantities. Consider the Legendre transformation which maps the total pressure onto the total energy density:
\begin{equation}
\rho=T\frac{dP}{dT}-P\,.
\end{equation}
The requirement of self-consistency implies that the pressure shows only explicit temperature dependence. Since the massive gauge modes obtain a temperature dependent mass, the condition can be formulated in terms of a dimensionless mass
$a=\beta m=\beta e|\phi|=2\pi e\lambda^{-3/2}$ as
\begin{equation}
\label{self-consistency}
\partial_a P=0\,.
\end{equation}
In \cite{Hofmann05} the one-loop expression for the pressure was calculated as
\begin{equation}
P(\lambda)=-\Lambda^4\left(\frac{2\lambda^4}{(2\pi)^6}
\left[2\bar{P}(0)+6\bar{P}(2a)\right]+2\lambda\right)\,,
\end{equation}
where the function $\bar{P}(a)$ is defined as
\begin{equation}
\bar{P}(a)=\int\limits_0^\infty dx\,x^2 \ln\left[1-\e^{-\sqrt{x^2+a^2}}\right]\,.
\end{equation}
Employing the self-consistency condition Eq\,.(\ref{self-consistency}) we get the evolution equation for $\lambda(a)$:
\begin{equation}
\partial_a\lambda=-\frac{24\lambda^4 a}{(2\pi)^6}\frac{D(2a)}
{1+\frac{24\lambda^3a^2}{(2\pi)^6}D(2a)}\,,
\end{equation}
where the function $D(a)$ is defined as
\begin{equation}
D(a)=\int\limits_0^\infty dx\,\frac{x^2}{\sqrt{x^2+a^2}}\frac{1}{\e^{-\sqrt{x^2+a^2}}-1}\,.
\end{equation}
Inversion of the differential equation yields an evolution equation for the effective gauge coupling $\e(\lambda)$.

\subsection{Phase transitions: The preconfining and the confining phase}

In this section we summarize the essential features of the other phases in the theory of $SU(2)$ Yang-Mills thermodynamics. A comprehensive representation of the physics in these phases of the theory can be found in \cite{Hofmann05,
Hofmann-decon, Hofmann-precon, Hofmann-con}.

In the deconfining phase of $SU(2)$ Yang-Mills thermodynamics $e(\lambda)$ is mostly constant at a plateau value at $e=\sqrt{8}\pi$, except in the vicinity of the critical temperature at $\lambda_c=13.87$, where the gauge coupling diverges logarithmically
\begin{equation}
e(\lambda)\sim -\log(\lambda-\lambda_c)\,.
\end{equation}
At this point a phase transition to the preconfining (magnetic) phase occurs.

\underline{\textit{The preconfining phase:}} The deconfining-preconfining phase transition is  of second-order kind. Large-holonomy calorons dissociate into constituent BPS monopoles, whose masses $(M_m)=\frac{4\pi}{e\beta}$ approach zero due to the logarithmic divergence of the electric gauge coupling. Consequently, a new ground state emerges as a magnetic monopole condensate. This ground state is responsible for a dynamical breaking of the dual $U(1)$ gauge symmetry. The monopole condensate is described by an inert complex scalar field $\phi_M$. The gauge mode $\gamma$ becomes Meissner massive:
\begin{equation}
m_\gamma=g(t)|\phi_M|\,.
\end{equation}

The preconfining character of the magnetic phase is reflected in the  fact, that the phase does allow for the propagation of massive dual gauge modes, while fundamental, heavy fermions are confined by the monopole condensate.

Due to the genesis of the Meissner mass of $\gamma$, a longitudinal polarization arises additional to the two transversal polarizations. This causes a discontinuity $\rho_{c,E}\rightarrow\rho_{c,M}$ in the energy density for $T\searrow T_{c,E}$, since the excitation of this additional agree requires energy.

In the preconfining phase the ground state equation is given as
\begin{equation}
\rho^{gs}=-P^{gs}=\pi\Lambda_M^3 T\,.
\end{equation}
In contrast to the deconfining phase, the negativity of the ground state pressure is not a result of monopole-antimonopole attraction, but arises from collapsing and re-created center-vortex loops (Abrikosov-Nielsen-Olesen vortices).

\underline{\textit{The confining phase:}} There is a second phase transition from the magnetic to the center phase. The center phase is of confining character.

If the temperature approaches the critical temperature of the preconfining phase, $T_{c,M}=0.83 T_{c,E}$, from above, the magnetic gauge coupling diverges logarithmically at this phase boundary. The excitation $\gamma$ becomes infinitely massive and a phase transition of Hagedorn type occurs.

A condensate of massless, single center-vortex loops is generated, due to the decay of the monopole condensate. The local magnetic center symmetry is dynamically broken. Massive excitations are generated by twisting a vortex loop at a self-interaction point. A fascinating property of the center phase is that in contrast to the previous phases this vortex loop excitations are spin-$\frac{1}{2}$ fermions.

In the confining phase the ground state pressure and the ground state energy density are evaluated to be precisely zero.

%% file: energydensity.tex
\chapter{Correlation of energy
density in deconfining SU(2) Yang-Mills thermodynamics}
\label{coldclouds}

The main aim of this chapter is the calculation of the two-point correlation
function of the canonical energy density $\bra\Theta_{00}(x)\Theta_{00}(y)\ket$ at different locations $\vec{x}$ and $\vec{y}$ and at equal time. This quantity is associated with the energy transfer between the two locations $\vec{x}$ and $\vec{y}$. At first, we calculate this object in a pure, thermalized $U(1)$ gauge theory, where the propagating gauge field $A_\mu$ is equivalent to the photon of the SM. Our results
will serve as a benchmark for the more involved calculation of the case
when this object is embedded into a deconfining SU(2) Yang-Mills
theory, where the photon is identified with the massless mode surviving the dynamical gauge symmetry breaking $SU(2)\rightarrow U(1)$ in the deconfining phase of $SU(2)$ Yang-Mills thermodynamics. We investigate how the energy transfer is affected as mediated by the massless mode interacting with the non-trivial massive excitations at low temperatures. Finally, we will propose a possible explanation concerning the stability of old, innergalactic, cold clouds of atomic hydrogen. The results presented in this chapter are summarized in \cite{Hofmann/Keller}.

\section{General strategy in Coulomb gauge}

The two-point correlation of the energy density is computed
by letting derivative operators, associated with the structure of the
energy-momentum tensor $\Theta_{\mu\nu}$, act on the real-time
propagator of the $U(1)$ gauge field.

Recall that the traceless and symmetric (Belinfante) energy-momentum tensor of a pure
$U(1)$ gauge theory is given as
\begin{equation}
\label{energymomentumtensor}
\Theta_{\mu\nu}=-F_{\mu}\!^{\lambda}F_{\nu\lambda}+\frac{1}{4}g_{\mu\nu}F^{\kappa\lambda}F_{\kappa\lambda}\,,
\end{equation}
where $F_{\mu\nu}\equiv\pd_\mu A_\nu-\pd_\nu A_\mu$, and $A_\mu$ denotes the
$U(1)$ gauge field. In the deconfining phase of $SU(2)$ Yang-Mills
thermodynamics the effective theory contains an inert, adjoint Higgs
field $\phi$ \cite{Hofmann05, Hofmann/Herbst04}, and the field strength
$F_{\mu\nu}$ of the abelian theory can in Eq.\,(\ref{energymomentumtensor}) be replaced by the 't Hooft
tensor ${\cal F}_{\mu\nu}$ to define an $SU(2)$ gauge invariant
energy-momentum tensor. This tensor is defined as \cite{'tHooft74}
\begin{equation}
\label{'tHooft}
{\cal F}_{\mu\nu}\equiv \frac{1}{|\phi|}\phi_a
G_{\mu\nu a}-\frac{1}{e |\phi|^3}\epsilon_{abc}\phi_a(D_\mu\phi)_b(D_\nu\phi)_c\,,
\end{equation}
where $G^a_{\mu\nu}$ is the $SU(2)$ field strength of topologically
trivial, coarse-grained fluctuations, $D_\mu$ denotes the adjoint
covariant derivative, and $e$ is the effective gauge
coupling. Obviously, the quantity defined by the right-hand side of
Eq.\,(\ref{'tHooft}) is $SU(2)$ gauge invariant. In
unitary gauge $\phi_a=\delta_{a3}|\phi|$ the 't Hooft tensor reduces to the
abelian tensor $F_{\mu\nu}$ defined on the massless gauge field
$A_\mu^3$.

We evaluate the connected correlation
function $\la\Theta_{00}(x)\Theta_{00}(y)\ra$ in four-dimensional
Minkowskian spacetime ($g_{00}=1$). As described in the theory part the energy-momentum tensor $\Theta^{\mu\nu}$ is a gauge invariant quantity of composite fields. We use Wick's theorem in order to decompose the 2-point-function of the composite tensor into a sum of products of 2-point-functions involving the elementary gauge fields, i.e.
of the propagator $D_{\mu\nu}$. Remember that in real-time quantum field theory at finite temperature the propagator consists of a vacuum and a thermal contribution \cite{Le Bellac, Kapusta, Landsman/vanWeert, Rothe}. In Coulomb gauge and momentum space $D_{\mu\nu}$ is given as
\begin{eqnarray}
\label{U1propagator}
D_{\mu\nu}(p,T)&=&D^{\tiny\mbox{vac}}_{\mu\nu}(p,T)
+D^{\tiny\mbox{th}}_{\mu\nu}(p,T)\nonumber\\
&=&-P^T_{\mu\nu}(p)\frac{i}{p^2+i\epsilon}+i\frac{u_\mu
  u_\nu}{\vec{p}^2}-P^T_{\mu\nu}(p)2\pi\delta(p^2)n_B(\beta|p_0|)\,,
\end{eqnarray}
where
\begin{eqnarray}
P^T_{00}(p)&\equiv&P^T_{0i}(p)=P^T_{i0}(p)=0,\nonumber\\
P^T_{ij}(p)&\equiv&\delta_{ij}-\frac{p_i p_j}{\vec{p}^2}\,.
\end{eqnarray}
In this propagator $u_\mu=(1,0,0,0)$ can be interpreted as velocity of the heat-bath, $\beta=\frac{1}{T}$ denotes the inverse temperature and $n_B(x)\equiv\frac{1}{e^x-1}$ represents the Bose-Einstein distribution function.
By virtue of Eq.\,(\ref{energymomentumtensor}) one obtains
\begin{eqnarray}
\label{Wick decomposition}
\bra\Theta_{00}(x)\Theta_{00}(y)\ket &=&
    2\bra\partial_{x^0}A^{\lambda}(x)\partial_{y^0}A^{\tau}(y)\ket
    \bra\partial_{x^0}A_{\lambda}(x)\partial_{y^0}A_{\tau}(y)\ket\nonumber\\
&&-g_{00}\bra\partial_{x^0}A^{\lambda}(x)\partial_{y^{\sigma}}A^{\tau}(y)\ket
    \bra\partial_{x^0}A_{\lambda}(x)\partial_{y_\sigma}A_{\tau}(y)\ket\nonumber\\
&&+g_{00}\bra\partial_{x^0}A^{\lambda}(x)\partial_{y_\sigma}A^{\tau}(y)\ket
    \bra\partial_{x^0}A_{\lambda}(x)\partial_{y^\tau}A_{\sigma}(y)\ket\nonumber\\
&&-g_{00}\bra\partial_{x_\kappa}A^{\lambda}(x)\partial_{y^0}A^{\tau}(y)\ket
    \bra\partial_{x^\kappa}A_{\lambda}(x)\partial_{y^0}A_{\tau}(y)\ket\nonumber\\
&&+g_{00}\bra\partial_{x_\kappa}A^{\lambda}(x)\partial_{y^0}A^{\tau}(y)\ket
    \bra\partial_{x^\lambda}A_{\kappa}(x)\partial_{y^0}A_{\tau}(y)\ket\nonumber\\
&&+\frac{g_{00}^2}{2}
    \bra\partial_{x_\kappa}A^{\lambda}(x)\partial_{y_\sigma}A^{\tau}(y)\ket
    \bra\partial_{x^\kappa}A_{\lambda}(x)\partial_{y^\sigma}A_{\tau}(y)\ket\nonumber\\
&&-\frac{g_{00}^2}{2}
    \bra\partial_{x_\kappa}A^{\lambda}(x)\partial_{y_\sigma}A^{\tau}(y)\ket
    \bra\partial_{x^\kappa}A_{\lambda}(x)\partial_{y^\tau}A_{\sigma}(y)\ket\nonumber\\
&&-\frac{g_{00}^2}{2}
    \bra\partial_{x_\kappa}A^{\lambda}(x)\partial_{y_\sigma}A^{\tau}(y)\ket
    \bra\partial_{x^\lambda}A_{\kappa}(x)\partial_{y^\sigma}A_{\tau}(y)\ket\nonumber\\
&&+\frac{g_{00}^2}{2}
    \bra\partial_{x_\kappa}A^{\lambda}(x)\partial_{y_\tau}A^{\sigma}(y)\ket
    \bra\partial_{x^\lambda}A_{\kappa}(x)\partial_{y^\sigma}A_{\tau}(y)\ket\nonumber\\
&&+2\bra\partial_{x^0}A^0(x)\partial_{y^0}A^0(y)\ket
    \bra\partial_{x^0}A_0(x)\partial_{y^0}A_0(y)\ket\nonumber\\
&&-2\bra\partial_{x^0}A^0(x)\partial_{y^\tau}A^0(y)\ket
    \bra\partial_{x^0}A_0(x)\partial_{y_\tau}A_0(y)\ket\nonumber\\
&&+2\bra\partial_{x^\tau}A^0(x)\partial_{y^\sigma}A^0(y)\ket
    \bra\partial_{x_\tau}A_0(x)\partial_{y_\sigma}A_0(y)\ket\nonumber\\
&&-2g_{00}\bra\partial_{x^0}A^0(x)\partial_{y^0}A^0(y)\ket
    \bra\partial_{x^0}A_0(x)\partial_{y^0}A_0(y)\ket\nonumber\\
&&+4g_{00}\bra\partial_{x^0}A^0(x)\partial_{y^\tau}A^0(y)\ket
    \bra\partial_{x^0}A_0(x)\partial_{y_\tau}A_0(y)\ket\nonumber\\
&&-2g_{00}\bra\partial_{x^\tau}A^0(x)\partial_{y^\sigma}A^0(y)\ket
    \bra\partial_{x_\tau}A_0(x)\partial_{y_\sigma}A_0(y)\ket\nonumber\\
&&+\frac{g_{00}^2}{2}\bra\partial_{x^0}A^0(x)\partial_{y^0}A^0(y)\ket
    \bra\partial_{x^0}A_0(x)\partial_{y^0}A_0(y)\ket\nonumber\\
&&-g_{00}^2\bra\partial_{x^0}A^0(x)\partial_{y^\tau}A^0(y)\ket
    \bra\partial_{x^0}A_0(x)\partial_{y_\tau}A_0(y)\ket\nonumber\\
&&+\frac{g_{00}^2}{2}\bra\partial_{x^\tau}A^0(x)\partial_{y^\sigma}A^0(y)\ket
    \bra\partial_{x_\tau}A_0(x)\partial_{y_\sigma}A_0(y)\ket\,.\nonumber\\
\end{eqnarray}
The last nine lines in Eq.\,(\ref{Wick decomposition}) arise
from the term $\propto u_\mu u_\nu$ in the propagator, see
Eq.\,(\ref{U1propagator}).

\section{Two-point correlation of energy density in thermal U(1) gauge
  theory}
  \label{correlator u1}
\begin{figure}
\begin{center}
\vspace{5.3cm}
\includegraphics{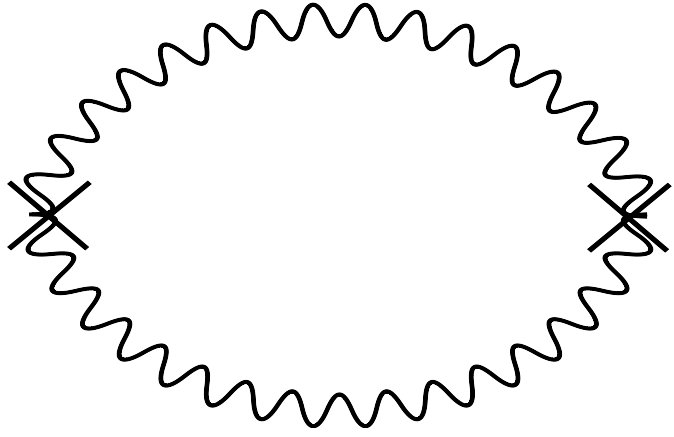}
\end{center}
\caption{Feynman diagram for the correlator
  $\bra\Theta_{00}(x)\Theta_{00}(y)\ket$ in a pure U(1) gauge
theory. Crosses denote the insertion of the composite operator $\Theta_{00}$.\label{Fig-1}}
\end{figure}

In evaluating the expression in Eq.\,(\ref{Wick decomposition})
the derivative operators are taken out of the expectation, and
Eq.\,(\ref{U1propagator}) is inserted. By momentum conservation the expression
Eq.\,(\ref{Wick decomposition}) separates into purely thermal and
purely vacuum contributions, see Fig.\,\ref{Fig-1}.
\subsection{The thermal contribution}
In this section we consider the thermal part of the two-point energy density correlator. We evaluate the derivative operators in Eq.\,(\ref{Wick decomposition}) and obtain\\
\newline
$\bra\Theta_{00}(x)\Theta_{00}(y)\ket^{\tiny\mbox{th}}=$
\begin{eqnarray}
\label{thermalpart}
&=&2\int\!\!\frac{d^4p}{(2\pi)^4}\int\frac{d^4k}{(2\pi)^4}
       P^{\lambda\tau}(p)P_{\lambda\tau}(k)p_0^2k_0^2(2\pi)^2\delta(p^2)\delta(k^2)
       n_B(\beta|p_0|)n_B(\beta|k_0|)e^{-ip(x-y)}e^{-ik(x-y)}\nonumber\\
&-&g_{00}\int\!\!\frac{d^4p}{(2\pi)^4}\int\frac{d^4k}{(2\pi)^4}
       P^{\lambda\tau}(p)P_{\lambda\tau}(k)p_0 p_\sigma k_0 k^\sigma(2\pi)^2
       \delta(p^2)\delta(k^2)n_B(\beta|p_0|)n_B(\beta|k_0|)
       e^{-ip(x-y)}e^{-ik(x-y)}\nonumber\\
&+&g_{00}\int\!\!\frac{d^4p}{(2\pi)^4}\int\frac{d^4k}{(2\pi)^4}
       P^{\lambda\tau}(p)P_{\lambda\sigma}(k)p_0 p^\sigma k_0 k_\tau (2\pi)^2
       \delta(p^2)\delta(k^2)n_B(\beta|p_0|)n_B(\beta|k_0|)
       e^{-ip(x-y)}e^{-ik(x-y)}\nonumber\\
&-&g_{00}\int\!\!\frac{d^4p}{(2\pi)^4}\int\frac{d^4k}{(2\pi)^4}
       P^{\lambda\tau}(p)P_{\lambda\tau}(k)p^\kappa p_0 k_\kappa k_0(2\pi)^2
       \delta(p^2)\delta(k^2)n_B(\beta|p_0|)n_B(\beta|k_0|)
       e^{-ip(x-y)}e^{-ik(x-y)}\nonumber\\
&+&g_{00}\int\!\!\frac{d^4p}{(2\pi)^4}\int\frac{d^4k}{(2\pi)^4}
       P^{\lambda\tau}(p)P_{\kappa\tau}(k)p^\kappa p_0 k_\lambda k_0(2\pi)^2
       \delta(p^2)\delta(k^2)n_B(\beta|p_0|)n_B(\beta|k_0|)
       e^{-ip(x-y)}e^{-ik(x-y)}\nonumber\\
&+&\frac{g_{00}^2}{2}\int\!\!\frac{d^4p}{(2\pi)^4}\int\frac{d^4k}{(2\pi)^4}
       P^{\lambda\tau}(p)P_{\lambda\tau}(k)p^\kappa p^\sigma k_\kappa k_\sigma(2\pi)^2
       \delta(p^2)\delta(k^2)n_B(\beta|p_0|)n_B(\beta|k_0|)
       e^{-ip(x-y)}e^{-ik(x-y)}\nonumber\\
&-&\frac{g_{00}^2}{2}\int\!\!\frac{d^4p}{(2\pi)^4}\int\frac{d^4k}{(2\pi)^4}
       P^{\lambda\tau}(p)P_{\lambda\sigma}(k)p^\kappa p^\sigma k_\kappa k_\tau (2\pi)^2
       \delta(p^2)\delta(k^2)n_B(\beta|p_0|)n_B(\beta|k_0|)
       e^{-ip(x-y)}e^{-ik(x-y)}\nonumber\\
&-&\frac{g_{00}^2}{2}\int\!\!\frac{d^4p}{(2\pi)^4}\int\frac{d^4k}{(2\pi)^4}
       P^{\lambda\tau}(p)P_{\kappa\tau}(k)p^\kappa p^\sigma k_\lambda k_\sigma(2\pi)^2
       \delta(p^2)\delta(k^2)n_B(\beta|p_0|)n_B(\beta|k_0|)
       e^{-ip(x-y)}e^{-ik(x-y)}\nonumber\\
&+&\frac{g_{00}^2}{2}\int\!\!\frac{d^4p}{(2\pi)^4}\int\frac{d^4k}{(2\pi)^4}
       P^{\lambda\sigma}(p)P_{\kappa\tau}(k)p^\kappa p^\tau k_\lambda k_\sigma(2\pi)^2
       \delta(p^2)\delta(k^2)n_B(\beta|p_0|)n_B(\beta|k_0|)
       e^{-ip(x-y)}e^{-ik(x-y)}\,.\nonumber\\
\end{eqnarray}
The tensor contractions are given as
\begin{eqnarray}
\label{tensorcontraction1}
P^{T\,\lambda\tau}(p)P^{T}_{\lambda\tau}(k)
&=&1+\frac{(\vec{p}\cdot
  \vec{k})^2}{\vec{p}^2\vec{k}^2}\,,\\
\label{tensorcontraction2}
P^{T}_{\lambda\sigma}(p)k^\lambda
k^\sigma&=&\vec{k}^2-\frac{(\vec{p}\cdot
  \vec{k})^2}{\vec{p}^2}\,,\\
\label{tensorcontraction3}
P^{T\,\kappa\tau}(p)P^{T}_{\kappa\sigma}(k)p^\sigma k_\tau
&=&-(\vec{p}\cdot \vec{k})
\left(\frac{(\vec{p}\cdot \vec{k})^2}{\vec{p}^2 \vec{k}^2}-1\right)\,.
\end{eqnarray}

The Dirac $\delta$-Distribution with quadratic argument can be written in the following way:
\begin{equation}
\delta(p^2)=\delta\left(p_0^2-\sqrt{\textbf{p}^2}^2\right)=
\frac{\delta(p_0-\sqrt{\textbf{p}^2})+\delta(p_0+\sqrt{\textbf{p}^2})}
{2\sqrt{\textbf{p}^2}}\,.
\end{equation}
This decomposition proves to be useful for the calculation of the integrals concerning the zero-coordinate.
We use the following property of the $\delta$-distribution:
\begin{equation}
\label{convolution}
\int dx f(x) \delta(x-\eta)=f(\eta)\,.
\end{equation}
The integration yields four summands:
\begin{eqnarray}
\label{Delta decomposition}
& &(\delta(p_0-\sqrt{\textbf{p}^2})+\delta(p_0+\sqrt{\textbf{p}^2})]\cdot
    [\delta(k_0-\sqrt{\textbf{k}^2})+\delta(k_0+\sqrt{\textbf{k}^2})]=\nonumber\\
&=&\delta(p_0-\sqrt{\textbf{p}^2})\delta(k_0-\sqrt{\textbf{k}^2})
   +\delta(p_0-\sqrt{\textbf{p}^2})\delta(k_0+\sqrt{\textbf{k}^2})+\nonumber\\
& &+\delta(p_0+\sqrt{\textbf{p}^2})\delta(k_0-\sqrt{\textbf{k}^2})
   +\delta(p_0+\sqrt{\textbf{p}^2})\delta(k_0+\sqrt{\textbf{k}^2})\\.
\nonumber
\end{eqnarray}
Performing the (trivial) integration over the zero-coordinates
leads us to
\begin{eqnarray}
\label{thermalintegralu1}
\bra\Theta_{00}(\vec{x})\Theta_{00}(\vec{y})\ket^{\tiny\mbox{th}}
&=&\left(\int\!\!\frac{d^3p}{(2\pi)^3}
        |\vec{p}|\,n_B(\beta|\vec{p}|)
        \,\e^{i\vec{p}\vec{z}}\right)^2\nonumber\\
&+&\int\!\!\frac{d^3p}{(2\pi)^3}\int\!\!\frac{d^3k}{(2\pi)^3}
        \left(\frac{\vec{p}\vec{k}}{|\vec{p}||\vec{k}|}\right)^2
        |\vec{p}||\vec{k}|\,n_B(\beta|\vec{p}|)n_B(\beta|\vec{k}|)
        \,\e^{i\vec{p}\vec{z}}\,\e^{i\vec{k}\vec{z}}\nonumber\\
\end{eqnarray}
where $\vec{z}\equiv \vec{x}-\vec{y}$. In deriving Eq.\,(\ref{thermalintegralu1}) we have set $x^0=y^0$ thus neglecting oscillatory terms. This prescription should reflect the time-averaged energy transport and is technically much easier to handle. An expression for the two-point correlator of energy density including the oscillatory terms can be found in Sec.\,(\ref{appendix thermal}).

In order to evaluate the integrals, we introduce rescaled momenta and rescaled coordinates subject to $\tilde{p}_i=\beta p_i$, $\tilde{k}_i=\beta k_i$ and $\tilde{z}_i= z_i/\beta$, respectively. Subsequently, we introduce $3D$ spherical coordinates
\begin{eqnarray}
   (\tilde{p}_x,\tilde{p}_y,\tilde{p}_z)&\rightarrow&
   (|\tilde{\vec{p}}|\sin\theta_{|\tilde{\vec{p}}|}\cos\phi_{|\tilde{\vec{p}}|},
   |\tilde{\vec{p}}|\sin\theta_{|\tilde{\vec{p}}|}\sin\phi_{|\tilde{\vec{p}}|},
   |\tilde{\vec{p}}|\cos\theta_{|\tilde{\vec{p}}|})\,,\\
   (\tilde{k}_x,\tilde{k}_y,\tilde{k}_z)&\rightarrow&
   (|\tilde{\vec{k}}|\sin\theta_{\tilde{\vec{k}}}\cos\phi_{\tilde{\vec{k}}},
   |\tilde{\vec{k}}|\sin\theta_{\tilde{\vec{k}}}\sin\phi_{\tilde{\vec{k}}},
   |\tilde{\vec{k}}|\cos\theta_{\tilde{\vec{k}}})\,,
\end{eqnarray}
and express the spatial scalar product between the momenta $\tilde{\vec{p}}$ and $\tilde{\vec{k}}$ as function of the polar and azimuthal angles
\begin{eqnarray}
\frac{\tilde{\vec{p}}\tilde{\vec{k}}}{|\tilde{\vec{p}}||\tilde{\vec{k}}|} &=&
  \sin\theta_{\tilde{\vec{p}}}\cos\phi_{\tilde{\vec{p}}}\sin\theta_{\tilde{\vec{k}}}
  \cos\phi_{\tilde{\vec{k}}}
  +\sin\theta_{\tilde{\vec{p}}}\sin\phi_{\tilde{\vec{p}}}\sin\theta_{\tilde{\vec{k}}}
  \sin\phi_{\tilde{\vec{k}}}+\cos\theta_{\tilde{\vec{p}}}\cos\theta_{\tilde{\vec{k}}}
  =\nonumber\\
&=&\sin\theta_{\tilde{\vec{p}}}\sin\theta_{\tilde{\vec{k}}}
  \cos(\phi_{\tilde{\vec{p}}}-\phi_{\tilde{\vec{k}}})
  +\cos\theta_{\tilde{\vec{p}}}\cos\theta_{\tilde{\vec{k}}}\,.
\end{eqnarray}
Without constraining on generality, we can set $\phi_k$ equal to zero.
At first, we evaluate the integration over the azimuthal angles $\phi_p$ and $\phi_k$. Thus the integrals in the remaining variables $|\vec{p}|$ and $\theta$ factorize.
\newpage
\noindent
We obtain\\

\begin{eqnarray}
\bra\Theta_{00}(\vec{x})\Theta_{00}(\vec{y})\ket^{\tiny\mbox{th}}
&=&\frac{1}{(2\pi)^6\beta^8}\left(
4\pi^2\cdot
         \left(\int\limits_0^{\infty}\!d|\tilde{\vec{p}}|
         \int\limits_{-1}^{+1}\!d\cos\theta
         \,\frac{|\tilde{\vec{p}}|^3}{(\e^{|\tilde{\vec{p}}|}-1)}
         \,\e^{i|\tilde{\vec{p}}||\vec{z}|\cos\theta}\right)^2\nonumber\right.\\
&+&2\pi^2\cdot
         \left(\int\limits_0^{\infty}\!d|\tilde{\vec{p}}|
         \int\limits_{-1}^{+1}\!d\cos\theta
         \,\sin^2\theta\frac{|\tilde{\vec{p}}|^3}{(\e^{|\tilde{\vec{p}}|}-1)}
         \,\e^{i|\tilde{\vec{p}}||\vec{z}| \cos\theta}\right)^2\nonumber\\
&+&4\pi^2\cdot
         \left(\int\limits_0^{\infty}\!d|\tilde{\vec{p}}| \int\limits_{-1}^{+1}\!d\cos\theta
         \,\cos^2\theta\frac{|\tilde{\vec{p}}|^3}{(\e^{|\tilde{\vec{p}}|}-1)}
         \,\e^{i|\tilde{\vec{p}}||\vec{z}|\cos\theta}\right)^2\,.
\end{eqnarray}

In order to evaluate the $\theta$-integration, we write the spatial Fourier transform in a more convenient way.
We are allowed to choose the distance  vector $\vec{z}$ to point into the three direction. Thereby, the scalar product of momenta $\textbf{q}$ and distance $\vec{z}$ in the exponential can be written as $\textbf{q}\cdot\vec{z}=|\vec{q}||\vec{z}|\cos\theta$.
Finally we use the expansion of a plane wave into spherical harmonics. It holds that
\begin{equation}
\label{sphericalexpansion3D}
e^{i|\vec{\tilde{q}}||\vec{\tilde{\vec{z}}}|\cos\theta}=
\sum\limits_{l=0}^{\infty} i^l (2l+1) j_l(|\vec{\tilde{q}}||\vec{\tilde{\vec{z}}}|) P_l(\cos\theta)\,,
\end{equation}
where $\theta\equiv\angle({\vec{q},\vec{\zeta}})$, $q=p,k$, $j_l$
denotes a spherical Bessel function of the first kind, and $P_l(\cos\theta)$ is a Legendre polynomial in $\cos\theta$. An important fact is that the Legendre polynomials obey an orthogonality relation defined on the interval $(-1,1)$:\\
\begin{equation}
\label{Legendre orthogonality}
\int\limits_{-1}^{+1}\!dx\, P_m(x) P_n(x)=\frac{2}{2m+1}\delta_{mn}\,.
\end{equation}
Subsequently, we write the thermal part of the two-point correlator in terms of linear combinations of Legendre polynomials with the argument $\cos\theta$ and exploit the orthogonality relation(\ref{Legendre orthogonality}). Consequently, the infinite sum in the plane wave expansion reduces to a finite amount of terms. The orthogonality relation of the Legendre polynomials fixes the degree of the involved spherical Bessel function.

\noindent
We arrive at\\

\begin{eqnarray}
\bra\Theta_{00}(\vec{x})\Theta_{00}(\vec{y})\ket^{\tiny\mbox{th}}
&=& \frac{1}{(2\pi)^6\beta^8}\cdot4\pi^2\cdot
    \left(2\int\limits_0^{\infty}d|\tilde{\vec{p}}|
    \frac{|\tilde{\vec{p}}|^3}{e^{|\tilde{\vec{p}}|}-1}
    (j_0(|\tilde{\vec{p}}||\tilde{\vec{z}}|))\right)^2\nonumber\\
&&+\quad \frac{1}{(2\pi)^6\beta^8}\cdot2\pi^2\cdot
   \left(\frac{4}{3}\int\limits_0^{\infty}d|\tilde{\vec{p}}|
   \frac{|\tilde{\vec{p}}|^3}{e^{|\tilde{\vec{p}}|}-1}
   (j_0(|\tilde{\vec{p}}||\tilde{\vec{z}}|))
   +\frac{4}{3}\int\limits_0^{\infty}d|\tilde{\vec{p}}|
   \frac{|\tilde{\vec{p}}|^3}{e^{|\tilde{\vec{p}}|}-1}
   (j_2(|\tilde{\vec{p}}||\tilde{\vec{z}}|))\right)^2\nonumber\\
&&+\quad \frac{1}{(2\pi)^6\beta^8}\cdot4\pi^2\cdot
   \left(\frac{2}{3}\int\limits_0^{\infty}d|\tilde{\vec{p}}|
   \frac{|\tilde{\vec{p}}|^3}{e^{|\tilde{\vec{p}}|}-1}
   (j_0(|\tilde{\vec{p}}||\tilde{\vec{z}}|))
   -\frac{4}{3}\int\limits_0^{\infty}d|\tilde{\vec{p}}|
   \frac{|\tilde{\vec{p}}|^3}{e^{|\tilde{\vec{p}}|}-1}
   (j_2(|\tilde{\vec{p}}||\tilde{\vec{z}}|))\right)^2\,.\nonumber\\
\end{eqnarray}
\newpage
After calculating the binomials and arranging integrals of the same type together, we obtain
\begin{eqnarray}
\label{thermalu1bessel}
\bra\Theta_{00}(\vec{x})\Theta_{00}(\vec{y})\ket^{\tiny\mbox{th}}
&=&\frac{1}{(2\pi)^6\beta^8}\left(\frac{64\pi^2}{3}\cdot
      \left(\int\limits_0^{\infty}d|\tilde{\vec{p}}|\,
      \frac{|\tilde{\vec{p}}|^3}
      {\e^{|\tilde{\vec{p}}|}-1} j_0(|\tilde{\vec{p}}||\tilde{\vec{z}}|)\right)^2\right.\nonumber\\
&&+\frac{32\pi^2}{3}\cdot
      \left.\left(\int\limits_0^{\infty}d|\tilde{\vec{p}}|\,
      \frac{|\tilde{\vec{p}}|^3}
      {\e^{|\tilde{\vec{p}}|}-1} j_2(|\tilde{\vec{p}}||\tilde{\vec{z}}|)\right)^2\right)\,.
\end{eqnarray}
We remark, that only partial waves of an even degree contribute to the result, modes with odd parity vanish for reasons of symmetry.

The remaining integral concerning the radial momentum are solvable in an analytical way by implementing the following definite integrals \cite{Gradshteyn}:
\begin{eqnarray}
\int\limits_0^{\infty} dx \frac{x^{2m}\sin(bx)}{e^x-1} &=& (-1)^m\frac{\partial^{2m}}{\partial b^{2m}}
               \left[\frac{\pi}{2}\coth(\pi b)-\frac{1}{2b}\right]\,,\\
\int\limits_0^{\infty} dx \frac{x^{2m+1}\cos(bx)}{e^x-1} &=& \frac{\partial^{2m+1}}{\partial b^{2m+1}}
               \left[\frac{\pi}{2}\coth(\pi b)-\frac{1}{2b}\right]\,,
\end{eqnarray}
where $b>0$, $m>0$ and $m\in\mathbb{Z}$ is required. Evaluating the last integration, we arrive at the final result for the thermal contribution of the two-point correlation of energy density in a $U(1)$ gauge theory:
\begin{eqnarray}
\label{thermalu1 finalresult}
&&\bra\Theta_{00}(\vec{x})\Theta_{00}(\vec{y})\ket^{\tiny\mbox{th}}=\nonumber\\
&&\frac{1}{(2\pi)^6\beta^8}\Bigg(\frac{64\pi^2}{3}\left(\frac{1}{|\tilde{\vec{z}}|^4}
     -\frac{\pi^3\coth(\pi|\tilde{\vec{z}}|)\textrm{cosech}^2(\pi|\tilde{\vec{z}}|)}
     {|\tilde{\vec{z}}|}\right)^2\nonumber\\
&&\left.+\frac{32\pi^2}{3}\left(\frac{-8+\pi|\tilde{\vec{z}}|(3\coth(\pi|\tilde{\vec{z}}|)
     +\pi|\tilde{\vec{z}}|(3+2\pi|\tilde{\vec{z}}|\coth(\pi|\tilde{\vec{z}}|))
     \,\textrm{cosech}^2(\pi|\tilde{\vec{z}}|))}
     {2|\tilde{\vec{z}}|^4}\right)^2\right)\nonumber\\
&&=\frac{1}{24\pi^4|\tilde{\vec{z}}|^8}\bigg(8\big(-1+\pi^3|\tilde{\vec{z}}|^3
     \coth(\pi|\tilde{\vec{z}}|)
     \,\textrm{cosech}^2(\pi|\tilde{\vec{z}}|)\big)^2\nonumber\\
     &&\hspace{30mm}
     +\big(-8 + \pi |\tilde{\vec{z}}| (3\coth(\pi |\tilde{\vec{z}}|) + \pi |\tilde{\vec{z}}| (3 +
         2 \pi |\tilde{\vec{z}}| \coth(\pi|\tilde{\vec{z}}|)
     \,\textrm{cosech}^2(\pi|\tilde{\vec{z}}|)))\big)^2\bigg)\,.\nonumber\\
\end{eqnarray}

\subsection{The vacuum contribution}
In this section we evaluate the quantum contribution to the two-point correlation function \begin{equation}
\bra\Theta_{00}(\textbf{x})\Theta_{00}(\textbf{y})\ket\,.
\end{equation}
We start again with Eq.\,(\ref{Wick decomposition}), but we replace the thermal part of the propagator by the zero-temperature quantum propagator.
We expect $\bra\Theta_{00}(x)\Theta_{00}(y)\ket^{\tiny\mbox{vac}}$ to be
a negligible correction to $\bra\Theta_{00}(\textbf{x})\Theta_{00}(\textbf{y})\ket^{\tiny\mbox{th}}$
for $|\tilde{\vec{z}}|>1$.
and actually the result will show a significant suppression in comparison to the thermal part.
Inserting the propagator yields\\
\begin{eqnarray}
\bra\Theta_{00}(\vec{x})\Theta_{00}(\vec{y})\ket^{\tiny\mbox{vac}}
&=&-2\int\!\frac{d^4p}{(2\pi)^4}\int\!\frac{d^4p}{(2\pi)^4}
   P^{\lambda\tau} P_{\lambda\tau} \frac{p_0^2}{p^2} \frac{k_0^2}{k^2}
   \e^{-ip\zeta} \e^{-ik\zeta}\nonumber\\
&&+2g_{00}\int\!\frac{d^4p}{(2\pi)^4}\int\!\frac{d^4p}{(2\pi)^4}
   P^{\lambda\tau} P_{\lambda\tau}
   \frac{p_0 p^\sigma}{p^2} \frac{k_0 k_\sigma}{k^2}
   \e^{-ip\zeta} \e^{-ik\zeta}\nonumber\\
&&-2g_{00}\int\!\frac{d^4p}{(2\pi)^4}\int\!\frac{d^4p}{(2\pi)^4}
   P^{\lambda\tau} P_{\lambda\sigma}
   \frac{p_0 p^\sigma}{p^2} \frac{k_0 k_\tau}{k^2}
   \e^{-ip\zeta} \e^{-ik\zeta}\nonumber\\
&&-\frac{g_{00}^2}{2}\int\!\frac{d^4p}{(2\pi)^4}\int\!\frac{d^4p}{(2\pi)^4}
   P^{\lambda\tau} P_{\lambda\tau}
   \frac{p^\kappa p^\sigma}{p^2} \frac{k_\kappa k_\sigma}{k^2}
   \e^{-ip\zeta} \e^{-ik\zeta}\nonumber\\
&&+g_{00}^2\int\!\frac{d^4p}{(2\pi)^4}\int\!\frac{d^4p}{(2\pi)^4}
   P^{\lambda\tau} P_{\lambda\sigma}
   \frac{p_\kappa p^\sigma}{p^2} \frac{k_\kappa k_\tau}{k^2}
   \e^{-ip\zeta} \e^{-ik\zeta}\nonumber\\
&&-\frac{g_{00}^2}{2}\int\!\frac{d^4p}{(2\pi)^4}\int\!\frac{d^4p}{(2\pi)^4}
   P^{\lambda\tau} P_{\kappa\sigma}
   \frac{p^\kappa p^\sigma}{p^2} \frac{k_\lambda k_\tau}{k^2}
   \e^{-ip\zeta} \e^{-ik\zeta}\nonumber\\
\hspace{25mm}&&-2\int\!\frac{d^4p}{(2\pi)^4}\int\!\frac{d^4p}{(2\pi)^4}
   (p_0^2 k_0^2-p_0 k_0 p_\tau k^\tau+p_\lambda k^\lambda p_\tau k^\tau)
   \frac{\e^{-ip\zeta}}{\vec{p}^2} \frac{\e^{-ik\zeta}}{\vec{k}^2}\nonumber\\
&&+g_{00}\int\!\frac{d^4p}{(2\pi)^4}\int\!\frac{d^4p}{(2\pi)^4}
   (p_0^2 k_0^2-2p_0 k_0 p_\tau k^\tau+p_\lambda k^\lambda p_\tau k^\tau)
   \frac{\e^{-ip\zeta}}{\vec{p}^2}
   \frac{\e^{-ik\zeta}}{\vec{k}^2}\nonumber\\
&&-\frac{g_{00}^2}{2}\int\!\frac{d^4p}{(2\pi)^4}\int\!\frac{d^4p}{(2\pi)^4}
   (p_0^2 k_0^2-2p_0 k_0 p_\tau k^\tau+p_\lambda k^\lambda p_\tau k^\tau)
   \frac{\e^{-ip\zeta}}{\vec{p}^2}
   \frac{\e^{-ik\zeta}}{\vec{k}^2}\nonumber\\
\end{eqnarray}
The last three summands are a consequence of the Coulomb-Term in the propagator (\ref{U1propagator}), which describes the 'propagation' of the $A^3_0$ field.
In first instance we use tensor contractions similar to Eqs.\,(\ref{tensorcontraction1}), (\ref{tensorcontraction2}), (\ref{tensorcontraction3}). Now we accomplish an analytic continuation and rotate to euclidean signature ($p_0,k_0\to ip_0,ik_0$, $x^0,y^0\to -ix^0,-iy^0$, $g_{\mu\nu}\to -\delta_{\mu\nu}$).

We obtain the to Eq.\,(\ref{thermalintegralu1}) corresponding part resulting from the zero-temperature propagator
\begin{eqnarray}
\label{vacuumintegralu1}
&=&\frac{9}{2}\left(\int\frac{d^4p}{(2\pi)^4}
   \,p_{0}^2\,\frac{\e^{ip\zeta}}{p^2}\right)^2
   +\frac{1}{2}\left(\int\frac{d^4p}{(2\pi)^4}
   \,|\vec{p}|^2\,\frac{\e^{ip\zeta}}{p^2}\right)^2\nonumber\\
&&+6\int\frac{d^4p}{(2\pi)^4}\int\frac{d^4k}{(2\pi)^4}
   \left(\frac{\vec{p}\vec{k}}{|\vec{p}||\vec{k}|}\right)
   p_0 k_0 |\vec{p}| |\vec{k}|
   \,\frac{\e^{ip\zeta}}{p^2}\,\frac{\e^{ik\zeta}}{k^2}\nonumber\\
&&+\frac{1}{2}\int\frac{d^4p}{(2\pi)^4}\int\frac{d^4k}{(2\pi)^4}
   \left(\frac{\vec{p}\vec{k}}{|\vec{p}||\vec{k}|}\right)^2
   \left(9p_0^2 k_0^2 + |\vec{p}|^2 |\vec{k}|^2\right)
   \,\frac{\e^{ip\zeta}}{p^2}\,\frac{\e^{ik\zeta}}{k^2}\nonumber\\
&&+2\left(\int\frac{d^4p}{(2\pi)^4}\,p_{0}^2
   \,\frac{\e^{ip\zeta}}{\vec{p}^2}\right)^2\nonumber\\
&&+2\int\frac{d^4p}{(2\pi)^4}\int\frac{d^4k}{(2\pi)^4}
   \left(\frac{\vec{p}\vec{k}}{|\vec{p}||\vec{k}|}\right)
   p_0 k_0 |\vec{p}| |\vec{k}|
   \frac{\e^{ip\zeta}}{\vec{p}^2}\,\frac{\e^{ik\zeta}}{\vec{k}^2}\nonumber\\
&&+\frac{9}{2}\int\frac{d^4p}{(2\pi)^4}\int\frac{d^4k}{(2\pi)^4}
   \left(\frac{\vec{p}\vec{k}}{|\vec{p}||\vec{k}|}\right)^2
   |\vec{p}|^2 |\vec{k}|^2
   \,\frac{\e^{ip\zeta}}{\vec{p}^2}\,\frac{\e^{ik\zeta}}{\vec{k}^2}\,.
   \nonumber\\
\end{eqnarray}
where $\zeta\equiv x-y$ and $p\zeta=p_\mu\zeta_\mu, k\zeta=k_\mu\zeta_\mu,
p^2=p_\mu p_\mu,$ and $k^2=k_\mu k_\mu$. The last three lines in
in the equation above arise
from the term $\propto u_\mu u_\nu$ in the propagator, see Eq.\,(\ref{U1propagator}).

In analogy to the calculation concerning the thermal contribution we rescale the variables dimensionless as $\tilde{p}_\mu\equiv \beta p_\mu,$ and $\tilde{k}_\mu\equiv\beta k_\mu$. The strategy is now to express the integrals in Eq.\,(\ref{vacuumintegralu1}) in terms of four-dimensional hyperspherical coordinates. We restrict to $x_0=y_0$ to be able to compare $\bra\Theta_{00}(x)\Theta_{00}(y)\ket^{\tiny\mbox{vac}}$ with
$\bra\Theta_{00}(\textbf{x})\Theta_{00}(\textbf{y})\ket^{\tiny\mbox{th}}$.\\
We define
\begin{eqnarray}
\tilde{p}=
|\tilde{p}|\left(
   \begin{array}{c} \cos\psi\\
                    \sin\psi\sin\theta\cos\phi\\
                    \sin\psi\sin\theta\sin\phi\\
                    \sin\psi\cos\theta
   \end{array}
\right)
\qquad\mbox{with}\quad
   \begin{array}{c} 0\leq\psi,\theta\leq\pi\\
                    0\leq\phi\leq2\pi\\
                    |\tilde{p}|=\sqrt{x_1^2+x_2^2+x_3^2+x_4^2}
   \end{array}\,,
\end{eqnarray}
and the integration measure follows as
$d^4x=|\tilde{\vec{p}}|^3\sin^2\psi\sin\theta\ d|\tilde{\vec{p}}| d\psi d\theta d\phi$. We proceed now in a similar way as in the thermal case.
The integration over the azimuthal angles is practicable without any difficulties and performed in total analogy to the thermal case. Consequently, the integrals over the remaining variables
factorize in two identical contributions for each term in Eq.\,(\ref{vacuumintegralu1}).
In order to evaluate the integrals over the first polar angle $\theta$, we determine the spatial distance vector $z$ to be collinear to the 3-direction and expand the exponential into partial waves. Due to the definition of the 4D spherical coordinates the following expansion is valid
\begin{equation}
\label{sphericalexpansion4D}
e^{i|\tilde{q}||\vec{\tilde{\vec{z}}}|\sin\psi\,\cos\theta}=
\sum\limits_{l=0}^{\infty} i^l (2l+1) j_l(|\tilde{q}||\vec{\tilde{\vec{z}}}|\sin\psi) P_l(\cos\theta)\,.
\end{equation}

After expressing
polynomial factors in $\cos\theta$ in terms of a linear combination of
Legendre polynomials $P_l(\cos\theta)$, we are enabled to exploit the orthogonality relation (\ref{Legendre orthogonality}).
Performing the integration over the second
polar angle $\theta$ yields
\begin{eqnarray}
\label{vacuummomentumpsiu1}
&&\bra\Theta_{00}(\vec{x})\Theta_{00}(\vec{y})\ket^{\tiny\mbox{vac}}
     =\frac{1}{(2\pi)^8\beta^8}\left(96\pi^2
     \left(\int\limits_0^{\infty}\!\!d|\tilde{p}|\,|\tilde{p}|^3
     \int\limits_0^{\pi}\!\!d\psi\,\sin^2\psi \cos^2\psi
     j_0(|\tilde{p}||\tilde{\vec{z}}|\sin\psi) \right)^2\right. \nonumber\\
&&\hspace{30mm}+\left.48\pi^2
     \left(\int\limits_0^{\infty}\!\!d|\tilde{p}|\,|\tilde{p}|^3
     \int\limits_0^{\pi}\!\!d\psi\,\sin^2\psi \cos^2\psi j_2(|\tilde{p}||\tilde{\vec{z}}|\sin\psi) \right)^2\right. \nonumber\\
&&\hspace{30mm}+\left.96\pi^2
     \left(\int\limits_0^{\infty}\!\!d|\tilde{p}|\,|\tilde{p}|^3
     \int\limits_0^{\pi}\!\!d\psi\,\sin^3\psi \cos\psi
     j_1(|\tilde{p}||\tilde{\vec{z}}|\sin\psi) \right)^2 \nonumber\right. \\
&&\hspace{30mm}+\left.\frac{32\pi^2}{3}
     \left(\int\limits_0^{\infty}\!\!d|\tilde{p}|\,|\tilde{p}|^3
     \int\limits_0^{\pi}\!\!d\psi\,\sin^4\psi j_0(|\tilde{p}||\tilde{\vec{z}}|\sin\psi) \right)^2\right.  \nonumber\\
&&\hspace{30mm}+\left.\frac{16\pi^2}{3}
     \left(\int\limits_0^{\infty}\!\!d|\tilde{p}|\,|\tilde{p}|^3
     \int\limits_0^{\pi}\!\!d\psi\,\sin^4\psi j_2(|\tilde{p}||\tilde{\vec{z}}|\sin\psi) \right)^2\right.  \nonumber\\
&&\hspace{30mm}+\left.32\pi^2
     \left(\int\limits_0^{\infty}\!\!d|\tilde{p}|\,|\tilde{p}|^3
     \int\limits_0^{\pi}\!\!d\psi\,\cos^2\psi j_0(|\tilde{p}||\tilde{\vec{z}}|\sin\psi) \right)^2\right.  \nonumber\\
&&\hspace{30mm}+\left.32\pi^2
     \left(\int\limits_0^{\infty}\!\!d|\tilde{p}|\,|\tilde{p}|^3
     \int\limits_0^{\pi}\!\!d\psi\,\sin\psi \cos\psi j_1(|\tilde{p}||\tilde{\vec{z}}|\sin\psi) \right)^2\right.\nonumber\\
&&\hspace{30mm}+\left.24\pi^2
     \left(\int\limits_0^{\infty}\!\!d|\tilde{p}|\,|\tilde{p}|^3
     \int\limits_0^{\pi}\!\!d\psi\,\sin^2\psi j_0(|\tilde{p}||\tilde{\vec{z}}|\sin\psi) \right)^2\right.\nonumber\\
&&\hspace{30mm}+\left.48\pi^2
     \left(\int\limits_0^{\infty}\!\!d|\tilde{p}|\,|\tilde{p}|^3
     \int\limits_0^{\pi}\!\!d\psi\,\sin^2\psi j_2(|\tilde{p}||\tilde{\vec{z}}|\sin\psi)\right)^2\right)\,,\nonumber\\
&&
\end{eqnarray}
In Eq.\,(\ref{vacuummomentumpsiu1}) the last four terms result from the term $\propto u_\mu u_\nu$ in the propagator, see Eq.\,(\ref{U1propagator}).
(Details of this straight-forward calculation can be regarded in the appendix).
Subsequently, we consider the integration over the second polar angle $\theta$. Merely the terms containing spherical Bessel functions of even degree contribute, obviously the third and the seventh term disappear for reason of symmetry.

The integrals
can be evaluated by applying Sonine's first integral formula
\begin{equation}
\int\limits_0^{\pi/2}\!d\psi\,J_\mu(a \sin\psi)\sin^{\mu+1}\psi\cos^{2\rho+1}\psi=
2^{\rho+1}\Gamma(\rho+1)a^{-\rho-1}J_{\rho+\mu+1}(a)
\end{equation}
where $J_\mu$ denotes an ordinary Bessel function of the first kind,
and the generalization
\begin{eqnarray}
\int\limits_0^{\pi}\!\!d\psi \sin^{\mu}\psi \cos^{\nu}\psi J_{\rho}(a\sin\psi)&=&
         2^{-1-\rho}(1+(-1)^{\nu})a^{\rho}\Gamma\left(\frac{1+\nu}{2}\right)
         \Gamma\left(\frac{1+\mu+\rho}{2}\right)\nonumber\\
         &&\quad\cdot\,\,_1\tilde{F}_2\left(\frac{1+\mu+\rho}{2};
         \frac{2+\mu+\nu+\rho}{2},\frac{1+\rho}{2};\frac{-a^2}{4}\right)\,,
\end{eqnarray}
provided that $\mathfrak{Re}\nu>-1, \mathfrak{Re}\rho>-1, \mathfrak{Re}(\mu+\rho)>-1\, \mbox{and}\,a\geq0$).
We get\\
\newline
$\bra\Theta_{00}(\vec{x})\Theta_{00}(\vec{y})\ket^{\tiny\mbox{vac}}$
\begin{eqnarray}
\label{vacuum momentum}
&=&\frac{1}{(2\pi)^8\beta^8}\left(96\pi^2\cdot
   \left(\int\limits_0^{\infty}\!\!d|\tilde{p}|
   \frac{\pi|\tilde{p}|}{|\tilde{\vec{z}}|^2}
   J_2(|\tilde{p}||\tilde{\vec{z}}|)\right)^2\right. \nonumber\\
&&\hspace{5mm}+48\pi^2\cdot
   \Bigg(\int\limits_0^{\infty}\!\!d|\tilde{p}|
   \frac{\pi}{2|\tilde{\vec{z}}|^3}
   \left(-6J_1(|\tilde{p}||\tilde{\vec{z}}|)
   -2|\tilde{p}||\tilde{\vec{z}}| J_2(|\tilde{p}||\tilde{\vec{z}}|)
   \right.\nonumber\\
&&\left.\hspace{60mm}+3|\tilde{p}||\tilde{\vec{z}}|
         \,_1F_2\left(\frac{1}{2};\frac{3}{2},2;-\frac{(|\tilde{p}||\tilde{\vec{z}}|)
         ^2}{4}\right) \right)\Bigg)^2 \nonumber\\
&&\hspace{5mm}+\frac{32\pi^2}{3}\cdot
   \left(\int\limits_0^{\infty}\!\!d|\tilde{p}|
   \frac{\pi|\tilde{p}|}{|\tilde{\vec{z}}|^2}
   \Big(3J_2(|\tilde{p}||\tilde{\vec{z}}|)
   -|\tilde{p}||\tilde{\vec{z}}| J_3(|\tilde{p}||\tilde{\vec{z}}|)\Big)\right)^2\nonumber\\
&&\hspace{5mm}+\frac{16\pi^2}{3}\cdot
   \left(\int\limits_0^{\infty}\!\!d|\tilde{p}| \frac{\pi|\tilde{p}|^2}{|\tilde{\vec{z}}|}
   J_3(|\tilde{p}||\tilde{\vec{z}}|) \right)^2 \nonumber\\
&&\hspace{5mm}+32\pi^2\cdot
   \left(\int\limits_0^{\infty}\!\!d|\tilde{p}|\,
   \frac{\pi|\tilde{p}|^3}{2}\,
   _1F_2\left(\frac{1}{2};\frac{3}{2},2;-\frac{|\tilde{p}|^2 |\tilde{\vec{z}}|^2}{4}\right)\right)^2\nonumber\\
&&\hspace{5mm}+24\pi^2\cdot
   \left(\int\limits_0^{\infty}\!\!d|\tilde{p}|\,
   \frac{\pi |\tilde{p}|^2}{|\tilde{\vec{z}}|}
   J_1(|\tilde{p}||\tilde{\vec{z}}|)\right)^2\nonumber\\
&&\hspace{5mm}+\left.48\pi^2\cdot
   \left(-\int\limits_0^{\infty}\!\!d|\tilde{p}|\,
   \frac{\pi|\tilde{p}|}{|\tilde{\vec{z}}|^2}
   \left(3J_0(|\tilde{p}||\tilde{\vec{z}}|)+|\tilde{p}||\tilde{\vec{z}}|
   \left(J_1(|\tilde{p}||\tilde{\vec{z}}|)
   -\frac{3}{|\tilde{p}||\tilde{\vec{z}}|}\,
   _1\tilde{F}_2\left(\frac{1}{2};1,\frac{3}{2},
   -\frac{(|\tilde{p}||\tilde{\vec{z}}|)^2}{4}\right)
   \right)\right)\right)^2\right)\nonumber\\
&&\nonumber\\
\end{eqnarray}
The integration over the radial momentum is still remaining. Evaluation of these integrals\footnote{For a detailed calculation consult Sec.\,(\ref{appendix vacuum}).}
yields the final result for the vacuum contribution of the two-point correlation of the energy density in a thermalized $U(1)$ gauge theory:
\begin{eqnarray}
\label{vacuumu1 finalresult}
\bra\Theta_{00}(\vec{x})\Theta_{00}(\vec{y})\ket^{\tiny\mbox{vac}}
&=&\frac{1}{(2\pi)^8\beta^8}\cdot
   \left(96\pi^2\cdot
   \left(\frac{2\pi}{|\tilde{\vec{z}}|^4}\right)^2
   +48\pi^2\cdot
   \left(\frac{-8\pi}{|\tilde{\vec{z}}|^4}\right)^2\right.\nonumber\\
&&\left.+\frac{32\pi^2}{3}\cdot
   \left(\frac{-2\pi}{|\tilde{\vec{z}}|^4}\right)^2
   +\frac{16\pi^2}{3}\cdot
   \left(\frac{8\pi}{|\tilde{\vec{z}}|^4}\right)^2\right)\nonumber\\
&=&\frac{15}{\pi^4\beta^8\,|\tilde{{\vec{z}}}|^8}
  =\frac{0.15399}{|\vec{z}|^8}\,.
\end{eqnarray}
\begin{figure}
\begin{center}
\vspace{5.3cm}
\includegraphics{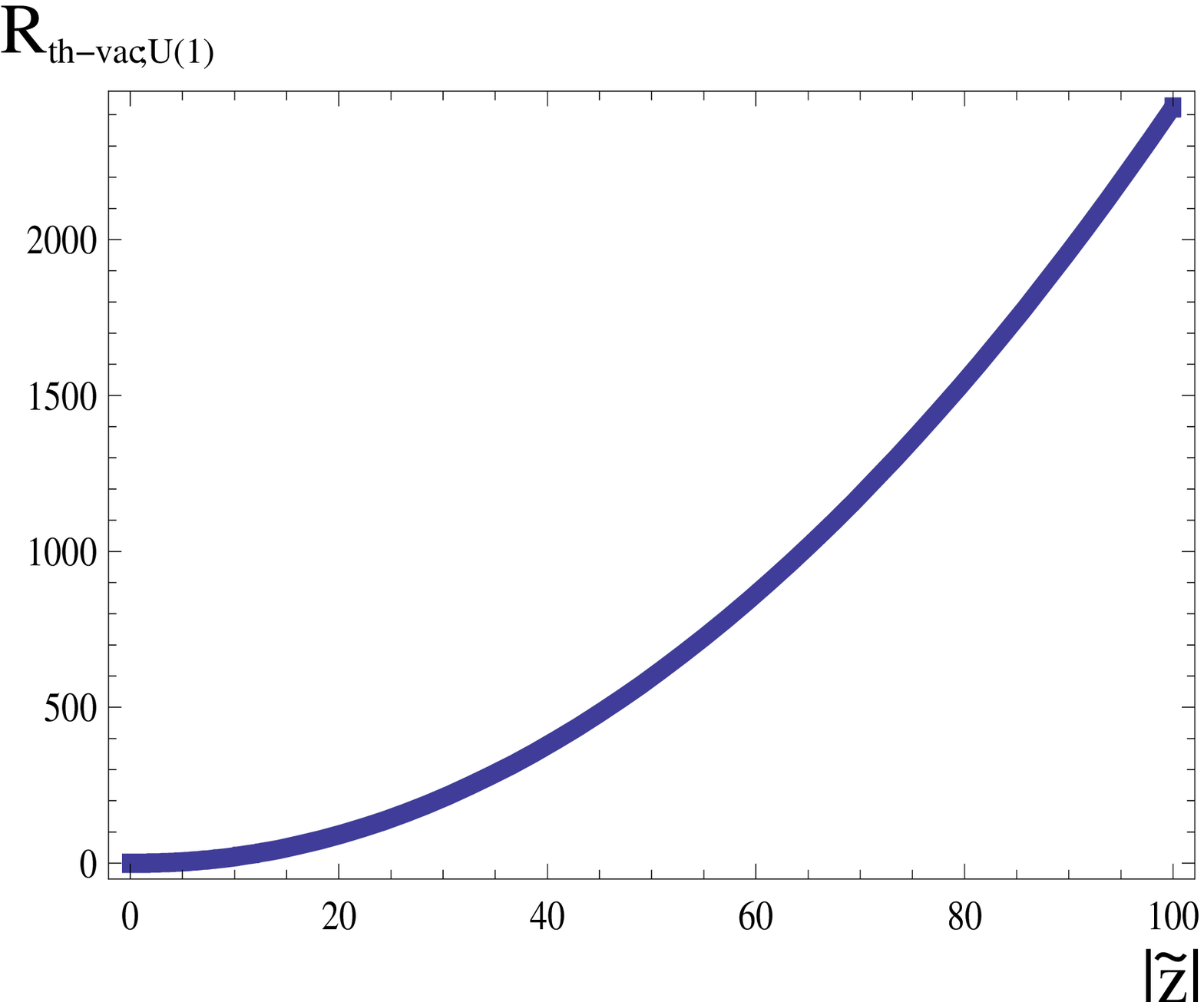}
\end{center}
\caption{The ratio
$R_{\tiny\mbox{th}-\tiny\mbox{vac};\tiny\mbox{U(1)}}\equiv\frac{\bra\Theta_{00}(\vec{x})\Theta_{00}(\vec{y})\ket^{\tiny\mbox{th}}}
{\bra\Theta_{00}(\vec{x})\Theta_{00}(\vec{y})\ket^{\tiny\mbox{vac}}}$
  as a function of $|\tilde{\vec{z}}|$ for a thermalized pure U(1) gauge theory.\label{Fig-2}}
\end{figure}
The contribution resulting from Coulomb part of the propagator (\ref{U1propagator})
in Eq.\,(\ref{vacuum momentum}) vanishes; this observation is in accordance with the fact that no energy transfer between points $\vec{x}$ and $\vec{y}$ is mediated
by the Coulomb part of the photon propagator. In Fig.\,\ref{Fig-2} the ratio
$R_{\tiny\mbox{th}-\tiny\mbox{vac};\tiny\mbox{U(1)}}\equiv\frac{\bra\Theta_{00}(\vec{x})\Theta_{00}(\vec{y})\ket^{\tiny\mbox{th}}}
{\bra\Theta_{00}(\vec{x})\Theta_{00}(\vec{y})\ket^{\tiny\mbox{vac}}}$
is shown as a function of $|\tilde{\vec{z}}|$. For example, at $T=5.5, 8.2,
10.9\,$K a distance $|\vec{z}|$ of 1\,cm corresponds to
$|\tilde{\vec{z}}|\sim 24, 36, 48$, respectively. As a consequence, the
thermal part of $\bra\Theta_{00}(\vec{x})\Theta_{00}(\vec{y})\ket$
dominates the vacuum part by at least a factor of a hundred.

\section{Two-point correlation of energy density in deconfining thermal
SU(2) Yang-Mills thermodynamics}
\label{correlator su2}
In this section we consider now the energy transfer mediated by the massless mode surviving the dynamical gauge symmetry
breaking $SU(2)\to U(1)$ in the deconfining phase of $SU(2)$ Yang-Mills
thermodynamics. As already described in sec.\,(\ref{Deconfining YMTD}), this symmetry breaking is a consequence of the
nontrivial thermal ground state composed of interacting calorons and
anticalorons. Upon a unique spatial coarse-graining this macroscopic ground
state is described by a spatially homogeneous field configuration, in particular a topological trivial pure-gauge solution $a_\mu^{gs}$ with an inert, adjoint scalar field $\phi$ as static background
\cite{Hofmann05}. Due to the adjoint Higgs mechanism Two out of three directions in the $SU(2)$
algebra dynamically acquire a temperature dependent mass. Working in unitary-Coulomb gauge, where the adjoint
Higgs field $\phi^a$ is given as  $\phi^a=\delta^{a3}|\phi|$
($a=1,2,3$), the tree-level massless, coarse-grained, topologically
trivial gauge field is $A_\mu^3$.\\
Since the gauge group is of non-abelian kind, the ground state is not of trivial nature (as in the case of an Abelian gauge group like $U(1)$) and interactions between the massless mode $\gamma$ and the
two massive excitations $V^\pm$ have to be considered. We investigate how the energy transfer between points $\vec{x}$ and $\vec{y}$ as
mediated by this mode, is affected by this effects induced by the non-trivial vacuum . This energy transfer is characterized by the
two-point correlator
$\bra\Theta_{00}(\textbf{x})\Theta_{00}(\textbf{y})\ket^{\tiny\mbox{th}}$
where $\Theta_{00}$ is now calculated as in
Eq.\,(\ref{energymomentumtensor}) replacing\footnote{This can be made manifestly $SU(2)$
  gauge invariant by substituting the 't Hooft tensor \cite{'tHooft74} for the field
  strength into Eq.\,(\ref{energymomentumtensor}).} $A_\mu$ by $A_\mu^3$.
We expect that the energy transfer in regions of interest for the evolution physics of cold, innergalactic clouds is sizeable suppressed in comparison to the counterpart in the standard thermalized $U(1)$-gauge theory of quantum electrodynamics.

\subsection{The thermal contribution}
\begin{figure}
\begin{center}
\vspace{6.5cm}
\includegraphics{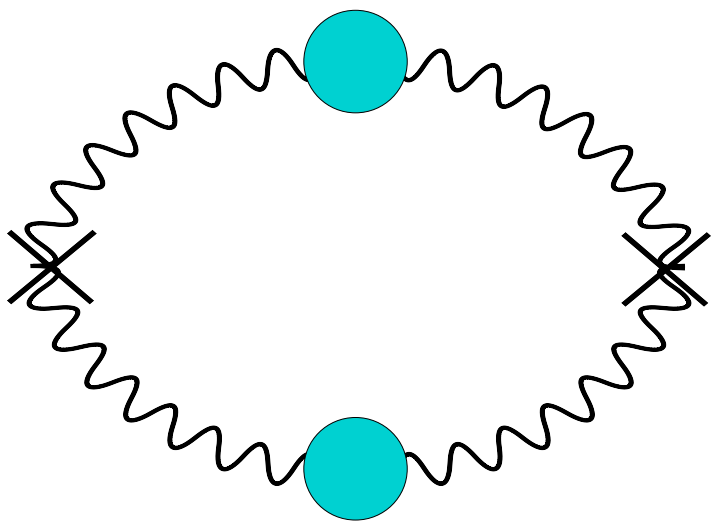}
\end{center}
\caption{Feynman diagram for the correlator
  $\bra\Theta_{00}(x)\Theta_{00}(y)\ket$ in a thermalized, deconfining $SU(2)$ gauge
theory including radiative corrections to lowest order. (Blobs signal a
resummation of the one-loop polarization for the massless mode.)
Crosses denote the insertion of the composite operator $\Theta_{00}$.\label{Fig-4}}
\end{figure}

As already explained in Sec.\,(\ref{1looppolarization}), the dispersion law of the (tree-level) massless gauge field is modified. Since we consider now photon propagation through the non-trivial vacuum of a non-abelian theory, we have to take interaction with this medium into account. This non-abelian effects manifests in the polarization tensor $\Pi_{\mu\nu}$. This object is calculated as a summation of one-loop self-energies for the on-shell massless mode \cite{Schwarz, Hofmann/Schwarz/Giacosa06-1}, see Sec.\,\ref{1looppolarization}. In the effective theory it is totally sufficient to account for radiative
corrections in the correlator
$\bra\Theta_{00}(\vec{x})\Theta_{00}(\vec{y})\ket^{\tiny\mbox{th}}$ in
terms of a resummation of the one-loop polarization tensor $\Pi_{\mu\nu}$ for the
massless mode only, see Fig.\,\ref{Fig-4}, since higher irreducible loop diagrams yield merely insignificant contributions to the total result \cite{Hofmann/Kaviani07}. According to \cite{Hofmann/Schwarz/Giacosa06-1} this polarization tensor leads to a modification of the
dispersion law in comparison to standard model quantum electrodynamics
\begin{equation}
\label{modified dispersion law}
p_0^2=\vec{p}^2 \to p_0^2=\vec{p}^2+G(T,|\vec{p}|,\Lambda)\,,
\end{equation}
where $\Lambda$
denotes the Yang-Mills scale related to the critical temperature $T_c$
for the deconfining-preconfining phase transition as
$T_c=\frac{\lambda_c}{2\pi}\,\Lambda=\frac{13.87}{2\pi}\,\Lambda$. Photon propagation is then governed by Eq.\,(\ref{modified dispersion law}). As described in Sec.\,(\ref{1looppolarization}), the function $G$ acquires relevance  for temperatures not much larger than
$T_c$: There is a regime of
antiscreening ($G<0$) for spatial momenta larger than
$|\vec{p}_{\tiny\mbox{as}}|\sim 0.2\,T$. This effect, however, dies off
exponentially fast with increasing momenta. For momenta smaller than
$|\vec{p}_{\tiny\mbox{high}}|\sim 0.1\,T$ and larger than
$|\vec{p}_{\tiny\mbox{low}}|\sim 0.02\,T$ the function $G$ is so strongly
positive that the propagation of the associated modes
is forbidden (total screening).\\
\begin{figure}
\begin{center}
\vspace{6.5cm}
\includegraphics{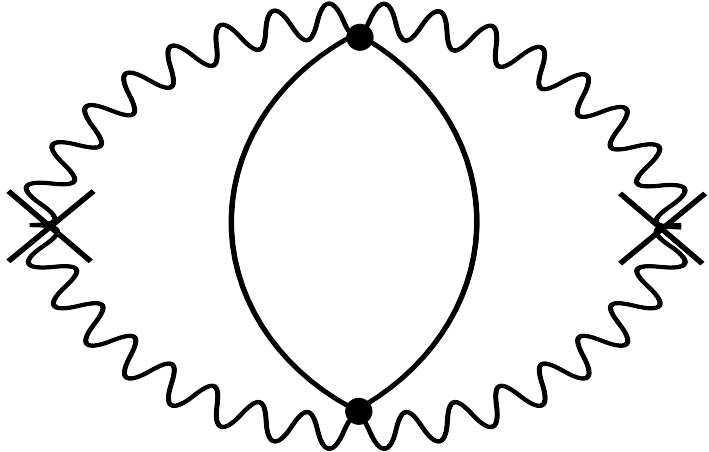}
\end{center}
\caption{Vanishing irreducible three-loop diagram for the correlator
  $\bra\Theta_{00}(x)\Theta_{00}(y)\ket$ in a thermalized, deconfining $SU(2)$ gauge
theory. A wavy (solid) line is associated with
the propagator of the massless (massive) mode.
Crosses denote the insertion of the composite operator $\Theta_{00}$.\label{Fig-3a}}
\end{figure}
Initially we derive the photon propagator in deconfining $SU(2)$ Yang-Mills thermodynamics. Considering the modification in the dispersion law, Eq.\,(\ref{modified dispersion law}), the propagator follows as
\begin{equation}
\label{SU2propagator}
D_{\mu\nu}(p,T)
=-P^T_{\mu\nu}(p)\left(\frac{i}{p^2-G(T,|\vec{p}|,\Lambda)+i\epsilon}+i\frac{u_\mu
u_\nu}{\vec{p}^2}-2\pi\delta(p^2-G(T,|\vec{p}|,\Lambda))n_B(\beta|p_0|)\right)\,.
\end{equation}
After inserting the thermal part of Eq.\,(\ref{SU2propagator}) into Eq.\,(\ref{Wick decomposition})and implanting the same tensor contractions Eqs.\,(\ref{tensorcontraction1}, \ref{tensorcontraction2}, \ref{tensorcontraction3}) like in the $U(1)$ case, we decompose the $\delta$-distribution similar to Eq.\,(\ref{Delta decomposition}).
Performing the integration over the zero-coordinate,
the thermal part $\bra\Theta_{00}(\textbf{x})\Theta_{00}(\textbf{y})\ket^{\tiny\mbox{th}}$
calculates in analogy to Eq.\,(\ref{thermalintegralu1}) as\\

\begin{eqnarray}
\label{thermalintegralsu2}
\bra\Theta_{00}(\vec{x})\Theta_{00}(\vec{y})\ket^{\tiny\mbox{th}}&=&
\frac{1}{2}\left(\int\!\!\frac{d^3p}{(2\pi)^3}
        \sqrt{\vec{p}^2+G}
        \,n_B(\beta\sqrt{\vec{p}^2+G})
        \,\e^{i\vec{p}(\vec{x}-\vec{y})}\right)^2\nonumber\\
& &+\frac{1}{2}\left(\int\!\!\frac{d^3p}{(2\pi)^3}
        \frac{\vec{p}^2}{\sqrt{\vec{p}^2+G}}
        \,n_B(\beta\sqrt{\vec{p}^2+G})
        \,\e^{i\vec{p}(\vec{x}-\vec{y})}\right)^2\nonumber\\
& &+\frac{1}{2}\int\!\!\frac{d^3p}{(2\pi)^3}\int\!\!\frac{d^3k}{(2\pi)^3}
        \left(\frac{\vec{p}\vec{k}}{|\vec{p}||\vec{k}|}\right)^2
        \sqrt{\vec{p}^2+G}\sqrt{\vec{k}^2+G}\times\nonumber\\
       & & \ \ \ \ \ \ n_B(\beta\sqrt{\vec{p}^2+G})n_B(\beta\sqrt{\vec{k}^2+G})
        \,\e^{i\vec{p}(\vec{x}-\vec{y})}\e^{i\vec{k}(\vec{x}-\vec{y})}
        \nonumber\\
& &+\frac{1}{2}\int\!\!\frac{d^3p}{(2\pi)^3}\int\!\!\frac{d^3k}{(2\pi)^3}
        \left(\frac{\vec{p}\vec{k}}{|\vec{p}||\vec{k}|}\right)^2
        \frac{\vec{p}^2}{\sqrt{\vec{p}^2+G}}
        \frac{\vec{k}^2}{\sqrt{\vec{k}^2+G}}\times\nonumber\\
        & &\ \ \ \ \ \ n_B(\beta\sqrt{\vec{p}^2+G})n_B(\beta\sqrt{\vec{k}^2+G})
        \,\e^{i\vec{p}(\vec{x}-\vec{y})}\e^{i\vec{k}(\vec{x}-\vec{y})}
        \nonumber\\
\end{eqnarray}
The integration over the $p_0$- and $k_0$-coordinates is executed
without constraints and
just fixes the dispersion law (\ref{modified dispersion law}). We proceed now in exactly the same manner as in the $U(1)$ case, since we
introduce spherical coordinates and evaluate
the integrals over the angles analytically. Upon performing the azimuthal integrations for
each summand in Eq.\,(\ref{thermalintegralsu2}) the respective expression
reduces to a square of two integrals over the polar angle and the modulus of spatial momentum, respectively. The integral concerning the polar angle is computed by expanding the exponential in spherical harmonics, writing polynomial factors in terms of linear combinations of Legendre polynomials and exploiting their orthogonality relation, Eq.\,(\ref{Legendre orthogonality}). The calculation of integral over the modulus of the spatial momentum is a more complex task, since we have to consider the screening effects induced by the one-loop polarization tensor. We continue the evaluation in analogy
to the derivation of the black-body spectrum in \cite{Hofmann/Schwarz/Giacosa06-2}.
Employing the modified dispersion law, the integral over
momentum-modulus is replaced in favor of an integral over frequency
$\omega$. Those values of $\omega$, which yield an
imaginary modulus of the spatial momentum due to strong screening, are excluded from the domain
of integration. This regime of strong screening is in the range
$\omega_1<\omega<\omega_2$, where $\omega_1, \omega_2$ denote the
solutions of the equation
\begin{equation}
\label{roots}
\omega^2-G(\omega,T,\Lambda)=0\,.
\end{equation}
Finally, we
introduce a dimensionless frequency $\tilde{\omega}\equiv\beta\omega$
and a dimensionless screening function $\tilde{G}\equiv\beta^2 G$ and arrive at
\begin{eqnarray}
\label{thermalsu2 finalresult}
&&\bra\Theta_{00}(\vec{x})\Theta_{00}(\vec{y})\ket^{\tiny\mbox{th}}=\nonumber\\
&&\frac{1}{(2\pi)^6\beta^8}
    \left(\frac{32\pi^2}{3}
    \left(\int\limits_{0\leq\tilde{\omega}\leq\tilde{\omega}_1,
    \atop\tilde{\omega}_2\leq\tilde{\omega}\leq\infty}
    d\tilde{\omega}\left(\tilde{\omega} -\frac{1}{2}\frac{d\tilde{G}}{d\tilde{\omega}}\right)\,\tilde{\omega}
    \sqrt{\tilde{\omega}^2-\tilde{G}}\,
    \frac{j_0(\sqrt{\tilde{\omega}^2-\tilde{G}}\,|\tilde{\vec{z}}|)}
    {e^{\tilde{\omega}}-1}\right)^2\right.\nonumber\\
&+&\left.\frac{32\pi^2}{3}\left(\int\limits_{0\leq\tilde{\omega}\leq\tilde{\omega}_1,
    \atop\tilde{\omega}_2\leq\tilde{\omega}\leq\infty}
    d\tilde{\omega}\left(\tilde{\omega}
      -\frac{1}{2}\frac{d\tilde{G}}{d\tilde{\omega}}\right)\,
    \frac{(\sqrt{\tilde{\omega}^2-\tilde{G}})^3}{\tilde{\omega}}\,
    \frac{j_0(\sqrt{\tilde{\omega}^2-\tilde{G}}\,|\tilde{\vec{z}}|)}
    {e^{\tilde{\omega}}-1}\right)^2\right.\nonumber\\
&+&\left.\frac{16\pi^2}{3}\left(\int\limits_{0\leq\tilde{\omega}\leq\tilde{\omega}_1,
    \atop\tilde{\omega}_2\leq\tilde{\omega}\leq\infty}
    d\tilde{\omega}\left(\tilde{\omega}
      -\frac{1}{2}\frac{d\tilde{G}}{d\tilde{\omega}}\right)\,\,
    \tilde{\omega}\sqrt{\tilde{\omega}^2-\tilde{G}}\,
    \frac{j_2(\sqrt{\tilde{\omega}^2-\tilde{G}}\,|\tilde{\vec{z}}|)}
    {e^{\tilde{\omega}}-1}\right)^2\right.\nonumber\\
&+&\left.\frac{16\pi^2}{3}\left(\int\limits_{0\leq\tilde{\omega}\leq\tilde{\omega}_1,
    \atop\tilde{\omega}_2\leq\tilde{\omega}\leq\infty}
    d\tilde{\omega}\left(\tilde{\omega} -\frac{1}{2}\frac{d\tilde{G}}{d\tilde{\omega}}\right)\,
    \frac{(\sqrt{\tilde{\omega}^2-\tilde{G}})^3}{\tilde{\omega}}\,
    \frac{j_2(\sqrt{\tilde{\omega}^2-\tilde{G}}\,|\tilde{\vec{z}}|)}
    {e^{\tilde{\omega}}-1}\right)^2\right.\,.\nonumber\\
\end{eqnarray}
The frequency integral is calculated numerically. The results for different temperatures and conclusive implications are analyzed in Sec.\,(\ref{Discussion correlator}).

\subsection{Estimate for the vacuum contribution}

A this point we would like to obtain an order-of-magnitude estimate
for the vacuum part of the two-point correlation of the energy density concerning the
massless mode in deconfining SU(2) Yang-Mills thermodynamics. To do this we
ignore the modification of the dispersion law in
Eq.\,(\ref{modified dispersion law}). Anyways, the function $G$ has so far only been
computed for external momentum $p$ with $p^2=0$. The difference as
compared to the $U(1)$ case is then a restriction of the (euclidean)
four-momentum $p$ as $p^2\le\phi^2$ due to the existence
of a scale of maximal resolution $|\phi|$ in the effective
theory. This compositeness constraint , Eq.\,(\ref{compositeness1}) originates from spatial coarse-graining.
Again, we will see that
$\bra\Theta_{00}(\vec{x})\Theta_{00}(\vec{y})\ket^{\tiny\mbox{vac}}$ is
a negligible correction to
$\bra\Theta_{00}(\vec{x})\Theta_{00}(\vec{y})\ket^{\tiny\mbox{th}}$
for physically interesting distances.

In case of
$\bra\Theta_{00}(\vec{x})\Theta_{00}(\vec{y})\ket^{\tiny\mbox{vac}}$ we restrict to $x_0=y_0$ to be able to compare with
$\bra\Theta_{00}(\textbf{x})\Theta_{00}(\textbf{y})\ket^{\tiny\mbox{th}}$.
Introducing the dimensionless
modulus of $\phi$ as $\tilde{\phi}\equiv\beta\phi$, proceeding in a
way analogous to the derivation of Eq.\,(\ref{vacuummomentumpsiu1}), and
performing the $\psi$- and $|\tilde{p}|$-integrations\footnote{Concerning a detailed calculation of the integrals we refer to the appendix Sec.\,\ref{appendix vacuum}},
we arrive at
\newpage
\begin{eqnarray}
\label{vacuum estimate su2}
&&\bra\Theta_{00}(\vec{x})\Theta_{00}(\vec{y})\ket^{\tiny\mbox{vac}}
   \sim\nonumber\\
&&\frac{1}{(2\pi)^8\beta^8}
   \Bigg(96\pi^2
   \cdot\frac{\pi^2}{|\vec{\tilde{z}}|^8}
   \left(2J_0(|\vec{\tilde{z}}||\tilde{\phi}|)
   +|\vec{\tilde{z}}||\tilde{\phi}|J_1(|\vec{\tilde{z}}||\tilde{\phi}|)
   -2\right)^2 \nonumber\\
&+&\left.48\pi^2\cdot\frac{\pi^2}{4|\vec{\tilde{z}}|^8}
   \left(2(8J_0(|\vec{\tilde{z}}||\tilde{\phi}|)+
   |\vec{\tilde{z}}||\tilde{\phi}|J_1(|\vec{\tilde{z}}||\tilde{\phi}|)-8)
   +3|\vec{\tilde{z}}|^2|\tilde{\phi}|^2\,\,_1F_2\left(\frac{1}{2}
   ;\frac{3}{2},2;-\frac{|\vec{\tilde{z}}|^2|\tilde{\phi}|^2}{4}\right)
   \right)^2\right.\nonumber\\
&+&\left.\frac{32\pi^2}{3}\cdot\frac{\pi^2}{|\vec{\tilde{z}}|^8}
   \left((|\vec{\tilde{z}}|^2|\tilde{\phi}|^2-2)
   J_0(|\vec{\tilde{z}}||\tilde{\phi}|)-3|\vec{\tilde{z}}||\tilde{\phi}|
   J_1(|\vec{\tilde{z}}||\tilde{\phi}|)+2\right)^2\right.  \nonumber\\
&+&\left.\frac{16\pi^2}{3}\cdot\frac{\pi^2}{|\vec{\tilde{z}}|^8}
   \left((|\vec{\tilde{z}}|^2|\tilde{\phi}|^2-8)
   J_0(|\vec{\tilde{z}}||\tilde{\phi}|)
   -6|\vec{\tilde{z}}||\tilde{\phi}|J_1(|\vec{\tilde{z}}||\tilde{\phi}|)
   +8\right)^2\right.  \nonumber\\
&+&\left.32\pi^2\cdot\frac{\pi^2|\tilde{\phi}|^8}{64}
   \left(\,\,_1F_2\left(\frac{1}{2};\frac{3}{2},3;
   -\frac{|\vec{\tilde{z}}|^2|\tilde{\phi}|^2}{4}\right)\right)^2\right.\nonumber\\
&+&\left.24\pi^2\cdot\frac{\pi^2|\tilde{\phi}|^4}
   {|\vec{\tilde{z}}|^4}\left(J_2(|\vec{\tilde{z}}||\tilde{\phi}|)\right)^2
   \right.  \nonumber\\
&+&48\pi^2\cdot\frac{\pi^2|\tilde{\phi}|^2}
   {4|\vec{\tilde{z}}|^6}\left(6J_1(|\vec{\tilde{z}}||\tilde{\phi}|)+
   2|\vec{\tilde{z}}||\tilde{\phi}|
   J_2(|\vec{\tilde{z}}||\tilde{\phi}|)-3|\vec{\tilde{z}}||\tilde{\phi}|
   \,\,_1F_2\left(\frac{1}{2};\frac{3}{2},2;
   -\frac{|\vec{\tilde{z}}|^2|\tilde{\phi}|^2}{4}\right)\right)^2\Bigg)
   \,,\nonumber\\
\end{eqnarray}
where $\,_1F_2$ is a hypergeometric function and $J_0, J_1, J_2$ are Bessel
functions of the first kind (conventions as in \cite{Gradshteyn}).

\section{Numerical results and discussion}
\label{Discussion correlator}
According to sec.\,(\ref{correlator u1}) it is evident that if photon propagation is governed by a $U(1)$ gauge theory, neglecting the vacuum contribution in comparison to the thermal part of the two-point correlation of energy density is well justified. In the following we turn to the case of energy transfer mediated by the massless gauge boson in deconfining Yang-Mills thermodynamics.

The frequency integral in Eq.\,(\ref{thermalsu2 finalresult}) is evaluated numerically.
We compare the estimate for the vacuum contribution in Eq.\,(\ref{vacuum estimate su2}) with the thermal part of the correlator
$\bra\Theta_{00}(\vec{x})\Theta_{00}(\vec{y})\ket$. To make contact
with $SU(2)_{\tiny\mbox{CMB}}$, whose Yang-Mills scale is
$\Lambda=1.065\times10^{-4}$\,eV \cite{Hofmann/Schwarz/Giacosa06-1, Hofmann/Schwarz/Giacosa06-2},
we relate at a given temperature the
dimensionless distance $|\tilde{\vec{z}}|$ to the physical distance in
centimeters\footnote{We can convert the values from Planck units to SI units by applying the relation
\begin{equation}
\hbar c=1\equiv197\,\mbox{Mev}\,\mbox{fm}
\end{equation}}.

\begin{figure}
\begin{center}
\vspace{5.3cm}
\includegraphics{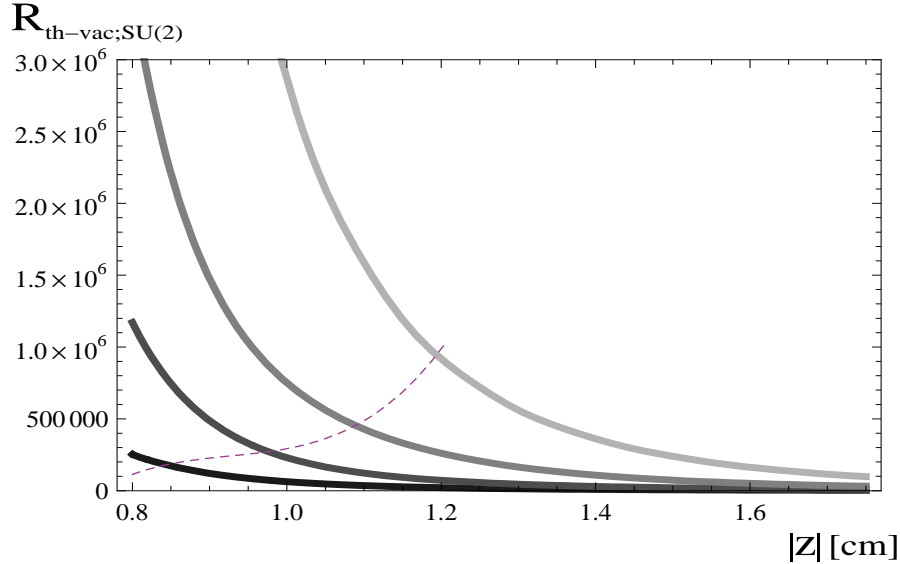}
\end{center}
\caption{\protect{\label{Fig-5}The function
  $R_{\tiny\mbox{th}-\tiny\mbox{vac};\tiny\mbox{SU(2)}}(|\vec{z}|)$,
  defined as in Eq.\,(\ref{ratio su2thvac}), when specializing to
the case of SU(2)$_{\tiny\mbox{CMB}}$ ($T_c=2.73\,$K), for
various temperatures: black curve ($T=1.5\,T_c$), dark grey curve
($T=2.0\,T_c$), grey curve ($T=2.5\,T_c$), light grey curve ($T=3.0\,T_c$). The dashed line separates distances
smaller than $|\phi|^{-1}$ from those that are larger than $|\phi|^{-1}$.}}
\end{figure}

We define
\begin{equation}
\label{ratio su2thvac}
R_{\tiny\mbox{th}-\tiny\mbox{vac};\tiny\mbox{SU(2)}}(|\vec{z}|)\equiv
\frac{\bra\Theta_{00}(\vec{x})\Theta_{00}(\vec{y})\ket^{\tiny\mbox{th}}}
{\left|\bra\Theta_{00}(\vec{x})\Theta_{00}(\vec{y})\ket^{\tiny\mbox{vac}}\right|}\,,
\end{equation}
where we use the estimate in Eq.\,(\ref{ratio su2thvac}) for
$\left|\bra\Theta_{00}(\vec{x})\Theta_{00}(\vec{y})\ket^{\tiny\mbox{vac}}\right|$.
In Fig.\,\ref{Fig-5} the quantity
$R_{\tiny\mbox{th}-\tiny\mbox{vac};\tiny\mbox{SU(2)}}(|\vec{z}|)$ is
depicted for various temperatures specializing to the case
of $SU(2)_{\tiny\mbox{CMB}}$. Notice the strong dominance of the
thermal part. Notice also that although this result resembles
qualitatively the result of Fig.\,\ref{Fig-2} for fixed distance and
varying temperature this is not true for fixed temperature and varying
distance. Namely, the existence of a nontrivial thermal ground
state in deconfining $SU(2)$ Yang-Mills thermodynamics constrains quantum
fluctuations of the massless mode to be softer than the compositeness scale
$|\phi|$. Owing to its trivial ground state,
no such constraint exists in a thermalized $U(1)$ gauge theory. As a
consequence and in accord with Fig.\,\ref{Fig-2}, quantum fluctuations dominate thermal
fluctuations at small distances in such a theory.

\begin{figure}
\begin{center}
\vspace{5.3cm}
\includegraphics{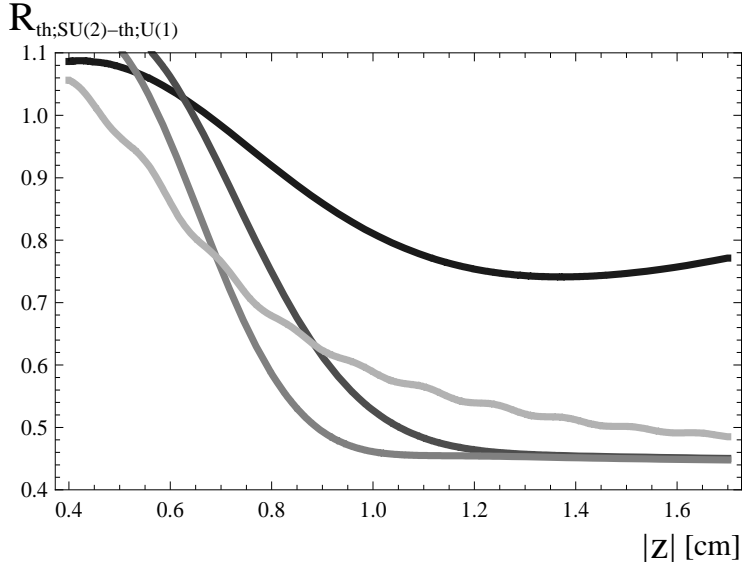}
\end{center}
\caption{\protect{\label{Fig-6}The function
  $R_{\tiny\mbox{th,SU(2)}-\tiny\mbox{th,U(1)}}(|\vec{z}|)$,
as defined in Eq.\,(\ref{ratio su2u1th}), when specializing to
the case of SU(2)$_{\tiny\mbox{CMB}}$ ($T_c=2.73\,$K), for
various temperatures: black curve ($T=1.5\,T_c$), dark grey
($T=2.0\,T_c$), grey ($T=2.5\,T_c$), light grey ($T=3.0\,T_c$). Notice the regime of antiscreening for small
$|z|$. For $T=3.0\,T_c$, where the approximation $p^2=0$ deviates sizably from the full result for the screening function $G$,
we have used $G$ as obtained 1-loop selfconsistently \cite{LH2008}.}}
\end{figure}

\begin{figure}
\begin{center}
\vspace{5.3cm}
\includegraphics{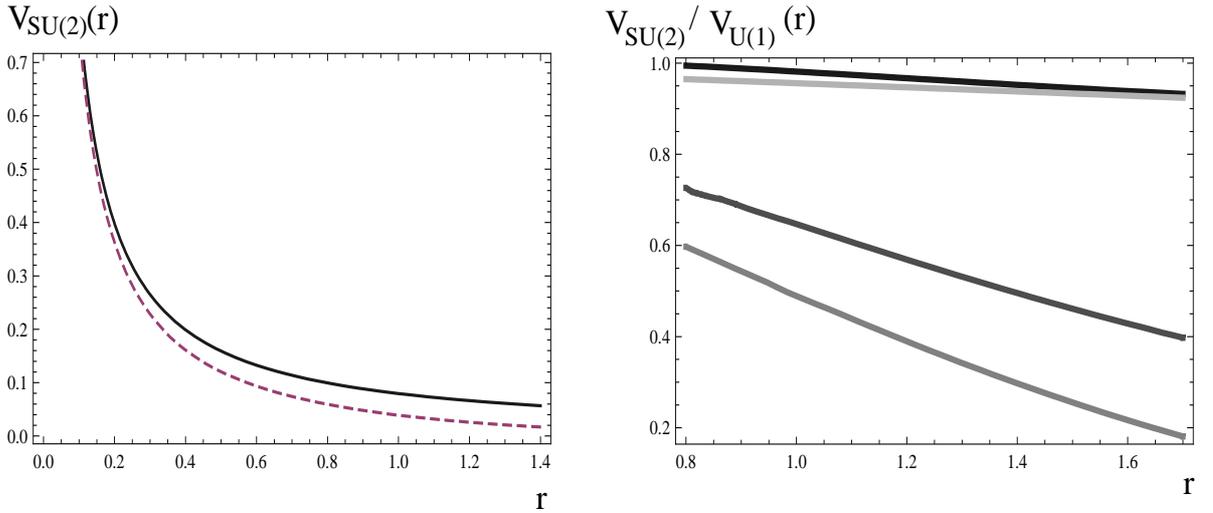}
\end{center}
\caption{\protect{\label{Fig-7} Left panel: Plot of the potentials
$V_{\tiny\mbox{U(1)}}(r)$ (dashed line) and
$V_{\tiny\mbox{SU(2)}_{\tiny\mbox{CMB}}}(r)$ (solid line)
as a function of $r$ at $T=2.5\,T_{\tiny\mbox{CMB}}\sim 6.8$\,K. Right panel: Plot of the
ratio
$V_{\tiny\mbox{SU(2)}_{\tiny\mbox{CMB}}}(r)/V_{\tiny\mbox{U(1)}}(r)$ as
a function of $r$. The temperature is set to $T=1.5\,T_c$ (black),
$T=2.0\,T_c$ (dark grey), $T=2.5\,T_c$ (grey), and $T=3.0\,T_c$ (light
grey). For $T=3.0\,T_c$, where the approximation $p^2=0$ deviates sizably from the full result for the screening function $G$,
we have used $G$ as obtained 1-loop selfconsistently \cite{LH2008}. Notice how close to unity the curve at $T=3.0\,T_c$
is as compared to the curve at $T=2.5\,T_c$ for the right-hand side panel.}}
\end{figure}

By virtue of the results shown in
Fig.\,\ref{Fig-2} and Fig.\,\ref{Fig-5}
we neglect the vacuum contribution to
$\bra\Theta_{00}(\vec{x})\Theta_{00}(\vec{y})\ket$ in the following.\\
Let us now turn to the interesting question of how much suppression
there is in the two-point correlation of the photon energy density
in the case of $SU(2)_{\tiny\mbox{CMB}}$ as compared to the conventional
$U(1)$ case. In Fig.\,\ref{Fig-6} the ratio
$R_{\tiny\mbox{th,SU(2)}-\tiny\mbox{th,U(1)}}(|\vec{z}|)$,
defined as
\begin{equation}
\label{ratio su2u1th}
R_{\tiny\mbox{th,SU(2)}-\tiny\mbox{th,U(1)}}(|\vec{z}|)\equiv
\frac{\bra\Theta_{00}(\vec{x})\Theta_{00}(\vec{y})\ket^{\tiny\mbox{th,SU(2)}}}
{\bra\Theta_{00}(\vec{x})\Theta_{00}(\vec{y})\ket^{\tiny\mbox{th,U(1)}}}\,,
\end{equation}
is depicted for various temperatures as a function of distance in
centimeters.
Notice the significant suppression of the correlation between photon energy
densities in the case of an underlying $SU(2)$ gauge symmetry as compared
to the case of the conventional $U(1)$ up to a factor of two. Since the two-point correlator $\bra\Theta_{00}(\vec{x})\Theta_{00}(\vec{y})\ket$
is a measure for the energy transfer between the spatial points
$\vec{x}$ and $\vec{y}$ in thermal equilibrium and hence a measure
for the interaction of microscopic objects emitting and absorbing
radiation, we conclude that this interaction is, as compared to the
conventional theory, suppressed on distances
$\sim 1\,$cm if photon propagation is subject to an $SU(2)$ gauge
principle.
To make
the situation even more explicit we have computed the
Coulomb potential $V(r)$ ($r\equiv|\vec{x}|$) of a heavy point charge in case of photons being
described by a $U(1)$ and an $SU(2)$ gauge theory. This potential is given in a $U(1)$
theory as
\begin{eqnarray}
\label{CoulombU1}
V_{\tiny\mbox{U(1)}}(r)&=&\frac{1}{(2\pi)^3}\int
d^3p\,\frac{\e^{-i\vec{p}\cdot\vec{x}}}{\vec{p}^2}=\frac{1}{2\pi^2}\int_0^\infty
dp\,\frac{\sin pr}{pr}=\frac{1}{2\pi^2r}\int_0^\infty
d\xi\,\frac{\sin\xi}{\xi}\nonumber\\
&=&\frac{1}{4\pi r}\,.
\end{eqnarray}
Going from $U(1)$ to $SU(2)$, we take into
account the resummed one-loop polarization \cite{Hofmann/Schwarz/Giacosa06-1}
by letting $\vec{p}^2\to\vec{p}^2+G$ in the denominator of the integrand
in Eq.\,(\ref{CoulombU1}). Here $G$ is the same function as discussed in
Sec.\,(\ref{1looppolarization}). The physical situation is a heavy point charge
immersed into the $SU(2)$ plasma with the photons associated with it being
(anti)screened by nonabelian, thermal fluctuations. Although the
function $G$ is known for on-shell photons $\omega^2=\vec{p}^2$
only this recipe should work well for sufficiently large distances since the
off-shellness of the photon, which does not mediate any energy transfer
from the source, then is sufficiently small. Thus for
$SU(2)$ we approximately\footnote{A point charge immersed into the plasma
  locally distorts the latter's ground state, and, strictly speaking, the
theory to describe this distortion yet needs to be worked out.
But measuring the (anti)screening of the potential sufficiently far away from the location of
the charge should still be describable in terms
of unadulterated SU(2) Yang-Mills thermodynamics.}
have
\begin{eqnarray}
\label{SU2}
V_{\tiny\mbox{SU(2)}}(r)&=&\frac{1}{(2\pi^3)}\int
d^3p\,\frac{\e^{-i\vec{p}\cdot\vec{x}}}{\vec{p}^2+G(T,|\vec{p}|,\Lambda)}\nonumber\\
&=&\frac{1}{2\pi^2r}\int_0^\infty
dp\,\frac{p}{p^2+G(T,p,\Lambda)}\sin pr\,,
\end{eqnarray}
where the last integral is performed numerically. In Fig.\,\ref{Fig-7} both
potentials, $V_{\tiny\mbox{U(1)}}(r)$ and $V_{\tiny\mbox{SU(2)}_{\tiny\mbox{CMB}}}(r)$, as
well as their ratio are plotted as functions of $r$.
Notice that the dimensionless quantity
$V_{\tiny\mbox{SU(2)}_{\tiny\mbox{CMB}}}(r)/V_{\tiny\mbox{U(1)}}(r)$
shows similar suppression as the dimensionless quantity
$R_{\tiny\mbox{th,SU(2)}-\tiny\mbox{th,U(1)}}(|\vec{z}|)$ depicted in
Fig.\,\ref{Fig-6}. This seems to confirm the validity of
our above approximation.

The following tabular lists the maximal resolution $|\phi|^{-1}$ as dimensionless quantity as well as the conversion into the physical interesting unit [cm] for various temperatures.
\begin{center}
\begin{tabular}{l||l|l}
Dimensionless temperature $\lambda$ & $|\tilde{\phi}|^{-1}$ & $|\phi|^{-1}$\,\,[cm]\\
\hline\hline\\
1.5$\times\lambda_c$=20,81 & 15.10 & 0.84 \\
\hline
2.0$\times\lambda_c$=27.74 & 23.25 & 0.98 \\
\hline
2.5$\times\lambda_c$=34.68 & 32.50 & 1.09 \\
\hline
3.0$\times\lambda_c$=41.61 & 42.72 & 1.19 \\
\hline
3.5$\times\lambda_c$=48.55 & 53.83 & 1.28 \\
\end{tabular}
\end{center}
\section{Stability of clouds of atomic hydrogen in the Milky Way}
\label{MilkyWay}

A large fraction of the interstellar medium in our galaxy consists of warm ($\sim 10^3-10^4 K$) and cold ($\sim 50-100 K$) atomic hydrogen gas. The standard theory of the interstellar medium implies that the colder the region, the higher the affinity of atomic hydrogen to form molecular hydrogen This leads to the assumption that extrem cold regions are dominated by dense clouds of molecular hydrogen.

Initially, we should give a short review concerning the evolution of such cloud structures \cite{Dickey01}. In the galactic center hot ionized hydrogen gas is emitted. This gas escapes the main galactic disk and forms a giant 'mushroom' structure above it. In the course of time the ion gas expands, governed by gravitational effects, cools down, the ions recombine and trickle down onto the galactic disk in to the voids between the spiral arms of the galaxy, where clouds of atomic hydrogens are formed. The clouds continue to cool down to $10-25 K$. At the same time neutral atoms are assumed to combine to molecular hydrogen HI via dipole interaction (i.e. SM QED). Numerical simulations of the cloud evolution predict a time scale of less than $10^7 y$ \cite{Kavars, Goldsmith} for the process of generating a substantial fraction of molecular hydrogen.

But in 2001, the discovery of an arc-like structure located in-between the spiral arms of the Milky Way  and consisting of atomic hydrogen HI, designated as GSH139-03-69, was reported in \cite{Brunt/Knee01, Dickey01}. This object GSH139-03-69 is impressive, with a size $\sim 2 kpc$, a mass of about $1.9\cdot10^7$ solar masses,and an atomic number density of $\sim 1.5 cm^{-3}$. The cold cloud exhibits a mean brightness temperature $T_B\sim 20 K$ with cold regions of $T_B\sim 5\dots10 K$. Although GSH139-03-69 is as cold as a molecular cloud, it consists mostly of atomic hydrogen. Regarding the observed data, one expects that the age of GSH139-03-69 is at most $10^6 y$. However, the astonishing fact is that this object possesses an estimated age of about $5\cdot 10^7 y$, which is an obvious contradiction to the SM of the interstellar medium and hence not conform with the SM of particle physics.

The results obtained in Sec.\,(\ref{correlator su2}) concerning the electromagnetic interaction of microscopic objects yield immediate implications for the
gradual metamorphosis of cold ($T=5\dots 10\,$K) astrophysical objects such as the
hydrogen cloud GSH139-03-69. Although a quantitative estimate of the
increased stability of atomic hydrogen clouds due to the $SU(2)$ effects in thermalized photon propagation is beyond the scope of the present work,
Fig.\,\ref{Fig-6} clearly expresses that the mean energy
transfer in the photon gas of $T=5\cdots 10\,$K is suppressed by up to a
factor of two at interatomic distances in the hydrogen cloud GSH139-03-69.
 It would be interesting to see how the
simulation of the cloud evolution, taking into account the effects as
expressed in Fig.\,\ref{Fig-6}, would increase the estimate for its age
as compared to the standard picture.

As already pointed out in \cite{Hofmann/Schwarz/Giacosa06-1}, the propagation of the 21-cm line, which thermalizes structures such as GSH139-03-69, is unscreened even when subjecting photons to $SU(2)_{\tiny\mbox{CMB}}$. A recent full
calculation of $G$ shows, however, that this result is an artefact of
the approximation $p^2=0$: Thermalization of the hydrogen cloud
thus takes place solely via the coupling of the photon to the nontrivial
thermal ground state \cite{GHKL2008}.

%% file: stringtension.tex
\chapter{The spatial string tension in deconfining SU(2) Yang Mills Thermodynamics}

\section{The Wegner-Wilson loop and magnetic screening effects}

In this section we investigate a gauge-invariant construction called Wegner-Wilson loop. As already mentioned in the Introduction the gauge theories in their strongly-coupled regime
are not considerable in the framework of perturbation theory, standard Feynman diagrammatic techniques are not applicable. Thus in the second half of the 70's the investigation and description of gauge theories with a large coupling (for instance QCD in the infrared) demanded the development of nonperturbative methods involving for example the Wegner-Wilson loop.
This quantity is defined as the expectation value of a path-ordered exponential of a gauge field $A_\mu$
\begin{equation}
\textrm{tr}\,\mathcal{P}\exp\left(ig\oint dz^\mu A_\mu\right)\,,
\end{equation}
where the contour integral is evaluated along a rectangular path around the origin. Gauge-invariance is guaranteed by the cyclicity of the trace.

This object was originally motivated in lattice gauge theories as an attempt at a nonperturbative formulation of QCD in its strongly-coupled region. Nevertheless, the asymptotic behavior of the Wegner-Wilson loop is an appropriate criterion to distinguish between the phases of deconfinement and confinement. We focus on another aspect of the Wegner-Wilson loop which monitores screening effects in a plasma originating from the existence of magnetic quasi-particles (monopoles) at finite temperature.

Consider a hot plasma in the framework of an abelian gauge theory. It is a well-known fact, that the existence of long-range static magnetic fields is possible, and no magnetic screening effects occur \cite{Gross/Pisarski/Yaffe81}. In contrast to these long-range electric fields are absent due to Debye screening. However, in the non-abelian scenario of an $SU(N)$ gauge theory, apart from Debye screening in the electric sector, additional screening effects in the magnetic sector emerge. This observation suggests the existence of magnetic monopoles in the plasma. This is mathematically corroborated in the spatial Wegner-Wilson loop,
which measures the magnetic flux induced by magnetic quasi-particles through the enclosed area of the loop, due to Stokes' theorem.
A result from lattice simulations is that the thermal average of the Wegner-Wilson loop obeys an area-law, which corresponds to a gas of free screened magnetic monopoles. For more details concerning the screening effects resulting from magnetic activity due to monopoles at high temperature we refer to \cite{KorthalsAltes, Giovannangeli}.

The spatial string tension is defined as the negative logarithm of the thermal average of the Wegner-Wilson loop divided by the enclosed area $A$ of the loop:
\begin{equation}
\sigma\equiv-\frac{\ln\bra \textrm{tr}\mathcal{P}\exp ie \oint_C dz_\mu A_\mu\ket}{A}\,,
\end{equation}
where $A_\mu$ denotes the gauge field and $e$ the (effective) coupling constant.
The integration contour $C$ is given as the edge of a square with sidelength $L$ in the limit $L\rightarrow\infty$.

We are interested in the evaluation of the spatial string tension in the deconfining phase of $SU(2)$ Yang-Mills thermodynamics using the effective theory \cite{Hofmann05}. In this context, we assume magnetic monopoles liberated from dissociated calorons to be responsible for an area-law.

If the spatial string tension shows area-law behavior, we will obtain the following relation (for $T\gg T_C$):
\begin{equation}
\lim\limits_{A\rightarrow\infty}\,\frac{1}{A}\ln \textrm{tr}\,\mathcal{P}\exp \left(ie\oint\limits_C dz^\mu A^\mu\right)=F(T)\,\propto\,T^2\,.
\end{equation}
Otherwise, if the spatial string tension is governed by an perimeter-law, we will get
\begin{equation}
\lim\limits_{L\rightarrow\infty}\,\frac{1}{L}\,
\ln \textrm{tr}\,\mathcal{P}\exp \left(ie\oint\limits_C dz^\mu A^\mu\right)
=f(T)\,\propto\,T^2\,.
\end{equation}
In the following the Wegner-Wilson loop will be denoted as $W[C]$

\section{Calculation of the spatial string tension including the on-shell polarization effects}

\begin{figure}
\begin{center}
\vspace{5.3cm}
\includegraphics{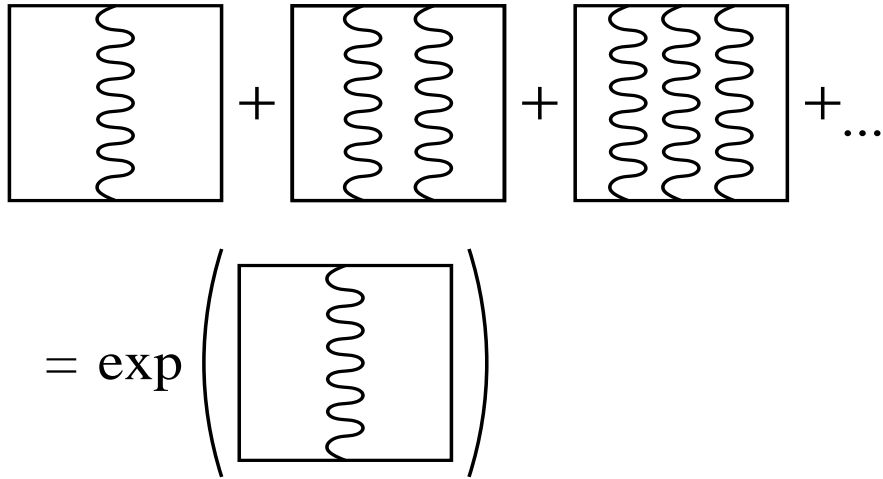}
\end{center}
\caption{\protect{\label{Wilson}Illustration of the diagrammatic approach. Every summand in this formula represents a whole class of $N$-gluons exchange diagrams, endowed with an combinatoric factor.}}
\end{figure}
In order to calculate the spatial string tension, we use an expansion into loops in the effective theory \cite{Bassetto}. A resummation of all possibly contributing $N$-gluon exchange diagrams can be rewritten in form of an exponential of the one-gluon exchange diagram. This fact is illustrated in Fig.\,\ref{Wilson}. Thus we can write for the logarithm of $W[C]$
\begin{equation}
\ln W[C]=-\frac{1}{2}C_F \oint\!dx_\mu dy_\nu\,D_{\mu\nu}(x-y)\,,
\end{equation}
where $C_F$ denotes the Dynkin index in the fundamental representation, defined as the normalization factor of the generators of the algebra (i.e. $C_F\equiv\frac{1}{2}$), and
\begin{equation}
D_{\mu\nu}=\sum\limits_{a=1}^3 D_{\mu\nu}^{(a)}
\end{equation}
is the sum of the tree-level propagators (\ref{PropagatorTLH},\,\ref{PropagatorTLM}) of the three gauge modes in deconfining, effective Yang-Mills thermodynamics. On-shell polarization effects concerning the tree-level massless gauge mode are taken into account by including the resummed one-loop polarization tensor into the propagator of the massless gauge mode (\ref{PropagatorTLM}). We constrain ourselves to the magnetic part of the polarization tensor (screening function $G(p^0,\vec{p})$), electric sreening effects (screening function $G(p^0,\vec{p})$) do not concern the spatial Wegner-Wilson loop. We have to emphasize that we only consider on-shell polarization effects for the massless gauge mode, radiative corrections for the massive gauge modes are not taken into account. Beyond this fact we should remark that this on-shell polarization tensor for the massless mode is calculated exactly up to one-loop order, since higher loop contributions vanish identically in the effec
 tive theory \cite{Hofmann/Kaviani07}, see Sec.\,\ref{1looppolarization}.

The tree-level propagators in position space are given as the Fourier transformations of their momentum space counterparts
\begin{equation}
D_{\mu\nu}^{1,2}(\beta,x-y)
=-\int\!\frac{d^4p}{(2\pi)^4}\,\e^{-ip(x-y)}\left(g_{\mu\nu}-\frac{p_\mu p_\nu}{m^2}
\right)\left[\frac{i}{p^2-m^2+i\varepsilon}
+2\pi\delta(p^2-m^2)n_B(\beta|p_0|)\right]\,,\\
\end{equation}
and
\begin{equation}
D_{\mu\nu}^{3}(\beta,x-y)
=\int\!\frac{d^4p}{(2\pi)^4}\,\e^{-ip(x-y)}
\left[P^T_{\mu\nu}\left(\frac{i}{p^2-G(p^0,\vec{p})+i\varepsilon}
+2\pi\delta(p^2-G(p^0,\vec{p})n_B(\beta|p_0|)\right)-i\frac{u_\mu u_\nu}{\vec{p}^2}\right]\,,
\end{equation}
where the transversal projection operator is given as
\begin{eqnarray}
P^T_{00}(p)&=&P^T_{0i}(p)=P^T_{i0}(p)=0\nonumber\\
P^T_{ij}(p)&=&\delta_{ij}-\frac{p_i p_j}{\vec{p}^2}.
\end{eqnarray}
We calculate the contour integral in the 1-2-plane, that is $x_0=y_0=x_0=y_3=0$,
and consider at first an arbitrary tensor structure $D_{\mu\nu}(p)$. Later we will insert the explicit tensor structure of $D_{\mu\nu}^{1,2}$ for the massive gauge modes ($V^\pm$ gauge modes) and $D_{\mu\nu}^{3}$ for the massless one ($\gamma$ mode).
We obtain
\begin{eqnarray}
\label{konturintegral}
&&\oint\oint dx_\mu dy_\nu\int\!\frac{d^4p}{(2\pi)^4}\,
  D_{\mu\nu}(p)\,\e^{-ip(x-y)}\left.\right|_{x_0=y_0=x_3=y_3=0}=\nonumber\\
&=&\int\!\frac{d^4p}{(2\pi)^4}\oint\oint dx_i dy_j\,D_{ij}(p)\,\e^{-ip(x-y)}\nonumber\\
&=&\int\!\frac{d^4p}{(2\pi)^4}\oint dx_i\,\e^{i\vec{p}\vec{x}}
  \left(D_{i1}\,\int\limits_{-\frac{L}{2}}^{\frac{L}{2}}\!dy_1\left(\e^{-i(p_1 y_1+p_2\frac{(-L)}{2})}-\e^{-i(p_1 y_1+p_2\frac{L}{2})}\right)\right.\nonumber\\
&&\left.\hspace{30mm}
  +D_{i2}\,\int\limits_{-\frac{L}{2}}^{\frac{L}{2}}\!dy_2\left(\e^{-i(p_1 \frac{L}{2} +p_2 y_2)}-\e^{-i(p_1 \frac{(-L)}{2}+p_2 y_2)}\right)\right)\nonumber\\
&=&\int\!\frac{d^4p}{(2\pi)^4}\oint dx_i\,\e^{i\vec{p}\vec{x}}
  \left(2i\sin\left(\frac{p_2L}{2}\right)\,D_{i1}\int\limits_{-\frac{L}{2}}^{\frac{L}{2}}
  \!dy_1\,\e^{-ip_1y_1}-2i\sin\left(\frac{p_1L}{2}\right)\,D_{i2}\int
  \limits_{-\frac{L}{2}}^{\frac{L}{2}}\!dy_2\,\e^{-ip_2y_2}\right)\nonumber\\
&=&4i\int\!\frac{d^4p}{(2\pi)^4}\sin\left(\frac{p_1L}{2}\right)\sin\left(\frac{p_2L}{2}\right)
  \oint dx_i\left(\frac{D_{i1}}{p_1}-\frac{D_{i2}}{p_2}\right)\e^{i\vec{p}\vec{x}}
  \nonumber\\
&=&4i\int\!\frac{d^4p}{(2\pi)^4}\sin\left(\frac{p_1L}{2}\right)\sin\left(\frac{p_2L}{2}\right)
  \left(\left(\frac{D_{11}}{p_1}-\frac{D_{12}}{p_2}\right)\int
  \limits_{-\frac{L}{2}}^{\frac{L}{2}} dx_1\!\left(\e^{i(p_1x_1+p_2\frac{(-L)}{2})}-
  \e^{i(p_1x_1+p_2\frac{L}{2})}\right)\right.\nonumber\\
&&\left.\hspace{30mm}
  +\left(\frac{D_{21}}{p_1}-\frac{D_{22}}{p_2}\right)\int
  \limits_{-\frac{L}{2}}^{\frac{L}{2}}dx_2\,\left(\e^{i(p_1\frac{L}{2}+p_2x_2)}-
  \e^{i(p_1\frac{(-L)}{2}+p_2x_2)}\right)\right)\nonumber\\
\nonumber
\end{eqnarray}
\newpage
\begin{eqnarray}
&=&4i\int\!\frac{d^4p}{(2\pi)^4}\sin\left(\frac{p_1L}{2}\right)\sin\left(\frac{p_2L}{2}\right)
  \left(-2i\sin\left(\frac{p_2L}{2}\right)
  \left(\frac{D_{11}}{p_1}-\frac{D_{12}}{p_2}\right)\int
  \limits_{\frac{-L}{2}}^{\frac{L}{2}}\!dx_1\,\e^{ip_1x_1}\right.
  \nonumber\\
&&\left.\hspace{30mm}
  +2i\sin\left(\frac{p_1L}{2}\right)
  \left(\frac{D_{21}}{p_1}-\frac{D_{22}}{p_2}\right)\int
  \limits_{\frac{-L}{2}}^{\frac{L}{2}}\!dx_1\,\e^{ip_2x_2}\right)\nonumber\\
&=&16\int\!\frac{d^4p}{(2\pi)^4}\sin^2\left(\frac{p_1L}{2}\right)\sin^2\left(\frac{p_2L}{2}\right)
  \left(\frac{D_{11}}{p_1^2}-\frac{D_{12}}{p_1p_2}
  -\frac{D_{21}}{p_1p_2}+\frac{D_{22}}{p_2^2}\right)\,.
\end{eqnarray}
We arrive at
\begin{eqnarray}
\ln W[C]&=&
-\frac{1}{4}\left.\oint\oint\!dx_\mu dy_\nu\,D_{\mu\nu}(p)
\e{-ip(x-y)}\right|_{x_0=y_0=x_3=y_3=0}\nonumber\\
&=&-4\int\!\frac{d^4p}{(2\pi)^4}\sin^2\left(\frac{p_1L}{2}\right)\sin^2\left(\frac{p_2L}{2}\right)
  \left(\frac{D_{11}}{p_1^2}-\frac{D_{12}}{p_1p_2}
  -\frac{D_{21}}{p_1p_2}+\frac{D_{22}}{p_2^2}\right)\nonumber\,,
\end{eqnarray}
At this point we insert the propagators for the $V^\pm$ and th  $\gamma$ mode. In case of the massive gauge modes we only have to consider the thermal part of Eq.\,(\ref{PropagatorTLH}), since due to the constraint on the maximal off shellness Eq.\,(\ref{compositeness1}) the vacuum part is vanishing. Subsequently, we perform the integration over the $p_0$-coordinate in the thermal contributions.
Thus we obtain
\begin{eqnarray}
\label{Wilsondimension}
\ln W[C]&=&\ln W[C]_{V^\pm}^{\tiny\mbox{th}}
             +\ln W[C]_{\gamma}^{\tiny\mbox{th}}
             +\ln W[C]_{\gamma}^{\tiny\mbox{vac}}\nonumber\\
&=&\frac{1}{\pi^3}\int\!d^3p\,
   \frac{\sin^2\left(\frac{p_1L}{2}\right)\sin^2\left(\frac{p_2L}{2}\right)}
   {\sqrt{p_1^2+p_2^2+p_3^2+m^2}}\,n_B\left(\beta{\sqrt{p_1^2+p_2^2+p_3^2+m^2}}\right)\,
   \left(\frac{1}{p_1^2}+\frac{1}{p_2^2}\right)\nonumber\\
 &-&\frac{1}{2\pi^3}\int\!d^3p\,
   \frac{\sin^2\left(\frac{p_1L}{2}\right)\sin^2\left(\frac{p_2L}{2}\right)}
   {\sqrt{p_1^2+p_2^2+p_3^2+G}}\,n_B\left(\beta{\sqrt{p_1^2+p_2^2+p_3^2+G}}\right)\,
   \left(\frac{1}{p_1^2}+\frac{1}{p_2^2}\right)\nonumber\\
 &-&\frac{i}{4\pi^4}\int\!d^4p\,
   \sin^2\left(\frac{p_1L}{2}\right)\sin^2\left(\frac{p_2L}{2}\right)\,
   \frac{1}{p^2-G+i\varepsilon}\,
   \left(\frac{1}{p_1^2}+\frac{1}{p_2^2}\right)\nonumber\\
\end{eqnarray}

We rescale the momenta and the squared gauge boson mass to dimensionless variables in terms of temperature and the loop length:
\begin{equation}
\hat{p_i}=p_i\cdot L\,
\end{equation}
and with $e=\sqrt{8}\pi$
\begin{equation}
\hat{m}^2=\frac{m^2}{T^2}=\frac{(2e)^2}{T^2}\frac{\Lambda}{2\pi T}=\frac{128\pi^4}{\lambda^3}\,.
\end{equation}
The screening function $G$ (as dynamically generated mass) is rescaled in the same manner
\begin{equation}
\hat{G}=\frac{G}{T^2}\,.
\end{equation}
Subsequently, in order to control the limit of growing loop length $L\rightarrow\infty$, we  introduce the dimensionless parameter $\tau$ into the integral (\ref{Wilsondimension}) via the definition
\begin{equation}
\tau=T\cdot L\,.
\end{equation}
Finally, Eq.\,(\ref{Wilsondimension}) is recast as
\begin{eqnarray}
\label{Wilsondimensionless}
\ln W[C]
&=&\frac{1}{\pi^3}\int\!d^3\hat{p}\,
   \frac{\sin^2\left(\frac{\hat{p}_1L}{2}\right)\sin^2\left(\frac{\hat{p}_2L}{2}\right)}
   {\sqrt{\hat{p}_1^2+\hat{p}_2^2+\hat{p}_3^2+\frac{128\pi^4}{\lambda^3}\tau^2}}\,
   n_B\left(\frac{{\sqrt{\hat{p}_1^2+\hat{p}_2^2+\hat{p}_3^2+\frac{128\pi^4}{\lambda^3}\tau^2}}}
   {\tau}\right)\,
   \left(\frac{1}{\hat{p}_1^2}+\frac{1}{\hat{p}_2^2}\right)\nonumber\\
 &-&\frac{1}{2\pi^3}\int\!d^3\hat{p}\,
   \frac{\sin^2\left(\frac{\hat{p}_1L}{2}\right)\sin^2\left(\frac{\hat{p}_2L}{2}\right)}
   {\sqrt{\hat{p}_1^2+\hat{p}_2^2+\hat{p}_3^2+G\tau^2}}\,
   n_B\left(\frac{\sqrt{\hat{p}_1^2+\hat{p}_2^2+\hat{p}_3^2+G\tau^2}}{\tau}\right)\,
   \left(\frac{1}{\hat{p}_1^2}+\frac{1}{\hat{p}_2^2}\right)\nonumber\\
 &-&\frac{i}{4\pi^4}\int\!d^4\hat{p}\,
   \sin^2\left(\frac{\hat{p}_1L}{2}\right)\sin^2\left(\frac{\hat{p}_2L}{2}\right)\,
   \frac{1}{\hat{p}^2-G\tau^2+i\varepsilon}\,
   \left(\frac{1}{\hat{p}_1^2}+\frac{1}{\hat{p}_2^2}\right)\nonumber\\
\end{eqnarray}
An area-law behavior of the spatial string tension corresponds to the existence of a nonvanishing limit
\begin{eqnarray}
\sigma=-\lim_{\tau\rightarrow\infty}\frac{\ln W[C]}{\tau^2}\cdot T^2\,,
\end{eqnarray}
a perimeter-law is given, if a finite limit according to
\begin{eqnarray}
\sigma=-\lim_{\tau\rightarrow\infty}\frac{\ln W[C]}{\tau}\cdot T^2\,
\end{eqnarray}
exists.
\begin{figure}
\begin{center}
\vspace{5.3cm}
\includegraphics{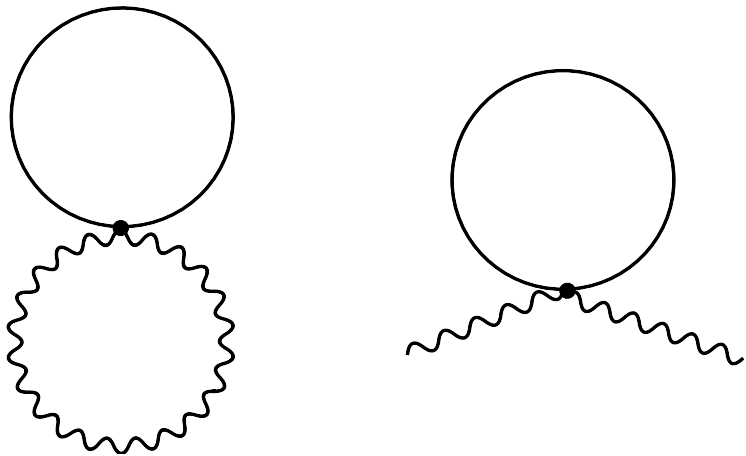}
\end{center}
\caption{\protect{\label{Pressure}Two-loop contribution to the pressure. Cutting of the massless line yields exactly on-shell one-loop polarization tensor for the massless gauge mode.}}
\end{figure}
Let us analyze the asymptotic behavior of Eq.\,(\ref{Wilsondimensionless}) at high temperature and low momenta for $\tau\rightarrow\infty$:
It is obvious, that in this limit the only contribution results from the massless mode, the contribution of the massive modes vanishes. This can be seen by an expansion of the integrand of $\ln W[C]_{V^\pm}$, in particular the Bose-Einstein distribution, for the momenta. In the limit $\tau\rightarrow\infty$ the integral is proportional $\tau^{-1}$. Concerning the thermal contribution of the massless mode we observe a perimeter-law  for each temperature in the limit $\tau\rightarrow\infty$. However, for values of $\tau$ below $\tau^*$, determined such that $L\cdot|\phi|^{-1}$, we are able to observe curvature in $\ln W[C]_{\gamma}^{\tiny\mbox{th}}$. Concerning the vacuum part the integrand can be estimated by neglecting the screening effects. We observe that $\ln W[C]_{\gamma}^{\tiny\mbox{vac}}$ represents a $\tau$-independent, infinite constant, and thus does not contribute to the force associated with the potential of a fundamental quark and its antiquark in the (2+1)-dimensional situation described by the spatial Wilson loop. A detailed discussion is beyond the scope of this thesis and will be part of future work \cite{LKGH2008}.

\begin{figure}
\begin{center}
\leavevmode
\leavevmode
\vspace{6.8cm}
\includegraphics{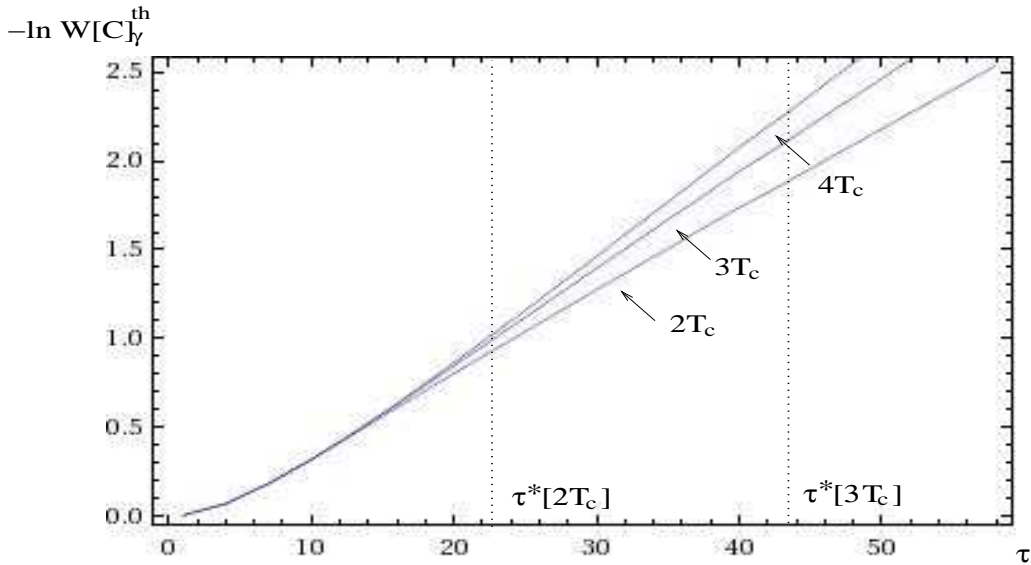}
\end{center}
\caption{\protect{\label{Fig-lnWC}} Plots of
$-\ln W[C]_\gamma^{\tiny\mbox{th}}$ as a function of $\tau=T\cdot L$. The dotted
lines correspond to the value of $L$ coinciding with the minimal length
scale $|\phi|^{-1}$ in the effective theory at a given temperature.
The left line is for $T=2\,T_c$, the right line for $T=3\,T_c$, and the
line for $T=4\,T_c$ would be at $\tau^*[4\,T_c]=65.77$ and thus is not
contained in the figure.}
\end{figure}

The negative logarithm of the Wegner-Wilson-loop $W[C]$ is depicted as function of the parameter $\tau$ in Fig.\,\ref{Fig-lnWC}.
Including merely the on-shell polarization tensor for the massless mode, we are obviously not able to reproduce the generally accepted results from lattice simulations. This leads to the idea that the lattice-obtained high-temperature behavior of the string tension is a consequence of off-shell contributions to the polarization tensor. As mentioned above, an area-law of the string tension requires the magnetic activity of free screened magnetic monopoles. These isolated magnetic charges are liberated during the dissociation of calorons. The likelihood for the decay of a large-holonomy caloron into a monopole-antimonopole pair is enormously Boltzmann-suppressed, $\left(\e^{-\beta M_{m/a}}\sim\e^{-8\pi^2}\right)$. However, isolated magnetic monopoles and antimonopoles can be generated from small-holonomy calorons by dissociation. A small-holonomy caloron interacts with an off-shell TLM-mode. In this process an amount of energy $\Delta E$ is transferred to the bound system of th
 e monopole-antimonopole pair. If $\Delta E$ is large enough to exceed the attractive potential induced by quantum fluctuations \cite{Diakonov/Gromov/Slizovskiy04}, the pair of constituents can split up into isolated magnetic charges.

Another aspect of this scenario concerns the two-loop contribution to the pressure $\Delta P^{HH}_{tt}$. This quantity was computed in \cite{Hofmann/Schwarz/Giacosa06-1, Hofmann/Herbst/Rohrer}. After scattering at the caloron, the TLM-mode yielding the energy to decouple the monopole-antimonopole pair possesses diminished energy than in the initial situation. As a consequence, the thermodynamical pressure is diminished and isolated magnetic monopoles are liberated. Since the pressure is proportional to $-T^4$, we can conclude that the density of isolated magnetic monopoles grows with increasing temperature. Consider the two-loop Feynman diagram concerning the pressure, where the massless line is cut. As depicted in Fig.\,\ref{Pressure} we obtain exactly the polarization tensor for the massless mode. Thus we can interpret the effects summarized in the one-loop polarization tensor on a microscopic level as scattering of massless modes on calorons. This leads to isolated magneti
 c charges and thus to magnetic flux in the plasma, measured by the spatial string tension. Finally, we have to emphasize that the off-shell polarization tensor contains an additional contribution (resulting from diagram A, see Fig.\,\ref{Fig-poltensor}, in contrast to its on-shell counterpart. This fact let us assume that in our effective theory a reproduction of the area-law behavior of the spatial string tension at high temperature possibly requires a consideration of the off-shell polarization tensor.

%% file: summary.tex
\chapter{Summary and Outlook}

In this thesis we have considered two aspects of $SU(2)$ Yang-Mills thermodynamics in the deconfining phase.
At first, we have calculated the two-point correlation function for the energy density of the photon, followed by an evaluation of the spatial string tension including on-shell polarization effects for the massless mode in the effective theory. This investigation was motivated by the postulate that the $U(1)_Y$ gauge group of the SM is not the progenitor to the gauge group of QED. Alternatively, the $U(1)_Y$ gauge group is originating from a more fundamental $SU(2)$ gauge group, named $SU(2)_{\tiny\mbox{CMB}}$, which dynamically breaks down to $U(1)_Y$.

In this effective theory the thermal ground state is composed of interacting calorons and anticalorons with trivial holonomy and characterized by an adjoint scalar field $\phi$ and a pure-gauge configuration $a_\mu^{gs}$. The macroscopic scalar field $\phi$ emerges as a result of spatial coarse-graining and acts as a Higgs field, responsible for the dynamical symmetry breaking, with two out of the three gauge modes acquiring a temperature dependent mass. The remaining massless gauge boson is identified with the photon. Distorted photon propagation according to the postulated $SU(2)$ gauge principle manifests itself in the polarization effects due to interactions with the massive excitations. As a consequence, the dispersion law is modified.

In Chapter three we have calculated the two-point correlator of a composite object, the energy density of the photon, in the conventional $U(1)$ gauge theory of electromagnetism and in deconfining $SU(2)$ Yang-Mills thermodynamics. This correlation function can be interpreted as a measure for the energy transfer between two spatial points. Apart from the fact that the computation is interesting from a technical point of view, we have considered a new feature of photon propagation governed by the $SU(2)_{\tiny\mbox{CMB}}$ gauge theory.  The visible suppression of $\bra\Theta_{00}(\vec{x})\Theta_{00}(\vec{y})\ket$ in case of photon propagation underlying a $SU(2)$ gauge theory in comparison to the conventional SM-QED implies a possible explanation for the unexpected stability of large, cold, innergalactic clouds consisting of atomic hydrogen.

Subsequently, we have considered the spatial string tension in deconfining $SU(2)$ Yang-Mills thermodynamics. This quantity can be regarded as a measure for the magnetic flux induced by magnetic monopoles at finite temperature through the area enclosed by a spatial contour. Including only on-shell polarization effects for the photon, we were not able to reproduce the lattice-obtained area-law. This has led us to the conclusion that the area-law is a consequence of off-shell contributions to the one-loop polarization tensor, which is natural since the Coulomb-field of a monopole is composed of off-shell modes. A calculation of the off-shell radiative corrections and thus a possible verification of our statement about the origin of the generally accepted area-law for the spatial string tension would be an interesting challenge for future works.

%% file: FTFT.tex
\chapter{Thermal field theory}
\label{FTFT}
In this section we summarize the fundamental concepts of finite temperature field theory. The analogy of quantum statistical mechanics and quantum field theory is developed, in order to transfer methods of zero-temperature quantum field theory to thermal field theory. The imaginary-time and the real-time formalism in thermal field theory are reviewed, and the free propagator in a thermalized theory is derived.
After that, the Feynman rules in QFT at finite temperature are stated, which are necessary prerequisites for the evaluation of Feynman diagrams. For a detailed introduction into the topic of finite temperature field theory and its applications in high energy physics, we refer to the literature \cite{Le Bellac, Kapusta, Landsman/vanWeert, Rothe}.

\section{The partition function as functional integral}

We restrict ourselves to the bosonic case.

At first, we remind the connection of quantum statistical mechanics and quantum field theory. Consider the amplitude $A(\phi_f,t_t;\phi_i,t_i)$ for the evolution of a system from an initial Heisenberg state $|\phi_i\ket$ at time $t_i$ to a final state $|\phi_f\ket$ at time $t_f$. The dynamics of the system is governed by the time-independent Hamiltonian
\begin{equation}
H=\int\!d^3x\,\mathcal{H}(\phi,\pi)=\int d^3x\,(\pi\partial_0\phi-\mathcal{L}(\phi,\pi))\,,
\end{equation}
where $\mathcal{H}$ and $\mathcal{L}$ denote the Hamiltonian density, respectively the Lagrangian density, as a functional of the field $\phi$ and its conjugated momentum $\pi$. If we introduce the time-evolution operator $\e^{-iH(t_f-t_i)}$, we are able to write the transition amplitude as
\begin{equation}
A(\phi_f,t_t;\phi_i,t_i)=\bra\phi_f|\e^{-iH(t_f-t_i)}|\phi_i\ket\,.
\end{equation}
In the path integral formulation of quantum field theory this object is calculated via functional integration in phase space:
\begin{equation}
A(\phi_f,t_f;\phi_i,t_i)=\int\limits_{\phi(\vec{x}_i,t_i)}^{\phi(\vec{x}_f,t_f)}\!
\mathcal{D}\phi\int\mathcal{D}\pi\,\exp\left({i\int\limits_{t_i}^{t_f}dt\int d^3x
                \left(\pi(\vec{x},t)\frac{\partial\phi(\vec{x},t)}{\partial t}-
                \mathcal{H}(\phi(\vec{x},t),\pi(\vec{x},t))\right)}\right)\,.
\end{equation}
In the following, we assume that $\mathcal{H}$ depends only quadratic on the conjugated momentum. In this case one can complete the square in the exponential, thus the $\pi$-integral is of gaussian type and can be evaluated analytically. This leads us to
\begin{equation}
A(\phi_f,t_f;\phi_i,t_i)=\mathcal{N}\int\limits_{\phi(\vec{x}_i,t_i)}^{\phi(\vec{x}_f,t_f)}
\!\mathcal{D}\phi\,\e^{iS[\phi]}\,,
\end{equation}
with the action functional $S[\phi]=\int d^4x\,\mathcal{L(\phi)}$ and a normalization constant $\mathcal{N}$ (which we can absorb in the integration measure).
The integration considers every possible field configuration on which the system can evolve from the initial to the final state, weighed with a phase factor proportional to the action of the field configuration.

In several cases it is convenient to evaluate path integrals in euclidean instead of Minkowskian spacetime. Especially in quantum field theory at non-zero temperature euclidean path integration is a common method of calculation. We obtain the path integral in euclidean formulation by performing an analytical continuation from real to imaginary time, \begin{equation}
t\rightarrow-i\tau\,,\qquad\mbox{or}\qquad x^0\rightarrow-ix_4\,,
\end{equation}
where $\tau$ real.
This transformation is called Wick-rotation and changes the signature of our metric. Since in Minkowskian signature we have $x^\mu x_\mu=t^2-\vec{x}^2$ we obtain $x^\mu x_\mu=x_\mu x_\mu=t^2+\vec{x}^2$ in euclidean signature. The transition amplitude for the field is then $\phi(\vec{x},t)=\phi(\vec{x},-i\tau)$ is then transformed into
\begin{equation}
A(\phi_f,\tau_f;\phi_i,\tau_i)=\bra\phi_f|e^{-H(\tau_f-\tau_i)}|\ket
                              =\mathcal{N}\int\!\mathcal{D}\phi\,\e^{-S_E[\phi]}\,.
\end{equation}
Here $S_E[\phi]$ denotes the action functional in euclidean spacetime, the boundary conditions concerning the paths are given as $\phi_i(\vec{x})=\phi(\vec{x},\tau_i)$ and $\phi_f(\vec{x})=\phi(\vec{x},\tau_f)$. (Without constraining generality, we set $\tau_i=t_i=0$.)

Now we look at quantum statistics in the canonical ensemble. Our intention is to transfer
the concepts of functional integration in quantum field theory to statistical mechanics with the aim to develop a path integral formulation of the partition function. The partition function occupies a central role in the canonical ensemble of statistical mechanics and is given as
\begin{equation}
Z=\textrm{tr}\,\e^{-\beta H}=\sum\limits_n\bra\phi_n|\e^{-\beta H}|\phi_n\ket\,.
\end{equation}
The sum is to evaluate over a complete set of states of the system with Hamiltonian $H$ and inverse temperature $\beta=\frac{1}{T}$. The similarity of the partition function and the euclidean path integral formulation of the transition amplitude is therefore obvious.
Formally, we are allowed to write
\begin{equation}
Z=\sum\limits_n A(\phi_n,-i\tau;\phi_n,0)\,.
\end{equation}
If we compare this formula with the euclidean expression for the transition amplitude, we can write the canonical partition function as an functional integral, since we adjust the following requirements. We identify finite temperature with imaginary time compactified on a circle with circumference $\beta$ and interpret $\e^{-\beta H}$ as imaginary time evolution operator. As consequence of the trace operator in the partition function, the physical fields have to satisfy $\beta$-periodic boundary conditions in euclidean time, the functional integration is constrained to periodic fields, $\phi(\vec{x},0)=\phi(\vec{x},\beta)$. With these conditions we obtain
\begin{equation}
\label{partition function}
Z=\int\limits_{\phi(\vec{x},0)=\phi(\vec{x},\beta)}\!\mathcal{D}\phi\,\e^{-S_E[\phi]}\,.
\end{equation}

\section{Derivation of the propagator for the free scalar field}

Now we receive the propagator by applying a standard functional method in quantum field theory. We define a generating functional $Z(\beta,J)$ with a source term $J$ through
\begin{equation}
\label{generating functional}
Z(\beta,J)=\int\mathcal{D}\!\phi\,\e^{-S_E(\beta)+\int_0^\beta\!d^4x\,J(x)\phi(x)}\,.
\end{equation}
It is obvious that for a vanishing source term, the generating functional is equal to the canonical partition function Eq.\,(\ref{partition function}).

Functional differentiation of Eq.\,(\ref{generating functional}) yields the propagator in euclidean, that is imaginary time:
\begin{equation}
\frac{1}{Z(\beta)}\left.\frac{\delta^2Z(\beta,J)}{\delta J(x)\delta J(y)}\right|_{J=0}=
\frac{1}{Z(\beta)}\int\!\mathcal{D}\phi(\tau)\,\phi(x)\phi(y)\e^{-S_E(\beta)}\,.
\end{equation}
We specialize to the case of a free bosonic field. The free generating functional in euclidean formulation is the given as
\begin{eqnarray}
Z_F(\beta,J)&=&\int\!\mathcal{D}\phi\,\exp\left(-\int\limits_0^\beta\!d\tau\int\!d^3x \left(\phi(x)\frac{1}{2}\left[-\frac{\partial^2}{\partial\tau^2}-\nabla^2+m^2\right]
\phi(x)-J(x)\phi(x)\right)\right)\nonumber\\
&=&Z_F(\beta)\,\exp\left(\frac{1}{2}\int\limits_0^\beta\!d^4x\,d^4y\,J(x)\Delta_F(x-y) J(y)\right)\,.
\end{eqnarray}
The two-point Greens function $\Delta_F(x-y)$ solves the partial differential equation
\footnote{We should mention, that $\delta(\tau_x-\tau_y)$ represents a periodic $\delta$-Distribution
\begin{equation}
\delta(\tau)=\frac{1}{\beta}\sum\limits_n \e^{i\omega_n\tau}\,.
\end{equation}}
\begin{equation}
\left(-\square^2+m^2\right)\Delta_F(x-y)^{(\beta)}=\left(-\frac{\partial^2}{\partial\tau^2}
-\nabla^2+m^2\right)\Delta_F(x-y)^{(\beta)}=\delta(\tau_x-\tau_y)\delta^3(\vec{x}-\vec{y})\,.
\end{equation}
As solution in Fourier space, we obtain the imaginary-time propagator, also called Matsubara propagator
\begin{equation}
\tilde{\Delta}_F(x-y)^{(\beta)}(\omega_n,k)=\frac{1}{\omega_n^2+\vec{k}^2+m^2}
=\frac{1}{\omega_n^2+\omega_k^2}\,,
\end{equation}
where the $\omega_n$ denote the discrete Matsubara frequencies $\omega_n=
\frac{2\pi n}{\beta}$ and $\omega_k$ is defined as $\omega_k=\sqrt{\vec{k}^2+m^2}$.
The corresponding expression in position space is given as
\begin{equation}
\Delta_F(x-y)^{(\beta)}=\frac{1}{\beta}\sum\limits_n\int\!\frac{d^3k}{(2\pi)^3}
\,\frac{\e^{i\omega_n\tau+i\vec{k}\vec{z}}}{\omega_n^2+k^2+m^2}\,.
\end{equation}
A comparison with the zero-temperature counterparts in euclidean metric
\begin{eqnarray}
\tilde{\Delta}(k)&=&\frac{1}{k^2+m^2}\,,\\
\Delta(x-y)&=&\int\frac{d^4k}{(2\pi)^4}\frac{\e^{ik(x-y)}}{k^2+m^2}\,,
\end{eqnarray}
where $k\equiv(k_4,\vec{k}), k^2=k_4^2+\vec{k}^2$,
yields transformation rules from zero to finite temperature:
\begin{eqnarray}
k_4&\rightarrow&\omega_n\,,\\
\int\!\frac{dk_4}{2\pi}f(k_4)&\rightarrow&\frac{1}{\beta}\sum\limits_n\omega_n\,,\\
\int d^4x&\rightarrow&\int\limits_0^\beta\!d\tau\int\!d^3x\,.
\end{eqnarray}
The temperature dependence is contained in the infinite sums over the temperature dependent, discrete Matsubara frequencies. As a consequence of the frequency summation, the finite temperature propagator $\Delta^{(\beta)}(x-y)$ consists of a zero-temperature and a finite-temperature contribution.

\begin{figure}
\begin{center}
\vspace{5.3cm}
\includegraphics{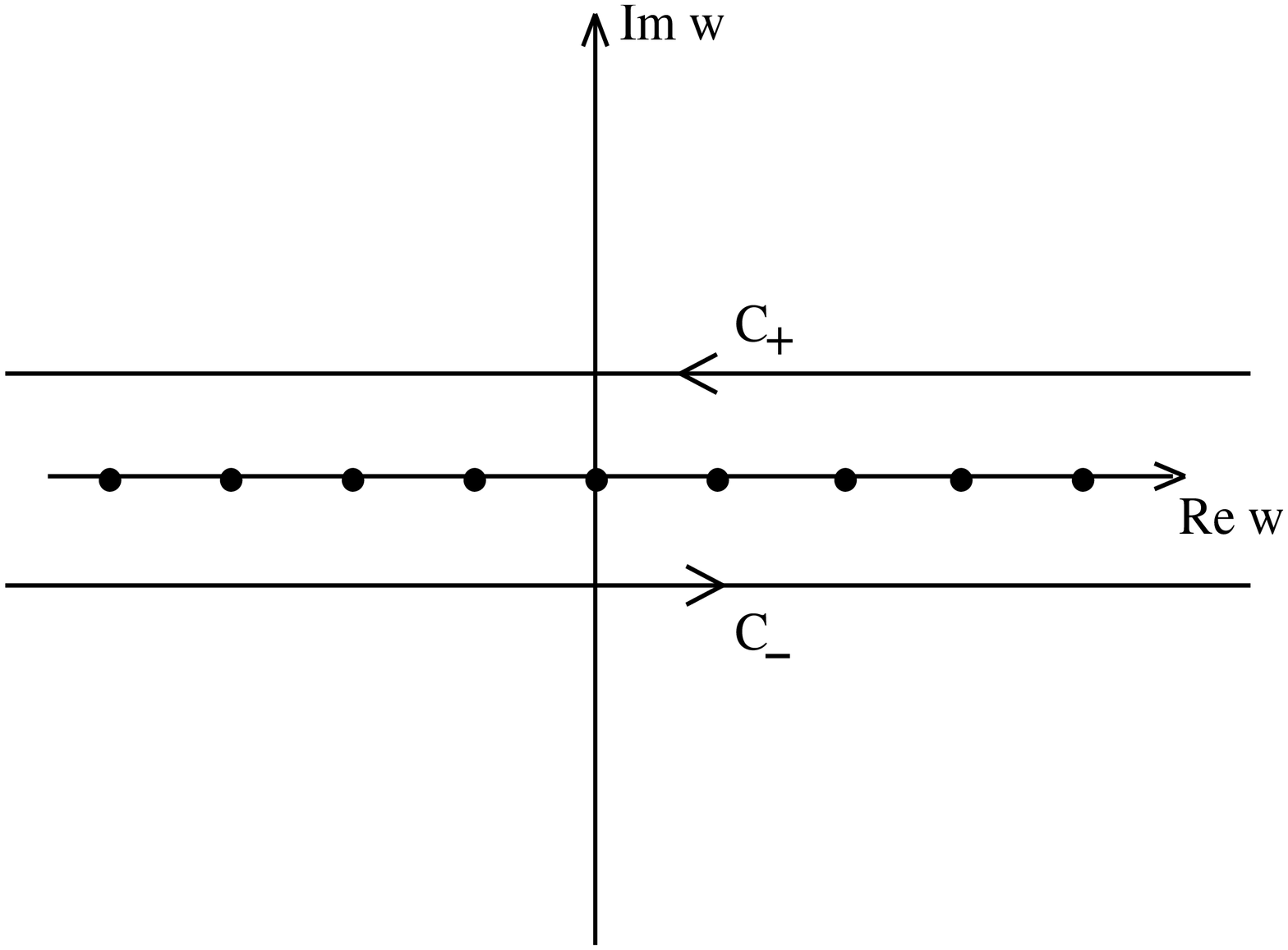}
\end{center}
\caption{\protect{\label{Kontur}Integration contour.}}
\end{figure}

A frequency sum $\frac{1}{\beta}\sum_n f(\omega_n, K)$, where conventionally $K$ represents the remaining variables on which $f$ depends on, can be executed with methods of complex analysis. We substitute the discrete values $\omega_n$ by a continuous variable $\omega$ and define a function $f(\omega,K)$, which we assume to have no singularities on the real axis. Additionally, we define the complex function
\begin{equation}
h(\omega)=\frac{i\beta}{\e^{i\beta\omega}-1}=-\frac{i\beta}{\e^{-i\beta\omega}-1}-i\beta\,,
\end{equation}
whose poles are located at $\omega=\frac{2\pi n}{\beta}$ with unit residue. Employing the residue theorem, we are able to write the sum as follows,
\begin{equation}
\label{contour propagator}
F(K)=\frac{1}{2\pi i\beta}\int\limits_\alpha \!d\omega\,h(\omega)f(\omega,K)\,,
\end{equation}
where $\alpha$ designates the contour shown in Fig.\,\ref{Kontur}. With respect to the behavior of $h(\omega)$ for large imaginary parts, Eq.\,(\ref{contour propagator}) can be decomposed into three integrals
\begin{equation}
F(K)=\frac{1}{2\pi}\int\limits_{-\infty}^{\infty}\!d\omega\,f(\omega,K)
+\frac{1}{2\pi}\int\limits_{-\infty-i\epsilon}^{\infty-i\epsilon}\!d\omega\,
\frac{f(\omega,K)}{\e^{i\beta\omega}-1}
+\frac{1}{2\pi}\int\limits_{-\infty+i\epsilon}^{\infty+i\epsilon}\!d\omega\,
\frac{f(\omega,K)}{\e^{-i\beta\omega}-1}\,.
\end{equation}
If we provide that $f(\omega,K)$ is meromorphic as a function of $\omega$ and well-behaved for $|\omega|\rightarrow\infty$, the integration contours concerning the second and third integrals can be closed in the lower, respectively upper half plane at infinity, and we obtain
\begin{equation}
\label{frequency sum}
F(K)=\frac{1}{2\pi}\int\limits_{-\infty}^{\infty}d\omega\,f(\omega,K)
+i\sum\limits_{Im\,\bar{\omega}_i>0}
\frac{Res_f(\bar{\omega}_i)}{\e^{i\beta\bar{\omega}_i}-1}
-i\sum\limits_{Im\,\bar{\omega}_i<0}
\frac{Res_f(\bar{\omega}_i)}{\e^{-i\beta\bar{\omega}_i}-1}\,,
\end{equation}
where $Res_f(\bar{\omega}_i)$ denotes the residuum of $f(\omega,K)$ at the pole located at $\bar{\omega}_i$.

An expression for the propagator in Minkowskian spacetime can be calculated as analytic continuation of the euclidean propagator to real times. In this concrete case we insert
\begin{equation}
f(\omega,K)=\frac{\exp(i\omega\tau)}{\omega^2+\vec{k^2}+m^2}
\end{equation}
into the frequency summation formula, Eq.\,(\ref{frequency sum}). Since $\tau$ is restrained to the finite interval $[0,\beta]$, the integration contours can be closed appropriate to Eq.\,(\ref{frequency sum}) in the complex $\omega$-plane. Accordingly, we get
\begin{eqnarray}
\Delta^{(\beta)}(x-y)&=&\int\!\frac{d^4k}{(2\pi)^4}
\,\frac{\e^{ik_4\tau+i\vec{k}(\vec{x}-\vec{y})}}{k^2+m^2}
+\int\!\frac{d^3k}{(2\pi)^3}\,\frac{1}{2E}n_B(\beta E)
(\e^{-E\tau}+\e^{E\tau})\,\e^{i\vec{k}(\vec{x}-\vec{y})}\nonumber\\
&=&\int\!\frac{d^4k}{(2\pi)^4}
\,\frac{\e^{ik_4\tau+i\vec{k}(\vec{x}-\vec{y})}}{k^2+m^2}
+\int\!\frac{d^3k}{(2\pi)^3}\,\frac{1}{E}n_B(\beta E)
\cosh(E\tau)\e^{i\vec{k}(\vec{x}-\vec{y})}\,,
\end{eqnarray}
where $E=\sqrt{\vec{k^2}+m^2}$, $\omega=k_4$, and $n_B(\beta E)=(\e^{\beta E}-1)^{-1}$ denotes the Bose-Einstein distribution. We rotate back to real time values by setting $\tau\rightarrow it$, $k_4\rightarrow ik^0$ and arrive at
\begin{eqnarray}
\left.\Delta^{(\beta)}(x-y)\right|_{\tau=it}&=&
i\int\!\frac{d^4k}{(2\pi)^4}\,\frac{\e^{-ik(x-y)}}{k^2-m^2+i\varepsilon}
+2\pi\int\!\frac{d^4k}{(2\pi)^4}\,\frac{n_B(\beta E)}{2E}
\left(\delta(k^0-E)+\delta(k^0+E)\right)\e^{-ik(x-y)}\,.\nonumber\\
\end{eqnarray}
Inserting
\begin{equation}
\delta(k^2-m^2)=\frac{1}{2E}\left(\delta(k^0-E)+\delta(k^0+E)\right)
\end{equation}
we arrive at the final result for the propagator of a scalar field in Minkowskian spacetime:\footnote{For completeness we mention that in case of a fermionic particle the propagator is given as
\begin{eqnarray}
S^{(\beta)}_{Mink}(x-y)&=&\int\!\frac{d^4p}{(2\pi)^4}\,
\tilde{S}^{(\beta)}_{Mink}(p)\e^{-ip(x-y)}\,,\\
\tilde{S}^{(\beta)}_{Mink}(p)&=&\frac{i}{\slashed{p}-m+i\varepsilon}+2\pi n_F(\beta E)
(\slashed{p}+m)\delta(p^2-m^2)\,,
\end{eqnarray}
where $n_F(x)$ denotes the Fermi-Dirac distribution function $n_F=\frac{1}{\e^x+1}$}
\begin{eqnarray}
\Delta^{(\beta)}_{Mink}(x-y)&=&\int\!\frac{d^4k}{(2\pi)^4}\,
\tilde{\Delta}^{(\beta)}_{Mink}(k)\,\e^{-ik(x-y)}\,,\\
\tilde{\Delta}^{(\beta)}_{Mink}(k)&=&\frac{i}{k^2-m^2+i\varepsilon}+2\pi n_B(\beta E)
\delta(k^2-m^2)\,.
\end{eqnarray}
We have to emphasize that the derivation of the propagator above is merely valid in a free theory. In an interacting theory the finite-temperature real-time Feynman rules are more complicated and a doubling of degrees of freedom is required to obtain the right perturbative expansion. The corresponding propagator is then given in form of a $2\times2$-matrix.

\section{Feynman rules in real-time finite temperature field theory}

In this subsection we state the Feynman rules for quantum field theory at finite temperature.

\begin{figure}
\begin{center}
\vspace{5.9cm}
\includegraphics{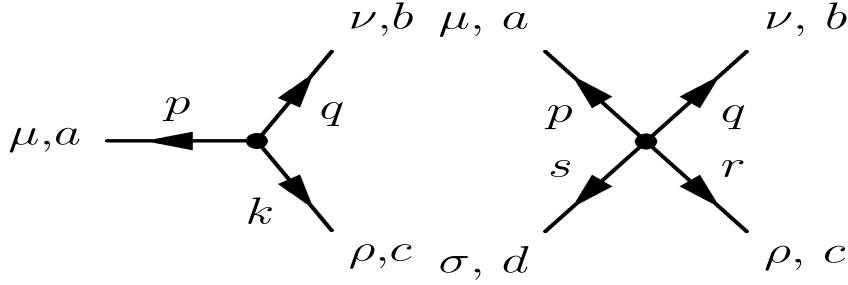}
\caption{\protect{\label{Fig-vertex}}The four-boson-vertex and the three-boson-vertex}
\end{center}
\end{figure}

In unitary Coulomb gauge, $\phi$ is diagonal and the pure-gauge ground state field $a^{gs}$ vanishes as mentioned in a previous chapter (unitarity). This physical gauge fixing does not require the introduction of Faddeev-Popov ghost fields. Subsequently, the remaining gauge freedom provides transversality of the massless gauge mode $\partial_i\delta a_i^{TLM}=0$ (Coulomb). A calculation in the real-time formalism of thermal field theory is advantageous, since the propagator splits up into a pure thermal and a pure vacuum part.
In the following we list the Feynman rules:

The tree-level propagator for a free TLM mode is given as
\begin{equation}
\label{PropagatorTLM}
D_{\mu\nu,a b}^{TLM}(p,\beta)
=-\delta_{a b}\left(P^T_{\mu\nu}\left(\frac{i}{p^2+i\varepsilon}
+2\pi\delta(p^2)n_B(\beta|p_0|)\right)-i\frac{u_\mu u_\nu}{\vec{p}^2}\right)\,,
\end{equation}
where $n_B(x)=\frac{1}{\e^x-1}$ denotes the Bose-Einstein distribution function and the projector
\begin{eqnarray}
\label{transversalprojector}
P^T_{00}(p)&=&P^T_{0i}(p)=P^T_{i0}(p)=0\nonumber\\
P^T_{ij}(p)&=&\delta_{ij}-\frac{p_i p_j}{\vec{p}^2}.
\end{eqnarray}
The tree-level propagator for a free TLH mode is given as
\begin{equation}
\label{PropagatorTLH}
D_{\mu\nu,a b}^{TLH}(p,\beta)
=-\delta_{a b}\tilde{D}_{\mu\nu}(p)\left(\frac{i}{p^2-m^2+i\varepsilon}
+2\pi\delta(p^2-m^2)n_B(\beta|p_0|)\right)\,,
\end{equation}
with the tensor structure
\begin{eqnarray}
\tilde{D}_{\mu\nu}(p)=\left(g_{\mu\nu}-\frac{p_\mu p_\nu}{m^2}\right)\,.
\end{eqnarray}
The three-gauge-boson-vertex and the four-gauge-boson-vertex are the conventional ones
\begin{eqnarray}
\Gamma_{[3]\ a b c}^{\mu\nu\rho}&=&e(2\pi)^4\delta(p+q+k)f_{a b c}
\left[g^{\mu\nu}(q-p)^\rho+g^{\nu\rho}(k-q)^\mu+g^{\rho\mu}(p-k)^\nu\right]\,,\\
\Gamma_{[4]\ a b c d}^{\mu\nu\rho\sigma}&=&-ie^2(2\pi)^4\delta(p+q+r+s)
\left[f_{a b e}f_{c d e}(g^{\mu\rho}g^{\nu\sigma}-g^{\mu\sigma}g^{\nu\rho})\right.\nonumber\\
&&\left.\hspace{5mm}+f_{a c e}f_{b d e}
(g^{\mu\nu}g^{\rho\sigma}-g^{\mu\sigma}g^{\nu\rho})
+f_{a d e}f_{b d e}(g^{\mu\rho}g^{\rho\sigma}-g^{\mu\rho}g^{\nu\sigma})\right]\,.
\end{eqnarray}
The four-momenta and color indices are defined in figure (\ref{Fig-vertex}).
Finally, one has to divide every loop diagram by $i$ and the number of its vertices \cite{Landsman/vanWeert}.

\section{One-loop polarization tensor of the massless gauge mode}
\label{1looppolarization}

In this section we consider the one-loop polarization tensor for the massless gauge mode in $SU(2)$ Yang-Mills thermodynamics. This object was first calculated in \cite{Hofmann/Schwarz/Giacosa06-1} and led to some astonishing implications \cite{Schwarz}.
It is evident, that if than elementary particle propagates in a medium, its properties are modified caused by interaction with this medium. As we will see, these effects dynamically generate an effective mass. In case of $SU(2)$ Yang-Mills thermodynamics this medium is given as the nontrivial vacuum, the propagating gauge boson is the TLM-mode $\gamma$, interacting with the TLH-modes $V^\pm$. All these non-Abelian effects are summarized in the polarization tensor $\Pi^{\mu\nu}$.

The polarization tensor of the TLM-mode is transversal for any value of $p^2$,
\begin{equation}
p_\mu\Pi^{\mu\nu}=0\,.
\end{equation}
Thus the following decomposition in a transversal and a longitudinal part is valid:
\begin{equation}
\Pi^{\mu\nu}=G(p_0,\vec{p})P_T^{\mu\nu}+F(p_0,\vec{p})P_L^{\mu\nu}\,,
\end{equation}
where the projection operators are related via
\begin{equation}
P_L^{\mu\nu}=\frac{p^\mu p^\nu}{p^2}-g^{\mu\nu}-P_T^{\mu\nu}\,,
\end{equation}
and $P_L^{\mu\nu}$ defined in Eq.\,(\ref{transversalprojector}).

In euclidean signature the full propagator of the massless gauge mode results as
\begin{equation}
\label{screened propagator}
D_{\mu \nu,a b}^{TLM}(p)=-\delta_{a b}\left(P^T_{\mu\nu}\,\left[\frac{1}{G-p^2}+2\pi\delta(p^2)n_B(\beta|p_0|)\right]
+\frac{p^2}{\vec{p}^2}\frac{1}{F-p^2}u_\mu u_\nu\right)\,.
\end{equation}
If photon propagation would not be affected by interactions with the virtual massive gauge modes, the screening functions F and G would vanish and the propagator (\ref{screened propagator}) transforms into Eq.\,(\ref{PropagatorTLM}).

According to \cite{Linde80/Polyakov75} the $00$-component of the polarization tensor $\Pi_{\mu\nu}(p)$ in the limit $p_0=0,\vec{p}\rightarrow 0$ corresponds to an electric screening mass (also called Debye mass) $m_{el}$. It holds
\begin{equation}
F(p_0=0,\vec{p}\rightarrow 0)=-\Pi_{00}(p_0=0,\vec{p}\rightarrow 0)=m_{el}^2\,.
\end{equation}
This quantity is diverging in this limit, what implies that static electric fields are completely screened by calorons. On the other side it is obvious that in the same limit $G(p_0=0,\vec{p}\rightarrow 0)$ disappears, so that static magnetic fields are not screened. Furthermore, we realize that on-shell $\Pi_{00}(p^2\rightarrow 0)$ remains finite, due to Bose-Einstein suppression, with the consequence that the screening function $F(p^2\rightarrow 0)$ vanishes in that limit.

Relevant for this thesis is the screening function $G(p_0,\vec{p})$, since G modifies the dispersion law for the TLM-mode. Instead of
\begin{equation}
\omega^2=\vec{p}^2,
\end{equation}
now
\begin{eqnarray}
\label{SU2 dispersion}
\omega^2(\vec{p})&=&\vec{p}^2+\mathfrak{Re}\,G(\omega(\vec{p}),\vec{p},\Lambda,T)\,,\\
\gamma(\vec{p})&=&-\frac{1}{2\omega}\mathfrak{Im}\,G(\omega(\vec{p}),\vec{p},\Lambda,T)\\
\nonumber
\end{eqnarray}
holds. In this context we can identify the square root of the real part of G as a dynamical mass.

The polarization tensor $\Pi^{\mu\nu}$ was computed in \cite{Schwarz} and \cite{Hofmann/Schwarz/Giacosa06-1} in on-shell approximation $p^2\rightarrow 0$. We should shortly render the essentials of the calculation as well as the results.
\begin{figure}
\label{DiagramPolarisation}
\begin{center}
\vspace{5.9cm}
\includegraphics{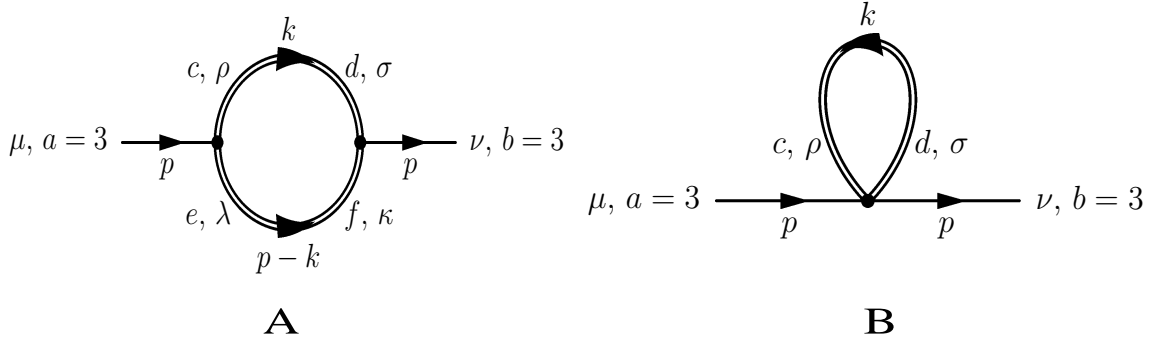}
\caption{\protect{\label{Fig-poltensor}}Diagrams contributing to the one-loop polarization tensor of the massless mode.}
\end{center}
\end{figure}

\begin{figure}
\begin{center}
\vspace{5.0cm}
\includegraphics{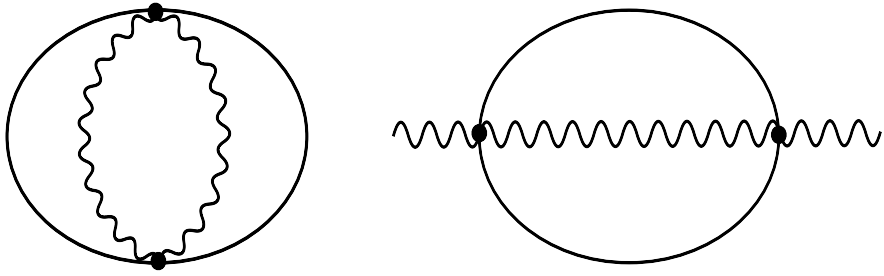}
\caption{\protect{\label{Higherloop}}Cutting of one of the massless lines in the three-loop diagram for the pressure yields the irreducible two-loop contribution for the on-shell polarization tensor.}
\end{center}
\end{figure}

The one-loop polarization tensor
is evaluated as the sum of the two Feynman diagrams depicted in Fig.\,\ref{Fig-poltensor}. An important fact is that for thermal massless modes this on-shell polarization tensor is exactly calculated in one-loop order. This is resulting from \cite{Hofmann/Kaviani07, Hofmann-loop}. In this work it was proved that the irreducible three-loop contribution to the pressure vanishes identically. If we cut one of the massless lines in this three-loop diagram we obtain the irreducible two-loop contribution to the on-shell polarization tensor of the massless mode, see Fig.\,\ref{Higherloop}. Thus the two-loop contribution is exactly nil, due to the vanishing three-loop diagram.

At first we have to emphasize that exclusively the thermal part of the propagator contributes to the polarization tensor, each vacuum part is exactly nil due to the constraint on the maximal off-shellness of the massive gauge mode, Eq.\,(\ref{compositeness1}), in combination with the fact that in the deconfining phase $e\geq\sqrt{8}\pi$ holds. Furthermore, the calculation shows that the contribution of diagram A is identically zero in the on-shell limit by virtue of momentum conservation.

\begin{figure}
\begin{center}
\vspace{7.5cm}
\includegraphics{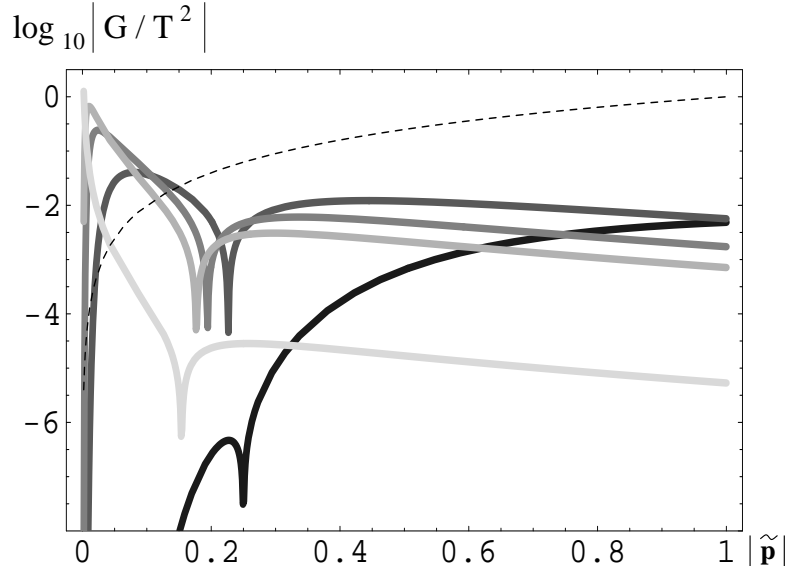}
\caption{\protect{\label{Fig-3}}$\log_{10}\left|\frac{G}{T^2}\right|$ as
  a function of
$|\tilde{\vec{p}}|$
for $\lambda=1.12\,\lambda_{c}$ (black), $\lambda=2\,\lambda_{c}$ (dark grey),
$\lambda=3\,\lambda_{c}$ (grey), $\lambda=4\,\lambda_{c}$ (light grey), $\lambda=20\,\lambda_{c}$
(very light grey). This result is obtained by appealing to the
approximation $\tilde{p}^2=0$. The full calculation shows similar results for
finite $|\tilde{\vec{p}}|$. However, there we have
$\lim_{|\tilde{\vec{p}}|\to 0}\left|\frac{G}{T^2}\right|>0$ in contrast to
the here-indicated result. The dashed curve is a
plot of the function
$f(|\tilde{\vec{p}}|)=2\log_{10}|\tilde{\vec{p}}|$. Here $\lambda\equiv
13.87\frac{T}{T_c}=\frac{2\pi T}{\Lambda}$. Photons are strongly screened at
$|\tilde{\vec{p}}|$-values for which $\log_{10}\left|\frac{G}{T^2}\right|>f(|\tilde{\vec{p}}|)$, that is, to the left
of the dashed line. The dips correspond to the zeros of $G$.}
\end{center}
\end{figure}

The tensor structure of the tadpole diagram B evaluates to \cite{Schwarz}
\begin{equation}
\Pi^{\mu\nu}(p)=4e^2\int\!\frac{d^4k}{(2\pi)^4}\,
\left[g^{\mu\nu}\left(3-\frac{k^2}{m^2}\right)+\frac{k^\mu k^\nu}{m^2}\right]
\cdot2\pi\delta(k^2-m^2)n_B(\beta|k_0|)\,,
\end{equation} where the four-vertex constraint (\ref{compositeness2}) implies severe restrictions on the limits of integration.  Without constraints on generality, \vec{p} is assumed to point into the three-direction. Then the diagonal components $\Pi_{11}$ and $\Pi_{22}$ are equal to G. Momenta and temperature are rescaled to dimensionless variables subject to
\begin{equation}
X=\frac{|\vec{p}|}{T},\qquad \vec{y}=\frac{\vec{k}}{|\phi|},
\qquad \lambda=\frac{2\pi T}{\Lambda}\,.
\end{equation}
Introducing cylindrical coordinates for $\vec{y}$,
\begin{equation}
y_1=\rho\cos\phi,\qquad y_2=\rho\sin\phi,\qquad y_3=\xi,
\end{equation}
the screening function G is calculated as
\begin{equation}
G=\left[\int\limits_{-\infty}^{\xi_m}d\xi\int\limits_{\rho_m}^{\rho_M}d\rho
+\int\limits_{\xi_m}^{\xi_M}d\xi\int\limits_{0}^{\rho_M}d\rho\right]
\frac{e^2}{\beta^2\lambda^3}\left(-4+\frac{\rho^2}{4e^2}\right)
\frac{\rho}{\sqrt{\rho^2+\xi^2+4e^2}}
n_B\left(\frac{2\pi}{\lambda^{3/2}}\sqrt{\rho^2+\xi^2+4e^2}\right)\,
\end{equation}
where the integration limits are given as
\begin{eqnarray}
\rho_m&=&\sqrt{\left(\frac{\pi}{X}\right)^2\frac{(4e^2-1)^2}{\lambda^3}
-\frac{2\pi}{X}\frac{4e^2-1}{\lambda^{3/2}}\xi-4e^2}\,,\nonumber\\
\rho_M&=&\sqrt{\left(\frac{\pi}{X}\right)^2\frac{(4e^2+1)^2}{\lambda^3}
-\frac{2\pi}{X}\frac{4e^2+1}{\lambda^{3/2}}\xi-4e^2}\,,\nonumber\\
\xi_m&=&\frac{\pi}{2X}\frac{4e^2-1}{\lambda^{3/2}}
-2\frac{X}{\pi}\lambda^{3/2}\frac{e^2}{4e^2-1}\,,\nonumber\\
\xi_M&=&\frac{\pi}{2X}\frac{4e^2+1}{\lambda^{3/2}}
-2\frac{X}{\pi}\lambda^{3/2}\frac{e^2}{4e^2+1}\,.\nonumber\\
\end{eqnarray}
The screening function G was calculated numerically in \cite{Schwarz} and is drawn in Fig.\,\ref{Fig-3} for various temperatures.

As a consequence of the modified dispersion law (\ref{SU2 dispersion}) photon propagation is affected by screening effects arising from the polarization tensor. This effects manifests for temperatures a few times the critical temperature. For momenta to the right of the sign change in the screening function (associated with the dips in the curves) G is negative and possesses a small modulus. Thus the energy of the propagating TLM-mode is reduced in comparison to the free case. This effect is known as antiscreening
\footnote{Nevertheless the influence of the tree-level massive modes $V^\pm$ is very weak due to the strong restrictions onto the maximal off-shellness.}. However, for momenta to the left of the sign change, we enter the regime of screening, G is positive and reaches sizeable values $\sim|\vec{p}|^2$. The photon acquires a screening mass. Thus the photon can travel through the plasma only up to a distance $G^{-1/2}$. The situation is summarized in
Fig.\,\ref{Fig-3}.

\begin{figure}
\begin{center}
\vspace{5.9cm}
\includegraphics{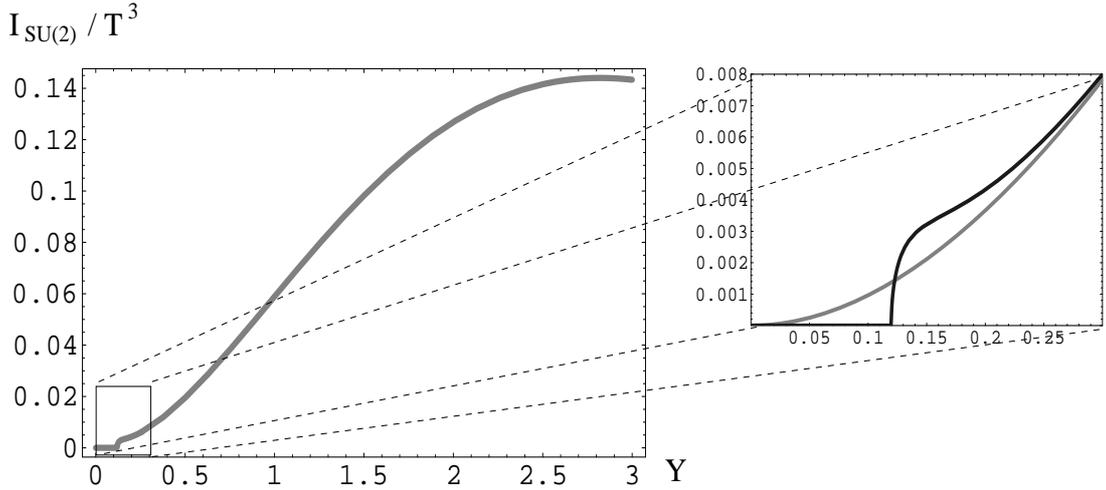}
\caption{\protect{\label{Fig-gap}}Dimensionless black-body spectral power $\frac{I_{SU(2)}}{T^3}$ as a function of the dimensionless
frequency $Y\equiv\frac{\omega}{T}$. The black curve in the magnified region depicts the
modification of the spectrum as compared to $\frac{I_{U(1)}}{T^3}$ (grey curve) for T = 10K \cite{Hofmann/Schwarz/Giacosa06-2}.}
\end{center}
\end{figure}

Subsequently, fascinating implications concerning the black body spectrum are arising from the modified photon propagation properties. Photon propagation subject to an $U(1)$ gauge symmetry leads to the well-known Planck law for the energy density of the black body radiation
\begin{equation}
\rho_{U(1)}=\int\!dp\,\frac{n_B(\beta\omega)}{\pi^2}\vec{p}^2\omega(\vec{p})\,.
\end{equation}
If photon propagation is considered in the framework of $SU(2)_{\tiny\mbox{CMB}}$ the energy density modifies to
\begin{equation}
\rho_{SU(2)}=\int\!d\omega\,\frac{n_B(\beta\omega)}{\pi^2}
\omega\left(\omega-\frac{1}{2}\frac{dG}{d\omega}\right)
\sqrt{\omega^2-G(\omega)}\,,
\end{equation}
where $\omega_1$ and $\omega_2$ denote the solutions of the equation $\omega^2-G(\omega)=0$. Photons of frequency $\omega_1\leq\omega\leq\omega_2$ possess an imaginary momentum modulus and can not propagate. This regime of strong screening causes a gap in the black body spectrum at low temperature \cite{Hofmann/Schwarz/Giacosa06-2}, depicted in Fig.\,\ref{Fig-gap}.

%% file: details1.tex
\chapter{Technical details concerning Chapter 3}

\section{Wick decomposition of the correlator}
This section contains the Wick decomposition of $\bra\Theta_{00}(x)\Theta_{00}(y)\ket$ into a sum of products two-point functions of the elementary gauge fields  $A_\mu(x)$. The evaluation is not a mathematical challenge, but nevertheless a cumbersome task. We emphasize that our focus is on the connected correlations, local diagrams are neglected.

\begin{flushleft}
$\bra\Theta_{00}(x)\Theta_{00}(y)\ket=$
\end{flushleft}
\begin{eqnarray}
&=& \bra (-F_{0\lambda}(x)F_0\,^\lambda(x)
          +\frac{g_{00}}{4}F_{\kappa\lambda}(x)F^{\kappa\lambda}(x))
          (-F_{0\tau}(y)F_0\,^\tau(y)
          +\frac{g_{00}}{4}F_{\sigma\tau}(y)F^{\sigma\tau}(y)) \ket
          \nonumber\\
&=& \bra F_{0\lambda}(x)F_0\,^\lambda(x)F_{0\tau}(y)F_0\,^\tau(y) \ket
   -\frac{g_{00}}{4}\bra F_{0\lambda}(x)F_0\,^\lambda(x)
    F_{\sigma\tau}(y)F^{\sigma\tau}(y)) \ket\nonumber\\
&& -\frac{g_{00}}{4}\bra F_{\kappa\lambda}(x)F^{\kappa\lambda}(x))
    F_{0\tau}(y)F_0\,^\tau(y) \ket
   +\frac{g_{00}^2}{16}\bra F_{\kappa\lambda}(x)F^{\kappa\lambda}(x))
    F_{\sigma\tau}(y)F^{\sigma\tau}(y)) \ket\nonumber\\
&=& \bra(\partial_{x^0}A^{\lambda}(x))(\partial_{x^0}A_{\lambda}(x))
              (\partial_{y^0}A^{\tau}(y))(\partial_{y^0}A_{\tau}(y))\ket
   -\bra(\partial_{x^0}A^{\lambda}(x))(\partial_{x^0}A_{\lambda}(x))
              (\partial_{y^0}A^{\tau}(y))(\partial_{y^\tau}A_{0}(y))\ket\nonumber\\
& &-\bra(\partial_{x^0}A^{\lambda}(x))(\partial_{x^0}A_{\lambda}(x))
              (\partial_{y_\tau}A_{0}(y))(\partial_{y^0}A_{\tau}(y))\ket
   +\bra(\partial_{x^0}A^{\lambda}(x))(\partial_{x^0}A_{\lambda}(x))
              (\partial_{y_\tau}A_{0}(y))(\partial_{y^\tau}A_{0}(y))\ket\nonumber\\
& &-\bra(\partial_{x^0}A^{\lambda}(x))(\partial_{x^\lambda}A_{0}(x))
              (\partial_{y^0}A^{\tau}(y))(\partial_{y^0}A_{\tau}(y))\ket
   +\bra(\partial_{x^0}A^{\lambda}(x))(\partial_{x^\lambda}A_{0}(x))
              (\partial_{y^0}A^{\tau}(y))(\partial_{y^\tau}A_{0}(y))\ket\nonumber\\
& &+\bra(\partial_{x^0}A^{\lambda}(x))(\partial_{x^\lambda}A_{0}(x))
              (\partial_{y_\tau}A_{0}(y))(\partial_{y^0}A_{\tau}(y))\ket
   -\bra(\partial_{x^0}A^{\lambda}(x))(\partial_{x^\lambda}A_{0}(x))
              (\partial_{y_\tau}A_{0}(y))(\partial_{y^\tau}A_{0}(y))\ket\nonumber\\
& &-\bra(\partial_{x_\lambda}A_{0}(x))(\partial_{x^0}A_{\lambda}(x))
              (\partial_{y^0}A^{\tau}(y))(\partial_{y^0}A_{\tau}(y))\ket
   +\bra(\partial_{x_\lambda}A_{0}(x))(\partial_{x^0}A_{\lambda}(x))
              (\partial_{y^0}A^{\tau}(y))(\partial_{y^\tau}A_{0}(y))\ket\nonumber\\
& &+\bra(\partial_{x_\lambda}A_{0}(x))(\partial_{x^0}A_{\lambda}(x))
              (\partial_{y_\tau}A_{0}(y))(\partial_{y^0}A_{\tau}(y))\ket
   -\bra(\partial_{x_\lambda}A_{0}(x))(\partial_{x^0}A_{\lambda}(x))
              (\partial_{y_\tau}A_{0}(y))(\partial_{y^\tau}A_{0}(y))\ket\nonumber\\
& &+\bra(\partial_{x_\lambda}A_{0}(x))(\partial_{x^\lambda}A_{0}(x))
              (\partial_{y^0}A^{\tau}(y))(\partial_{y^0}A_{\tau}(y))\ket
   -\bra(\partial_{x_\lambda}A_{0}(x))(\partial_{x^\lambda}A_{0}(x))
              (\partial_{y^0}A^{\tau}(y))(\partial_{y^\tau}A_{0}(y))\ket\nonumber\\
& &-\bra(\partial_{x_\lambda}A_{0}(x))(\partial_{x^\lambda}A_{0}(x))
              (\partial_{y_\tau}A_{0}(y))(\partial_{y^0}A_{\tau}(y))\ket
   +\bra(\partial_{x_\lambda}A_{0}(x))(\partial_{x^\lambda}A_{0}(x))
              (\partial_{y_\tau}A_{0}(y))(\partial_{y^\tau}A_{0}(y))\ket\nonumber\\
&+&\left(\frac{-g_{00}}{4}\right)\nonumber\\
& &\times[\bra(\partial_{x^0}A^{\lambda}(x))(\partial_{x^0}A_{\lambda}(x))
        (\partial_{y_\sigma}A^{\tau}(y))(\partial_{y^\sigma}A_{\tau}(y))\ket
   -\bra(\partial_{x^0}A^{\lambda}(x))(\partial_{x^0}A_{\lambda}(x))
        (\partial_{y_\sigma}A^{\tau}(y))(\partial_{y^\tau}A_{\sigma}(y))\ket\nonumber\\
& &-\bra(\partial_{x^0}A^{\lambda}(x))(\partial_{x^0}A_{\lambda}(x))
        (\partial_{y_\tau}A^{\sigma}(y))(\partial_{y^\sigma}A_{\tau}(y))\ket
   +\bra(\partial_{x^0}A^{\lambda}(x))(\partial_{x^0}A_{\lambda}(x))
        (\partial_{y_\tau}A^{\sigma}(y))(\partial_{y^\tau}A_{\sigma}(y))\ket\nonumber\\
& &-\bra(\partial_{x^0}A^{\lambda}(x))(\partial_{x^\lambda}A_{0}(x))
        (\partial_{y_\sigma}A^{\tau}(y))(\partial_{y^\sigma}A_{\tau}(y))\ket
   +\bra(\partial_{x^0}A^{\lambda}(x))(\partial_{x^\lambda}A_{0}(x))
        (\partial_{y_\sigma}A^{\tau}(y))(\partial_{y^\tau}A_{\sigma}(y))\ket\nonumber\\
& &+\bra(\partial_{x^0}A^{\lambda}(x))(\partial_{x^\lambda}A_{0}(x))
        (\partial_{y_\tau}A^{\sigma}(y))(\partial_{y^\sigma}A_{\tau}(y))\ket
   -\bra(\partial_{x^0}A^{\lambda}(x))(\partial_{x^\lambda}A_{0}(x))
        (\partial_{y_\tau}A^{\sigma}(y))(\partial_{y^\tau}A_{\sigma}(y))\ket\nonumber\\
& &-\bra(\partial_{x_\lambda}A_{0}(x))(\partial_{x^0}A_{\lambda}(x))
        (\partial_{y_\sigma}A_{\tau}(y))(\partial_{y^\sigma}A_{\tau}(y))\ket
   +\bra(\partial_{x_\lambda}A_{0}(x))(\partial_{x^0}A_{\lambda}(x))
        (\partial_{y_\sigma}A_{\tau}(y))(\partial_{y^\tau}A_{\sigma}(y))\ket\nonumber\\
& &+\bra(\partial_{x_\lambda}A_{0}(x))(\partial_{x^0}A_{\lambda}(x))
        (\partial_{y_\tau}A_{\sigma}(y))(\partial_{y^\sigma}A_{\tau}(y))\ket
   -\bra(\partial_{x_\lambda}A_{0}(x))(\partial_{x^0}A_{\lambda}(x))
        (\partial_{y_\tau}A_{\sigma}(y))(\partial_{y^\tau}A_{\sigma}(y))\ket\nonumber\\
& &+\bra(\partial_{x_\lambda}A_{0}(x))(\partial_{x^\lambda}A_{0}(x))
        (\partial_{y_\sigma}A_{\tau}(y))(\partial_{y^\sigma}A_{\tau}(y))\ket
   -\bra(\partial_{x_\lambda}A_{0}(x))(\partial_{x^\lambda}A_{0}(x))
        (\partial_{y_\sigma}A_{\tau}(y))(\partial_{y^\tau}A_{\sigma}(y))\ket\nonumber\\
& &-\bra(\partial_{x_\lambda}A_{0}(x))(\partial_{x^\lambda}A_{0}(x))
        (\partial_{y_\tau}A_{\sigma}(y))(\partial_{y^\sigma}A_{\tau}(y))\ket
   +\bra(\partial_{x_\lambda}A_{0}(x))(\partial_{x^\lambda}A_{0}(x))
        (\partial_{y_\tau}A_{\sigma}(y))(\partial_{y^\tau}A_{\sigma}(y))\ket]\nonumber
\end{eqnarray}
\newpage
\begin{eqnarray}
&+&\left(\frac{-g_{00}}{4}\right)\nonumber\\
& &\times[\bra(\partial_{x_\kappa}A^{\lambda}(x))(\partial_{x^\kappa}A_{\lambda}(x))
        (\partial_{y^0}A^{\tau}(y))(\partial_{y^0}A_{\tau}(y))\ket
   -\bra(\partial_{x_\kappa}A^{\lambda}(x))(\partial_{x^\kappa}A_{\lambda}(x))
        (\partial_{y^0}A^{\tau}(y))(\partial_{y^\tau}A_{0}(y))\ket\nonumber\\
& &-\bra(\partial_{x_\kappa}A^{\lambda}(x))(\partial_{x^\kappa}A_{\lambda}(x))
        (\partial_{y_\tau}A_{0}(y))(\partial_{y^0}A_{\tau}(y))\ket
   +\bra(\partial_{x_\kappa}A^{\lambda}(x))(\partial_{x^\kappa}A_{\lambda}(x))
        (\partial_{y_\tau}A_{0}(y))(\partial_{y^\tau}A_{0}(y))\ket\nonumber\\
& &-\bra(\partial_{x_\kappa}A^{\lambda}(x))(\partial_{x^\lambda}A_{\kappa}(x))
        (\partial_{y^0}A^{\tau}(y))(\partial_{y^0}A_{\tau}(y))\ket
   +\bra(\partial_{x_\kappa}A^{\lambda}(x))(\partial_{x^\lambda}A_{\kappa}(x))
        (\partial_{y^0}A^{\tau}(y))(\partial_{y^\tau}A_{0}(y))\ket\nonumber\\
& &+\bra(\partial_{x_\kappa}A^{\lambda}(x))(\partial_{x^\lambda}A_{\kappa}(x))
        (\partial_{y_\tau}A_{0}(y))(\partial_{y^0}A_{\tau}(y))\ket
   -\bra(\partial_{x_\kappa}A^{\lambda}(x))(\partial_{x^\lambda}A_{\kappa}(x))
        (\partial_{y_\tau}A_{0}(y))(\partial_{y^\tau}A_{0}(y))\ket\nonumber\\
& &-\bra(\partial_{x_\lambda}A^{\kappa}(x))(\partial_{x^\kappa}A_{\lambda}(x))
        (\partial_{y^0}A^{\tau}(y))(\partial_{y^0}A_{\tau}(y))\ket
   +\bra(\partial_{x_\lambda}A^{\kappa}(x))(\partial_{x^\kappa}A_{\lambda}(x))
        (\partial_{y^0}A^{\tau}(y))(\partial_{y^\tau}A_{0}(y))\ket\nonumber\\
& &+\bra(\partial_{x_\lambda}A^{\kappa}(x))(\partial_{x^\kappa}A_{\lambda}(x))
        (\partial_{y_\tau}A_{0}(y))(\partial_{y^0}A_{\tau}(y))\ket
   -\bra(\partial_{x_\lambda}A^{\kappa}(x))(\partial_{x^\kappa}A_{\lambda}(x))
        (\partial_{y_\tau}A_{0}(y))(\partial_{y^\tau}A_{0}(y))\ket\nonumber\\
& &+\bra(\partial_{x_\lambda}A^{\kappa}(x))(\partial_{x^\lambda}A_{\kappa}(x))
        (\partial_{y^0}A^{\tau}(y))(\partial_{y^0}A_{\tau}(y))\ket
   -\bra(\partial_{x_\lambda}A^{\kappa}(x))(\partial_{x^\lambda}A_{\kappa}(x))
        (\partial_{y^0}A^{\tau}(y))(\partial_{y^\tau}A_{0}(y))\ket\nonumber\\
& &-\bra(\partial_{x_\lambda}A^{\kappa}(x))(\partial_{x^\lambda}A_{\kappa}(x))
        (\partial_{y_\tau}A_{0}(y))(\partial_{y^0}A_{\tau}(y))\ket
   +\bra(\partial_{x_\lambda}A^{\kappa}(x))(\partial_{x^\lambda}A_{\kappa}(x))
        (\partial_{y_\tau}A_{0}(y))(\partial_{y^\tau}A_{0}(y))\ket]\nonumber\\
&+&\left(\frac{g_{00}^2}{16}\right)\nonumber\\
& &\times[\bra(\partial_{x_\kappa}A^{\lambda}(x))(\partial_{x^\kappa}A_{\lambda}(x))
        (\partial_{y_\sigma}A^{\tau}(y))(\partial_{y^\sigma}A_{\tau}(y))\ket
   -\bra(\partial_{x_\kappa}A^{\lambda}(x))(\partial_{x^\kappa}A_{\lambda}(x))
        (\partial_{y_\sigma}A^{\tau}(y))(\partial_{y^\tau}A_{\sigma}(y))\ket\nonumber\\
& &-\bra(\partial_{x_\kappa}A^{\lambda}(x))(\partial_{x^\kappa}A_{\lambda}(x))
        (\partial_{y_\tau}A^{\sigma}(y))(\partial_{y^\sigma}A_{\tau}(y))\ket
   +\bra(\partial_{x_\kappa}A^{\lambda}(x))(\partial_{x^\kappa}A_{\lambda}(x))
        (\partial_{y_\tau}A^{\sigma}(y))(\partial_{y^\tau}A_{\sigma}(y))\ket\nonumber\\
& &-\bra(\partial_{x_\kappa}A^{\lambda}(x))(\partial_{x^\lambda}A_{\kappa}(x))
        (\partial_{y_\sigma}A^{\tau}(y))(\partial_{y^\sigma}A_{\tau}(y))\ket
   +\bra(\partial_{x_\kappa}A^{\lambda}(x))(\partial_{x^\lambda}A_{\kappa}(x))
        (\partial_{y_\sigma}A^{\tau}(y))(\partial_{y^\tau}A_{\sigma}(y))\ket\nonumber\\
& &+\bra(\partial_{x_\kappa}A^{\lambda}(x))(\partial_{x^\lambda}A_{\kappa}(x))
        (\partial_{y_\tau}A^{\sigma}(y))(\partial_{y^\sigma}A_{\tau}(y))\ket
   -\bra(\partial_{x_\kappa}A^{\lambda}(x))(\partial_{x^\lambda}A_{\kappa}(x))
        (\partial_{y_\tau}A^{\sigma}(y))(\partial_{y^\tau}A_{\sigma}(y))\ket\nonumber\\
& &-\bra(\partial_{x_\lambda}A^{\kappa}(x))(\partial_{x^\kappa}A_{\lambda}(x))
        (\partial_{y_\sigma}A^{\tau}(y))(\partial_{y^\sigma}A_{\tau}(y))\ket
   +\bra(\partial_{x_\lambda}A^{\kappa}(x))(\partial_{x^\kappa}A_{\lambda}(x))
        (\partial_{y_\sigma}A^{\tau}(y))(\partial_{y^\tau}A_{\sigma}(y))\ket\nonumber\\
& &+\bra(\partial_{x_\lambda}A^{\kappa}(x))(\partial_{x^\kappa}A_{\lambda}(x))
        (\partial_{y_\tau}A^{\sigma}(y))(\partial_{y^\sigma}A_{\tau}(y))\ket
   -\bra(\partial_{x_\lambda}A^{\kappa}(x))(\partial_{x^\kappa}A_{\lambda}(x))
        (\partial_{y_\tau}A^{\sigma}(y))(\partial_{y^\tau}A_{\sigma}(y))\ket\nonumber\\
& &+\bra(\partial_{x_\lambda}A^{\kappa}(x))(\partial_{x^\lambda}A_{\kappa}(x))
        (\partial_{y_\sigma}A^{\tau}(y))(\partial_{y^\sigma}A_{\tau}(y))\ket
   -\bra(\partial_{x_\lambda}A^{\kappa}(x))(\partial_{x^\lambda}A_{\kappa}(x))
        (\partial_{y_\sigma}A^{\tau}(y))(\partial_{y^\tau}A_{\sigma}(y))\ket\nonumber\\
& &-\bra(\partial_{x_\lambda}A^{\kappa}(x))(\partial_{x^\lambda}A_{\kappa}(x))
        (\partial_{y_\tau}A^{\sigma}(y))(\partial_{y^\sigma}A_{\tau}(y))\ket
   +\bra(\partial_{x_\lambda}A^{\kappa}(x))(\partial_{x^\lambda}A_{\kappa}(x))
        (\partial_{y_\tau}A^{\sigma}(y))(\partial_{y^\tau}A_{\sigma}(y))\ket]\nonumber\\
\end{eqnarray}
\begin{eqnarray*}
&=&\bra(\partial_{x^0}A^{\lambda}(x))(\partial_{x^0}A_{\lambda}(x))\ket
   \bra(\partial_{y^0}A^{\tau}(y))(\partial_{y^0}A_{\tau}(y))\ket
   +2\bra(\partial_{x^0}A^{\lambda}(x))(\partial_{y^0}A^{\tau}(y))\ket
   \bra(\partial_{x^0}A_{\lambda}(x))(\partial_{y^0}A_{\tau}(y))\ket\\
& &-\bra(\partial_{x^0}A^{\lambda}(x))(\partial_{x^0}A_{\lambda}(x))\ket
   \bra(\partial_{y^0}A^{\tau}(y))(\partial_{y_\tau}A_{0}(y))\ket
   -4\bra(\partial_{x^0}A^{\lambda}(x))(\partial_{y^0}A^{\tau}(y))\ket
   \bra(\partial_{x^0}A_{\lambda}(x))(\partial_{y^\tau}A_{0}(y))\ket\\
& &-\bra(\partial_{x^0}A^{\lambda}(x))(\partial_{x^0}A_{\lambda}(x))\ket
   \bra(\partial_{y_\tau}A_{0}(y))(\partial_{y^0}A_{\tau}(y))\ket\\
& &+\bra(\partial_{x^0}A^{\lambda}(x))(\partial_{x^0}A_{\lambda}(x))\ket
   \bra(\partial_{y_\tau}A_{0}(y))(\partial_{y^\tau}A_{0}(y))\ket
   +2\bra(\partial_{x^0}A^{\lambda}(x))(\partial_{y_\tau}A_{0}(y))\ket
   \bra(\partial_{x^0}A_{\lambda}(x))(\partial_{y^\tau}A_{0}(y))\ket\\
& &-\bra(\partial_{x^0}A^{\lambda}(x))(\partial_{x^\lambda}A_{0}(x))\ket
   \bra(\partial_{y^0}A^{\tau}(y))(\partial_{y^0}A_{\tau}(y))\ket
   -2\bra(\partial_{x^0}A^{\lambda}(x))(\partial_{y^0}A^{\tau}(y))\ket
   \bra(\partial_{x^\lambda}A_{0}(x))(\partial_{y^0}A_{\tau}(y))\ket\\
& &+\bra(\partial_{x^0}A^{\lambda}(x))(\partial_{x^\lambda}A_{0}(x))\ket
   \bra(\partial_{y^0}A^{\tau}(y))(\partial_{y^\tau}A_{0}(y))\ket
   +2\bra(\partial_{x^0}A^{\lambda}(x))(\partial_{y^0}A_{\tau}(y))\ket
   \bra(\partial_{x^\lambda}A_{0}(x))(\partial_{y^\tau}A_{0}(y))\ket\\
& &+\bra(\partial_{x^0}A^{\lambda}(x))(\partial_{x^\lambda}A_{0}(x))\ket
   \bra(\partial_{y_\tau}A_{0}(y))(\partial_{y^0}A_{\tau}(y))\ket
   +2\bra(\partial_{x^0}A^{\lambda}(x))(\partial_{y_\tau}A_{0}(y))\ket
   \bra(\partial_{x^\lambda}A_{0}(x))(\partial_{y^0}A_{\tau}(y))\ket\\
& &-\bra(\partial_{x^0}A^{\lambda}(x))(\partial_{x^\lambda}A_{0}(x))\ket
   \bra(\partial_{y_\tau}A_{0}(y))(\partial_{y^\tau}A_{0}(y))\ket
   -2\bra(\partial_{x^0}A^{\lambda}(x))(\partial_{y_\tau}A_{0}(y))\ket
   \bra(\partial_{x^\lambda}A_{0}(x))(\partial_{y^\tau}A_{0}(y))\ket\\
& &-\bra(\partial_{x_\lambda}A_{0}(x))(\partial_{x^0}A_{\lambda}(x))\ket
   \bra(\partial_{y^0}A^{\tau}(y))(\partial_{y^0}A_{\tau}(y)\ket
   -2\bra(\partial_{x_\lambda}A_{0}(x))(\partial_{y^0}A^{\tau}(y))\ket
   \bra(\partial_{x^0}A_{\lambda}(x))(\partial_{y^0}A_{\tau}(y))\ket\\
& &+\bra(\partial_{x_\lambda}A_{0}(x))(\partial_{x^0}A_{\lambda}(x))\ket
   \bra(\partial_{y^0}A^{\tau}(y))(\partial_{y^\tau}A_{0}(y)\ket
   +2\bra(\partial_{x_\lambda}A_{0}(x))(\partial_{y_\tau}A_{0}(y))\ket
   \bra(\partial_{x^0}A_{\lambda}(x))(\partial_{y^0}A_{\tau}(y))\ket\\
& &+\bra(\partial_{x_\lambda}A_{0}(x))(\partial_{x^0}A_{\lambda}(x))\ket
   \bra(\partial_{y_\tau}A_{0}(y))(\partial_{y^0}A_{\tau}(y)\ket
   +2\bra(\partial_{x_\lambda}A_{0}(x))(\partial_{y^0}A^{\tau}(y))\ket
   \bra(\partial_{x^0}A_{\lambda}(x))(\partial_{y^\tau}A_{0}(y))\ket\\
& &-\bra(\partial_{x_\lambda}A_{0}(x))(\partial_{x^0}A_{\lambda}(x))\ket
   \bra(\partial_{y_\tau}A_{0}(y))(\partial_{y^\tau}A_{0}(y)\ket
   -2\bra(\partial_{x_\lambda}A_{0}(x))(\partial_{y_\tau}A_{0}(y))\ket
   \bra(\partial_{x^0}A_{\lambda}(x))(\partial_{y^\tau}A_{0}(y))\ket\\
& &+\bra(\partial_{x_\lambda}A_{0}(x))(\partial_{x^\lambda}A_{0}(x))\ket
   \bra(\partial_{y^0}A^{\tau}(y))(\partial_{y^0}A_{\tau}(y)\ket
   +2\bra(\partial_{x_\lambda}A_{0}(x))(\partial_{y^0}A^{\tau}(y))\ket
   \bra(\partial_{x^\lambda}A_{0}(x))(\partial_{y^0}A_{\tau}(y))\ket\\
& &-\bra(\partial_{x_\lambda}A_{0}(x))(\partial_{x^\lambda}A_{0}(x))\ket
   \bra(\partial_{y^0}A^{\tau}(y))(\partial_{y^\tau}A_{0}(y))\ket
   -4\bra(\partial_{x_\lambda}A_{0}(x))(\partial_{y^0}A^{\tau}(y))\ket
   \bra(\partial_{x^\lambda}A_{0}(x))(\partial_{y^\tau}A_{0}(y))\ket\\
& &-\bra(\partial_{x_\lambda}A_{0}(x))(\partial_{x^\lambda}A_{0}(x))\ket
   \bra(\partial_{y_\tau}A_{0}(y))(\partial_{y^0}A_{\tau}(y))\ket\\
& &+\bra(\partial_{x_\lambda}A_{0}(x))(\partial_{x^\lambda}A_{0}(x))\ket
   \bra(\partial_{y_\tau}A_{0}(y))(\partial_{y^\tau}A_{0}(y))\ket
   +2\bra(\partial_{x_\lambda}A_{0}(x))(\partial_{y_\tau}A_{0}(y))\ket
   \bra(\partial_{x^\lambda}A_{0}(x))(\partial_{y^\tau}A_{0}(y))\ket\\
\end{eqnarray*}
\newpage
\begin{eqnarray*}
&-&\left(\frac{g_{00}}{4}\right)\\
& &\times\left[\bra(\partial_{x^0}A^{\lambda}(x))(\partial_{x^0}A_{\lambda}(x))\ket
   \bra(\partial_{y_\sigma}A^{\tau}(y))(\partial_{y^\sigma}A_{\tau}(y))\ket
   +2\bra(\partial_{x^0}A^{\lambda}(x))(\partial_{y_\sigma}A^{\tau}(y))\ket
   \bra(\partial_{x^0}A_{\lambda}(x))(\partial_{y^\sigma}A_{\tau}(y))\ket\right.\\
& &-\bra(\partial_{x^0}A^{\lambda}(x))(\partial_{x^0}A_{\lambda}(x))\ket
   \bra(\partial_{y_\sigma}A^{\tau}(y))(\partial_{y^\tau}A_{\sigma}(y))\ket
   -4\bra(\partial_{x^0}A^{\lambda}(x))(\partial_{y_\tau}A^{\sigma}(y))\ket
   \bra(\partial_{x^0}A_{\lambda}(x))(\partial_{y^\sigma}A_{\tau}(y))\ket\\
& &-\bra(\partial_{x^0}A^{\lambda}(x))(\partial_{x^0}A_{\lambda}(x))\ket
   \bra(\partial_{y_\tau}A^{\sigma}(y))(\partial_{y^\sigma}A_{\tau}(y))\ket\\
& &+\bra(\partial_{x^0}A^{\lambda}(x))(\partial_{x^0}A_{\lambda}(x))\ket
   \bra(\partial_{y_\tau}A^{\sigma}(y))(\partial_{y^\tau}A_{\sigma}(y))\ket
   +2\bra(\partial_{x^0}A^{\lambda}(x))(\partial_{y_\tau}A^{\sigma}(y))\ket
   \bra(\partial_{x^0}A^{\lambda}(x))(\partial_{y^\tau}A_{\sigma}(y))\ket\\
& &-\bra(\partial_{x^0}A^{\lambda}(x))(\partial_{x^\lambda}A_{0}(x))\ket
   \bra(\partial_{y_\sigma}A^{\tau}(y))(\partial_{y^\sigma}A_{\tau}(y))\ket
   -2\bra(\partial_{x^0}A^{\lambda}(x))(\partial_{y_\sigma}A^{\tau}(y))\ket
   \bra(\partial_{x^\lambda}A_{0}(x))(\partial_{y^\sigma}A_{\tau}(y))\ket\\
& &+\bra(\partial_{x^0}A^{\lambda}(x))(\partial_{x^\lambda}A_{0}(x))\ket
   \bra(\partial_{y_\sigma}A^{\tau}(y))(\partial_{y^\tau}A_{\sigma}(y))\ket
   +2\bra(\partial_{x^0}A^{\lambda}(x))(\partial_{y_\tau}A^{\sigma}(y))\ket
   \bra(\partial_{x^\lambda}A_{0}(x))(\partial_{y^\sigma}A_{\tau}(y))\ket\\
& &+\bra(\partial_{x^0}A^{\lambda}(x))(\partial_{x^\lambda}A_{0}(x))\ket
   \bra(\partial_{y_\tau}A^{\sigma}(y))(\partial_{y^\sigma}A_{\tau}(y))\ket
   +2\bra(\partial_{x^0}A^{\lambda}(x))(\partial_{y_\sigma}A^{\tau}(y))\ket
   \bra(\partial_{x^\lambda}A_{0}(x))(\partial_{y^\tau}A_{\sigma}(y))\ket\\
& &-\bra(\partial_{x^0}A^{\lambda}(x))(\partial_{x^\lambda}A_{0}(x))\ket
   \bra(\partial_{y_\tau}A^{\sigma}(y))(\partial_{y^\tau}A_{\sigma}(y))\ket
   -2\bra(\partial_{x^0}A_{\lambda}(x))(\partial_{y_\tau}A^{\sigma}(y))\ket
   \bra(\partial_{x^\lambda}A_{0}(x))(\partial_{y^\tau}A_{\sigma}(y))\ket\\
& &-\bra(\partial_{x_\lambda}A_{0}(x))(\partial_{x^0}A_{\lambda}(x))\ket
   \bra(\partial_{y_\sigma}A^{\tau}(y))(\partial_{y^\sigma}A_{\tau}(y))\ket
   -2\bra(\partial_{x_\lambda}A_{0}(x))(\partial_{y_\sigma}A^{\tau}(y))\ket
   \bra(\partial_{x^0}A_{\lambda}(x))(\partial_{y^\sigma}A_{\tau}(y))\ket\\
& &+\bra(\partial_{x_\lambda}A_{0}(x))(\partial_{x^0}A_{\lambda}(x))\ket
   \bra(\partial_{y_\sigma}A^{\tau}(y))(\partial_{y^\tau}A_{\sigma}(y))\ket
   +2\bra(\partial_{x_\lambda}A_{0}(x))(\partial_{y_\tau}A^{\sigma}(y))\ket
   \bra(\partial_{x^0}A_{\lambda}(x))(\partial_{y^\sigma}A_{\tau}(y))\ket\\
& &+\bra(\partial_{x_\lambda}A_{0}(x))(\partial_{x^0}A_{\lambda}(x))\ket
   \bra(\partial_{y_\tau}A^{\sigma}(y))(\partial_{y^\sigma}A_{\tau}(y))\ket
   +2\bra(\partial_{x_\lambda}A_{0}(x))(\partial_{y\sigma}A^{\tau}(y))\ket
   \bra(\partial_{x^0}A_{\lambda}(x))(\partial_{y^\tau}A_{\sigma}(y))\ket\\
& &-\bra(\partial_{x_\lambda}A_{0}(x))(\partial_{x^0}A_{\lambda}(x))\ket
   \bra(\partial_{y_\tau}A^{\sigma}(y))(\partial_{y^\tau}A_{\sigma}(y))\ket
   -2\bra(\partial_{x_\lambda}A_{0}(x))(\partial_{y_\tau}A^{\sigma}(y))\ket
   \bra(\partial_{x^0}A_{\lambda}(x))(\partial_{y^\tau}A_{\sigma}(y))\ket\\
& &+\bra(\partial_{x_\lambda}A_{0}(x))(\partial_{x^\lambda}A_{0}(x))\ket
   \bra(\partial_{y_\sigma}A^{\tau}(y))(\partial_{y^\sigma}A_{\tau}(y))\ket
   +2\bra(\partial_{x_\lambda}A_{0}(x))(\partial_{y\sigma}A^{\tau}(y))\ket
   \bra(\partial_{x^\lambda}A_{0}(x))(\partial_{y^\sigma}A_{\tau}(y))\ket\\
& &-\bra(\partial_{x_\lambda}A_{0}(x))(\partial_{x^\lambda}A_{0}(x))\ket
   \bra(\partial_{y_\sigma}A_{\tau}(y))(\partial_{y^\tau}A_{\sigma}(y))\ket
   -4\bra(\partial_{x_\lambda}A_{0}(x))(\partial_{y_\tau}A^{\sigma}(y))\ket
   \bra(\partial_{x^\lambda}A_{0}(x))(\partial_{y^\sigma}A_{\tau}(y))\ket\\
& &-\bra(\partial_{x_\lambda}A_{0}(x))(\partial_{x^\lambda}A_{0}(x))\ket
   \bra(\partial_{y_\tau}A^{\sigma}(y))(\partial_{y^\sigma}A_{\tau}(y))\ket\\
& &\left.+\bra(\partial_{x_\lambda}A_{0}(x))(\partial_{x^\lambda}A_{0}(x))\ket
   \bra(\partial_{y_\tau}A^{\sigma}(y))(\partial_{y^\tau}A_{\sigma}(y))\ket
   +2\bra(\partial_{x_\lambda}A_{0}(x))(\partial_{y_\tau}A_{\sigma}(y))\ket
   \bra(\partial_{x^\lambda}A_{0}(x))(\partial_{y^\tau}A_{\sigma}(y))\ket\right]\\
&-&\left(\frac{g_{00}}{4}\right)\\
& &\times\left[\bra(\partial_{x_\kappa}A^{\lambda}(x))(\partial_{x^\kappa}A_{\lambda}(x))\ket
   \bra(\partial_{y^0}A^{\tau}(y))(\partial_{y^0}A_{\tau}(y))\ket
   +2\bra(\partial_{x_\kappa}A^{\lambda}(x))(\partial_{y^0}A^{\tau}(y))\ket
   \bra(\partial_{x^\kappa}A_{\lambda}(x))(\partial_{y^0}A_{\tau}(y))\ket\right.\\
& &-\bra(\partial_{x_\kappa}A^{\lambda}(x))(\partial_{x^\kappa}A_{\lambda}(x))\ket
   \bra(\partial_{y^0}A^{\tau}(y))(\partial_{y\tau}A_{0}(y))\ket
   -4\bra(\partial_{x\kappa}A^{\lambda}(x))(\partial_{y_\tau}A_{0}(y))\ket
   \bra(\partial_{x^\kappa}A_{\lambda}(x))(\partial_{y^0}A_{\tau}(y))\ket\\
& &-\bra(\partial_{x_\kappa}A^{\lambda}(x))(\partial_{x^\kappa}A_{\lambda}(x))\ket
   \bra(\partial_{y_\tau}A_{0}(y))(\partial_{y^0}A_{\tau}(y))\ket\\
& &+\bra(\partial_{x_\kappa}A^{\lambda}(x))(\partial_{x^\kappa}A_{\lambda}(x))\ket
   \bra(\partial_{y_\tau}A_{0}(y))(\partial_{y^\tau}A_{0}(y))\ket
   +2\bra(\partial_{x_\kappa}A^{\lambda}(x))(\partial_{y_\tau}A_{0}(y))\ket
   \bra(\partial_{x^\kappa}A_{\lambda}(x))(\partial_{y^\tau}A_{0}(y))\ket\\
& &-\bra(\partial_{x_\kappa}A^{\lambda}(x))(\partial_{x^\lambda}A_{\kappa}(x))\ket
   \bra(\partial_{y^0}A^{\tau}(y))(\partial_{y^0}A_{\tau}(y))\ket
   -2\bra(\partial_{x_\kappa}A^{\lambda}(x))(\partial_{y^0}A^{\tau}(y))\ket
   \bra(\partial_{x^\lambda}A_{\kappa}(x))(\partial_{y^0}A_{\tau}(y))\ket\\
& &+\bra(\partial_{x_\kappa}A^{\lambda}(x))(\partial_{x^\lambda}A_{\kappa}(x))\ket
   \bra(\partial_{y^0}A^{\tau}(y))(\partial_{y^\tau}A_{0}(y))\ket
   +2\bra(\partial_{x_\kappa}A^{\lambda}(x))(\partial_{y_\tau}A_{0}(y))\ket
   \bra(\partial_{x^\lambda}A_{\kappa}(x))(\partial_{y^0}A_{\tau}(y))\ket\\
& &+\bra(\partial_{x_\kappa}A^{\lambda}(x))(\partial_{x^\lambda}A_{\kappa}(x))\ket
   \bra(\partial_{y_\tau}A_{0}(y))(\partial_{y^0}A_{\tau}(y))\ket
   +2\bra(\partial_{x\kappa}A^{\lambda}(x))(\partial_{y^0}A^{\tau}(y))\ket
   \bra(\partial_{x^\lambda}A_{\kappa}(x))(\partial_{y^\tau}A_{0}(y))\ket\\
& &-\bra(\partial_{x_\kappa}A^{\lambda}(x))(\partial_{x^\lambda}A_{\kappa}(x))\ket
   \bra(\partial_{y_\tau}A_{0}(y))(\partial_{y^\tau}A_{0}(y))\ket
   -2\bra(\partial_{x_\kappa}A^{\lambda}(x))(\partial_{y\tau}A_{0}(y))\ket
   \bra(\partial_{x^\lambda}A_{\kappa}(x))(\partial_{y^\tau}A_{0}(y))\ket\\
& &-\bra(\partial_{x_\lambda}A^{\kappa}(x))(\partial_{x^\kappa}A_{\lambda}(x))\ket
   \bra(\partial_{y^0}A^{\tau}(y))(\partial_{y^0}A_{\tau}(y))\ket
   -2\bra(\partial_{x_\lambda}A^{\kappa}(x))(\partial_{y^0}A^{\tau}(y))\ket
   \bra(\partial_{x^\kappa}A_{\lambda}(x))(\partial_{y^0}A_{\tau}(y))\ket\\
& &+\bra(\partial_{x_\lambda}A^{\kappa}(x))(\partial_{x^\kappa}A_{\lambda}(x))\ket
   \bra(\partial_{y^0}A^{\tau}(y))(\partial_{y^\tau}A_{0}(y))\ket
   +2\bra(\partial_{x_\lambda}A^{\kappa}(x))(\partial_{y_\tau}A_{0}(y))\ket
   \bra(\partial_{x^0}A_{\lambda}(x))(\partial_{y^0}A_{\tau}(y))\ket\\
& &+\bra(\partial_{x_\lambda}A^{\kappa}(x))(\partial_{x^\kappa}A_{\lambda}(x))\ket
   \bra(\partial_{y_\tau}A_{0}(y))(\partial_{y^0}A_{\tau}(y))\ket
   +2\bra(\partial_{x_\lambda}A^{\kappa}(x))(\partial_{y^0}A^{\tau}(y))\ket
   \bra(\partial_{x^0}A_{\lambda}(x))(\partial_{y^\tau}A_{0}(y))\nonumber\ket\\
& &-\bra(\partial_{x_\lambda}A^{\kappa}(x))(\partial_{x^\kappa}A_{\lambda}(x))\ket
   \bra(\partial_{y_\tau}A_{0}(y))(\partial_{y^\tau}A_{0}(y))\ket
   -2\bra(\partial_{x_\lambda}A^{\kappa}(x))(\partial_{y_\tau}A_{0}(y))\ket
   \bra(\partial_{x^0}A_{\lambda}(x))(\partial_{y^\tau}A_{0}(y))\ket\nonumber\\
& &+\bra(\partial_{x_\lambda}A^{\kappa}(x))(\partial_{x^\lambda}A_{\kappa}(x))\ket
   \bra(\partial_{y^0}A^{\tau}(y))(\partial_{y^0}A_{\tau}(y))\ket
   +2\bra(\partial_{x_\lambda}A^{\kappa}(x))(\partial_{y^0}A^{\tau}(y))\ket
   \bra(\partial_{x^\lambda}A_{\kappa}(x))(\partial_{y^0}A_{\tau}(y))\ket\nonumber\\
& &-\bra(\partial_{x_\lambda}A^{\kappa}(x))(\partial_{x^\lambda}A_{\kappa}(x))\ket
   \bra(\partial_{y^0}A^{\tau}(y))(\partial_{y^\tau}A_{0}(y))\ket
   -4\bra(\partial_{x_\lambda}A^{\kappa}(x))(\partial_{y^0}A^{\tau}(y))\ket
   \bra(\partial_{x^\lambda}A_{\kappa}(x))(\partial_{y^\tau}A_{0}(y))\ket\nonumber\\
& &-\bra(\partial_{x_\lambda}A^{\kappa}(x))(\partial_{x^\lambda}A_{\kappa}(x))\ket\
   \bra(\partial_{y_\tau}A_{0}(y))(\partial_{y^0}A_{\tau}(y))\ket\nonumber\\
& &\left.+\bra(\partial_{x_\lambda}A^{\kappa}(x))(\partial_{x^\lambda}A_{\kappa}(x))\ket
   \bra(\partial_{y_\tau}A_{0}(y))(\partial_{y^\tau}A_{0}(y))\ket
   +2\bra(\partial_{x_\lambda}A^{\kappa}(x))(\partial_{y_\tau}A_{0}(y))\ket
   \bra(\partial_{x^\lambda}A_{\kappa}(x))(\partial_{y^\tau}A_{0}(y))\ket\right]
   \nonumber\\
&+&\left(\frac{g_{00}^2}{16}\right)\nonumber\\
&&\times\left[\bra(\partial_{x_\kappa}A^{\lambda}(x))(\partial_{x^\kappa}A_{\lambda}(x))\ket
   \bra(\partial_{y_\sigma}A^{\tau}(y))(\partial_{y^\sigma}A_{\tau}(y))\ket
   +2\bra(\partial_{x_\kappa}A^{\lambda}(x))(\partial_{y_\sigma}A^{\tau}(y))\ket
   \bra(\partial_{x^\kappa}A_{\lambda}(x))(\partial_{y^\sigma}A_{\tau}(y))\ket\right.\nonumber\\
& &-\bra(\partial_{x_\kappa}A^{\lambda}(x))(\partial_{x^\kappa}A_{\lambda}(x))\ket
   \bra(\partial_{y_\sigma}A^{\tau}(y))(\partial_{y^\tau}A_{\sigma}(y))\ket
   -4\bra(\partial_{x_\kappa}A^{\lambda}(x))(\partial_{y_\tau}A^{\sigma}(y))\ket
   \bra(\partial_{x^\kappa}A_{\lambda}(x))(\partial_{y^\sigma}A_{\tau}(y))\ket\nonumber\\
& &-\bra(\partial_{x_\kappa}A^{\lambda}(x))(\partial_{x^\kappa}A_{\lambda}(x))\ket
   \bra(\partial_{y_\tau}A^{\sigma}(y))(\partial_{y^\sigma}A_{\tau}(y))\ket\nonumber\\
\end{eqnarray*}
\newpage
\begin{eqnarray}
\label{WICK}
& &+\bra(\partial_{x_\kappa}A^{\lambda}(x))(\partial_{x^\kappa}A_{\lambda}(x))\ket
   \bra(\partial_{y_\tau}A^{\sigma}(y))(\partial_{y^\tau}A_{\sigma}(y))\ket
   +2\bra(\partial_{x_\kappa}A^{\lambda}(x))(\partial_{y_\tau}A^{\sigma}(y))\ket
   \bra(\partial_{x^\kappa}A_{\lambda}(x))(\partial_{y^\tau}A_{\sigma}(y))\ket\nonumber\\
& &-\bra(\partial_{x_\kappa}A^{\lambda}(x))(\partial_{x^\lambda}A_{\kappa}(x))\ket
   \bra(\partial_{y_\sigma}A^{\tau}(y))(\partial_{y^\sigma}A_{\tau}(y))\ket
   -2\bra(\partial_{x_\kappa}A^{\lambda}(x))(\partial_{y_\sigma}A^{\tau}(y))\ket
   \bra(\partial_{x^\lambda}A_{\kappa}(x))(\partial_{y^\sigma}A_{\tau}(y))\ket\nonumber\\
& &+\bra(\partial_{x_\kappa}A^{\lambda}(x))(\partial_{x^\lambda}A_{\kappa}(x))\ket
   \bra(\partial_{y_\sigma}A^{\tau}(y))(\partial_{y^\tau}A_{\sigma}(y))\ket
   +2\bra(\partial_{x_\kappa}A^{\lambda}(x))(\partial_{y_\tau}A^{\sigma}(y))\ket
   \bra(\partial_{x^\lambda}A_{\kappa}(x))(\partial_{y^\sigma}A_{\tau}(y))\ket\nonumber\\
& &+\bra(\partial_{x_\kappa}A^{\lambda}(x))(\partial_{x^\lambda}A_{\kappa}(x))\ket
   \bra(\partial_{y_\tau}A^{\sigma}(y))(\partial_{y^\sigma}A_{\tau}(y))\ket
   +2\bra(\partial_{x_\kappa}A^{\lambda}(x))(\partial_{y_\sigma}A^{\tau}(y))\ket
   \bra(\partial_{x^\lambda}A_{\kappa}(x))(\partial_{y^\tau}A_{\sigma}(y))\ket\nonumber\\
& &-\bra(\partial_{x_\kappa}A^{\lambda}(x))(\partial_{x^\lambda}A_{\kappa}(x))\ket
   \bra(\partial_{y_\tau}A^{\sigma}(y))(\partial_{y^\tau}A_{\sigma}(y))\ket
   -2\bra(\partial_{x_\kappa}A^{\lambda}(x))(\partial_{y_\tau}A^{\sigma}(y))\ket
   \bra(\partial_{x^\lambda}A_{\kappa}(x))(\partial_{y^\tau}A_{\sigma}(y))\ket\nonumber\\
& &-\bra(\partial_{x_\lambda}A^{\kappa}(x))(\partial_{x^\kappa}A_{\lambda}(x))\ket
   \bra(\partial_{y_\sigma}A^{\tau}(y))(\partial_{y^\sigma}A_{\tau}(y))\ket
   -2\bra(\partial_{x_\lambda}A^{\kappa}(x))(\partial_{y_\sigma}A^{\tau}(y))\ket
   \bra(\partial_{x^\kappa}A_{\lambda}(x))(\partial_{y^\sigma}A_{\tau}(y))\ket\nonumber\\
& &+\bra(\partial_{x_\lambda}A^{\kappa}(x))(\partial_{x^\kappa}A^{\lambda}(x))\ket
   \bra(\partial_{y_\sigma}A^{\tau}(y))(\partial_{y^\tau}A_{\sigma}(y))\ket
   +2\bra(\partial_{x_\lambda}A^{\kappa}(x))(\partial_{y\tau}A^{\sigma}(y))\ket
   \bra(\partial_{x^\kappa}A_{\lambda}(x))(\partial_{y^\sigma}A_{\tau}(y))\ket\nonumber\\
& &+\bra(\partial_{x_\lambda}A^{\kappa}(x))(\partial_{x^\kappa}A^{\lambda}(x))\ket
   \bra(\partial_{y_\tau}A^{\sigma}(y))(\partial_{y^\sigma}A_{\tau}(y))\ket
   +2\bra(\partial_{x_\lambda}A^{\kappa}(x))(\partial_{y_\sigma}A^{\tau}(y))\ket
   \bra(\partial_{x^\kappa}A_{\lambda}(x))(\partial_{y^\tau}A_{\sigma}(y))\ket\nonumber\\
& &-\bra(\partial_{x_\lambda}A^{\kappa}(x))(\partial_{x^\kappa}A_{\lambda}(x))\ket
   \bra(\partial_{y_\tau}A^{\sigma}(y))(\partial_{y^\tau}A_{\sigma}(y))\ket
   -2\bra(\partial_{x_\lambda}A^{\kappa}(x))(\partial_{y_\tau}A^{\sigma}(y))\ket
   \bra(\partial_{x^\kappa}A_{\lambda}(x))(\partial_{y^\tau}A_{\sigma}(y))\ket\nonumber\\
& &+\bra(\partial_{x_\lambda}A^{\kappa}(x))(\partial_{x^\lambda}A_{\kappa}(x))\ket
   \bra(\partial_{y_\sigma}A^{\tau}(y))(\partial_{y^\sigma}A_{\tau}(y))\ket
   +2\bra(\partial_{x_\lambda}A^{\kappa}(x))(\partial_{y_\sigma}A^{\tau}(y))\ket
   \bra(\partial_{x^\lambda}A_{\kappa}(x))(\partial_{y^\sigma}A_{\tau}(y))\ket\nonumber\\
& &-\bra(\partial_{x_\lambda}A^{\kappa}(x))(\partial_{x^\lambda}A_{\kappa}(x))\ket
   \bra(\partial_{y_\sigma}A^{\tau}(y))(\partial_{y^\tau}A_{\sigma}(y))\ket
   -4\bra(\partial_{x_\lambda}A^{\kappa}(x))(\partial_{y_\tau}A^{\sigma}(y))\ket
   \bra(\partial_{x^\lambda}A_{\kappa}(x))(\partial_{y^\sigma}A_{\tau}(y))\ket\nonumber\\
& &-\bra(\partial_{x_\lambda}A^{\kappa}(x))(\partial_{x^\lambda}A_{\kappa}(x))\ket
   \bra(\partial_{y_\tau}A^{\sigma}(y))(\partial_{y^\sigma}A_{\tau}(y))\ket\nonumber\\
& &+\underbrace{\bra(\partial_{x_\lambda}A^{\kappa}(x))(\partial_{x^\lambda}A_{\kappa}(x))\ket
   \bra(\partial_{y_\tau}A^{\sigma}(y))(\partial_{y^\tau}A_{\sigma}(y))\ket}_{\textrm{local diagrams}}
   +2\underbrace{\bra(\partial_{x_\lambda}A^{\kappa}(x))(\partial_{y_\tau}A^{\sigma}(y))\ket
   \bra(\partial_{x^\lambda}A_{\kappa}(x))(\partial_{y^\tau}A_{\sigma}(y))\ket}_
   {\textrm{real correlations}}\left.\right]\,.\nonumber\\
\end{eqnarray}

We are interested in the connected two-point correlation function, thus we neglect the local diagrams in Eq.\,(\ref{WICK}). Inserting the propagator for the massless gauge mode in Coulomb gauge and summarizing the terms of the same structure leads us to Eq.\,(\ref{Wick decomposition}):
\begin{eqnarray}
\bra\Theta_{00}(x)\Theta_{00}(y)\ket &=&
    2\bra\partial_{x^0}A^{\lambda}(x)\partial_{y^0}A^{\tau}(y)\ket
    \bra\partial_{x^0}A_{\lambda}(x)\partial_{y^0}A_{\tau}(y)\ket\nonumber\\
&&-g_{00}\bra\partial_{x^0}A^{\lambda}(x)\partial_{y^{\sigma}}A^{\tau}(y)\ket
    \bra\partial_{x^0}A_{\lambda}(x)\partial_{y_\sigma}A_{\tau}(y)\ket\nonumber\\
&&+g_{00}\bra\partial_{x^0}A^{\lambda}(x)\partial_{y_\sigma}A^{\tau}(y)\ket
    \bra\partial_{x^0}A_{\lambda}(x)\partial_{y^\tau}A_{\sigma}(y)\ket\nonumber\\
&&-g_{00}\bra\partial_{x_\kappa}A^{\lambda}(x)\partial_{y^0}A^{\tau}(y)\ket
    \bra\partial_{x^\kappa}A_{\lambda}(x)\partial_{y^0}A_{\tau}(y)\ket\nonumber\\
&&+g_{00}\bra\partial_{x_\kappa}A^{\lambda}(x)\partial_{y^0}A^{\tau}(y)\ket
    \bra\partial_{x^\lambda}A_{\kappa}(x)\partial_{y^0}A_{\tau}(y)\ket\nonumber\\
&&+\frac{g_{00}^2}{2}
    \bra\partial_{x_\kappa}A^{\lambda}(x)\partial_{y_\sigma}A^{\tau}(y)\ket
    \bra\partial_{x^\kappa}A_{\lambda}(x)\partial_{y^\sigma}A_{\tau}(y)\ket\nonumber\\
&&-\frac{g_{00}^2}{2}
    \bra\partial_{x_\kappa}A^{\lambda}(x)\partial_{y_\sigma}A^{\tau}(y)\ket
    \bra\partial_{x^\kappa}A_{\lambda}(x)\partial_{y^\tau}A_{\sigma}(y)\ket\nonumber\\
&&-\frac{g_{00}^2}{2}
    \bra\partial_{x_\kappa}A^{\lambda}(x)\partial_{y_\sigma}A^{\tau}(y)\ket
    \bra\partial_{x^\lambda}A_{\kappa}(x)\partial_{y^\sigma}A_{\tau}(y)\ket\nonumber\\
&&+\frac{g_{00}^2}{2}
    \bra\partial_{x_\kappa}A^{\lambda}(x)\partial_{y_\tau}A^{\sigma}(y)\ket
    \bra\partial_{x^\lambda}A_{\kappa}(x)\partial_{y^\sigma}A_{\tau}(y)\ket\nonumber\\
&&+2\bra\partial_{x^0}A^0(x)\partial_{y^0}A^0(y)\ket
    \bra\partial_{x^0}A_0(x)\partial_{y^0}A_0(y)\ket\nonumber\\
&&-2\bra\partial_{x^0}A^0(x)\partial_{y^\tau}A^0(y)\ket
    \bra\partial_{x^0}A_0(x)\partial_{y_\tau}A_0(y)\ket\nonumber\\
&&+2\bra\partial_{x^\tau}A^0(x)\partial_{y^\sigma}A^0(y)\ket
    \bra\partial_{x_\tau}A_0(x)\partial_{y_\sigma}A_0(y)\ket\nonumber\\
&&-2g_{00}\bra\partial_{x^0}A^0(x)\partial_{y^0}A^0(y)\ket
    \bra\partial_{x^0}A_0(x)\partial_{y^0}A_0(y)\ket\nonumber\\
&&+4g_{00}\bra\partial_{x^0}A^0(x)\partial_{y^\tau}A^0(y)\ket
    \bra\partial_{x^0}A_0(x)\partial_{y_\tau}A_0(y)\ket\nonumber\\
&&-2g_{00}\bra\partial_{x^\tau}A^0(x)\partial_{y^\sigma}A^0(y)\ket
    \bra\partial_{x_\tau}A_0(x)\partial_{y_\sigma}A_0(y)\ket\nonumber\\
&&+\frac{g_{00}^2}{2}\bra\partial_{x^0}A^0(x)\partial_{y^0}A^0(y)\ket
    \bra\partial_{x^0}A_0(x)\partial_{y^0}A_0(y)\ket\nonumber\\
&&-g_{00}^2\bra\partial_{x^0}A^0(x)\partial_{y^\tau}A^0(y)\ket
    \bra\partial_{x^0}A_0(x)\partial_{y_\tau}A_0(y)\ket\nonumber\\
&&+\frac{g_{00}^2}{2}\bra\partial_{x^\tau}A^0(x)\partial_{y^\sigma}A^0(y)\ket
    \bra\partial_{x_\tau}A_0(x)\partial_{y_\sigma}A_0(y)\ket\,.\nonumber\\
\end{eqnarray}

Subsequently, we take the partial derivatives out of the expectation value and insert the propa\-gator for the massless mode, Eq.\,(\ref{PropagatorTLM}).

%% file: details2.tex
\section{Calculation of the thermal contribution in Coulomb gauge}
\label{appendix thermal}
\subsection{The entire thermal part including oscillatory terms}
The evaluation of the integrals over the zero-components of the momenta eliminates the Dirac $\delta$-distribution and yields additional oscillatory terms. Since we are focussed on a time-averaged expression for the energy transport, we are setting $x^0=y^0$ for the subsequent calculation. At this point we state the time-dependent two-point correlator after performing the integration over $p_0$ and $k_0$ for reasons of completeness.
Starting with Eq.\,(\ref{thermalpart}) we insert the decomposition Eq.\,(\ref{Delta decomposition}) and perform the integration by employing the convolution property Eq.\,(\ref{convolution}) of the $\delta$-distribution.\\
\newline

$\bra\Theta_{00}(x)\Theta_{00}(y)\ket^{\tiny\mbox{th}}$
\begin{eqnarray}
&=&2\int\!dp_0\int\!dk_0\int\!\frac{d^3p}{(2\pi)^3 2|\vec{p}|}\int\!\frac{d^3k}{(2\pi)^3 2|\vec{k}|}
  \left(1+\left(\frac{\vec{p}\vec{k}}{|\vec{p}||\vec{k}|}\right)^2\right)
  p_0^2 k_0^2 \cdot n_B(\beta |p_0|)\,n_B(\beta |k_0|)\cdot\e^{-ip(x-y)}\e^{-ik(x-y)}\nonumber\\
  &&\hspace{20mm}\cdot
  (\delta(p_0-|\vec{p}|)+\delta(p_0+|\vec{p}|))(\delta(k_0-|\vec{k}|)+\delta(k_0+|\vec{k}|))\nonumber\\
&-&g_{00}
  \int\!dp_0\int\!dk_0\int\!\frac{d^3p}{(2\pi)^3 2|\vec{p}|}\int\!\frac{d^3k}{(2\pi)^3 2|\vec{k}|}
  \left(1+\left(\frac{\vec{p}\vec{k}}{|\vec{p}||\vec{k}|}\right)^2\right)
  p_0 k_0 p k \cdot n_B(\beta|p_0|)\,n_B(\beta|k_0|)\cdot\e^{-ip(x-y)}\e^{-ik(x-y)}\nonumber\\
  &&\hspace{20mm}\cdot
  (\delta(p_0-|\vec{p}|)+\delta(p_0+|\vec{p}|))(\delta(k_0-|\vec{k}|)+\delta(k_0+|\vec{k}|))\nonumber\\
&+&g_{00}
  \int\!dp_0\int\!dk_0\int\!\frac{d^3p}{(2\pi)^3 2|\vec{p}|}\int\!\frac{d^3k}{(2\pi)^3 2|\vec{k}|}
  \left(\left(\frac{\vec{p}\vec{k}}{|\vec{p}||\vec{k}|}\right)^2-1\right)
  p_0 k_0 \vec{p}\vec{k} \cdot n_B(\beta|p_0|)\,n_B(\beta|k_0|)\cdot\e^{-ip(x-y)}\e^{-ik(x-y)}\nonumber\\
  &&\hspace{20mm}\cdot
  (\delta(p_0-|\vec{p}|)+\delta(p_0+|\vec{p}|))(\delta(k_0-|\vec{k}|)+\delta(k_0+|\vec{k}|))\nonumber\\
&-&g_{00}
  \int\!dp_0\int\!dk_0\int\!\frac{d^3p}{(2\pi)^3 2|\vec{p}|}\int\!\frac{d^3k}{(2\pi)^3 2|\vec{k}|}
  \left(1+\left(\frac{\vec{p}\vec{k}}{|\vec{p}||\vec{k}|}\right)^2\right)
  p_0 k_0 p k \cdot n_B(\beta|p_0|)\,n_B(\beta|k_0|)\cdot\e^{-ip(x-y)}\e^{-ik(x-y)}\nonumber\\
  &&\hspace{20mm}\cdot
  (\delta(p_0-|\vec{p}|)+\delta(p_0+|\vec{p}|))(\delta(k_0-|\vec{k}|)+\delta(k_0+|\vec{k}|))\nonumber\\
&+&g_{00}
  \int\!dp_0\int\!dk_0\int\!\frac{d^3p}{(2\pi)^3 2|\vec{p}|}\int\!\frac{d^3k}{(2\pi)^3 2|\vec{k}|}
  \left(\left(\frac{\vec{p}\vec{k}}{|\vec{p}||\vec{k}|}\right)^2-1\right)
  p_0 k_0 \vec{p}\vec{k} \cdot n_B(\beta|p_0|)\,n_B(\beta|k_0|)\cdot\e^{-ip(x-y)}\e^{-ik(x-y)}\nonumber\\
  &&\hspace{20mm}\cdot
  (\delta(p_0-|\vec{p}|)+\delta(p_0+|\vec{p}|))(\delta(k_0-|\vec{k}|)+\delta(k_0+|\vec{k}|))\nonumber\\
&+&\frac{g_{00}^2}{2}
  \int\!dp_0\int\!dk_0\int\!\frac{d^3p}{(2\pi)^3 2|\vec{p}|}\int\!\frac{d^3k}{(2\pi)^3 2|\vec{k}|}
  \left(1+\left(\frac{\vec{p}\vec{k}}{|\vec{p}||\vec{k}|}\right)^2\right)
  (p k)^2 \cdot n_B(\beta|p_0|)\,n_B(\beta|k_0|)\cdot\e^{-ip(x-y)}\e^{-ik(x-y)}\nonumber\\
  &&\hspace{20mm}\cdot
  (\delta(p_0-|\vec{p}|)+\delta(p_0+|\vec{p}|))(\delta(k_0-|\vec{k}|)+\delta(k_0+|\vec{k}|))\nonumber\\
&-&g_{00}^2
  \int\!dp_0\int\!dk_0\int\!\frac{d^3p}{(2\pi)^3 2|\vec{p}|}\int\!\frac{d^3k}{(2\pi)^3 2|\vec{k}|}
  \left(\left(\frac{\vec{p}\vec{k}}{|\vec{p}||\vec{k}|}\right)^2-1\right)
  \vec{p}\vec{k} p k \cdot n_B(\beta|p_0|)\,n_B(\beta|k_0|)\cdot\e^{-ip(x-y)}\e^{-ik(x-y)}\nonumber\\
  &&\hspace{20mm}\cdot
  (\delta(p_0-|\vec{p}|)+\delta(p_0+|\vec{p}|))(\delta(k_0-|\vec{k}|)+\delta(k_0+|\vec{k}|))\nonumber\\
&+&\frac{g_{00}^2}{2}
  \int\!dp_0\int\!dk_0\int\!\frac{d^3p}{(2\pi)^3 2|\vec{p}|}\int\!\frac{d^3k}{(2\pi)^3 2|\vec{k}|}
  \left(\left(\frac{\vec{p}\vec{k}}{|\vec{p}||\vec{k}|}\right)^2-1\right)^2
  \vec{p}^2\vec{k}^2 \cdot n_B(\beta|p_0|)\,n_B(\beta|k_0|)\cdot\e^{-ip(x-y)}\e^{-ik(x-y)}\nonumber\\
  &&\hspace{20mm}\cdot
  (\delta(p_0-|\vec{p}|)+\delta(p_0+|\vec{p}|))(\delta(k_0-|\vec{k}|)+\delta(k_0+|\vec{k}|))\nonumber\\
\end{eqnarray}
\newpage
\begin{eqnarray}
&=&2\int\!\frac{d^3p}{(2\pi)^3}\int\!\frac{d^3k}{(2\pi)^3}
  \left(1+\left(\frac{\vec{p}\vec{k}}{|\vec{p}||\vec{k}|}\right)^2\right)
  |\vec{p}||\vec{k}|
  \cdot n_B(\beta|\vec{p}|)\,n_B(\beta|\vec{k}|)
  \cdot\e^{i\vec{p}\vec{(x-y)}}\,\e^{i\vec{k}\vec{(x-y)}}\nonumber\\
 &&\hspace{50mm}
  \cdot (\cos((|\vec{p}|+|\vec{k}|)(x-y)_0)+\cos((|\vec{p}|-|\vec{k}|)(x-y)_0))
  \nonumber\\
&&-\int\!\frac{d^3p}{(2\pi)^3}\int\!\frac{d^3k}{(2\pi)^3}
  \left(1+\left(\frac{\vec{p}\vec{k}}{|\vec{p}||\vec{k}|}\right)^2\right)
  |\vec{p}||\vec{k}|
  \cdot n_B(\beta|\vec{p}|)\,n_B(\beta|\vec{k}|)
  \cdot\e^{i\vec{p}\vec{(x-y)}}\,\e^{i\vec{k}\vec{(x-y)}}
  \nonumber\\
 &&\hspace{50mm}
  \cdot (\cos((|\vec{p}|+|\vec{k}|)(x-y)_0)+\cos((|\vec{p}|-|\vec{k}|)(x-y)_0))
  \nonumber\\
 &&\hspace{5mm}
  -\int\!\frac{d^3p}{(2\pi)^3}\int\!\frac{d^3k}{(2\pi)^3}
  \left(\left(\frac{\vec{p}\vec{k}}{|\vec{p}||\vec{k}|}\right)
  +\left(\frac{\vec{p}\vec{k}}{|\vec{p}||\vec{k}|}\right)^3\right)
  |\vec{p}||\vec{k}|
  \cdot n_B(\beta|\vec{p}|)\,n_B(\beta|\vec{k}|)
  \cdot\e^{i\vec{p}\vec{(x-y)}}\,\e^{i\vec{k}\vec{(x-y)}}
  \nonumber\\
 &&\hspace{50mm}
  \cdot (\cos((|\vec{p}|-|\vec{k}|)(x-y)_0)-\cos((|\vec{p}|+|\vec{k}|)(x-y)_0))
  \nonumber\\
&&+\int\!\frac{d^3p}{(2\pi)^3}\int\!\frac{d^3k}{(2\pi)^3}
  \left(\left(\frac{\vec{p}\vec{k}}{|\vec{p}||\vec{k}|}\right)^3
  -\left(\frac{\vec{p}\vec{k}}{|\vec{p}||\vec{k}|}\right)\right)
  |\vec{p}||\vec{k}|
  \cdot n_B(\beta|\vec{p}|)\,n_B(\beta|\vec{k}|)
  \cdot\e^{i\vec{p}\vec{(x-y)}}\,\e^{i\vec{k}\vec{(x-y)}}
  \nonumber\\
 &&\hspace{50mm}
  \cdot (\cos((|\vec{p}|+|\vec{k}|)(x-y)_0)-\cos((|\vec{p}|-|\vec{k}|)(x-y)_0))
  \nonumber\\
&&+\frac{1}{4}\int\!\frac{d^3p}{(2\pi)^3}\int\!\frac{d^3k}{(2\pi)^3}
  \left(1+2\left(\frac{\vec{p}\vec{k}}{|\vec{p}||\vec{k}|}\right)^2
  +\left(\frac{\vec{p}\vec{k}}{|\vec{p}||\vec{k}|}\right)^4\right)
  |\vec{p}||\vec{k}|
  \cdot n_B(\beta|\vec{p}|)\,n_B(\beta|\vec{k}|)
  \cdot\e^{i\vec{p}\vec{(x-y)}}\,\e^{i\vec{k}\vec{(x-y)}}
  \nonumber\\
 &&\hspace{50mm}
  \cdot (\cos((|\vec{p}|+|\vec{k}|)(x-y)_0)+\cos((|\vec{p}|-|\vec{k}|)(x-y)_0))
  \nonumber\\
 &&\hspace{5mm}
  +\frac{1}{4}\int\!\frac{d^3p}{(2\pi)^3}\int\!\frac{d^3k}{(2\pi)^3}
  \left(2\left(\frac{\vec{p}\vec{k}}{|\vec{p}||\vec{k}|}\right)
  +2\left(\frac{\vec{p}\vec{k}}{|\vec{p}||\vec{k}|}\right)^3\right)
  |\vec{p}||\vec{k}|
  \cdot n_B(\beta|\vec{p}|)\,n_B(\beta|\vec{k}|)
  \cdot\e^{i\vec{p}\vec{(x-y)}}\,\e^{i\vec{k}\vec{(x-y)}}
  \nonumber\\
 &&\hspace{50mm}
  \cdot(\cos((|\vec{p}|-|\vec{k}|)(x-y)_0)-\cos((|\vec{p}|+|\vec{k}|)(x-y)_0))
  \nonumber\\
&&-\frac{1}{2}\int\!\frac{d^3p}{(2\pi)^3}\int\!\frac{d^3k}{(2\pi)^3}
  \left(\left(\frac{\vec{p}\vec{k}}{|\vec{p}||\vec{k}|}\right)^2
  -\left(\frac{\vec{p}\vec{k}}{|\vec{p}||\vec{k}|}\right)^4\right)
  |\vec{p}||\vec{k}|
  \cdot n_B(\beta|\vec{p}|)\,n_B(\beta|\vec{k}|)
  \cdot\e^{i\vec{p}\vec{(x-y)}}\,\e^{i\vec{k}\vec{(x-y)}}
  \nonumber\\
 &&\hspace{50mm}
  \cdot (\cos((|\vec{p}|+|\vec{k}|)(x-y)_0)+\cos((|\vec{p}|-|\vec{k}|)(x-y)_0))
  \nonumber\\
 &&\hspace{5mm}
  -\frac{1}{2}\int\!\frac{d^3p}{(2\pi)^3}\int\!\frac{d^3k}{(2\pi)^3}
  \left(\left(\frac{\vec{p}\vec{k}}{|\vec{p}||\vec{k}|}\right)^3
  -\left(\frac{\vec{p}\vec{k}}{|\vec{p}||\vec{k}|}\right)\right)
  |\vec{p}||\vec{k}|
  \cdot n_B(\beta|\vec{p}|)\,n_B(\beta|\vec{k}|)
  \cdot\e^{i\vec{p}\vec{(x-y)}}\,\e^{i\vec{k}\vec{(x-y)}}
  \nonumber\\
 &&\hspace{50mm}
  \cdot (\cos((|\vec{p}|+|\vec{k}|)(x-y)_0)-\cos((|\vec{p}|-|\vec{k}|)(x-y)_0))
  \nonumber\\
&&+\frac{1}{4}\int\!\frac{d^3p}{(2\pi)^3}\int\!\frac{d^3k}{(2\pi)^3}
  \left(1-2\left(\frac{\vec{p}\vec{k}}{|\vec{p}||\vec{k}|}\right)^2
  +\left(\frac{\vec{p}\vec{k}}{|\vec{p}||\vec{k}|}\right)^4\right)
  |\vec{p}||\vec{k}|
  \cdot n_B(\beta|\vec{p}|)\,n_B(\beta|\vec{k}|)
  \cdot\e^{i\vec{p}\vec{(x-y)}}\,\e^{i\vec{k}\vec{(x-y)}}
  \nonumber\\
 &&\hspace{50mm}
  \cdot (\cos((|\vec{p}|+|\vec{k}|)(x-y)_0)+\cos((|\vec{p}|-|\vec{k}|)(x-y)_0))\,.
  \nonumber\\
\end{eqnarray}
\noindent
Restricting to equal-time and summarizing the terms of identical structure leads us to Eq.\,(\ref{thermalintegralu1})
\begin{eqnarray}
\bra\Theta_{00}(\vec{x})\Theta_{00}(\vec{y})\ket^{\tiny\mbox{th}}
&=&\left(\int\!\!\frac{d^3p}{(2\pi)^3}
        |\vec{p}|\,n_B(\beta|\vec{p}|)
        \,\e^{i\vec{p}\vec{z}}\right)^2\nonumber\\
&+&\int\!\!\frac{d^3p}{(2\pi)^3}\int\!\!\frac{d^3k}{(2\pi)^3}
        \left(\frac{\vec{p}\vec{k}}{|\vec{p}||\vec{k}|}\right)^2
        |\vec{p}||\vec{k}|\,n_B(\beta|\vec{p}|)n_B(\beta|\vec{k}|)
        \,\e^{i\vec{p}\vec{z}}\e^{i\vec{k}\vec{z}}\,.\nonumber\\
\end{eqnarray}

\subsection{Integration over azimuthal and polar angle}

In this subsection concerning the thermal contribution we proceed with the detailed calculation of the integrals over the azimuthal angle $\phi$ and the polar angle $\theta$. Introducing three-dimensional spherical coordinates for the momenta $\vec{p}$ and $\vec{k}$ we formulate the spatial scalar product $\vec{p}\vec{k}$ in terms of the azimuthal and polar angles.
We evaluate the integration over the azimuthal angles $\phi_p$ and $\phi_k$. As a consequence the integrals factorize in the remaining variables radial momentum-modulus and polar angle. We obtain\\
\newline
$\bra\Theta_{00}(\vec{x})\Theta_{00}(\vec{y})\ket^{\tiny\mbox{th}}$
\begin{eqnarray}
&=&\frac{1}{(2\pi)^6\beta^8}\Bigg(4\pi^2\cdot
   \left(\int\limits_0^{\infty}\!d|\tilde{\vec{p}}|
   \int\limits_{-1}^{+1}\!d\cos\theta_p
   \frac{|\tilde{\vec{p}}|^3}{(\e^{|\tilde{\vec{p}}|}-1)}
   \,\e^{i|\tilde{\vec{p}}||\tilde{\vec{z}}|\cos\theta_p}\right)^2
   \nonumber\\
&&\hspace{15mm}+2\pi\cdot
   \int\limits_0^{\infty}\!d|\tilde{\vec{p}}|
   \int\limits_{-1}^{+1}\!d\cos\theta_p
   \int\limits_0^{\infty}\!d|\tilde{\vec{k}}|
   \int\limits_{-1}^{+1}\!d\cos\theta_k
   \frac{|\tilde{\vec{p}}|^3}{(\e^{|\tilde{\vec{p}}|}-1)}
   \frac{|\tilde{\vec{k}}|^3}{(\e^{|\tilde{\vec{k}}|}-1)}
   \,\e^{i|\tilde{\vec{p}}||\tilde{\vec{z}}|\cos\theta_p}
   \,\e^{i|\tilde{\vec{k}}||\tilde{\vec{z}}|\cos\theta_k}\nonumber\\
   &&\hspace{40mm}\cdot\;
   \int\limits_0^{2\pi}\!d\phi_p
   \left(\sin\theta_p\sin\theta_k\cos\phi_p+\cos\theta_p\cos\theta_k\right)^2\Bigg)
   \nonumber\\
&=&\frac{1}{(2\pi)^6\beta^8}\left(4\pi^2
         \left(\int\limits_0^{\infty}\!d|\tilde{\vec{p}}|
         \int\limits_{-1}^{+1}\!d\cos\theta
         \frac{|\tilde{\vec{p}}|^3}{(e^{|\tilde{\vec{p}}|}-1)}
         e^{i|\tilde{\vec{p}}||\tilde{\vec{z}}|\cos\theta}\right)^2\right.\nonumber\\
&&+2\pi^2
         \left(\int\limits_0^{\infty}\!d|\tilde{\vec{p}}|
         \int\limits_{-1}^{+1}\!d\cos\theta
         \sin^2\theta\frac{|\tilde{\vec{p}}|^3}{(e^{|\tilde{\vec{p}}|}-1)}
         e^{i|\tilde{\vec{p}}||\tilde{\vec{z}}| \cos\theta}\right)^2\nonumber\\
&&+4\pi^2
         \left.\left(\int\limits_0^{\infty}\!d|\tilde{\vec{p}}|
         \int\limits_{-1}^{+1}\!d\cos\theta
         \cos^2\theta\frac{|\tilde{\vec{p}}|^3}{(e^{|\tilde{\vec{p}}|}-1)}
         e^{i|\tilde{\vec{p}}||\tilde{\vec{z}}| \cos\theta}\right)^2\right)\nonumber\\
\end{eqnarray}

Subsequently, we expand the exponential into spherical harmonics and express the powers of $\cos\theta$ in terms of the first Legendre polynomials $P_l(\cos\theta)$. The Legendre polynomials up to the fourth order are given as:
\begin{eqnarray}
P_0(\cos\theta)&=&1\,,\nonumber\\
P_1(\cos\theta)&=&\cos\theta\,,\nonumber\\
P_2(\cos\theta)&=&\frac{1}{2}(3\cos^2\theta-1)\,,\nonumber\\
P_3(\cos\theta)&=&\frac{1}{2}(5\cos^3\theta-3\cos\theta)\,,\nonumber\\
P_4(\cos\theta)&=&\frac{1}{8}(35\cos^4\theta-30\cos^2\theta-3)\,.\nonumber\\
\end{eqnarray}
Employing their orthogonality relation
\begin{equation}
\label{Legendre orthogonality}
\int\limits_{-1}^{+1}\!dx\, P_m(x) P_n(x)=\frac{2}{2m+1}\delta_{mn}\,.
\end{equation}
leads us to\\
\newpage
$\bra\Theta_{00}(\vec{x})\Theta_{00}(\vec{y})\ket^{\tiny\mbox{th}}$
\begin{eqnarray}
&=& \frac{1}{(2\pi)^6\beta^8}\cdot\left(4\pi^2\cdot
          \left(\int\limits_0^{\infty}d|\tilde{\vec{p}}|\int\limits_{-1}^{+1}d\!\cos\theta\,
          \frac{|\tilde{\vec{p}}|^3}{e^{|\tilde{\vec{p}}|}-1}
          \sum\limits_{l=0}^{\infty} i^l (2l+1) j_l(|\tilde{\vec{z}}| |\tilde{\vec{p}}|)
          P_l(\cos\theta) P_0(\cos\theta) \right)^2\right.\nonumber\\
&&\hspace{8mm}+2\pi^2\cdot
          \left(\int\limits_0^{\infty}d|\tilde{\vec{p}}|\int\limits_{-1}^{+1}d\!\cos\theta\,
          \frac{|\tilde{\vec{p}}|^3}{e^{|\tilde{\vec{p}}|}-1}
          \sum\limits_{l=0}^{\infty} i^l (2l+1) j_l(|\tilde{\vec{z}}| |\tilde{\vec{p}}|)
          P_l(\cos\theta) P_0(\cos\theta) \right.\nonumber\\
&&\left.\hspace{10mm}-\frac{1}{3}
          \int\limits_0^{\infty}d|\tilde{\vec{p}}|\int\limits_{-1}^{+1}d\!\cos\theta\,
          \frac{|\tilde{\vec{p}}|^3}{e^{|\tilde{\vec{p}}|}-1}
          \sum\limits_{l=0}^{\infty} i^l (2l+1) j_l(|\tilde{\vec{z}}| |\tilde{\vec{p}}|)
          P_l(\cos\theta) P_0(\cos\theta) \right.\nonumber\\
&&\left.\hspace{10mm}-
          \frac{2}{3}\int\limits_0^{\infty}d|\tilde{\vec{p}}|\int\limits_{-1}^{+1}d\!\cos\theta\,
          \frac{|\tilde{\vec{p}}|^3}{e^{|\tilde{\vec{p}}|}-1}
          \sum\limits_{l=0}^{\infty} i^l (2l+1) j_l(|\tilde{\vec{z}}| |\tilde{\vec{p}}|)
          P_l(\cos\theta) P_2(\cos\theta) \right)^2\nonumber\\
&&\hspace{8mm}+4\pi^2\cdot
          \left(\frac{1}{3}
          \int\limits_0^{\infty}d|\tilde{\vec{p}}|\int\limits_{-1}^{+1}d\!\cos\theta\,
          \frac{|\tilde{\vec{p}}|^3}{e^{|\tilde{\vec{p}}|}-1}
          \sum\limits_{l=0}^{\infty} i^l (2l+1) j_l(|\tilde{\vec{z}}| |\tilde{\vec{p}}|)
          P_l(\cos\theta) P_0(\cos\theta) \right.\nonumber\\
&&\left.\hspace{10mm}+
          \frac{2}{3}\int\limits_0^{\infty}d|\tilde{\vec{p}}|\int\limits_{-1}^{+1}d\!\cos\theta\,
          \frac{|\tilde{\vec{p}}|^3}{e^{|\tilde{\vec{p}}|}-1}
          \sum\limits_{l=0}^{\infty} i^l (2l+1) j_l(|\tilde{\vec{z}}| |\tilde{\vec{p}}|)
          P_l(\cos\theta) P_0(\cos\theta) \right)\nonumber\\
&=& \frac{1}{(2\pi)^6\beta^8}\cdot\left(4\pi^2\cdot
   \left(2\int\limits_0^{\infty}d|\tilde{\vec{p}}|
   \frac{|\tilde{\vec{p}}|^3}{e^{|\tilde{\vec{p}}|}-1}
   j_0(|\tilde{\vec{p}}||\tilde{\vec{z}}|)\right)^2\right.\nonumber\\
&&+\quad 2\pi^2\cdot
   \left(\frac{4}{3}\int\limits_0^{\infty}d|\tilde{\vec{p}}|
   \frac{|\tilde{\vec{p}}|^3}{e^{|\tilde{\vec{p}}|}-1}
   j_0(|\tilde{\vec{p}}||\tilde{\vec{z}}|)
   +\frac{4}{3}\int\limits_0^{\infty}d|\tilde{\vec{p}}|
   \frac{|\tilde{\vec{p}}|^3}{e^{|\tilde{\vec{p}}|}-1}
   j_2(|\tilde{\vec{p}}||\tilde{\vec{z}}|)\right)^2\nonumber\\
&&+\quad 4\pi^2\cdot
   \left.\left(\frac{2}{3}\int\limits_0^{\infty}d|\tilde{\vec{p}}|
   \frac{|\tilde{\vec{p}}|^3}{e^{|\tilde{\vec{p}}|}-1}
   j_0(|\tilde{\vec{p}}||\tilde{\vec{z}}|)
   -\frac{4}{3}\int\limits_0^{\infty}d|\tilde{\vec{p}}|
   \frac{|\tilde{\vec{p}}|^3}{e^{|\tilde{\vec{p}}|}-1}
   j_2(|\tilde{\vec{p}}||\tilde{\vec{z}}|)\right)^2\right)\,.\nonumber\\
\end{eqnarray}
Summarizing integrals of the same structure let us finally arrive at Eq.\,(\ref{thermalu1bessel})
\begin{eqnarray}
\bra\Theta_{00}(\vec{x})\Theta_{00}(\vec{y})\ket^{\tiny\mbox{th}}
&=&\frac{1}{(2\pi)^6\beta^8}
      \left(\frac{64\pi^2}{3}\cdot
      \left(\int\limits_0^{\infty}d|\tilde{\vec{p}}|\frac{|\tilde{\vec{p}}|^3}
      {\e^{|\tilde{\vec{p}}|}-1}\right. j_0(|\tilde{\vec{p}}||\tilde{\vec{z}}|)\right)^2\nonumber\\
&&+\left.\frac{32\pi^2}{3}\cdot
      \left(\int\limits_0^{\infty}d|\tilde{\vec{p}}|\frac{|\tilde{\vec{p}}|^3}
      {\e^{|\tilde{\vec{p}}|}-1} j_2(|\tilde{\vec{p}}||\tilde{\vec{z}}|)\right)^2\right)\nonumber\\
\end{eqnarray}
\newpage

\subsection{Integration over momentum}
The remaining integrals involve spherical Bessel functions of the first kind. These function are related to the ordinary first kind Bessel functions of fractional order via
\begin{equation}
\label{ordinary-spherical}
j_\nu(z)=\sqrt{\frac{\pi}{2z}} J_{\nu+\frac{1}{2}}(z)\,.
\end{equation}
The (spherical) Bessel functions satisfy the following recurrence formula:
\begin{equation}
\label{Bessel recurrence}
j_{\nu+1}(z)=\frac{2\nu}{z}j_\nu(z)+j_{\nu-1}(z)\,.
\end{equation}
The spherical Bessel functions for $\nu=1,\dots,4$ are given as
\begin{eqnarray}
j_0(z)&=&\frac{\sin z}{z}\nonumber\,,\\
j_1(z)&=&\frac{\sin z}{z^2}-\frac{\cos z}{z}\nonumber\,,\\
j_2(z)&=&\left(\frac{3}{z^3}-\frac{1}{z}\right)\sin z-\frac{3}{z^2}\cos z\nonumber\,,\\
j_3(z)&=&\left(\frac{15}{z^4}-\frac{6}{z^2}\right)\sin z
           -\left(\frac{15}{z^3}-\frac{1}{z}\right)\cos z\nonumber\,,\\
j_4(z)&=&\left(\frac{105}{z^5}-\frac{45}{z^3}+\frac{1}{z}\right)\sin z
           -\left(\frac{105}{z^4}-\frac{10}{z^2}\right)\cos z\nonumber\,.\\
\end{eqnarray}

In case of the thermal contribution we employ the two integral formulas \cite{Gradshteyn}
\begin{eqnarray}
\int\limits_0^{\infty} dx \frac{x^{2m}\sin(bx)}{e^x-1} &=& (-1)^m\frac{\partial^{2m}}{\partial b^{2m}}
               \left[\frac{\pi}{2}\coth(\pi b)-\frac{1}{2b}\right]\,,\\
\int\limits_0^{\infty} dx \frac{x^{2m+1}\cos(bx)}{e^x-1} &=& \frac{\partial^{2m+1}}{\partial b^{2m+1}}
               \left[\frac{\pi}{2}\coth(\pi b)-\frac{1}{2b}\right]\,,
\end{eqnarray}
where $b>0$, $m>0$ and $m\in\mathbb{Z}$,
and obtain:
\begin{eqnarray}
\int\limits_0^\infty d|\tilde{\vec{p}}|\,
\frac{|\tilde{\vec{p}}|^3\,j_0(|\tilde{\vec{p}}||\tilde{\vec{z}}|)}
{\e^{|\tilde{\vec{p}}|}-1}&=&
\frac{1}{|\tilde{\vec{z}}|}
\int\limits_0^\infty d|\tilde{\vec{p}}|\,
\frac{|\tilde{\vec{p}}|^2\,\sin(|\tilde{\vec{p}}||\tilde{\vec{z}}|)}
{\e^{|\tilde{\vec{p}}|}-1}\nonumber\\
&=&
\frac{1}{|\tilde{\vec{z}}|^4}
   -\frac{\pi^3\coth(\pi|\tilde{\vec{z}}|)\textrm{cosech}^2(\pi|\tilde{\vec{z}}|)}
                                         {|\tilde{\vec{z}}|}\,,\\
\int\limits_0^\infty d|\tilde{\vec{p}}|\,
\frac{|\tilde{\vec{p}}|^3\,j_2(|\tilde{\vec{p}}||\tilde{\vec{z}}|)}
{\e^{|\tilde{\vec{p}}|}-1}&=&
\frac{3}{|\tilde{\vec{z}}|^3}
\int\limits_0^\infty d|\tilde{\vec{p}}|\,
\frac{\sin(|\tilde{\vec{p}}||\tilde{\vec{z}}|)}{\e^{|\tilde{\vec{p}}|}-1}
-\frac{1}{|\tilde{\vec{z}}|}
\int\limits_0^\infty d|\tilde{\vec{p}}|\,
\frac{|\tilde{\vec{p}}|^2\,\sin(|\tilde{\vec{p}}||\tilde{\vec{z}}|)}
{\e^{|\tilde{\vec{p}}|}-1}\nonumber\\
&&\qquad-\,\frac{3}{|\tilde{\vec{z}}|^2}
\int\limits_0^\infty d|\tilde{\vec{p}}|\,
\frac{|\tilde{\vec{p}}|\,\cos(|\tilde{\vec{p}}||\tilde{\vec{z}}|)}
{\e^{|\tilde{\vec{p}}|}-1}\nonumber\\
&=&
\frac{-8+\pi|\tilde{\vec{z}}|(3\coth(\pi|\tilde{\vec{z}}|)
   +\pi|\tilde{\vec{z}}|(3+2\pi|\tilde{\vec{z}}|\coth(\pi|\tilde{\vec{z}}|))
   \,\textrm{cosech}^2(\pi|\tilde{\vec{z}}|))}
   {2|\tilde{\vec{z}}|^4}\nonumber\\
\end{eqnarray}

%% file: details3.tex
\newpage
\section{Calculation of the vacuum contribution in Coulomb gauge}
\label{appendix vacuum}
\subsection{Integration over azimuthal and second polar angle}
As similar procedure is applied to the case of the vacuum propagator. Again Eq.\,(\ref{Wick decomposition}) provides a benchmark for the calculation. In contrast to the thermal contribution we employ a Wick-rotation to euclidean signature ($p_0,k_0\to -ip_0,-ik_0$, $x^0,y^0\to -ix^0,-iy^0$ and $g_{\mu\nu}\to -\delta_{\mu\nu}$). Thus we are able to transform the variables to 4D spherical coordinates and choose $\tilde{\zeta_\mu}$ to point into the 3-direction. In detail we have\\
\newline
\begin{eqnarray}
\bra\Theta_{00}(\vec{x})\Theta_{00}(\vec{y})\ket^{\tiny\mbox{vac}}
&=&-2\int\!\frac{d^4p}{(2\pi)^4}\int\!\frac{d^4p}{(2\pi)^4}
   P^{\lambda\tau} P_{\lambda\tau} \frac{p_0^2}{p^2} \frac{k_0^2}{k^2}
   \e^{-ip\zeta} \e^{-ik\zeta}\nonumber\\
&&+2g_{00}\int\!\frac{d^4p}{(2\pi)^4}\int\!\frac{d^4p}{(2\pi)^4}
   P^{\lambda\tau} P_{\lambda\tau}
   \frac{p_0 p^\sigma}{p^2} \frac{k_0 k_\sigma}{k^2}
   \e^{-ip\zeta} \e^{-ik\zeta}\nonumber\\
&&-2g_{00}\int\!\frac{d^4p}{(2\pi)^4}\int\!\frac{d^4p}{(2\pi)^4}
   P^{\lambda\tau} P_{\lambda\sigma}
   \frac{p_0 p^\sigma}{p^2} \frac{k_0 k_\tau}{k^2}
   \e^{-ip\zeta} \e^{-ik\zeta}\nonumber\\
&&-\frac{g_{00}^2}{2}\int\!\frac{d^4p}{(2\pi)^4}\int\!\frac{d^4p}{(2\pi)^4}
   P^{\lambda\tau} P_{\lambda\tau}
   \frac{p^\kappa p^\sigma}{p^2} \frac{k_\kappa k_\sigma}{k^2}
   \e^{-ip\zeta} \e^{-ik\zeta}\nonumber\\
&&+g_{00}^2\int\!\frac{d^4p}{(2\pi)^4}\int\!\frac{d^4p}{(2\pi)^4}
   P^{\lambda\tau} P_{\lambda\sigma}
   \frac{p_\kappa p^\sigma}{p^2} \frac{k_\kappa k_\tau}{k^2}
   \e^{-ip\zeta} \e^{-ik\zeta}\nonumber\\
&&-\frac{g_{00}^2}{2}\int\!\frac{d^4p}{(2\pi)^4}\int\!\frac{d^4p}{(2\pi)^4}
   P^{\lambda\tau} P_{\kappa\sigma}
   \frac{p^\kappa p^\sigma}{p^2} \frac{k_\lambda k_\tau}{k^2}
   \e^{-ip\zeta} \e^{-ik\zeta}\nonumber\\
&&-2\int\!\frac{d^4p}{(2\pi)^4}\int\!\frac{d^4p}{(2\pi)^4}
   (p_0^2 k_0^2-p_0 k_0 p_\tau k^\tau+p_\lambda k^\lambda p_\tau k^\tau)
   \frac{\e^{-ip\zeta}}{\vec{p}^2} \frac{\e^{-ik\zeta}}{\vec{k}^2}\nonumber\\
&&+g_{00}\int\!\frac{d^4p}{(2\pi)^4}\int\!\frac{d^4p}{(2\pi)^4}
   (p_0^2 k_0^2-2p_0 k_0 p_\tau k^\tau+p_\lambda k^\lambda p_\tau k^\tau)
   \frac{\e^{-ip\zeta}}{\vec{p}^2}
   \frac{\e^{-ik\zeta}}{\vec{k}^2}\nonumber\\
&&-\frac{g_{00}^2}{2}2\int\!\frac{d^4p}{(2\pi)^4}\int\!\frac{d^4p}{(2\pi)^4}
   (p_0^2 k_0^2-2p_0 k_0 p_\tau k^\tau+p_\lambda k^\lambda p_\tau k^\tau)
   \frac{\e^{-ip\zeta}}{\vec{p}^2}
   \frac{\e^{-ik\zeta}}{\vec{k}^2}\nonumber\\
&=&2\int\!\frac{d^4p}{(2\pi)^4}\int\!\frac{d^4p}{(2\pi)^4}
   P_{\lambda\tau} P_{\lambda\tau} \frac{p_0^2}{p^2} \frac{k_0^2}{k^2}
   \e^{ip\zeta} \e^{ik\zeta}\nonumber\\
&&+2\delta_{00}\int\!\frac{d^4p}{(2\pi)^4}\int\!\frac{d^4p}{(2\pi)^4}
   P_{\lambda\tau} P_{\lambda\tau}
   \frac{p_0 p_\sigma}{p^2} \frac{k_0 k_\sigma}{k^2}
   \e^{ip\zeta} \e^{ik\zeta}\nonumber\\
&&-2\delta_{00}\int\!\frac{d^4p}{(2\pi)^4}\int\!\frac{d^4p}{(2\pi)^4}
   P_{\lambda\tau} P_{\lambda\sigma}
   \frac{p_0 p_\sigma}{p^2} \frac{k_0 k_\tau}{k^2}
   \e^{ip\zeta} \e^{ik\zeta}\nonumber\\
&&+\frac{\delta_{00}^2}{2}\int\!\frac{d^4p}{(2\pi)^4}\int\!\frac{d^4p}{(2\pi)^4}
   P_{\lambda\tau} P_{\lambda\tau}
   \frac{p_\kappa p_\sigma}{p^2} \frac{k_\kappa k_\sigma}{k^2}
   \e^{ip\zeta} \e^{ik\zeta}\nonumber\\
&&-\delta_{00}^2\int\!\frac{d^4p}{(2\pi)^4}\int\!\frac{d^4p}{(2\pi)^4}
   P_{\lambda\tau} P_{\lambda\sigma}
   \frac{p_\kappa p^\sigma}{p^2} \frac{k_\kappa k_\tau}{k^2}
   \e^{ip\zeta} \e^{ik\zeta}\nonumber\\
&&+\frac{\delta_{00}^2}{2}\int\!\frac{d^4p}{(2\pi)^4}\int\!\frac{d^4p}{(2\pi)^4}
   P_{\lambda\tau} P_{\kappa\sigma}
   \frac{p_\kappa p_\sigma}{p^2} \frac{k_\lambda k_\tau}{k^2}
   \e^{ip\zeta} \e^{ik\zeta}\nonumber\\
&&+2\int\!\frac{d^4p}{(2\pi)^4}\int\!\frac{d^4p}{(2\pi)^4}
   (p_0^2 k_0^2-p_0 k_0 p_\tau k_\tau+p_\lambda k_\lambda p_\tau k_\tau)
   \frac{\e^{ip\zeta}}{\vec{p}^2} \frac{\e^{ik\zeta}}{\vec{k}^2}\nonumber\\
&&-\delta_{00}\int\!\frac{d^4p}{(2\pi)^4}\int\!\frac{d^4p}{(2\pi)^4}
   (p_0^2 k_0^2-2p_0 k_0 p_\tau k_\tau+p_\lambda k_\lambda p_\tau k_\tau)
   \frac{\e^{ip\zeta}}{\vec{p}^2}
   \frac{\e^{ik\zeta}}{\vec{k}^2}\nonumber\\
&&+\frac{\delta_{00}^2}{2}2\int\!\frac{d^4p}{(2\pi)^4}\int\!\frac{d^4p}{(2\pi)^4}
   (p_0^2 k_0^2-2p_0 k_0 p_\tau k_\tau+p_\lambda k_\lambda p_\tau k_\tau)
   \frac{\e^{ip\zeta}}{\vec{p}^2}
   \frac{\e^{ik\zeta}}{\vec{k}^2}\nonumber\\
\end{eqnarray}
\newpage
\begin{eqnarray}
&=&\frac{9}{2}\left(\int\frac{d^4p}{(2\pi)^4}
   \,p_{0}^2\,\frac{\e^{ip\zeta}}{p^2}\right)^2
   +\frac{1}{2}\left(\int\frac{d^4p}{(2\pi)^4}
   \,|\vec{p}|^2\,\frac{\e^{ip\zeta}}{p^2}\right)^2\nonumber\\
&&+6\int\frac{d^4p}{(2\pi)^4}\int\frac{d^4k}{(2\pi)^4}
   \left(\frac{\vec{p}\vec{k}}{|\vec{p}||\vec{k}|}\right)
   p_0 k_0 |\vec{p}| |\vec{k}|
   \,\frac{\e^{ip\zeta}}{p^2}\,\frac{\e^{ik\zeta}}{k^2}\nonumber\\
&&+\frac{1}{2}\int\frac{d^4p}{(2\pi)^4}\int\frac{d^4k}{(2\pi)^4}
   \left(\frac{\vec{p}\vec{k}}{|\vec{p}||\vec{k}|}\right)^2
   \left(9p_0^2 k_0^2 + |\vec{p}|^2 |\vec{k}|^2\right)
   \,\frac{\e^{ip\zeta}}{p^2}\,\frac{\e^{ik\zeta}}{k^2}\nonumber\\
&&+2\left(\int\frac{d^4p}{(2\pi)^4}\,p_{0}^2
   \,\frac{\e^{ip\zeta}}{\vec{p}^2}\right)^2\nonumber\\
&&+2\int\frac{d^4p}{(2\pi)^4}\int\frac{d^4k}{(2\pi)^4}
   \left(\frac{\vec{p}\vec{k}}{|\vec{p}||\vec{k}|}\right)
   p_0 k_0 |\vec{p}| |\vec{k}|
   \frac{\e^{ip\zeta}}{\vec{p}^2}\,\frac{\e^{ik\zeta}}{\vec{k}^2}\nonumber\\
&&+\frac{9}{2}\int\frac{d^4p}{(2\pi)^4}\int\frac{d^4k}{(2\pi)^4}
   \left(\frac{\vec{p}\vec{k}}{|\vec{p}||\vec{k}|}\right)^2
   |\vec{p}|^2 |\vec{k}|^2
   \,\frac{\e^{ip\zeta}}{\vec{p}^2}\,\frac{\e^{ik\zeta}}{\vec{k}^2}\,.
   \nonumber\\
\end{eqnarray}
Introduction of dimensionless variables via rescaling of the momenta and  transformation to hyperspherical coordinates subject to the convention
\begin{eqnarray}
\tilde{p}=
|\tilde{p}|\left(
   \begin{array}{c} \cos\psi\\
                    \sin\psi\sin\theta\cos\phi\\
                    \sin\psi\sin\theta\sin\phi\\
                    \sin\psi\cos\theta
   \end{array}
\right)
\qquad\mbox{with}\quad
   \begin{array}{c} 0\leq\psi,\theta\leq\pi\\
                    0\leq\phi\leq2\pi\\
                    |\tilde{p}|=\sqrt{x_1^2+x_2^2+x_3^2+x_4^2}
   \end{array}\,
\end{eqnarray}
yields\\
\newline
$\bra\Theta_{00}(\vec{x})\Theta_{00}(\vec{y})\ket^{\tiny\mbox{vac}}$
\begin{eqnarray*}
&=&\frac{1}{(2\pi)^8\beta^8}
         \Bigg(
         \frac{9}{2}\cdot 4\pi^2
         \left(\int\!d|\tilde{p}| d\psi_p d\theta_p
         |\tilde{p}|^3 \sin^2\psi_p \cos^2\psi_p\sin\theta_p
         \e^{i|\tilde{p}||\tilde{\zeta}|\sin\psi_p\cos\theta_p}
         \right)^2\nonumber\\
&&\hspace{20mm}+\frac{1}{2}\cdot 4\pi^2
         \left(\int d|\tilde{p}| d\psi_p d\theta_p
         |\tilde{p}|^3 \sin^4\psi_p\sin\theta_p
         \e^{i|\tilde{p}||\tilde{\zeta}|\sin\psi_p\cos\theta_p}
         \right)^2\nonumber\\
&&\hspace{20mm}+\frac{9}{2}\cdot 2\pi
         \int\!d|\tilde{p}| d\psi_p d\theta_p
         \int\!d|\tilde{k}| d\psi_k d\theta_k
         |\tilde{p}|^3 \sin^2\psi_p\cos^2\psi_p\sin\theta_p
         |\tilde{k}|^3 \sin^2\psi_k\cos^2\psi_k\sin\theta_k
         \nonumber\\
         &&\hspace{30mm}\cdot\,
         \e^{i|\tilde{p}||\tilde{\zeta}|\sin\psi_p\cos\theta_p}
         \e^{i|\tilde{k}||\tilde{\zeta}|\sin\psi_k\cos\theta_k}\nonumber\\
         &&\hspace{30mm}\cdot\,
         \int_0^{2\pi}\!d\phi_p
         \left(\sin^2\theta_p\sin^2\theta_k\cos^2\phi_p
         +2\sin\theta_p\sin\theta_k\cos\theta_p\cos\theta_k
         +\cos^2\theta_p\cos^2\theta_k\right)\nonumber\\
&&\hspace{20mm}+6\cdot 2\pi
         \int d|\tilde{p}| d\psi_p d\theta_p
         \int d|\tilde{k}| d\psi_k d\theta_k
         |\tilde{p}|^3 \sin^3\psi_p\cos\psi_p\sin\theta_p
         |\tilde{k}|^3 \sin^3\psi_k\cos\psi_k\sin\theta_k\nonumber\\
         &&\hspace{30mm}\cdot\,
         \e^{i|\tilde{p}||\tilde{\zeta}|\sin\psi_p\cos\theta_p}
         \e^{i|\tilde{k}||\tilde{\zeta}|\sin\psi_k\cos\theta_k}\nonumber\\
         &&\hspace{30mm}\cdot\,
         \int_0^{2\pi}\!d\phi_p
         (\sin\theta_p\sin\theta_k\cos\phi_p+\cos\theta_p\cos\theta_k)
         \nonumber\\
&&\hspace{20mm}+\frac{1}{2}\cdot 2\pi
         \int\!d|\tilde{p}| d\psi_p d\theta_p
         \int\!d|\tilde{k}| d\psi_k d\theta_k
         |\tilde{p}|^3 \sin^4\psi_p\sin\theta_p
         |\tilde{k}|^3 \sin^4\psi_k\sin\theta_k\nonumber\\
         &&\hspace{30mm}\cdot\,
         \e^{i|\tilde{p}||\tilde{\zeta}|\sin\psi_p\cos\theta_p}
         \e^{i|\tilde{k}||\tilde{\zeta}|\sin\psi_k\cos\theta_k}\nonumber\\
         &&\hspace{30mm}\cdot\,
         \int_0^{2\pi}\!d\phi_p
         \left(\sin^2\theta_p\sin^2\theta_k\cos^2\phi_p
         +2\sin\theta_p\sin\theta_k\cos\theta_p\cos\theta_k
         +\cos^2\theta_p\cos^2\theta_k\right)\nonumber\\
&&\hspace{20mm}+2\cdot 4\pi^2
         \left(\int\!d|\tilde{p}| d\psi_p d\theta_p
         |\tilde{p}|^3 \cos^2\psi_p\sin\theta_p
         \e^{i|\tilde{p}||\tilde{\zeta}|\sin\psi_p\cos\theta_p}
         \right)^2\nonumber\\
&&\hspace{20mm}+2\cdot 2\pi
         \int d|\tilde{p}| d\psi_p d\theta_p
         \int d|\tilde{k}| d\psi_k d\theta_k
         |\tilde{p}|^3 \sin\psi_p\cos\psi_p\sin\theta_p
         |\tilde{k}|^3 \sin\psi_k\cos\psi_k\sin\theta_k\nonumber\\
         \end{eqnarray*}
\newpage
\begin{eqnarray}
&&\hspace{30mm}\cdot\,
         \e^{i|\tilde{p}||\tilde{\zeta}|\sin\psi_p\cos\theta_p}
         \e^{i|\tilde{k}||\tilde{\zeta}|\sin\psi_k\cos\theta_k}\nonumber\\
         &&\hspace{30mm}\cdot\,
         \int_0^{2\pi}\!d\phi_p
         (\sin\theta_p\sin\theta_k\cos\phi_p+\cos\theta_p\cos\theta_k)
         \nonumber\\
&&\hspace{20mm}+\frac{9}{2}\cdot 2\pi
         \int\!d|\tilde{p}| d\psi_p d\theta_p
         \int\!d|\tilde{k}| d\psi_k d\theta_k
         |\tilde{p}|^3 \sin^2\psi_p\sin\theta_p
         |\tilde{k}|^3 \sin^2\psi_k\sin\theta_k\nonumber\\
         &&\hspace{30mm}\cdot\,
         \e^{i|\tilde{p}||\tilde{\zeta}|\sin\psi_p\cos\theta_p}
         \e^{i|\tilde{k}||\tilde{\zeta}|\sin\psi_k\cos\theta_k}\nonumber\\
         &&\hspace{30mm}\cdot\,
         \int_0^{2\pi}\!d\phi_p
         \left(\sin^2\theta_p\sin^2\theta_k\cos^2\phi_p
         +2\sin\theta_p\sin\theta_k\cos\theta_p\cos\theta_k
         +\cos^2\theta_p\cos^2\theta_k\right)
         \Bigg)\,.\nonumber\\
\end{eqnarray}
Again the expression above factorizes in two identical contributions for each momenta. Thus we are able to write
\newline
\begin{eqnarray}
\bra\Theta_{00}(x)\Theta_{00}(y)\ket^{\tiny\mbox{vac}}
&=&\frac{1}{(2\pi)^8\beta^8}\left(\frac{9}{2}\cdot4\pi^2
   \left(\int\limits_0^{\infty}\!\! d|\tilde{p}| \int\limits_0^{\pi}\!\! d\psi \int\limits_0^{\pi}\!\! d\theta\,
   |\tilde{p}|^3 \sin^2\psi \cos^2\psi \sin\theta \,\e^{i|\tilde{p}||\tilde{\zeta}|\sin\psi\cos\theta}\right)^2
   \right.\nonumber\\
& &\left.+\frac{1}{2}\cdot4\pi^2
   \left(\int\limits_0^{\infty}\!\! d|\tilde{p}| \int\limits_0^{\pi}\!\! d\psi \int\limits_0^{\pi}\!\! d\theta\,
   |\tilde{p}|^3 \sin^4\psi \sin^3\theta \,\e^{i|\tilde{p}||\tilde{\zeta}|\sin\psi\cos\theta}\right)^2
   \right.\nonumber\\
& &\left.+\frac{9}{2}\cdot2\pi^2
   \left(\int\limits_0^{\infty}\!\! d|\tilde{p}| \int\limits_0^{\pi}\!\! d\psi \int\limits_0^{\pi}\!\! d\theta\,
   |\tilde{p}|^3 \sin^2\psi \cos^2\psi \sin\theta \sin^3\theta \,\e^{i|\tilde{p}||\tilde{\zeta}|\sin\psi\cos\theta}\right)^2
   \right.\nonumber\\
& &\left.+\frac{9}{2}\cdot4\pi^2
   \left(\int\limits_0^{\infty}\!\! d|\tilde{p}| \int\limits_0^{\pi}\!\! d\psi \int\limits_0^{\pi}\!\! d\theta\,
   |\tilde{p}|^3 \sin^2\psi \cos^2\psi \sin\theta \cos^2\theta \,\e^{i|\tilde{p}||\tilde{\zeta}|\sin\psi\cos\theta}\right)^2
   \right.\nonumber\\
& &\left.+6\cdot4\pi^2
   \left(\int\limits_0^{\infty}\!\! d|\tilde{p}| \int\limits_0^{\pi}\!\! d\psi \int\limits_0^{\pi}\!\! d\theta\,
   |\tilde{p}|^3 \sin^3\psi \cos\psi \cos\theta \,\e^{i|\tilde{p}||\tilde{\zeta}|\sin\psi\cos\theta}\right)^2
   \right.\nonumber\\
& &\left.+\frac{1}{2}\cdot2\pi^2
   \left(\int\limits_0^{\infty}\!\! d|\tilde{p}| \int\limits_0^{\pi}\!\! d\psi \int\limits_0^{\pi}\!\! d\theta\,
   |\tilde{p}|^3 \sin^4\psi \sin^3\theta \,\e^{i|\tilde{p}||\tilde{\zeta}|\sin\psi\cos\theta}\right)^2
   \right.\nonumber\\
& &\left.+\frac{1}{2}\cdot4\pi^2
   \left(\int\limits_0^{\infty}\!\! d|\tilde{p}| \int\limits_0^{\pi}\!\! d\psi \int\limits_0^{\pi}\!\! d\theta\,
   |\tilde{p}|^3 \sin^4\psi \sin\theta \cos^2\theta \,\e^{i|\tilde{p}||\tilde{\zeta}|\sin\psi\cos\theta}\right)^2
   \right.\nonumber\\
& &\left.+2\cdot4\pi^2
   \left(\int\limits_0^{\infty}\!\! d|\tilde{p}| \int\limits_0^{\pi}\!\! d\psi \int\limits_0^{\pi}\!\! d\theta\,
   |\tilde{p}|^3 \cos^2\psi \sin\theta \,\e^{i|\tilde{p}||\tilde{\zeta}|\sin\psi\cos\theta}\right)^2
   \right.\nonumber\\
& &\left.+2\cdot4\pi^2
   \left(\int\limits_0^{\infty}\!\! d|\tilde{p}| \int\limits_0^{\pi}\!\! d\psi \int\limits_0^{\pi}\!\! d\theta\,
   |\tilde{p}|^3 \sin\psi \cos\psi \sin\theta \cos\theta \,\e^{i|\tilde{p}||\tilde{\zeta}|\sin\psi\cos\theta}\right)^2
   \right.\nonumber\\
& &\left.+\frac{9}{2}\cdot\pi^2
   \left(\int\limits_0^{\infty}\!\! d|\tilde{p}| \int\limits_0^{\pi}\!\! d\psi \int\limits_0^{\pi}\!\! d\theta\,
   |\tilde{p}|^3 \sin^2\psi \sin^3\theta \,\e^{i|\tilde{p}||\tilde{\zeta}|\sin\psi\cos\theta}\right)^2
   \right.\nonumber\\
& &\left.+\frac{9}{2}\cdot2\pi^2
   \left(\int\limits_0^{\infty}\!\! d|\tilde{p}| \int\limits_0^{\pi}\!\! d\psi \int\limits_0^{\pi}\!\! d\theta\,
   |\tilde{p}|^3 \sin^2\psi \sin\theta \cos^2\theta \,\e^{i|\tilde{p}||\tilde{\zeta}|\sin\psi\cos\theta}\right)^2
   \right)\,.\nonumber\\
\end{eqnarray}
\newpage
In order to proceed with the $\theta$-integration we expand the plane wave into spherical harmonics. We replace $\sin^2\theta$ by $(1-\cos^2\theta)$ and $-d\theta\sin\theta$ by $d\cos\theta$. The polynomials in $\cos\theta$ are written in terms of Legendre polynomials.\\
We obtain\\
\newline
$\bra\Theta_{00}(x)\Theta_{00}(y)\ket^{\tiny\mbox{vac}}$
\begin{eqnarray*}
   \nonumber
&=&\frac{1}{(2\pi)^8\beta^8}
   \left(18\pi^2
   \left(\int\limits_0^{\infty}\!\!d|\tilde{p}| \int\limits_0^{\pi}\!\!d\psi |\tilde{p}|^3 \sin^2\psi \cos^2\psi
   \int\limits_{-1}^{+1}\!\!d\cos\theta P_0(\cos\theta)
   \sum\limits_{l=0}^{\infty}i^l(2l+1)
   j_l(|\tilde{p}||\tilde{\zeta}|\sin\psi)P_l(\cos\theta)\right)^2
   \right.\nonumber\\
&&+2\pi^2
   \left(\int\limits_0^{\infty}\!\!d|\tilde{p}| \int\limits_0^{\pi}\!\!d\psi |\tilde{p}|^3 \sin^4\psi
   \int\limits_{-1}^{+1}\!\!d\cos\theta
   \sum\limits_{l=0}^{\infty}i^l(2l+1)
   j_l(|\tilde{p}||\tilde{\zeta}|\sin\psi)P_l(\cos\theta)\right)^2
   \nonumber\\
&&+9\pi^2
   \left(\int\limits_0^{\infty}\!\!d|\tilde{p}| \int\limits_0^{\pi}\!\!d\psi |\tilde{p}|^3 \sin^2\psi \cos^2\psi
   \int\limits_{-1}^{+1}\!\!d\cos\theta\right.
   \nonumber\\
  &&\left.\hspace{25mm}
   \left(\frac{2}{3}P_0(\cos\theta)-\frac{2}{3}P_2(\cos\theta)\right)
   \sum\limits_{l=0}^{\infty}i^l(2l+1)
   j_l(|\tilde{p}||\tilde{\zeta}|\sin\psi)P_l(\cos\theta)\right)^2
   \nonumber\\
&&+18\pi^2
   \left(\int\limits_0^{\infty}\!\!d|\tilde{p}| \int\limits_0^{\pi}\!\!d\psi |\tilde{p}|^3 \sin^2\psi \cos^2\psi
   \int\limits_{-1}^{+1}\!\!d\cos\theta
   \right.\nonumber\\
  &&\left.\hspace{25mm}
   \left(\frac{1}{3}P_0(\cos\theta)+\frac{2}{3}P_2(\cos\theta)\right)
   \sum\limits_{l=0}^{\infty}i^l(2l+1)
   j_l(|\tilde{p}||\tilde{\zeta}|\sin\psi)P_l(\cos\theta)\right)^2
   \nonumber\\
&&+24\pi^2
   \left(\int\limits_0^{\infty}\!\!d|\tilde{p}| \int\limits_0^{\pi}\!\!d\psi |\tilde{p}|^3 \sin^3\psi \cos\psi
   \int\limits_{-1}^{+1}\!\!d\cos\theta
   P_1(\cos\theta)
   \sum\limits_{l=0}^{\infty}i^l(2l+1)
   j_l(|\tilde{p}||\tilde{\zeta}|\sin\psi)P_l(\cos\theta)\right)^2
   \nonumber\\
&&+\pi^2
   \left(\int\limits_0^{\infty}\!\!d|\tilde{p}| \int\limits_0^{\pi}\!\!d\psi |\tilde{p}|^3 \sin^4\psi
   \int\limits_{-1}^{+1}\!\!d\cos\theta\right.
   \nonumber\\
  &&\left.\hspace{25mm}
   \left(\frac{2}{3}P_0(\cos\theta)-\frac{2}{3}P_2(\cos\theta)\right)
   \sum\limits_{l=0}^{\infty}i^l(2l+1)
   j_l(|\tilde{p}||\tilde{\zeta}|\sin\psi)P_l(\cos\theta)\right)^2
   \nonumber\\
&&+2\pi^2
   \left(\int\limits_0^{\infty}\!\!d|\tilde{p}| \int\limits_0^{\pi}\!\!d\psi |\tilde{p}|^3 \sin^4\psi
   \int\limits_{-1}^{+1}\!\!d\cos\theta
   \right.\nonumber\\
  &&\left.\hspace{25mm}
   \left(\frac{1}{3}P_0(\cos\theta)+\frac{2}{3}P_2(\cos\theta)\right)
   \sum\limits_{l=0}^{\infty}i^l(2l+1)
   j_l(|\tilde{p}||\tilde{\zeta}|\sin\psi)P_l(\cos\theta)\right)^2
   \nonumber\\
&&+8\pi^2
   \left(\int\limits_0^{\infty}\!\!d|\tilde{p}| \int\limits_0^{\pi}\!\!d\psi |\tilde{p}|^3 \cos^2\psi
   \int\limits_{-1}^{+1}\!\!d\cos\theta
   \sum\limits_{l=0}^{\infty}i^l(2l+1)
   j_l(|\tilde{p}||\tilde{\zeta}|\sin\psi)P_l(\cos\theta)\right)^2
   \nonumber\\
&&+8\pi^2
   \left(\int\limits_0^{\infty}\!\!d|\tilde{p}| \int\limits_0^{\pi}\!\!d\psi |\tilde{p}|^3 \sin\psi \cos\psi
   \int\limits_{-1}^{+1}\!\!d\cos\theta
   P_1(\cos\theta)
   \sum\limits_{l=0}^{\infty}i^l(2l+1)
   j_l(|\tilde{p}||\tilde{\zeta}|\sin\psi)P_l(\cos\theta)\right)^2
   \nonumber\\
\end{eqnarray*}
\newpage
\begin{eqnarray}
&&+9\pi^2
   \left(\int\limits_0^{\infty}\!\!d|\tilde{p}| \int\limits_0^{\pi}\!\!d\psi |\tilde{p}|^3 \sin^2\psi
   \int\limits_{-1}^{+1}\!\!d\cos\theta\right.
   \nonumber\\
  &&\left.\hspace{25mm}
   \left(\frac{2}{3}P_0(\cos\theta)-\frac{2}{3}P_2(\cos\theta)\right)
   \sum\limits_{l=0}^{\infty}i^l(2l+1)
   j_l(|\tilde{p}||\tilde{\zeta}|\sin\psi)P_l(\cos\theta)\right)^2
   \nonumber\\
&&+18\pi^2
   \left(\int\limits_0^{\infty}\!\!d|\tilde{p}| \int\limits_0^{\pi}\!\!d\psi |\tilde{p}|^3 \sin^2\psi
   \int\limits_{-1}^{+1}\!\!d\cos\theta
   \right.\nonumber\\
  &&\left.\left.\hspace{25mm}
   \left(\frac{1}{3}P_0(\cos\theta)+\frac{2}{3}P_2(\cos\theta)\right)
   \sum\limits_{l=0}^{\infty}i^l(2l+1)
   j_l(|\tilde{p}||\tilde{\zeta}|\sin\psi)P_l(\cos\theta)\right)^2\right)\nonumber\\
&=&
\frac{1}{(2\pi)^8\beta^8}
   \left(72\pi^2
   \left(\int\limits_0^{\infty}\!\!d|\tilde{p}| \int\limits_0^{\pi}\!\!d\psi
   |\tilde{p}|^3 \sin^2\psi \cos^2\psi
   j_0(|\tilde{p}||\tilde{\zeta}|\sin\psi)\right)^2
   \right.\nonumber\\
&&+8\pi^2
   \left(\int\limits_0^{\infty}\!\!d|\tilde{p}| \int\limits_0^{\pi}\!\!d\psi
   |\tilde{p}|^3 \sin^4\psi
   j_0(|\tilde{p}||\tilde{\zeta}|\sin\psi)
   \right)^2\nonumber\\
&&+36\pi^2
   \left(\int\limits_0^{\infty}\!\!d|\tilde{p}| \int\limits_0^{\pi}\!\!d\psi
   |\tilde{p}|^3 \sin^2\psi \cos^2\psi
   \left(\frac{2}{3}j_0(|\tilde{p}||\tilde{\zeta}|\sin\psi)
   +\frac{2}{3}j_2(|\tilde{p}||\tilde{\zeta}|\sin\psi)\right)\right)^2
   \nonumber\\
&&+72\pi^2
   \left(\int\limits_0^{\infty}\!\!d|\tilde{p}| \int\limits_0^{\pi}\!\!d\psi
   |\tilde{p}|^3 \sin^2\psi \cos^2\psi\,
   \left(\frac{1}{3}j_0(|\tilde{p}||\tilde{\zeta}|\sin\psi)
   -\frac{2}{3}j_2(|\tilde{p}||\tilde{\zeta}|\sin\psi)\right)\right)^2
   \nonumber\\
&&+96\pi^2
   \left(\int\limits_0^{\infty}\!\!d|\tilde{p}| \int\limits_0^{\pi}\!\!d\psi
   |\tilde{p}|^3 \sin^3\psi \cos\psi
   i\,j_1(|\tilde{p}||\tilde{\zeta}|\sin\psi)
   \right)^2\nonumber\\
&&+4\pi^2
   \left(\int\limits_0^{\infty}\!\!d|\tilde{p}| \int\limits_0^{\pi}\!\!d\psi
   |\tilde{p}|^3 \sin^4\psi
   \left(\frac{2}{3}j_0(|\tilde{p}||\tilde{\zeta}|\sin\psi)
   +\frac{2}{3}j_2(|\tilde{p}||\tilde{\zeta}|\sin\psi)\right)\right)^2
   \nonumber\\
&&+8\pi^2
   \left(\int\limits_0^{\infty}\!\!d|\tilde{p}| \int\limits_0^{\pi}\!\!d\psi
   |\tilde{p}|^3 \sin^2\psi
   \left(\frac{1}{3}j_0(|\tilde{p}||\tilde{\zeta}|\sin\psi)
   -\frac{2}{3}j_2(|\tilde{p}||\tilde{\zeta}|\sin\psi)\right)\right)^2
   \nonumber\\
&&+32\pi^2
   \left(\int\limits_0^{\infty}\!\!d|\tilde{p}| \int\limits_0^{\pi}\!\!d\psi
   |\tilde{p}|^3 \cos^2\psi
   j_0(|\tilde{p}||\tilde{\zeta}|\sin\psi)
   \right)^2\nonumber\\
&&+32\pi^2
   \left(\int\limits_0^{\infty}\!\!d|\tilde{p}| \int\limits_0^{\pi}\!\!d\psi
   |\tilde{p}|^3 \sin\psi \cos\psi
   i\,j_1(|\tilde{p}||\tilde{\zeta}|\sin\psi)
   \right)^2\nonumber\\
&&+36\pi^2
   \left(\int\limits_0^{\infty}\!\!d|\tilde{p}| \int\limits_0^{\pi}\!\!d\psi
   |\tilde{p}|^3 \sin^4\psi
   \left(\frac{2}{3}j_0(|\tilde{p}||\tilde{\zeta}|\sin\psi)
   +\frac{2}{3}j_2(|\tilde{p}||\tilde{\zeta}|\sin\psi)\right)\right)^2
   \nonumber\\
&&+72\pi^2
   \left(\int\limits_0^{\infty}\!\!d|\tilde{p}| \int\limits_0^{\pi}\!\!d\psi
   |\tilde{p}|^3 \sin^2\psi
   \left(\frac{1}{3}j_0(|\tilde{p}||\tilde{\zeta}|\sin\psi)
   -\frac{2}{3}j_2(|\tilde{p}||\tilde{\zeta}|\sin\psi)\right)\right)^2
\end{eqnarray}
\newpage

We summarize the integrals of the same type and obtain Eq.\,(\ref{vacuummomentumpsiu1}):
\begin{eqnarray}
&&\bra\Theta_{00}(\vec{x})\Theta_{00}(\vec{y})\ket^{\tiny\mbox{vac}}
     =\frac{1}{(2\pi)^8\beta^8}\left(96\pi^2
     \left(\int\limits_0^{\infty}\!\!d|\tilde{p}|\,|\tilde{p}|^3
     \int\limits_0^{\pi}\!\!d\psi\,\sin^2\psi \cos^2\psi
     j_0(|\tilde{p}||\tilde{\vec{z}}|\sin\psi) \right)^2\right. \nonumber\\
&&+\left.48\pi^2
     \left(\int\limits_0^{\infty}\!\!d|\tilde{p}|\,|\tilde{p}|^3
     \int\limits_0^{\pi}\!\!d\psi\,\sin^2\psi \cos^2\psi j_2(|\tilde{p}||\tilde{\vec{z}}|\sin\psi) \right)^2\right. \nonumber\\
&&+\left.96\pi^2
     \left(\int\limits_0^{\infty}\!\!d|\tilde{p}|\,|\tilde{p}|^3
     \int\limits_0^{\pi}\!\!d\psi\,\sin^3\psi \cos\psi
     j_1(|\tilde{p}||\tilde{\vec{z}}|\sin\psi) \right)^2 \nonumber\right. \\
&&+\left.\frac{32\pi^2}{3}
     \left(\int\limits_0^{\infty}\!\!d|\tilde{p}|\,|\tilde{p}|^3
     \int\limits_0^{\pi}\!\!d\psi\,\sin^4\psi j_0(|\tilde{p}||\tilde{\vec{z}}|\sin\psi) \right)^2\right.  \nonumber\\
&&+\left.\frac{16\pi^2}{3}
     \left(\int\limits_0^{\infty}\!\!d|\tilde{p}|\,|\tilde{p}|^3
     \int\limits_0^{\pi}\!\!d\psi\,\sin^4\psi j_2(|\tilde{p}||\tilde{\vec{z}}|\sin\psi) \right)^2\right.  \nonumber\\
&&+\left.32\pi^2
     \left(\int\limits_0^{\infty}\!\!d|\tilde{p}|\,|\tilde{p}|^3
     \int\limits_0^{\pi}\!\!d\psi\,\cos^2\psi j_0(|\tilde{p}||\tilde{\vec{z}}|\sin\psi) \right)^2\right.  \nonumber\\
&&+\left.32\pi^2
     \left(\int\limits_0^{\infty}\!\!d|\tilde{p}|\,|\tilde{p}|^3
     \int\limits_0^{\pi}\!\!d\psi\,\sin\psi \cos\psi j_1(|\tilde{p}||\tilde{\vec{z}}|\sin\psi) \right)^2\right.\nonumber\\
&&+\left.24\pi^2
     \left(\int\limits_0^{\infty}\!\!d|\tilde{p}|\,|\tilde{p}|^3
     \int\limits_0^{\pi}\!\!d\psi\,\sin^2\psi j_0(|\tilde{p}||\tilde{\vec{z}}|\sin\psi) \right)^2\right.\nonumber\\
&&+\left.48\pi^2
     \left(\int\limits_0^{\infty}\!\!d|\tilde{p}|\,|\tilde{p}|^3
     \int\limits_0^{\pi}\!\!d\psi\,\sin^2\psi j_2(|\tilde{p}||\tilde{\vec{z}}|\sin\psi)\right)^2\right)\,,\nonumber\\
&&
\end{eqnarray}
where now $|\tilde{p}|\equiv
\sqrt{\tilde{p}_0^2+\tilde{p}_1^2+\tilde{p}_2^2+\tilde{p}_3^2}$ is the modulus of euclidean four-momentum.
In Eq.\,(\ref{vacuummomentumpsiu1}) the last four lines arise
from the term $\propto u_\mu u_\nu$ in the propagator, see Eq.\,(\ref{U1propagator}). Again the spherical Bessel functions of the first kind $j_0, j_1, j_2$ are involved. The evaluation of integrals concerning functions of this kind with nontrivial argument is the subject of interest in the next subsection.

\subsection{Integration over first polar angle and momentum}
\label{appendix vacuum2}
In order to solve the remaining integrals emerging in the vacuum contribution we use the relations Eqs.\,(\ref{ordinary-spherical}), (\ref{Bessel recurrence}) and employ Sonine's first integral formula and its generalizations \cite{Watson}. It holds that
\begin{equation}
\int\limits_0^{\frac{\pi}{2}}\!d\psi\,J_\mu(a \sin\psi)\sin^{\mu+1}\psi\cos^{2\rho+1}\psi=
2^{\rho+1}\Gamma(\rho+1)a^{-\rho-1}J_{\rho+\mu+1}(a)
\end{equation}
where $\mathfrak{Re}\mu, \mathfrak{Re}\nu >-1, a\geq0$,
\begin{equation}
\int\limits_0^\pi\!d\psi\,J_{\mu+1}(a \sin\psi)\sin^\mu\psi=
\sqrt{\pi}2^{-2-\mu}a^{1+\mu}\Gamma(1+\mu)
\,_1\tilde{F}_2\left(1+\mu;2+\mu,\frac{3}{2}+\mu;\frac{-a^2}{4}\right)\,,
\end{equation}
where $\mathfrak{Re}\mu, \mathfrak{Re}\nu >-1, a\geq0$,
\begin{eqnarray}
\int\limits_0^{\pi}\!\!d\psi \sin^{\mu}\psi \cos^{\nu}\psi J_{\rho}(a\sin\psi)&=&
         2^{-1-\rho}(1+(-1)^{\nu})a^{\rho}\Gamma\left(\frac{1+\nu}{2}\right)
         \Gamma\left(\frac{1+\mu+\rho}{2}\right)\nonumber\\
         &&\quad\cdot\,\,_1\tilde{F}_2\left(\frac{1+\mu+\rho}{2};
         \frac{2+\mu+\nu+\rho}{2},1+\rho;\frac{-a^2}{4}\right)
\end{eqnarray}
provided that $\mathfrak{Re}\nu>-1, \mathfrak{Re}\rho>-1, \mathfrak{Re}(\mu+\rho)>-1\, \mbox{and}\,a\geq0$).
In these formulas $_p\tilde{F}_q$ denotes a regularized hypergeometric function, which is defined as
\begin{equation}
_p\tilde{F}_q(\alpha_1,\dots\alpha_p;\beta_1,\dots\beta_q;z)
=\frac{_pF_q(\alpha_1,\dots\alpha_p;\beta_1,\dots\beta_q;z)}
{\Gamma(\beta_1)\cdot\dots\cdot\Gamma(\beta_q)}\,.
\end{equation}
We observe that the integrals which involve the spherical Bessel functions $j_1$ (and in general spherical Bessel functions of uneven degree $j_{2\nu-1}$) vanish for reasons of symmetry:
\begin{eqnarray}
&&\int\limits_0^{\infty}\!\!
     d|\tilde{p}|\,|\tilde{p}|^3\int\limits_0^{\pi}\!\!
     d\psi\,\sin^3\psi \cos\psi j_1(|\tilde{p}||\tilde{\vec{z}}|\sin\psi)=0\,,\\
&&\int\limits_0^{\infty}\!\!
     d|\tilde{p}|\,|\tilde{p}|^3\int\limits_0^{\pi}\!\!
     d\psi\,\sin\psi\cos\psi j_1(|\tilde{p}||\tilde{\vec{z}}|\sin\psi)=0\,.
\end{eqnarray}

Employing the Sonine integral formulas yields:
\begin{eqnarray}
&&\int\limits_0^{\infty}\!\!
     d|\tilde{p}|\,|\tilde{p}|^3\int\limits_0^{\pi}\!\!d\psi\,\sin^2\psi \cos^2\psi
     j_0(|\tilde{p}||\tilde{\vec{z}}|\sin\psi)\nonumber\\
&&=\sqrt{\frac{2\pi}{|\tilde{\vec{z}}|}}
     \int\limits_0^{\infty}\!\!d|\tilde{p}|\,|\tilde{p}|^{\frac{5}{2}}
     \int\limits_0^{\frac{\pi}{2}}\!\!d\psi\,\sin^{\frac{3}{2}}\psi \cos^2\psi
     J_{\frac{1}{2}}(|\tilde{p}||\tilde{\vec{z}}|\sin\psi)\nonumber\\
&&=\sqrt{\frac{2\pi}{|\tilde{\vec{z}}|}}
     \int\limits_0^{\infty}\!\!d|\tilde{p}|\,|\tilde{p}|^{\frac{5}{2}}
     \left(2^{\frac{1}{2}}\Gamma\left(\frac{3}{2}\right)
     (|\tilde{p}||\tilde{\vec{z}}|)^{-\frac{3}{2}}
     J_2(|\tilde{p}||\tilde{\vec{z}}|)\right)\nonumber\\
&&=\frac{\pi}{|\tilde{\vec{z}}|^2}
     \int\limits_0^{\infty}\!\!d|\tilde{p}|\,|\tilde{p}|J_2(|\tilde{p}||\tilde{\vec{z}}|)
\end{eqnarray}
\begin{eqnarray*}
&&\int\limits_0^{\infty}\!\!
     d|\tilde{p}|\,|\tilde{p}|^3\int\limits_0^{\pi}\!\!d\psi\,\sin^2\psi \cos^2\psi
     j_2(|\tilde{p}||\tilde{\vec{z}}|\sin\psi)\nonumber\\
&&=\sqrt{\frac{\pi}{2|\tilde{\vec{z}}|}}
     \int\limits_0^{\infty}\!\!d|\tilde{p}|\,|\tilde{p}|^{\frac{5}{2}}
     \int\limits_0^{\pi}\!\!d\psi\,\sin^{\frac{3}{2}}\psi \cos^2\psi
     J_{\frac{5}{2}}(|\tilde{p}||\tilde{\vec{z}}|\sin\psi)\nonumber\\
&&=\sqrt{\frac{\pi}{2|\tilde{\vec{z}}|}}
     \int\limits_0^{\infty}\!\!d|\tilde{p}|\,|\tilde{p}|^{\frac{5}{2}}
     \int\limits_0^{\pi}\!\!d\psi\,\sin^{\frac{3}{2}}\psi \cos^2\psi
     \left(\frac{3}{|\tilde{p}||\tilde{\vec{z}}|\sin\psi}J_{\frac{3}{2}}(|\tilde{p}||\tilde{\vec{z}}|\sin\psi)
        -J_{\frac{1}{2}}(|\tilde{p}||\tilde{\vec{z}}|\sin\psi)\right)\nonumber\\
&&=3\sqrt{\frac{\pi}{2}}|\tilde{\vec{z}}|^{-\frac{3}{2}}
     \int\limits_0^{\infty}\!\!d|\tilde{p}|\,|\tilde{p}|^{\frac{3}{2}}
     \int\limits_0^{\pi}\!\!d\psi\,\sin^{\frac{1}{2}}\psi \cos^2\psi
     J_{\frac{3}{2}}(|\tilde{p}||\tilde{\vec{z}}|\sin\psi)\nonumber\\
  &&\hspace{20mm}-\sqrt{\frac{2\pi}{|\tilde{\vec{z}}|}}
     \int\limits_0^{\infty}\!\!d|\tilde{p}|\,|\tilde{p}|^{\frac{5}{2}}
     \int\limits_0^{\frac{\pi}{2}}\!\!d\psi\,\sin^{\frac{3}{2}}\psi \cos^2\psi
     J_{\frac{1}{2}}(|\tilde{p}||\tilde{\vec{z}}|\sin\psi)\nonumber
\end{eqnarray*}
\newpage
\begin{eqnarray}
&&=3\sqrt{\frac{\pi}{2}}|\tilde{\vec{z}}|^{-\frac{3}{2}}
     \int\limits_0^{\infty}\!\!d|\tilde{p}|\,|\tilde{p}|^{\frac{3}{2}}
     \int\limits_0^{\pi}\!\!d\psi\,\sin^{\frac{1}{2}}\psi \cos^2\psi
     \left(\frac{1}{|\tilde{p}||\tilde{\vec{z}}|\sin\psi}J_{\frac{1}{2}}(|\tilde{p}||\tilde{\vec{z}}|\sin\psi)
     -J_{-\frac{1}{2}}(|\tilde{p}||\tilde{\vec{z}}|\sin\psi)\right)\nonumber\\
  &&\hspace{20mm}-\sqrt{\frac{2\pi}{|\tilde{\vec{z}}|}}
     \int\limits_0^{\infty}\!\!d|\tilde{p}|\,|\tilde{p}|^{\frac{5}{2}}
     \left(2^{\frac{1}{2}}\Gamma\left(\frac{3}{2}\right)(|\tilde{p}||\tilde{\vec{z}}|)^{-\frac{3}{2}}
     J_2(|\tilde{p}||\tilde{\vec{z}}|)\right)\nonumber\\
&&=3\sqrt{\frac{\pi}{2}}|\tilde{\vec{z}}|^{-\frac{5}{2}}
     \int\limits_0^{\infty}\!\!d|\tilde{p}|\,|\tilde{p}|^{\frac{1}{2}}
     \int\limits_0^{\pi}\!\!d\psi\,\sin^{-\frac{1}{2}}\psi \cos^2\psi
     J_{\frac{1}{2}}(|\tilde{p}||\tilde{\vec{z}}|\sin\psi)\nonumber\\
  &&\hspace{10mm}-3\sqrt{2\pi}|\tilde{\vec{z}}|^{-\frac{3}{2}}
     \int\limits_0^{\infty}\!\!d|\tilde{p}|\,|\tilde{p}|^{\frac{3}{2}}
     \int\limits_0^{\frac{\pi}{2}}\!\!d\psi\,\sin^{\frac{1}{2}}\psi \cos^2\psi
     J_{-\frac{1}{2}}(|\tilde{p}||\tilde{\vec{z}}|\sin\psi)
     -\frac{\pi}{|\tilde{\vec{z}}|^2}
     \int\limits_0^{\infty}\!\!d|\tilde{p}|\,|\tilde{p}|
     J_2(|\tilde{p}||\tilde{\vec{z}}|)\nonumber\\
&&=3\sqrt{\frac{\pi}{2}}|\tilde{\vec{z}}|^{-\frac{5}{2}}
     \int\limits_0^{\infty}\!\!d|\tilde{p}|\,|\tilde{p}|^{\frac{1}{2}}
     \left(2^{-\frac{3}{2}}(1+(-1)^2)\Gamma\left(\frac{3}{2}\right)
     (|\tilde{p}||\tilde{\vec{z}}|)^{\frac{1}{2}}
     \Gamma\left(\frac{1}{2}\right)
     \frac{_1F_2\left(\frac{1}{2};2,\frac{3}{2};
     -\frac{(|\tilde{p}||\tilde{\vec{z}}|)^2}{4}\right)}
     {\Gamma(2)\Gamma\left(\frac{3}{2}\right)}\right)\nonumber\\
  &&\hspace{10mm}-3\sqrt{2\pi}|\tilde{\vec{z}}|^{-\frac{3}{2}}
     \int\limits_0^{\infty}\!\!d|\tilde{p}|\,|\tilde{p}|^{\frac{3}{2}}
     \left(2^{\frac{1}{2}}\Gamma\left(\frac{3}{2}\right)
     (|\tilde{p}||\tilde{\vec{z}}|)^{-\frac{3}{2}}
     J_1(|\tilde{p}||\tilde{\vec{z}}|)\right)
     -\frac{\pi}{|\tilde{\vec{z}}|^2}
     \int\limits_0^{\infty}\!\!d|\tilde{p}|\,|\tilde{p}|
     J_2(|\tilde{p}||\tilde{\vec{z}}|)\nonumber\\
&&=\int\limits_0^{\infty}\!\!d|\tilde{p}|
     \frac{\pi}{2|\tilde{\vec{z}}|^3}
     \left(-6J_1(|\tilde{p}||\tilde{\vec{z}}|)
     -2|\tilde{p}||\tilde{\vec{z}}| J_2(|\tilde{p}||\tilde{\vec{z}}|)
     +3|\tilde{p}||\tilde{\vec{z}}|
     \,_1F_2\left(\frac{1}{2};\frac{3}{2},2;-\frac{(|\tilde{p}||\tilde{\vec{z}}|)
     ^2}{4}\right) \right)\nonumber\\
\end{eqnarray}
\begin{eqnarray}
&&\int\limits_0^{\infty}\!\!
     d|\tilde{p}|\,|\tilde{p}|^3
     \int\limits_0^{\pi}\!\!d\psi\,\sin^4\psi j_0(|\tilde{p}||\tilde{\vec{z}}|\sin\psi)\nonumber\\
&&=\sqrt{\frac{\pi}{2|\tilde{\vec{z}}|}}
     \int\limits_0^{\infty}\!\!d|\tilde{p}|\,|\tilde{p}|^{\frac{5}{2}}
     \int\limits_0^{\pi}\!\!d\psi\,\sin^{\frac{7}{2}}\psi
     J_{\frac{1}{2}}(|\tilde{p}||\tilde{\vec{z}}|\sin\psi)\nonumber\\
&&=\sqrt{\frac{\pi}{2|\tilde{\vec{z}}|}}
     \int\limits_0^{\infty}\!\!d|\tilde{p}|\,|\tilde{p}|^{\frac{5}{2}}
     \int\limits_0^{\pi}\!\!d\psi\,\sin^{\frac{7}{2}}\psi
     \left(\frac{3}{|\tilde{p}||\tilde{\vec{z}}|\sin\psi}
     J_{\frac{3}{2}}(|\tilde{p}||\tilde{\vec{z}}|\sin\psi)
     -J_{\frac{5}{2}}(|\tilde{p}||\tilde{\vec{z}}|\sin\psi)\right)\nonumber\\
&&=3\sqrt{2\pi}|\tilde{\vec{z}}|^{-\frac{3}{2}}
     \int\limits_0^{\infty}\!\!d|\tilde{p}|\,|\tilde{p}|^{\frac{3}{2}}
     \int\limits_0^{\frac{\pi}{2}}\!\!d\psi\,\sin^{\frac{5}{2}}\psi
     J_{\frac{3}{2}}(|\tilde{p}||\tilde{\vec{z}}|\sin\psi)\nonumber\\
  &&\hspace{10mm}-\sqrt{2\pi}|\tilde{\vec{z}}|^{-\frac{3}{2}}
     \int\limits_0^{\infty}\!\!d|\tilde{p}|\,|\tilde{p}|^{\frac{5}{2}}
     \int\limits_0^{\frac{\pi}{2}}\!\!d\psi\,\sin^{\frac{7}{2}}\psi
     J_{\frac{5}{2}}(|\tilde{p}||\tilde{\vec{z}}|\sin\psi)\nonumber\\
&&=3\sqrt{2\pi}|\tilde{\vec{z}}|^{-\frac{3}{2}}
     \int\limits_0^{\infty}\!\!d|\tilde{p}|\,|\tilde{p}|^{\frac{3}{2}}
     \left(2^{-\frac{1}{2}}\Gamma\left(\frac{1}{2}\right)
     (|\tilde{p}||\tilde{\vec{z}}|)^{-\frac{1}{2}}
     J_2(|\tilde{p}||\tilde{\vec{z}}|)\right)\nonumber\\
  &&\hspace{20mm}-3\sqrt{\frac{2\pi}{|\tilde{\vec{z}}|}}
     \int\limits_0^{\infty}\!\!d|\tilde{p}|\,|\tilde{p}|^{\frac{5}{2}}
     \left(2^{-\frac{1}{2}}\Gamma\left(\frac{1}{2}\right)
     (|\tilde{p}||\tilde{\vec{z}}|)^{-\frac{1}{2}}
     J_2(|\tilde{p}||\tilde{\vec{z}}|)\right)\nonumber\\
&&=\frac{3\pi}{|\tilde{\vec{z}}|^2}
     \int\limits_0^{\infty}\!\!d|\tilde{p}|\,|\tilde{p}|
     J_2(|\tilde{p}||\tilde{\vec{z}}|)
     -\frac{\pi}{|\tilde{\vec{z}}|}
     \int\limits_0^{\infty}\!\!d|\tilde{p}|\,|\tilde{p}|^2
     J_3(|\tilde{p}||\tilde{\vec{z}}|)
\end{eqnarray}
\newpage
\begin{eqnarray}
&&\int\limits_0^{\infty}\!\!
     d|\tilde{p}|\,|\tilde{p}|^3\int\limits_0^{\pi}\!\!d\psi\,\sin^4\psi
     j_2(|\tilde{p}||\tilde{\vec{z}}|\sin\psi)\nonumber\\
&&=\sqrt{\frac{2\pi}{|\tilde{\vec{z}}|}}
     \int\limits_0^{\infty}\!\!d|\tilde{p}|\,|\tilde{p}|^{\frac{5}{2}}
     \int\limits_0^{\frac{\pi}{2}}\!\!d\psi\,\sin^{\frac{7}{2}}\psi
     J_{\frac{5}{2}}(|\tilde{p}||\tilde{\vec{z}}|\sin\psi)\nonumber\\
&&=\sqrt{\frac{2\pi}{|\tilde{\vec{z}}|}}
     \int\limits_0^{\infty}\!\!d|\tilde{p}|\,|\tilde{p}|^{\frac{5}{2}}
     \left(2^{-\frac{1}{2}}\Gamma\left(\frac{1}{2}\right)
     (|\tilde{p}||\tilde{\vec{z}}|)^{-\frac{1}{2}}
     J_3(|\tilde{p}||\tilde{\vec{z}}|)\right)\nonumber\\
&&=\frac{\pi}{|\tilde{\vec{z}}|}
     \int\limits_0^{\infty}\!\!d|\tilde{p}|\,|\tilde{p}|^2 J_3(|\tilde{p}||\tilde{\vec{z}}|)\\
&&\int\limits_0^{\infty}\!\!
     d|\tilde{p}|\,|\tilde{p}|^3\int\limits_0^{\pi}\!\!d\psi\,
     \cos^2\psi j_0(|\tilde{p}||\tilde{\vec{z}}|\sin\psi)\nonumber\\
&&=\sqrt{\frac{\pi}{2|\tilde{\vec{z}}|}}
     \int\limits_0^{\infty}\!\!d|\tilde{p}|\,|\tilde{p}|^{\frac{5}{2}}
     \int\limits_0^{\pi}\!\!d\psi\,\sin^{-\frac{1}{2}}\psi\cos^2\psi
     J_{\frac{1}{2}}(|\tilde{p}||\tilde{\vec{z}}|\sin\psi)\nonumber\\
&&=\sqrt{\frac{\pi}{2|\tilde{\vec{z}}|}}
     \int\limits_0^{\infty}\!\!d|\tilde{p}|\,|\tilde{p}|^{\frac{5}{2}}
     \left(2^{-\frac{3}{2}}(1+(-1)^2)\Gamma\left(\frac{3}{2}\right)
     \Gamma\left(\frac{1}{2}\right)
     \frac{_1F_2\left(\frac{1}{2};2,\frac{3}{2};
     -\frac{(|\tilde{p}||\tilde{\vec{z}}|)^2}{4}\right)}
     {\Gamma(2)\Gamma\left(\frac{3}{2}\right)}\right)\nonumber\\
&&=\frac{\pi}{2}
     \int\limits_0^{\infty}\!\!d|\tilde{p}|\,|\tilde{p}|^3
     \,_1F_2\left(\frac{1}{2};2,\frac{3}{2};
     -\frac{(|\tilde{p}||\tilde{\vec{z}}|)^2}{4}\right)\\
&&\int\limits_0^{\infty}\!\!
     d|\tilde{p}|\,|\tilde{p}|^3\int\limits_0^{\pi}\!\!d\psi\,
     \sin^2\psi j_0(|\tilde{p}||\tilde{\vec{z}}|\sin\psi)\nonumber\\
&&=\sqrt{\frac{2\pi}{|\tilde{\vec{z}}|}}
     \int\limits_0^{\infty}\!\!d|\tilde{p}|\,|\tilde{p}|^{\frac{5}{2}}
     \int\limits_0^{\frac{\pi}{2}}\!\!d\psi\,\sin^{\frac{3}{2}}\psi
     J_{\frac{1}{2}}(|\tilde{p}||\tilde{\vec{z}}|\sin\psi)\nonumber\\
&&=\sqrt{\frac{2\pi}{|\tilde{\vec{z}}|}}
     \int\limits_0^{\infty}\!\!d|\tilde{p}|\,|\tilde{p}|^{\frac{5}{2}}
     \left(2^{-\frac{1}{2}}\Gamma\left(\frac{1}{2}\right)
     (|\tilde{p}||\tilde{\vec{z}}|)^{-\frac{1}{2}}
     J_1(|\tilde{p}||\tilde{\vec{z}}|)\right)\nonumber\\
&&=\frac{\pi}{|\tilde{\vec{z}}|}
     \int\limits_0^{\infty}\!\!d|\tilde{p}|\,|\tilde{p}|^2
     J_1(|\tilde{p}||\tilde{\vec{z}}|)\\
&&\int\limits_0^{\infty}\!\!d|\tilde{p}|\,|\tilde{p}|^3
     \int\limits_0^{\pi}\!\!d\psi\,
     \sin^2\psi j_2(|\tilde{p}||\tilde{\vec{z}}|\sin\psi)\nonumber\\
&&=\sqrt{\frac{\pi}{2|\tilde{\vec{z}}|}}
     \int\limits_0^{\infty}\!\!d|\tilde{p}|\,|\tilde{p}|^{\frac{5}{2}}
     \int\limits_0^{\pi}\!\!d\psi\,\sin^{\frac{3}{2}}\psi
     J_{\frac{5}{2}}(|\tilde{p}||\tilde{\vec{z}}|\sin\psi)\nonumber\\
&&=\sqrt{\frac{\pi}{2|\tilde{\vec{z}}|}}
     \int\limits_0^{\infty}\!\!d|\tilde{p}|\,|\tilde{p}|^{\frac{5}{2}}
     \int\limits_0^{\pi}\!\!d\psi\,\sin^{\frac{3}{2}}\psi
     \left(\frac{3}{|\tilde{p}||\tilde{\vec{z}}|\sin\psi}
     J_{\frac{3}{2}}(|\tilde{p}||\tilde{\vec{z}}|\sin\psi)
     -J_{\frac{1}{2}}(|\tilde{p}||\tilde{\vec{z}}|\sin\psi)\right)\nonumber\\
&&=3\sqrt{\frac{\pi}{2}}|\tilde{\vec{z}}|^{-\frac{3}{2}}
     \int\limits_0^{\infty}\!\!d|\tilde{p}|\,|\tilde{p}|^{\frac{3}{2}}
     \int\limits_0^{\pi}\!\!d\psi\,\sin^{\frac{1}{2}}\psi
     \left(\frac{1}{|\tilde{p}||\tilde{\vec{z}}|\sin\psi}
     J_{\frac{1}{2}}(|\tilde{p}||\tilde{\vec{z}}|\sin\psi)
     -J_{-\frac{1}{2}}(|\tilde{p}||\tilde{\vec{z}}|)\right)\nonumber\\
  &&\hspace{10mm}-\sqrt{\frac{2\pi}{|\tilde{\vec{z}}|\sin\psi}}
     \int\limits_0^{\infty}\!\!d|\tilde{p}|\,|\tilde{p}|^{\frac{5}{2}}
     \int\limits_0^{\frac{\pi}{2}}\!\!d\psi\,\sin^{\frac{3}{2}}\psi
     J_{\frac{1}{2}}(|\tilde{p}||\tilde{\vec{z}}|\sin\psi)\nonumber
\end{eqnarray}
\newpage
\begin{eqnarray}
&&=3\sqrt{\frac{\pi}{2}}|\tilde{\vec{z}}|^{-\frac{5}{2}}
     \int\limits_0^{\infty}\!\!d|\tilde{p}|\,|\tilde{p}|^{\frac{1}{2}}
     \int\limits_0^{\pi}\!\!d\psi\,\sin^{-\frac{1}{2}}\psi
     J_{\frac{1}{2}}(|\tilde{p}||\tilde{\vec{z}}|\sin\psi)\nonumber\\
  &&\hspace{10mm}-3\sqrt{2\pi}|\tilde{\vec{z}}|^{-\frac{3}{2}}
     \int\limits_0^{\infty}\!\!d|\tilde{p}|\,|\tilde{p}|^{\frac{3}{2}}
     \int\limits_0^{\frac{\pi}{2}}\!\!d\psi\,\sin^{\frac{1}{2}}\psi
     J_{-\frac{1}{2}}(|\tilde{p}||\tilde{\vec{z}}|\sin\psi)\nonumber\\
  &&\hspace{10mm}-\sqrt{2\pi}|\tilde{\vec{z}}|^{-\frac{1}{2}}
     \int\limits_0^{\infty}\!\!d|\tilde{p}|\,|\tilde{p}|^{\frac{5}{2}}
     \left(2^{-\frac{1}{2}}\Gamma\left(\frac{1}{2}\right)
     (|\tilde{p}||\tilde{\vec{z}}|)^{-\frac{1}{2}}
     J_1(|\tilde{p}||\tilde{\vec{z}}|)\right)\nonumber\\
&&=3\sqrt{\frac{\pi}{2}}|\tilde{\vec{z}}|^{-\frac{5}{2}}
     \int\limits_0^{\infty}\!\!d|\tilde{p}|\,|\tilde{p}|^{-\frac{3}{2}}
     \left(2^{-\frac{3}{2}}(1+(-1)^0)
     (|\tilde{p}||\tilde{\vec{z}}|)^{\frac{1}{2}}
     \Gamma\left(\frac{1}{2}\right)
     \Gamma\left(\frac{1}{2}\right)
     \frac{_1F_2\left(\frac{1}{2};1,\frac{3}{2};
     -\frac{(|\tilde{p}||\tilde{\vec{z}}|)^2}{4}\right)}
     {\Gamma(1)\Gamma\left(\frac{3}{2}\right)}\right)\nonumber\\
  &&\hspace{10mm}-3\sqrt{2\pi}|\tilde{\vec{z}}|^{-\frac{3}{2}}
     \int\limits_0^{\infty}\!\!d|\tilde{p}|\,|\tilde{p}|^{\frac{3}{2}}
     \left(2^{-\frac{1}{2}}\Gamma\left(\frac{1}{2}\right)
     (|\tilde{p}||\tilde{\vec{z}}|)^{-\frac{1}{2}}
     J_0(|\tilde{p}||\tilde{\vec{z}}|)\right)\nonumber\\
  &&\hspace{10mm}-\frac{\pi}{|\tilde{\vec{z}}|}
     \int\limits_0^{\infty}\!\!d|\tilde{p}|\,|\tilde{p}|^2 J_1(|\tilde{p}||\tilde{\vec{z}}|)\nonumber\\
  &&=-\int\limits_0^{\infty}\!\!d|\tilde{p}|\,
   \frac{\pi|\tilde{p}|}{|\tilde{\vec{z}}|^2}
   \bigg(3J_0(|\tilde{p}||\tilde{\vec{z}}|)\nonumber\\
&&\left.\hspace{25mm}+|\tilde{p}||\tilde{\vec{z}}|
   \left(J_1(|\tilde{p}||\tilde{\vec{z}}|)
   -\frac{3}{|\tilde{p}||\tilde{\vec{z}}|}\,
   _1\tilde{F}_2\left(\frac{1}{2};1,\frac{3}{2},
   -\frac{(|\tilde{p}||\tilde{\vec{z}}|)^2}{4}\right)
   \right)\right)
\end{eqnarray}

The analytic evaluation of the integrals over the modulus of the four-momentum is a very sophisticated task. In the following we compile the necessary prerequisites for a detailed calculation of the remaining integrals.

At first we have to mention that the involved hypergeometric functions $_1F_2\left(\frac{1}{2};1,\frac{3}{2};z\right)$, $_1F_2\left(\frac{1}{2};\frac{3}{2},2;z\right)$,
$_1F_2\left(2;3,3;z\right)$ and
$_1F_2\left(3;4,4;z\right)$,
can be expressed in terms of elementary functions and Bessel functions. It holds that \cite{Mathworld}
\begin{equation}
\label{hypergeometric1}
_1F_2\left(\frac{1}{2};1,\frac{3}{2};z\right)=I_0(2\sqrt{z})+\frac{\pi}{2}
\left(I_0(2\sqrt{z}){\bf{L}}_1(2\sqrt{z})
-I_1(2\sqrt{z}){\bf{L}}_0(2\sqrt{z})\right)\,,
\end{equation}
\begin{eqnarray}
\label{hypergeometric2}
_1F_2\left(\frac{1}{2};\frac{3}{2},2;z\right)&=&2I_0(2\sqrt{z})+
\pi\left(I_0(2\sqrt{z}){\bf{L}}_1(2\sqrt{z})
-I_1(2\sqrt{z}){\bf{L}}_0(2\sqrt{z})\right)
-\frac{1}{\sqrt{z}}I_1(2\sqrt{z})\,,\nonumber\\
\end{eqnarray}
\begin{equation}
\label{hypergeometric3}
_1F_2\left(2;3,3;z\right)=\frac{4}{z^2}\left(1-I_0(2\sqrt{z})
+\sqrt{z}I_1(2\sqrt{z})\right)\,,
\end{equation}
\begin{equation}
\label{hypergeometric4}
_1F_2\left(2;3,3;z\right)=\frac{18}{z^3}\left((z+2)I_0(2\sqrt{z})
-3\sqrt{z}I_1(2\sqrt{z})-2\right)\,,
\end{equation}
where $\sqrt{z}=\pm\frac{i\chi}{2}$. $I_\nu$ denotes a modified Bessel function of the first kind and $\bf{L}_\nu$ represents a modified Struve function. These functions are related to their regular counterparts via
\begin{eqnarray}
\label{modified regular1}
I_\nu(iz)&=&i^\nu J_\nu(z)\,,\\
\label{modified regular2}
{\bf{L}}_\nu(iz)&=&i^{\nu+1}{\bf{H}}_\nu(z)\,.
\end{eqnarray}
The primary definition of these functions are given as power series
\begin{eqnarray}
J_\nu(z)&=&\sum\limits_{k=0}^\infty\frac{(-1)^k}{\Gamma(k+\nu+1)k!}
\left(\frac{z}{2}\right)^{2k+\nu}\,,\\
I_\nu(z)&=&\sum\limits_{k=0}^\infty\frac{1}{\Gamma(k+\nu+1)k!}
\left(\frac{z}{2}\right)^{2k+\nu}\,,\\
{\bf{H}}_\nu(z)&=&\left(\frac{z}{2}\right)^{\nu+1}
\sum\limits_{k=0}^\infty\frac{(-1)^k}
{\Gamma\left(k+\frac{3}{2}\right)\Gamma\left(k+\nu+\frac{3}{2}\right)k!}
\left(\frac{z}{2}\right)^{2k}\,,\\
{\bf{L}}_\nu(z)&=&\left(\frac{z}{2}\right)^{\nu+1}
\sum\limits_{k=0}^\infty\frac{1}
{\Gamma\left(k+\frac{3}{2}\right)\Gamma\left(k+\nu+\frac{3}{2}\right)k!}
\left(\frac{z}{2}\right)^{2k}\,.
\end{eqnarray}
The parity of Bessel and Struve functions is determined by the order $\nu$ in the following way:
\begin{eqnarray}
\label{Bessel parity1}
I_\nu(-z)&=&(-1)^\nu I_\nu(z)\,,\\
\label{Bessel parity2}
{\bf{L}}_\nu(-z)&=&(-1)^{\nu+1}{\bf{L}}_\nu(z)\,,\\
\label{Bessel parity3}
J_\nu(-z)&=&(-1)^\nu J_\nu(z)\,,\\
\label{Bessel parity4}
{\bf{H}}_\nu(-z)&=&(-1)^{\nu+1}{\bf{H}}_\nu(z)\,.
\end{eqnarray}
Eqs.\,(\ref{modified regular}), (\ref{modified regular2}) and (\ref{Bessel parity}) - (\ref{Bessel parity4}) show that the sign of the square root of $z$ has no influence on the final result in this concrete case, since the sign cancels out anyway. Thus we constrain ourselves to the positive sign for the remainder of the calculation. In the following $\chi$ will denote the product $|\tilde{p}||\tilde{\vec{z}}|$. As a consequence of the mentioned properties of $I_\nu$ and $\bf{L}_\nu$, or $J_\nu$ and $\bf{H}_\nu$, respectively, we can transform Eqs.\,(\ref{hypergeometric1}) and (\ref{hypergeometric2}) into
\begin{eqnarray}
\label{hypergeometric5}
_1F_2\left(\frac{1}{2};1,\frac{3}{2};\frac{-\chi^2}{4}\right)&=&J_0(\chi)
+\frac{\pi}{2}\left(-J_0(\chi){\bf{H}}_1(\chi)
+J_1(\chi){\bf{H}}_0(\chi)\right)\,,
\end{eqnarray}
\begin{eqnarray}
\label{hypergeometric6}
_1F_2\left(\frac{1}{2};\frac{3}{2},2;\frac{-\chi^2}{4}\right)&=&2J_0(\chi)+
\pi\left(-J_0(\chi){\bf{H}}_1(\chi)+J_1(\chi){\bf{H}}_0(\chi)\right)
-\frac{2}{\chi}J_1(\chi)\,,
\end{eqnarray}
\begin{eqnarray}
\label{hypergeometric7}
_1F_2\left(2;3,3;\frac{-\chi^2}{4}\right)&=&
\frac{64}{\chi^4}\left(1-J_0(\chi)-\frac{\chi}{2}J_1(\chi)\right)\,,
\end{eqnarray}
\begin{eqnarray}
\label{hypergeometric8}
_1F_2\left(3;4,4;\frac{-\chi^2}{4}\right)&=&
-\frac{1152}{\chi^6}\left(\left(-\frac{\chi^2}{4}+2\right)J_0(\chi)
+\frac{3\chi}{2}J_1(\chi)-2\right)\,.
\end{eqnarray}
Another important property of the Bessel functions is their normalization. It holds that
\begin{equation}
\int\limits_0^\infty\!dz\,J_\nu(z)=1\,,
\end{equation}
for positive integral and real $\nu$.

The strategy for the evaluation of the momentum integration is to reduce every integral to a generalized Lipschitz-Hankel integral. An investigation concerning these integrals of the structure
\begin{equation}
\label{Lipschitz Hankel}
\int\limits_0^\infty\!dz\,\e^{-\alpha z}J_\nu(\beta z)\,z^{\mu-1}
\end{equation}
is presented in \cite{Watson}. The interesting integrals are available by performing the limiting process $\alpha\rightarrow0^+$.
In \cite{Watson} this prototype integral (\ref{Lipschitz Hankel}) is evaluated as
\begin{eqnarray}
\int\limits_0^\infty\!dz\,\e^{-\alpha z}J_\nu(\beta z)\,z^{\mu-1}&=&
\frac{\left(\frac{\beta}{2\alpha}\right)^\nu\Gamma(\mu+\nu)}{\alpha^\mu\Gamma(\nu+1)}\,
_2F_1\left(\frac{\mu+\nu}{2},\frac{\mu+\nu+1}{2};\nu+1;-\frac{\beta^2}{\alpha^2}\right)
\nonumber\\
&=&
\frac{\left(\frac{\beta}{2\alpha}\right)^\nu\Gamma(\mu+\nu)}{\alpha^\mu\Gamma(\nu+1)}\,
\left(1+\frac{\beta^2}{\alpha^2}\right)^{\frac{1}{2}-\mu}\,
_2F_1\left(\frac{\nu-\mu+1}{2},\frac{\nu-\mu}{2}+1;\nu+1;-\frac{\beta^2}{\alpha^2}\right)
\nonumber\\
&=&
\frac{\left(\frac{\beta}{2}\right)^\nu\Gamma(\mu+\nu)}{(\alpha^2+\beta^2)
^{\frac{1}{2}(\mu+\nu)}\Gamma(\nu+1)}\,
_2F_1\left(\frac{\mu+\nu}{2},\frac{1-\mu+\nu}{2};\nu+1;
\frac{\beta^2}{\alpha^2+\beta^2}\right)
\nonumber\,,\\
\end{eqnarray}
provided that $\mathfrak{Re}(\mu+\nu)>0$ and  $\mathfrak{Re}(\mu)>|\mathfrak{Re}(\nu)|$ to secure convergence at the origin and $\mathfrak{Re}(\alpha\pm i\beta)>0$ to secure convergence at infinity. The hypergeometric function $_2F_1$ is called Gauss' confluent hypergeometric function. Relevant for our considerations are the cases where $\mu$ is taken equal to $\nu+1$ or $\nu+2$. Then we obtain
\begin{eqnarray}
\int\limits_0^\infty\!dz\,\e^{-\alpha z}J_\nu(\beta z)\,z^\nu&=&
\frac{(2\beta)^\nu\Gamma\left(\nu+\frac{1}{2}\right)}
{(\alpha^2+\beta^2)^{\nu+\frac{1}{2}}\sqrt{\pi}}\,,\\
\int\limits_0^\infty\!dz\,\e^{-\alpha z}J_\nu(\beta z)\,z^{\nu+1}&=&
(2\alpha)\frac{(2\beta)^\nu\Gamma\left(\nu+\frac{3}{2}\right)}
{(\alpha^2+\beta^2)^{\nu+\frac{3}{2}}\sqrt{\pi}}\,,
\end{eqnarray}
so long as $\mathfrak{Re}(\nu)>-\frac{1}{2}$, $\mathfrak{Re}(\nu)>-1$, respectively.
The two integrals above imply that for $\mu-\nu$ being an even integer, an integral of the Lipschitz-Hankel type vanishes in the limit $\alpha\rightarrow0^+$.

At this point we have collected the tools to proceed with the explicit calculation.
\begin{eqnarray}
&&\frac{\pi}{|\tilde{\vec{z}}|^2}\int\limits_0^\infty\!d|\tilde{p}|\,
|\tilde{p}|J_2(|\tilde{p}||\tilde{\vec{z}}|)
=\frac{\pi}{|\tilde{\vec{z}}|^4}\int\limits_0^\infty\!d\chi\,
\chi\left(\frac{2}{\chi}J_1(\chi)-J_0(\chi)\right)\nonumber\\
&&=\frac{\pi}{|\tilde{\vec{z}}|^4}
\left(2\underbrace{\int\limits_0^\infty\!d\chi\,J_1(\chi)}_{=1}
-\underbrace{\int\limits_0^\infty\!d\chi\,\chi J_0(\chi)}_{=0}\right)
=\frac{2\pi}{|\tilde{\vec{z}}|^4}
\end{eqnarray}
\begin{eqnarray*}
&&\frac{\pi}{2|\tilde{\vec{z}}|^3}\int\limits_0^\infty\!d|\tilde{p}|\,
\left(-6J_1(|\tilde{p}||\tilde{\vec{z}}|)
-2|\tilde{p}||\tilde{\vec{z}}|J_2(|\tilde{p}||\tilde{\vec{z}}|)
+3|\tilde{p}||\tilde{\vec{z}}|\,_1F_2\left(\frac{1}{2};\frac{3}{2},2;
-\frac{(|\tilde{p}||\tilde{\vec{z}}|)^2}{4}\right)\right)\nonumber\\
&&=-\frac{3\pi}{|\tilde{\vec{z}}|^4}\int\limits_0^\infty\!d\chi\,J_1(\chi)
-\frac{\pi}{|\tilde{\vec{z}}|^4}\int\limits_0^\infty\!d\chi\,\chi J_2(\chi)
+\frac{3\pi}{2|\tilde{\vec{z}}|^4}
\int\limits_0^\infty\!d\chi\,\chi
\,_1F_2\left(\frac{1}{2};\frac{3}{2},2;-\frac{\chi^2}{4}\right)\nonumber\\
&&=-\frac{3\pi}{|\tilde{\vec{z}}|^4}
\underbrace{\int\limits_0^\infty\!d\chi\,J_1(\chi)}_{=1}
-\frac{\pi}{|\tilde{\vec{z}}|^4}
\left(2\underbrace{\int\limits_0^\infty\!d\chi\,J_1(\chi)}_{=1}
-\underbrace{\int\limits_0^\infty\!d\chi\,\chi J_0(\chi)}_{=0}\right)\nonumber\\
  &&\qquad+\frac{3\pi}{2|\tilde{\vec{z}}|^4}
  \int\limits_0^\infty\!d\chi\,\chi
  \left(2J_0(\chi)+
  \pi\left(-J_0(\chi){\bf{H}}_1(\chi)+J_1(\chi){\bf{H}}_0(\chi)\right)
  -\frac{2}{\chi}J_1(\chi)\right)\nonumber\\
\end{eqnarray*}
\newpage
\begin{eqnarray}
&&=-\frac{5\pi}{|\tilde{\vec{z}}|^4}
+\frac{3\pi}{2|\tilde{\vec{z}}|^4}
\left(2\underbrace{\int\limits_0^\infty\!d\chi\,\chi J_0(\chi)}_{=0}
-\pi\int\limits_0^\infty\!d\chi\,\chi J_0(\chi){\bf{H}}_1(\chi)
+\pi\int\limits_0^\infty\!d\chi\,\chi J_1(\chi){\bf{H}}_0(\chi)
-2\underbrace{\int\limits_0^\infty\!d\chi\,J_1(\chi)}_{=1}\right)\nonumber\\
&&=-\frac{8\pi}{|\tilde{\vec{z}}|^4}
-\frac{3\pi}{2|\tilde{\vec{z}}|^4}\pi\sum\limits_{k=0}^\infty\frac{(-1)^k}
{\Gamma\left(k+\frac{3}{2}\right)\Gamma\left(k+\frac{5}{2}\right)k!}
\underbrace
{\int\limits_0^\infty\!d\chi\,\frac{\chi^{2k+3}}{2^{2k+2}} J_0(\chi)}_{=0}
\nonumber\\
  &&\hspace{30mm}
  +\frac{3\pi}{2|\tilde{\vec{z}}|^4}\pi\sum\limits_{k=0}^\infty\frac{(-1)^k}
  {\Gamma\left(k+\frac{3}{2}\right)\Gamma\left(k+\frac{3}{2}\right)k!}
  \underbrace
  {\int\limits_0^\infty\!d\chi\,\frac{\chi^{2k+2}}{2^{2k+1}} J_1(\chi)}_{=0}
=-\frac{8\pi}{|\tilde{\vec{z}}|^4}
\end{eqnarray}
\begin{eqnarray}
&&\frac{\pi}{|\tilde{\vec{z}}|^3}
\int\limits_0^\infty\!d|\tilde{p}|\,
\left(3|\tilde{p}||\tilde{\vec{z}}| J_2(|\tilde{p}||\tilde{\vec{z}}|)
-(|\tilde{p}||\tilde{\vec{z}}|)^2 J_3(|\tilde{p}||\tilde{\vec{z}}|)\right)\nonumber\\
&&=\frac{3\pi}{|\tilde{\vec{z}}|^4}\int\limits_0^\infty\!d\chi\,
\chi\left(\frac{2}{\chi}J_1(\chi)-J_0(\chi)\right)
-\frac{\pi}{|\tilde{\vec{z}}|^4}\int\limits_0^\infty\!d\chi\,
\chi^2\left(\frac{4}{\chi}J_2(\chi)-J_1(\chi)\right)\nonumber\\
&&=\frac{6\pi}{|\tilde{\vec{z}}|^4}
\underbrace{\int\limits_0^\infty\!d\chi\,J_1(\chi)}_{=1}
-\frac{3\pi}{|\tilde{\vec{z}}|^4}
\underbrace{\int\limits_0^\infty\!d\chi\,\chi J_0(\chi)}_{=0}
-\frac{4\pi}{|\tilde{\vec{z}}|^4}
\int\limits_0^\infty\!d\chi\,
\chi\left(\frac{2}{\chi}J_1(\chi)-J_0(\chi)\right)
+\frac{\pi}{|\tilde{\vec{z}}|^4}
\underbrace{\int\limits_0^\infty\!d\chi\,\chi^2 J_1(\chi)}_{=0}\nonumber\\
&&=\frac{6\pi}{|\tilde{\vec{z}}|^4}
-\frac{8\pi}{|\tilde{\vec{z}}|^4}
\underbrace{\int\limits_0^\infty\!d\chi\,J_1(\chi)}_{=1}
+\frac{4\pi}{|\tilde{\vec{z}}|^4}
\underbrace{\int\limits_0^\infty\!d\chi\,\chi J_0(\chi)}_{=0}
=-\frac{2\pi}{|\tilde{\vec{z}}|^4}
\end{eqnarray}
\begin{eqnarray}
&&\frac{\pi}{|\tilde{\vec{z}}|}\int\limits_0^\infty\!d|\tilde{p}|\,
|\tilde{p}|^2J_3(|\tilde{p}||\tilde{\vec{z}}|)
=\frac{\pi}{|\tilde{\vec{z}}|^4}\int\limits_0^\infty\!d\chi\,
\chi^2\left(\frac{4}{\chi}J_2(\chi)-J_1(\chi)\right)\nonumber\\
&&=\frac{4\pi}{|\tilde{\vec{z}}|^4}
\int\limits_0^\infty\!d\chi\,
\chi\left(\frac{2}{\chi}J_1(\chi)-J_0(\chi)\right)
-\frac{\pi}{|\tilde{\vec{z}}|^4}
\underbrace{\int\limits_0^\infty\!d\chi\,\chi^2 J_1(\chi)}_{=0}\nonumber\\
&&=\frac{8\pi}{|\tilde{\vec{z}}|^4}
\underbrace{\int\limits_0^\infty\!d\chi\,J_1(\chi)}_{=1}
-\frac{4\pi}{|\tilde{\vec{z}}|^4}
\underbrace{\int\limits_0^\infty\!d\chi\,\chi J_0(\chi)}_{=0}
=\frac{8\pi}{|\tilde{\vec{z}}|^4}
\end{eqnarray}
\newpage

\begin{eqnarray}
&&\frac{\pi}{2}\int\limits_0^\infty\!d|\tilde{p}|\,|\tilde{p}|^3
\,_1F_2\left(\frac{1}{2};\frac{3}{2},2;-\frac{(|\tilde{p}||\tilde{\vec{z}}|)^2}{4}\right)
\nonumber\\
&&=\frac{\pi}{2|\tilde{\vec{z}}|^4}\int\limits_0^\infty\!d\chi\,\chi^3
\,_1F_2\left(\frac{1}{2};\frac{3}{2},2;-\frac{\chi^2}{4}\right)\nonumber\\
&&=\frac{\pi}{2|\tilde{\vec{z}}|^4}\int\limits_0^\infty\!d\chi\,\chi^3
\left(2J_0(\chi)+\pi\left(-J_0(\chi){\bf{H}}_1(\chi)+J_1(\chi){\bf{H}}_0(\chi)\right)
-\frac{2}{\chi}J_1(\chi)\right)\nonumber\\
&&=\frac{\pi}{2|\tilde{\vec{z}}|^4}
\left(2\underbrace{\int\limits_0^\infty\!d\chi\,\chi^3 J_0(\chi)}_{=0}
-\pi\int\limits_0^\infty\!d\chi\,\chi^3 J_0(\chi){\bf{H}}_1(\chi)
+\pi\int\limits_0^\infty\!d\chi\,\chi^3 J_1(\chi){\bf{H}}_0(\chi)
-2\underbrace{\int\limits_0^\infty\!d\chi\,\chi^3 J_1(\chi)}_{=0}
\right)\nonumber\\
&&=-\frac{\pi}{2|\tilde{\vec{z}}|^4}\pi\sum\limits_{k=0}^\infty\frac{(-1)^k}
{\Gamma\left(k+\frac{3}{2}\right)\Gamma\left(k+\frac{5}{2}\right)k!}
\underbrace
{\int\limits_0^\infty\!d\chi\,\frac{\chi^{2k+5}}{2^{2k+2}} J_0(\chi)}_{=0}\nonumber\\
  &&\hspace{35mm}
  +\frac{\pi}{2|\tilde{\vec{z}}|^4}\pi\sum\limits_{k=0}^\infty\frac{(-1)^k}
  {\Gamma\left(k+\frac{3}{2}\right)\Gamma\left(k+\frac{3}{2}\right)k!}
  \underbrace
  {\int\limits_0^\infty\!d\chi\,\frac{\chi^{2k+4}}{2^{2k+1}} J_1(\chi)}_{=0}
=0
\end{eqnarray}
\begin{eqnarray}
\frac{\pi}{|\tilde{\vec{z}}|}
\int\limits_0^\infty\!d|\tilde{p}|\,
|\tilde{p}|^2 J_1(|\tilde{p}||\tilde{\vec{z}}|)=
\frac{\pi}{|\tilde{\vec{z}}|^4}
\underbrace{\int\limits_0^\infty\!d\chi\,\chi^2 J_1(|\tilde{p}||\tilde{\vec{z}}|)}_{=0}
=0
\end{eqnarray}
\begin{eqnarray}
&&-\frac{\pi}{|\tilde{\vec{z}}|^2}\int\limits_0^\infty\!d|\tilde{p}|\,
\left(3|\tilde{p}|J_0(|\tilde{p}||\tilde{\vec{z}}|)
+|\tilde{p}|^2|\tilde{\vec{z}}|J_1(|\tilde{p}||\tilde{\vec{z}}|)
-|\tilde{p}|\,_1F_2\left(\frac{1}{2};1,\frac{3}{2};
-\frac{(|\tilde{p}||\tilde{\vec{z}}|)^2}{4}\right)\right)\nonumber\\
&&=-\frac{3\pi}{|\tilde{\vec{z}}|^4}
\underbrace{\int\limits_0^\infty\!d\chi\,\chi J_0(\chi)}_{=0}
-\frac{\pi}{|\tilde{\vec{z}}|^4}
\underbrace{\int\limits_0^\infty\!d\chi\,\chi^2 J_1(\chi)}_{=0}
+\frac{\pi}{|\tilde{\vec{z}}|^4}\frac{1}{\Gamma(1)\Gamma\left(\frac{3}{2}\right)}
\int\limits_0^\infty\!d\chi\,\chi\,_1F_2\left(\frac{1}{2};1,\frac{3}{2};
-\frac{\chi^2}{4}\right)\nonumber\\
&&=\frac{\pi}{|\tilde{\vec{z}}|^4}
\int\limits_0^\infty\!d\chi\,\chi
\left(J_0(\chi)+\frac{\pi}{2}(-J_0(\chi){\bf{H}}_1(\chi)+J_1(\chi){\bf{H}}_0(\chi))
\right)\nonumber\\
&&=\frac{\pi}{|\tilde{\vec{z}}|^4}
\underbrace{\int\limits_0^\infty\!d\chi\,\chi J_0(\chi)}_{=0}
-\frac{\pi}{|\tilde{\vec{z}}|^4}\frac{\pi}{2}
\int\limits_0^\infty\!d\chi\,\chi J_0(\chi){\bf{H}}_1(\chi)
+\frac{\pi}{|\tilde{\vec{z}}|^4}\frac{\pi}{2}
\int\limits_0^\infty\!d\chi\,\chi J_1(\chi){\bf{H}}_0(\chi)\nonumber\\
&&=-\frac{\pi}{|\tilde{\vec{z}}|^4}\frac{\pi}{2}
\sum\limits_{k=0}^\infty\frac{(-1)^k}
{\Gamma\left(k+\frac{3}{2}\right)\Gamma\left(k+\frac{5}{2}\right)k!}
\underbrace
{\int\limits_0^\infty\!d\chi\,\frac{\chi^{2k+3}}{2^{2k+2}} J_0(\chi)}_{=0}
\nonumber\\
  &&\hspace{30mm}
  +\frac{\pi}{|\tilde{\vec{z}}|^4}
  \frac{\pi}{2}\sum\limits_{k=0}^\infty\frac{(-1)^k}
  {\Gamma\left(k+\frac{3}{2}\right)\Gamma\left(k+\frac{3}{2}\right)k!}
  \underbrace{\int\limits_0^\infty\!d\chi\,\frac{\chi^{2k+2}}{2^{2k+1}} J_1(\chi)}_{=0}
=0
\end{eqnarray}

Finally we calculate the order-of-magnitude estimate for the vacuum contribution to the two-point energy density correlator in a more detailed manner. Recall that the integrations are performed up to a limit determined by the maximal resolution $|\phi|$ of the effective theory. In principle, we obtain our results by employing the following master integrals (with the integration cut-off $\Lambda$) \cite{Mathworld}
\begin{equation}
\int\limits_0^\Lambda\!dp\,p^\alpha J_\beta(\gamma p)=\frac{2^{-\beta}\Lambda^{1+\alpha+\beta}\gamma^\beta
\,_1F_2\left(\frac{1+\alpha+\beta}{2};
\frac{3+\alpha+\beta}{2},1+\beta;-\frac{\gamma^2\Lambda^2}{4}\right)}
{(1+\alpha+\beta)\Gamma(1+\beta)}\,,\\
\end{equation}
where $\mathfrak{Re}(\alpha+\beta)\geq-1$ and $\gamma p\geq0$,
\begin{equation}
\int\limits_0^\Lambda\!dp\,p^\alpha\,_1F_2(a;b,c;\sigma p^2)=\frac{\Lambda^{1+\alpha}\,_2F_3
    \left(a,\frac{1+\alpha}{2};b,c,\frac{3+\alpha}{2};\sigma\Lambda^2\right)}{1+\alpha}\,,
\end{equation}
where $\mathfrak{Re}(\alpha)>-1$ and $\sigma \Lambda^2\in\mathbb{R}_{\leq0}$.
An important formula concerning integrals over a separate Bessel function is given by the sum
\begin{equation}
\int\limits_0^{\Lambda}\!dz\,J_\nu(z)=2\sum\limits_{k=0}^\infty J_{\nu+2k+1}(\Lambda)\,,
\end{equation}
provided that $\mathfrak{Re}(\nu)>-1$. The regular Bessel functions of the first kind of even order are resummable into the expression
\begin{equation}
\sum\limits_{k=1}^\infty J_{2k}(z)=\frac{1}{2}(1-J_0(z))\,.
\end{equation}
In several cases we will be confronted with integrals containing a product of Bessel and Struve functions. Then the following indefinite integral is employed \cite{Mathworld}:
\begin{eqnarray}
\int\!dz \frac{J_\nu(z){\bf{H}}_\mu(z)}{z}&=&
\frac{1}{(\mu-\nu)(\mu+\nu)}
\Bigg(J_\nu(z){\bf{H}}_{\mu-1}(z)z\nonumber\\
&&
-\frac{2^{-\mu-\nu}z^{\mu+\nu+1}\Gamma\left(\frac{\mu+\nu+1}{2}\right)}
{\sqrt{\pi}\Gamma\left(\mu+\frac{1}{2}\right)
\Gamma\left(\nu+1\right)\Gamma\left(\frac{\mu+\nu+3}{2}\right)}
\,_1F_2\left(\frac{\mu+\nu+1}{2};\nu+1,\frac{\mu+\nu+3}{2};-\frac{z^2}{4}\right)
\nonumber\\
&&
-(z J_{\nu-1}(z)+(\mu-\nu)J_\nu(z)){\bf{H}}_\mu(z)
\Bigg)\,.
\end{eqnarray}

Finally we should mention the primary definition of a generalized hypergeometric function. $_pF_q(a_1,\dots,a_p;b_1,\dots,b_q;z)$ is defined in terms of a power series:
\begin{equation}
_pF_q(a_1,\dots,a_p;b_1,\dots,b_q;z)=\sum\limits_{k=0}^\infty
\frac{\prod_{j=1}^p(a_j)_k}{\prod_{j=1}^q(b_j)_k}\frac{z^k}{k!}\,,
\end{equation}
where $(\alpha)_k$ denotes the Pochhammer symbol
\begin{equation}
(\alpha)_k=\frac{\Gamma(\alpha+k)}{\Gamma(\alpha)}\,.
\end{equation}
This definition implies the following property:
\begin{equation}
_pF_q(a_1,\dots,a_p;b_1,\dots,b_{q-1},a_p;z)
=_{p-1}F_{q-1}(a_1,\dots,a_{p-1};b_1,\dots,b_{q-1};z)\,.
\end{equation}

With these formulas we are able to proceed with a detailed calculation of the integrals.

\begin{eqnarray}
&&\frac{\pi}{|\tilde{\vec{z}}|^2}\int\limits_0^{|\tilde{\phi}|}\!d|\tilde{p}|\,
|\tilde{p}|J_2(|\tilde{p}||\tilde{\vec{z}}|)\nonumber\\
&&=\frac{\pi}{|\tilde{\vec{z}}|^2}
\frac{|\tilde{\phi}|^4|\tilde{\vec{z}}|^2}{32}
\,_1F_2\left(2;3,3;-\frac{|\tilde{\phi}|^2|\tilde{\vec{z}}|^2}{4}\right)\nonumber\\
&&=\frac{\pi}{|\tilde{\vec{z}}|^2}
\frac{|\tilde{\phi}|^4|\tilde{\vec{z}}|^2}{32}
\frac{64}{|\tilde{\phi}|^4|\tilde{\vec{z}}|^4}
\left(1-I_0(i|\tilde{\phi}||\tilde{\vec{z}}|)+\frac{i|\tilde{\phi}||\tilde{\vec{z}}|}{2}
I_1(i|\tilde{\phi}||\tilde{\vec{z}}|)\right)\nonumber\\
&&=\frac{\pi}{|\tilde{\vec{z}}|^4}
\left(2-2J_0(|\tilde{\phi}||\tilde{\vec{z}}|)
-|\tilde{\phi}||\tilde{\vec{z}}|J_1(|\tilde{\phi}||\tilde{\vec{z}}|)\right)
\end{eqnarray}
\begin{eqnarray}
&&\frac{\pi}{2|\tilde{\vec{z}}|^3}\int\limits_0^{|\tilde{\phi}|}\!d|\tilde{p}|\,
\left(-6J_1(|\tilde{p}||\tilde{\vec{z}}|)
-2|\tilde{p}||\tilde{\vec{z}}|J_2(|\tilde{p}||\tilde{\vec{z}}|)
+\,_1F_2\left(\frac{1}{2};\frac{3}{2},2;-\frac{|\tilde{p}|^2|\tilde{\vec{z}}|^2}{4}\right)
\right)\nonumber\\
&&=-\frac{3\pi}{|\tilde{\vec{z}}|^4}
\int\limits_0^{|\tilde{\phi}||\tilde{\vec{z}}|}\!d\chi\,J_1(\chi)
-\frac{\pi}{|\tilde{\vec{z}}|^2}\int\limits_0^{|\tilde{\phi}|}\!d|\tilde{p}|\,
|\tilde{p}| J_2(|\tilde{p}||\tilde{\vec{z}}|)
+\frac{3\pi}{2|\tilde{\vec{z}}|^2}\int\limits_0^{|\tilde{\phi}|}\!d|\tilde{p}|\,
\,_1F_2\left(\frac{1}{2};\frac{3}{2},2;-\frac{|\tilde{p}|^2|\tilde{\vec{z}}|^2}{4}\right)
\nonumber\\
&&=-\frac{3\pi}{|\tilde{\vec{z}}|^4}
2\sum\limits_{k=0}^\infty J_{2(k+1)}(\chi)
+\frac{\pi}{|\tilde{\vec{z}}|^4}
\left(2J_0(|\tilde{\phi}||\tilde{\vec{z}}|)
+|\tilde{\phi}||\tilde{\vec{z}}|J_1(|\tilde{\phi}||\tilde{\vec{z}}|)-2\right)\nonumber\\
&&\hspace{40mm}
 +\frac{3\pi}{2|\tilde{\vec{z}}|^2}\int\limits_0^{|\tilde{\phi}|}\!d|\tilde{p}|\,
\left(2\,_1F_2\left(\frac{1}{2};1,\frac{3}{2};
-\frac{|\tilde{p}|^2|\tilde{\vec{z}}|^2}{4}\right)
-\frac{2}{|\tilde{p}||\tilde{\vec{z}}|}J_1(|\tilde{p}||\tilde{\vec{z}}|)\right)
\nonumber\\
&&=-\frac{3\pi}{|\tilde{\vec{z}}|^4}\left(1-J_0(|\tilde{\phi}||\tilde{\vec{z}}|)\right)
+\frac{\pi}{|\tilde{\vec{z}}|^4}
\left(2J_0(|\tilde{\phi}||\tilde{\vec{z}}|)
+|\tilde{\phi}||\tilde{\vec{z}}|J_1(|\tilde{\phi}||\tilde{\vec{z}}|)-2\right)\nonumber\\
&&\hspace{40mm}
 +\frac{3\pi}{2|\tilde{\vec{z}}|^4}|\tilde{\phi}|^2|\tilde{\vec{z}}|^2
\,_2F_3\left(\frac{1}{2},1;1,\frac{3}{2},2;-\frac{|\tilde{\phi}|^2|\tilde{\vec{z}}|^2}{4}
\right)
-\frac{3\pi}{2|\tilde{\vec{z}}|^4}\left(1-J_0(|\tilde{\phi}||\tilde{\vec{z}}|)\right)
\nonumber\\
&&=\frac{\pi}{2|\tilde{\vec{z}}|^4}\left(2\left(8J_0(|\tilde{\phi}||\tilde{\vec{z}}|)
+|\tilde{\phi}||\tilde{\vec{z}}|J_1(|\tilde{\phi}||\tilde{\vec{z}}|)-8\right)
+3|\tilde{\phi}|^2|\tilde{\vec{z}}|^2\,_1F_2\left(\frac{1}{2};\frac{3}{2},2;
-\frac{|\tilde{\phi}|^2|\tilde{\vec{z}}|^2}{4}\right)\right)\nonumber\\
\end{eqnarray}
\begin{eqnarray}
&&\frac{\pi}{|\tilde{\vec{z}}|^2}\int\limits_0^{|\tilde{\phi}|}\!d|\tilde{p}|\,
\left(3|\tilde{p}|J_2(|\tilde{p}||\tilde{\vec{z}}|)
-|\tilde{\vec{z}}||\tilde{p}|^2J_3(|\tilde{p}||\tilde{\vec{z}}|)\right)\nonumber\\
&&=\frac{3\pi}{|\tilde{\vec{z}}|^2}\int\limits_0^{|\tilde{\phi}|}\!d|\tilde{p}|\,
|\tilde{p}|J_2(|\tilde{p}||\tilde{\vec{z}}|)
-\frac{\pi}{|\tilde{\vec{z}}|}\int\limits_0^{|\tilde{\phi}|}\!d|\tilde{p}|\,
|\tilde{p}|^2J_3(|\tilde{p}||\tilde{\vec{z}}|)\nonumber\\
&&=\frac{6\pi}{|\tilde{\vec{z}}|^4}
\left(1-J_0(|\tilde{\phi}||\tilde{\vec{z}}|)
-\frac{|\tilde{\phi}||\tilde{\vec{z}}|}{2}J_1(|\tilde{\phi}||\tilde{\vec{z}}|)\right)
-\frac{\pi}{|\tilde{\vec{z}}|}\frac{|\tilde{\phi}|^6|\tilde{\vec{z}}|^3}{288}
\,_1F_2\left(3;4,4;-\frac{|\tilde{\phi}|^2|\tilde{\vec{z}}|^2}{4}\right)\nonumber\\
&&=\frac{6\pi}{|\tilde{\vec{z}}|^4}
\left(1-J_0(|\tilde{\phi}||\tilde{\vec{z}}|)
-\frac{|\tilde{\phi}||\tilde{\vec{z}}|}{2}J_1(|\tilde{\phi}||\tilde{\vec{z}}|)\right)
\nonumber\\
 &&\hspace{30mm}
-\frac{\pi}{|\tilde{\vec{z}}|}\frac{|\tilde{\phi}|^6|\tilde{\vec{z}}|^3}{288}
\frac{1152}{(-|\tilde{\phi}|^6|\tilde{\vec{z}}|^6)}
\left(\left(-\frac{|\tilde{\phi}|^2|\tilde{\vec{z}}|^2}{4}
+2\right)J_0(|\tilde{\phi}||\tilde{\vec{z}}|)+\frac{3|\tilde{\phi}||\tilde{\vec{z}}|}{2}
J_1(|\tilde{\phi}||\tilde{\vec{z}}|)-2\right)\nonumber\\
&&=-\frac{\pi}{|\tilde{\vec{z}}|^4}\left((|\tilde{\phi}|^2|\tilde{\vec{z}}|^2-2)
J_0(|\tilde{\phi}||\tilde{\vec{z}}|)
+3|\tilde{\phi}||\tilde{\vec{z}}|J_1(|\tilde{\phi}||\tilde{\vec{z}}|)-2\right)
\nonumber\\
\end{eqnarray}
\begin{eqnarray}
&&\frac{\pi}{|\tilde{\vec{z}}|}\int\limits_0^{|\tilde{\phi}|}\!d|\tilde{p}|\,
|\tilde{p}|^2J_3(|\tilde{p}||\tilde{\vec{z}}|)\nonumber\\
&&=\frac{\pi}{|\tilde{\vec{z}}|^4}\left((|\tilde{\phi}|^2|\tilde{\vec{z}}|^2-2)
J_0(|\tilde{\phi}||\tilde{\vec{z}}|)-6|\tilde{\phi}||\tilde{\vec{z}}|J_1(|\tilde{\phi}||\tilde{\vec{z}}|)+2\right)
\end{eqnarray}
\begin{eqnarray}
&&\frac{\pi}{2}\int\limits_0^{|\tilde{\phi}|}\!d|\tilde{p}|\,
|\tilde{p}|^3\,_1F_2\left(\frac{1}{2};\frac{3}{2},2;
-\frac{|\tilde{p}|^2|\tilde{\vec{z}}|^2}{4}\right)\nonumber\\
&&=\frac{\pi}{2}\frac{|\tilde{\phi}|^4}{4}
\,_2F_3\left(\frac{1}{2},2;\frac{3}{2},2,3;-\frac{|\tilde{\phi}|^2|\tilde{\vec{z}}|^2}{4}
\right)\nonumber\\
&&=\frac{\pi}{8}\,_1F_2\left(\frac{1}{2};\frac{3}{2},3;
-\frac{|\tilde{\phi}|^2|\tilde{\vec{z}}|^2}{4}\right)
\end{eqnarray}
\begin{eqnarray}
&&\frac{\pi}{|\tilde{\vec{z}}|}\int\limits_0^{|\tilde{\phi}|}\!d|\tilde{p}|\,
|\tilde{p}|^2J_1(|\tilde{p}||\tilde{\vec{z}}|)\nonumber\\
&&=\frac{\pi}{|\tilde{\vec{z}}|^4}\int\limits_0^{|\tilde{\phi}||\tilde{\vec{z}}|}\!d\chi\,
\chi^2J_1(\chi)\nonumber\\
&&=\frac{\pi}{|\tilde{\vec{z}}|^4}
\left(|\tilde{\phi}|^2|\tilde{\vec{z}}|^2J_1(|\tilde{\phi}||\tilde{\vec{z}}|)\right)
\end{eqnarray}
\begin{eqnarray}
&&-\frac{\pi}{|\tilde{\vec{z}}|^2}\int\limits_0^{|\tilde{\phi}|}\!d|\tilde{p}|\,
\left(3|\tilde{p}|J_0(|\tilde{p}||\tilde{\vec{z}}|)
+|\tilde{p}|^2|\tilde{\vec{z}}|J_1(|\tilde{p}||\tilde{\vec{z}}|)
-3|\tilde{p}|\,_1F_2\left(\frac{1}{2};\frac{3}{2},2;
-\frac{|\tilde{p}|^2|\tilde{\vec{z}}|^2}{4}\right)\right)\nonumber\\
&&=-\frac{3\pi}{|\tilde{\vec{z}}|^4}\int\limits_0^{|\tilde{\phi}||\tilde{\vec{z}}|}\!d\chi\,
\chi J_0(\chi)
-\frac{\pi}{|\tilde{\vec{z}}|^4}\int\limits_0^{|\tilde{\phi}||\tilde{\vec{z}}|}\!d\chi\,
\chi^2 J_1(\chi)
+\frac{3\pi}{|\tilde{\vec{z}}|^2}\int\limits_0^{|\tilde{\phi}|}\!d|\tilde{p}|\,
_1F_2\left(\frac{1}{2};\frac{3}{2},2;
-\frac{|\tilde{p}|^2|\tilde{\vec{z}}|^2}{4}\right)\nonumber\\
&&=-\frac{3\pi}{|\tilde{\vec{z}}|^4}
|\tilde{\phi}||\tilde{\vec{z}}|J_1(|\tilde{\phi}||\tilde{\vec{z}}|)
-\frac{\pi}{|\tilde{\vec{z}}|^4}
|\tilde{\phi}|^2|\tilde{\vec{z}}|^2J_2(|\tilde{\phi}||\tilde{\vec{z}}|)
+\frac{3\pi}{|\tilde{\vec{z}}|^2}\,_2F_3\left(1,1;1,\frac{3}{2},2;
-\frac{|\tilde{\phi}|^2|\tilde{\vec{z}}|^2}{4}\right)\nonumber\\
&&=-\frac{\pi}{2|\tilde{\vec{z}}|^4}
\left(6|\tilde{\phi}||\tilde{\vec{z}}|J_1(|\tilde{\phi}||\tilde{\vec{z}}|)
+2|\tilde{\phi}|^2|\tilde{\vec{z}}|^2J_2(|\tilde{\phi}||\tilde{\vec{z}}|)
-3|\tilde{\phi}|^2|\tilde{\vec{z}}|^2\,_1F_2\left(\frac{1}{2};\frac{3}{2},2;
-\frac{|\tilde{\phi}|^2|\tilde{\vec{z}}|^2}{4}\right)\right)\nonumber\\
\end{eqnarray}
\newpage 

%% file: details4.tex
\section{Calculation of the vacuum and thermal contribution in Lorenz gauge}
We evaluate the two-point correlator of energy density again in Lorenz gauge in order to show the gauge invariance (in case of the vacuum contribution) of this quantity explicitly. As mentioned above, we have used Coulomb gauge $\bra\Theta_{00}(x)\Theta_{00}(y)\ket$ in the case of the $SU(2)$-Yang-Mills thermodynamics calculation, since this gauge is a physical gauge and the implementation of the constraints on the momenta is given as in Eqs,.\,(\ref{compositeness1}), (\ref{compositeness2}), and thus considerably easier than in the covariant gauge.

\subsection{Propagator of the massless gauge mode in different gauges}
For reasons of completeness we will compile the the propagators of the massless gauge mode in momentum space and position space in consideration of different gauge conditions. We will constrain ourselves to the cases of Lorenz and Coulomb gauge. We recall that in the real-time formalism of finite temperature field theory the propagator consists of a vacuum (zero-temperature) and a thermal contribution:
\begin{equation}
D_{\mu\nu}(p)=D^{\tiny\mbox{vac}}_{\mu\nu}(p,T)
+D^{\tiny\mbox{th}}_{\mu\nu}(p,T).
\end{equation}
For a more detailed discussion one could consult for instance \cite{Le Bellac}.\\
\noindent
\underline{Coulomb gauge:}\\
The momentum-space vacuum contribution is given as
\begin{equation}
D_{\mu\nu}^{\tiny{\mbox{vac}}}(p)
=-P^T_{\mu\nu}\frac{i}{p^2+i\varepsilon}
+i\frac{\delta_{\mu0} \delta_{\nu0}}{\vec{p}^2}
-\xi\frac{p_{\mu}p_{\nu}}{p_0^2}\,,
\end{equation}
with the gauge-fixing parameter. The quantity $\bra\Theta_{00}(x)\Theta_{00}(y)\ket$ does not depend on any value of the gauge parameter $\xi$. The usual Coulomb gauge is realized for the case $\xi=0$.
The contribution coupling to the zero-component of the electromagnetic current is the Fourier-transform of the Coulomb-potential.
In position-space we obtain
\begin{equation}
D_{\mu\nu}^{\tiny{\mbox{vac,Coulomb}}}(|\vec{x-y}|)
=-\delta(x_0-y_0)\frac{u_\mu u_\nu}{|\vec{x-y}|}\,,
\end{equation}
and describes an instantaneous static interaction.
The momentum-space thermal contribution is given as
\begin{equation}
D_{\mu\nu}^{\tiny{\mbox{th}}}(p,\beta)
=-P^T_{\mu\nu}(p)2\pi\delta(p^2)n_B(\beta|p_0|)\,.
\end{equation}
\noindent
\underline{Lorenz gauge:}
The momentum-space propagator for the massless gauge mode in covariant gauge is given as
\begin{equation}
D_{\mu\nu}(p)
=-g_{\mu\nu}\frac{i}{p^2+i\varepsilon}
+\frac{\xi-1}{\xi}\frac{p_{\mu}p_{\nu}}{(p^2+i\varepsilon)^2}\,
-g_{\mu\nu}(p)2\pi\delta(p^2)n_B(\beta|p_0|)\,.
\end{equation}
Here $\xi$ again denotes the so-called gauge-fixing parameter. A convenient choice is $\xi=1$, which is referred to as Feynman gauge, or the Landau gauge $\xi=\infty$. We calculate in Feynman gauge.
While apart from the tensor structure the thermal part is equal to the Coulomb gauge thermal part, the Lorenz gauge vacuum propagator in position gauge is then given as
\begin{equation}
\label{Lorenzgauge position}
D_{\mu\nu}^{\tiny{\mbox{vac}}}(x-y)=\bra T[A_\mu(x)A_\nu(y)]\ket\,=\frac{g_{\mu\nu}}{4\pi^2 (x-y)^2}\,.
\end{equation}
In one of the following subsections we will use this propagator Eq.\,(\ref{Lorenzgauge position}) in order to proove the above obtained result for the vacuum part by a straight-forward position-space calculation.

Via employing a Fourier transform we are able to determine position-space counterpart of the thermal propagator. We obtain the equal-time correlator
\begin{equation}
\label{thermal position space}
D_{\mu\nu}^{\tiny{\mbox{th}}}(x-y,\beta)\Big|_{x^0=y^0}=
-g_{\mu\nu}\cdot\frac{\beta-\pi(\vec{x-y})\coth\left(\frac{\pi(\vec{x-y})}{\beta}\right)}
{2\pi\beta(\vec{x-y})^2}\,.
\end{equation}
A more sophisticated task is the calculation of the nonequal-time thermal part of the propagator. Nevertheless by consulting \cite{Gradshteyn} we are enabled to evaluate the four-dimensional Fourier integral, which yields
\begin{eqnarray}
\label{thermal position space 2}
&&D_{\mu\nu}^{\tiny{\mbox{th}}}(x-y,\beta)\nonumber\\
&=&-\frac{g_{\mu\nu}}{2}\cdot
\left(\frac{\sinh\left(\frac{2\pi|\vec{x-y}|}{\beta}\right)}
{\beta|\vec{x-y}|\cosh\left(\frac{2\pi|\vec{x-y}|}{\beta}\right)
-\beta|\vec{x-y}|\cosh\left(\frac{2\pi|x_0-y_0|}{\beta}\right)}
-\frac{1}{\pi|\vec{x-y}|^2-\pi|x_0-y_0|^2}\right)\nonumber\\
\end{eqnarray}
In the limit $x^0\to y^0$ Eq.\,(\ref{thermal position space 2}) is equal to Eq.\,(\ref{thermal position space}).

\subsection{The thermal contribution:\,Calculation in position space}
The thermal part can also be computed in coordinate space. Instead of calculating integrals, we insert the propagator Eq.\,(\ref{thermal position space 2}) into Eq.\,(\ref{WICK}) and calculate derivatives of this structure. Finally, we set $x^0=y^0=0$ and obtain the equal-time two-point correlation of energy density.\\
Nevertheless, the derivatives of Eq.\,(\ref{thermal position space 2}) with respect to $x$ and $y$ are very large expressions and involve hyperbolic functions. Although these derivatives can be simplified by making use of several functional identities, a calculation in momentum space is considerably more convenient.\\
In order to compile the necessary ingredients for a momentum space calculation of
$\bra\Theta_{00}(x)\Theta_{00}(y)\ket^{\tiny\mbox{th}}$ we give the following formulas:
\begin{eqnarray}
\left.\partial_{x^0}\partial_{y^0}\,D_{\mu\nu}^{\tiny{\mbox{th}}}(z,\beta)\right|_{z^0=0}&=&
\frac{1}{\pi|\vec{z}|^4}
-\frac{\coth\left(\frac{\pi|\vec{z}|}{\beta}\right)
\textrm{cosech}^2\left(\frac{\pi|\vec{z}|}{\beta}\right)}
{\beta^3|\vec{z}|}\,,\\
\left.\partial_{x^i}\partial_{y^j}\,D_{\mu\nu}^{\tiny{\mbox{th}}}(z,\beta)\right|_{z^0=0}&=&
\frac{z_i z_j\, \textrm{cosech}^2\left(\frac{\pi|\vec{z}|}{\beta}\right)}{4\pi|\vec{z}|^6\beta^3}
\left(8\beta^3\left(1-\cosh\left(\frac{2\pi|\vec{z}|}{\beta}\right)\right)
-4\pi^3\beta^3\coth\left(\frac{\pi|\vec{z}|}{\beta}\right)\right.\nonumber\\
&&\left.\hspace{30mm}-6\pi^2|\vec{z}|^2\beta
-3\pi|\vec{z}|\beta^2\sinh\left(\frac{2\pi|\vec{z}|}{\beta}\right)\right)\nonumber\\
&=&\frac{z_i z_j\, \textrm{cosech}^2\left(\frac{\pi|\vec{z}|}{\beta}\right)}{4\pi|\vec{z}|^6\beta^3}
\left(16\beta^3\sinh^2\left(\frac{\pi|\vec{z}|}{\beta}\right)\right.\\
&&\left.\hspace{10mm}
-\left(4\pi^3|\vec{z}|^3+6\pi|\vec{z}|\beta^2\right)
\sinh\left(\frac{\pi|\vec{z}|}{\beta}\right)\cosh\left(\frac{\pi|\vec{z}|}{\beta}\right)
-6\pi^2|\vec{z}|^2\beta\right)\,,\nonumber\\
\left.\partial_{x^0}\partial_{y^j}\,D_{\mu\nu}^{\tiny{\mbox{th}}}(z,\beta)\right|_{z^0=0}&=&
\left.\partial_{x^i}\partial_{y^0}\,D_{\mu\nu}^{\tiny{\mbox{th}}}(z,\beta)\right|_{z^0=0}\,\,=\,\,0\,,
\end{eqnarray}
where $z$ stands for $x-y$.

\subsection{The thermal contribution:\,Calculation in momentum space}
Let us first consider the thermal part to the correlator. We insert the propagator
\begin{equation}
D^{\tiny\mbox{th}}_{\mu\nu}(p,T)=-g_{\mu\nu}(p)2\pi\delta(p^2)n_B(\beta|p_0|)
\end{equation}
into the Wick decomposition of the energy density two-point correlation function, Eq.\,(\ref{WICK}). Subsequently, we rearrange terms of the same structure and use the properties of the Delta distribution Eqs.\,(\ref{convolution}, \ref{Delta decomposition}) and integrate over the zero-coordinates. This yields

$\bra\Theta_{00}(x)\Theta_{00}(y)\ket^{\tiny\mbox{th}}$
\begin{eqnarray}
&=&4\int\!dp_0\int\!dk_0\int\!\frac{d^3p}{(2\pi)^3 2|\vec{p}|}\int\!\frac{d^3k}{(2\pi)^3 2|\vec{k}|}
  p_0^2 k_0^2 \cdot n_B(\beta |p_0|)\,n_B(\beta |k_0|)\cdot\e^{-ip(x-y)}\e^{-ik(x-y)}\nonumber\\
  &&\hspace{20mm}\cdot
  (\delta(p_0-|\vec{p}|)+\delta(p_0+|\vec{p}|))(\delta(k_0-|\vec{k}|)+\delta(k_0+|\vec{k}|))\nonumber\\
&-&4\int\!dp_0\int\!dk_0\int\!\frac{d^3p}{(2\pi)^3 2|\vec{p}|}\int\!\frac{d^3k}{(2\pi)^3 2|\vec{k}|}
  p_0 k_0 p k \cdot n_B(\beta |p_0|)\,n_B(\beta |k_0|)\cdot\e^{-ip(x-y)}\e^{-ik(x-y)}\nonumber\\
  &&\hspace{20mm}\cdot
  (\delta(p_0-|\vec{p}|)+\delta(p_0+|\vec{p}|))(\delta(k_0-|\vec{k}|)+\delta(k_0+|\vec{k}|))\nonumber\\
&+&2\int\!dp_0\int\!dk_0\int\!\frac{d^3p}{(2\pi)^3 2|\vec{p}|}\int\!\frac{d^3k}{(2\pi)^3 2|\vec{k}|}
  p_0^2 k^2 \cdot n_B(\beta |p_0|)\,n_B(\beta |k_0|)\cdot\e^{-ip(x-y)}\e^{-ik(x-y)}\nonumber\\
  &&\hspace{20mm}\cdot
  (\delta(p_0-|\vec{p}|)+\delta(p_0+|\vec{p}|))(\delta(k_0-|\vec{k}|)+\delta(k_0+|\vec{k}|))\nonumber\\
&+&2\int\!dp_0\int\!dk_0\int\!\frac{d^3p}{(2\pi)^3 2|\vec{p}|}\int\!\frac{d^3k}{(2\pi)^3 2|\vec{k}|}
  p^2 k_0^2 \cdot n_B(\beta |p_0|)\,n_B(\beta |k_0|)\cdot\e^{-ip(x-y)}\e^{-ik(x-y)}\nonumber\\
  &&\hspace{20mm}\cdot
  (\delta(p_0-|\vec{p}|)+\delta(p_0+|\vec{p}|))(\delta(k_0-|\vec{k}|)+\delta(k_0+|\vec{k}|))\nonumber\\
&+&2\int\!dp_0\int\!dk_0\int\!\frac{d^3p}{(2\pi)^3 2|\vec{p}|}\int\!\frac{d^3k}{(2\pi)^3 2|\vec{k}|}
  (p k)^2 \cdot n_B(\beta |p_0|)\,n_B(\beta |k_0|)\cdot\e^{-ip(x-y)}\e^{-ik(x-y)}\nonumber\\
  &&\hspace{20mm}\cdot
  (\delta(p_0-|\vec{p}|)+\delta(p_0+|\vec{p}|))(\delta(k_0-|\vec{k}|)+\delta(k_0+|\vec{k}|))\nonumber\\
&-&2g_{00}
  \int\!dp_0\int\!dk_0\int\!\frac{d^3p}{(2\pi)^3 2|\vec{p}|}\int\!\frac{d^3k}{(2\pi)^3 2|\vec{k}|}
  (p k)^2 \cdot n_B(\beta |p_0|)\,n_B(\beta |k_0|)\cdot\e^{-ip(x-y)}\e^{-ik(x-y)}\nonumber\\
  &&\hspace{20mm}\cdot
  (\delta(p_0-|\vec{p}|)+\delta(p_0+|\vec{p}|))(\delta(k_0-|\vec{k}|)+\delta(k_0+|\vec{k}|))\nonumber\\
&-&2g_{00}
  \int\!dp_0\int\!dk_0\int\!\frac{d^3p}{(2\pi)^3 2|\vec{p}|}\int\!\frac{d^3k}{(2\pi)^3 2|\vec{k}|}
  p_0^2 k^2 \cdot n_B(\beta |p_0|)\,n_B(\beta |k_0|)\cdot\e^{-ip(x-y)}\e^{-ik(x-y)}\nonumber\\
  &&\hspace{20mm}\cdot
  (\delta(p_0-|\vec{p}|)+\delta(p_0+|\vec{p}|))(\delta(k_0-|\vec{k}|)+\delta(k_0+|\vec{k}|))\nonumber\\
&-&2g_{00}
  \int\!dp_0\int\!dk_0\int\!\frac{d^3p}{(2\pi)^3 2|\vec{p}|}\int\!\frac{d^3k}{(2\pi)^3 2|\vec{k}|}
  p^2 k_0^2 \cdot n_B(\beta |p_0|)\,n_B(\beta |k_0|)\cdot\e^{-ip(x-y)}\e^{-ik(x-y)}\nonumber\\
  &&\hspace{20mm}\cdot
  (\delta(p_0-|\vec{p}|)+\delta(p_0+|\vec{p}|))(\delta(k_0-|\vec{k}|)+\delta(k_0+|\vec{k}|))\nonumber\\
&+&g_{00}^2
  \int\!dp_0\int\!dk_0\int\!\frac{d^3p}{(2\pi)^3 2|\vec{p}|}\int\!\frac{d^3k}{(2\pi)^3 2|\vec{k}|}
  (p k)^2 \cdot n_B(\beta |p_0|)\,n_B(\beta |k_0|)\cdot\e^{-ip(x-y)}\e^{-ik(x-y)}\nonumber\\
  &&\hspace{20mm}\cdot
  (\delta(p_0-|\vec{p}|)+\delta(p_0+|\vec{p}|))(\delta(k_0-|\vec{k}|)+\delta(k_0+|\vec{k}|))\nonumber\\
&+&\frac{g_{00}^2}{2}
  \int\!dp_0\int\!dk_0\int\!\frac{d^3p}{(2\pi)^3 2|\vec{p}|}\int\!\frac{d^3k}{(2\pi)^3 2|\vec{k}|}
  p^2 k^2 \cdot n_B(\beta |p_0|)\,n_B(\beta |k_0|)\cdot\e^{-ip(x-y)}\e^{-ik(x-y)}\nonumber\\
  &&\hspace{20mm}\cdot
  (\delta(p_0-|\vec{p}|)+\delta(p_0+|\vec{p}|))(\delta(k_0-|\vec{k}|)+\delta(k_0+|\vec{k}|))\nonumber\\
\end{eqnarray}
\newpage
\begin{eqnarray}
&=&2\int\!\frac{d^3p}{(2\pi)^3}\int\!\frac{d^3k}{(2\pi)^3}
  |\vec{p}||\vec{k}|\cdot n_B(\beta|\vec{p}|)\,n_B(\beta|\vec{k}|)
  \cdot\e^{i\vec{p}\vec{(x-y)}}\,\e^{i\vec{k}\vec{(x-y)}}\nonumber\\
  &&\hspace{50mm}\cdot (\cos((|\vec{p}|+|\vec{k}|)(x-y)_0)+\cos((|\vec{p}|-|\vec{k}|)(x-y)_0))\nonumber\\
&&-2\int\!\frac{d^3p}{(2\pi)^3}\int\!\frac{d^3k}{(2\pi)^3}
  |\vec{p}||\vec{k}|
  \cdot n_B(\beta|\vec{p}|)\,n_B(\beta|\vec{k}|)
  \cdot\e^{i\vec{p}\vec{(x-y)}}\,\e^{i\vec{k}\vec{(x-y)}}
  \nonumber\\
 &&\hspace{50mm}
  \cdot (\cos((|\vec{p}|+|\vec{k}|)(x-y)_0)+\cos((|\vec{p}|-|\vec{k}|)(x-y)_0))
  \nonumber\\
 &&\hspace{5mm}
  -2\int\!\frac{d^3p}{(2\pi)^3}\int\!\frac{d^3k}{(2\pi)^3}
  \left(\frac{\vec{p}\vec{k}}{|\vec{p}||\vec{k}|}\right)
  |\vec{p}||\vec{k}|
  \cdot n_B(\beta|\vec{p}|)\,n_B(\beta|\vec{k}|)
  \cdot\e^{i\vec{p}\vec{(x-y)}}\,\e^{i\vec{k}\vec{(x-y)}}
  \nonumber\\
 &&\hspace{50mm}
  \cdot (\cos((|\vec{p}|-|\vec{k}|)(x-y)_0)-\cos((|\vec{p}|+|\vec{k}|)(x-y)_0))
  \nonumber\\
&&+1\int\!\frac{d^3p}{(2\pi)^3}\int\!\frac{d^3k}{(2\pi)^3}
  \left(1+\left(\frac{\vec{p}\vec{k}}{|\vec{p}||\vec{k}|}\right)^2\right)
  |\vec{p}||\vec{k}|
  \cdot n_B(\beta|\vec{p}|)\,n_B(\beta|\vec{k}|)
  \cdot\e^{i\vec{p}\vec{(x-y)}}\,\e^{i\vec{k}\vec{(x-y)}}
  \nonumber\\
 &&\hspace{50mm}
  \cdot (\cos((|\vec{p}|+|\vec{k}|)(x-y)_0)+\cos((|\vec{p}|-|\vec{k}|)(x-y)_0))
  \nonumber\\
 &&\hspace{5mm}
  -\int\!\frac{d^3p}{(2\pi)^3}\int\!\frac{d^3k}{(2\pi)^3}
  \left(\frac{\vec{p}\vec{k}}{|\vec{p}||\vec{k}|}\right)
  |\vec{p}||\vec{k}|
  \cdot n_B(\beta|\vec{p}|)\,n_B(\beta|\vec{k}|)
  \cdot\e^{i\vec{p}\vec{(x-y)}}\,\e^{i\vec{k}\vec{(x-y)}}
  \nonumber\\
 &&\hspace{50mm}
  \cdot (\cos((|\vec{p}|-|\vec{k}|)(x-y)_0)-\cos((|\vec{p}|+|\vec{k}|)(x-y)_0))
  \nonumber\\
&&-g_{00}\int\!\frac{d^3p}{(2\pi)^3}\int\!\frac{d^3k}{(2\pi)^3}
  \left(1+\left(\frac{\vec{p}\vec{k}}{|\vec{p}||\vec{k}|}\right)^2\right)
  |\vec{p}||\vec{k}|
  \cdot n_B(\beta|\vec{p}|)\,n_B(\beta|\vec{k}|)
  \cdot\e^{i\vec{p}\vec{(x-y)}}\,\e^{i\vec{k}\vec{(x-y)}}
  \nonumber\\
 &&\hspace{50mm}
  \cdot (\cos((|\vec{p}|+|\vec{k}|)(x-y)_0)+\cos((|\vec{p}|-|\vec{k}|)(x-y)_0))
  \nonumber\\
 &&\hspace{5mm}
  +g_{00}\int\!\frac{d^3p}{(2\pi)^3}\int\!\frac{d^3k}{(2\pi)^3}
  \left(\frac{\vec{p}\vec{k}}{|\vec{p}||\vec{k}|}\right)
  |\vec{p}||\vec{k}|
  \cdot n_B(\beta|\vec{p}|)\,n_B(\beta|\vec{k}|)
  \cdot\e^{i\vec{p}\vec{(x-y)}}\,\e^{i\vec{k}\vec{(x-y)}}
  \nonumber\\
 &&\hspace{50mm}
  \cdot (\cos((|\vec{p}|-|\vec{k}|)(x-y)_0)-\cos((|\vec{p}|+|\vec{k}|)(x-y)_0))
  \nonumber\\
&&+\frac{g_{00}^2}{2}\int\!\frac{d^3p}{(2\pi)^3}\int\!\frac{d^3k}{(2\pi)^3}
  \left(1+\left(\frac{\vec{p}\vec{k}}{|\vec{p}||\vec{k}|}\right)^2\right)
  |\vec{p}||\vec{k}|
  \cdot n_B(\beta|\vec{p}|)\,n_B(\beta|\vec{k}|)
  \cdot\e^{i\vec{p}\vec{(x-y)}}\,\e^{i\vec{k}\vec{(x-y)}}
  \nonumber\\
 &&\hspace{50mm}
  \cdot (\cos((|\vec{p}|+|\vec{k}|)(x-y)_0)+\cos((|\vec{p}|-|\vec{k}|)(x-y)_0))
  \nonumber\\
 &&\hspace{5mm}
  -g_{00}^2\int\!\frac{d^3p}{(2\pi)^3}\int\!\frac{d^3k}{(2\pi)^3}
  \left(\frac{\vec{p}\vec{k}}{|\vec{p}||\vec{k}|}\right)
  |\vec{p}||\vec{k}|
  \cdot n_B(\beta|\vec{p}|)\,n_B(\beta|\vec{k}|)
  \cdot\e^{i\vec{p}\vec{(x-y)}}\,\e^{i\vec{k}\vec{(x-y)}}
  \nonumber\\
 &&\hspace{50mm}
  \cdot (\cos((|\vec{p}|-|\vec{k}|)(x-y)_0)-\cos((|\vec{p}|+|\vec{k}|)(x-y)_0))
  \nonumber\\
&=&g_{00}^2\int\!\frac{d^3p}{(2\pi)^3}\int\!\frac{d^3k}{(2\pi)^3}
  \left(1+\left(\frac{\vec{p}\vec{k}}{|\vec{p}||\vec{k}|}\right)^2\right)
  |\vec{p}||\vec{k}|
  \cdot n_B(\beta|\vec{p}|)\,n_B(\beta|\vec{k}|)
  \cdot\e^{i\vec{p}\vec{(x-y)}}\,\e^{i\vec{k}\vec{(x-y)}}\,.
  \nonumber\\
\end{eqnarray}
Last line is equal to the expression derived in Coulomb gauge, Eq.\,(\ref{thermalintegralu1}). Subsequently we can proceed in exactly the same manner as we have done above.

\subsection{The vacuum contribution:\,Calculation in position space}
If we would only were interested in the value of $\bra\Theta_{00}(x)\Theta_{00}(y)\ket^{\tiny\mbox{vac}}$ a calculation in position space and Lorenz gauge would have been more convenient. For the reasons mentioned above we have preferred a calculation in momentum-space, in order to present how our method works, particularly with regard to the more involved calculation, when photon propagation is governed by an $SU(2)$ gauge principle. At this point we briefly compile the fundamental elements of the position-space computation of the energy density two-point-correlator and surrender details, since the calculation is of straight-forward character and more a challenge on concentration.\\
Starting point for the calculation is again the Wick decomposition of the correlator, Eq\,.(\ref{WICK}). Instead of shifting to momentum-space, we directly insert the position-space propagator Eq\,.(\ref{Lorenzgauge position}). Differentiation with respect to the $x$- and $y$-coordinates and contraction of the tensor structure will yield us the final result.\\
Concerning the calculation we have to consider derivatives of the propagator. It holds that
\begin{eqnarray}
\label{derivation correlator}
\bra T[\partial_{x^\mu}A_\kappa(x)\partial_{x^\nu}A_\lambda(y)]\ket&=&
\partial_{x^\mu}\partial_{x^\nu} \bra T[A_\kappa(x)A_\lambda(y)]\ket\nonumber\\
&=&g_{\kappa\lambda}\cdot\underbrace{\frac{1}{4\pi^2}
\left(\frac{2\delta_{\mu\nu}(x-y)^4
-8(x-y)_\mu(x-y)_\nu(x-y)^2}{(x-y)^8}\right)}_{\mathcal{I}_{\mu\nu}(x-y)}\,,
\end{eqnarray}
where $T$ denotes the time-ordering symbol. Since negligence of time-ordering is well justified we omit the symbol in the following.
In principle, we have to evaluate two-point correlators of squared field strength tensors. It is not difficult to see that following relation holds:
\begin{equation}
\bra T[F^2(x)G^2(y)]\ket=2\,\bra T[F(x)G(y)]\ket
\end{equation}
The correlator of field strengths is given as:
\begin{equation}
\label{derivation correlator 2}
\bra T(F_{\mu\nu}(x)F_{\kappa\lambda}(y))\ket=\underbrace{
g_{\mu\kappa}\mathcal{I}_{\nu\lambda}(x-y)-g_{\mu\lambda}\mathcal{I}_{\nu\kappa}(x-y)
-g_{\nu\kappa}\mathcal{I}_{\mu\lambda}(x-y)+g_{\nu\lambda}\mathcal{I}_{\mu\kappa}(x-y)}
_{\mathcal{F}_{\mu\nu\kappa\lambda}}(x-y)\,.
\end{equation}
The energy density correlator is then given as the sum
\begin{eqnarray}
\bra\Theta_{00}(x)\Theta_{00}(y)\ket^{\tiny\mbox{vac}}=
&&2\mathcal{F}_0\,^\lambda\,_0\,^\tau(x-y)\mathcal{F}_{0\lambda0\tau}(x-y)
-\frac{g_{00}}{2}\mathcal{F}_0\,^{\lambda\sigma\tau}(x-y)
                 \mathcal{F}_{0\lambda\sigma\tau}(x-y)\nonumber\\
&&
-\frac{g_{00}}{2}\mathcal{F}^{\kappa\lambda}\,_0\,^\tau(x-y)
                 \mathcal{F}_{\kappa\lambda0\tau}(x-y)
-\frac{g_{00}^2}{8}\mathcal{F}^{\kappa\lambda\sigma\tau}(x-y)
                 \mathcal{F}_{\kappa\lambda\sigma\tau}(x-y)\,.\nonumber\\
\end{eqnarray}
Inserting Eqs.\,(\ref{derivation correlator}) and (\ref{derivation correlator 2}) and contracting contravariant and covariant tensor components let us finally arrive at
\begin{eqnarray}
\bra\Theta_{00}(x)\Theta_{00}(y)\ket^{\tiny\mbox{vac}}
&=&\frac{1}{2\pi^4 (x-y)^8}\left(3g_{00}^2-8g_{00}\frac{(x_0-y_0)^2}{(x-y)^2}
                       +32\frac{(x_0-y_0)^4}{(x-y)^4}\right)\nonumber\\
&+&\frac{1}{4\pi^4 (x-y)^8}\left(-23g_{00}^2-16g_{00}\frac{(x_0-y_0)^2}{(x-y)^2}
                       \right)\nonumber\\
&+&\frac{1}{4\pi^4 (x-y)^8}\left(-11g_{00}^2-40g_{00}\frac{(x_0-y_0)^2}{(x-y)^2}
                       \right)\nonumber\\
&+&\frac{1}{\pi^4 (x-y)^8}\left(22g_{00}^2-10g_{00}\frac{(x_0-y_0)^2}{(x-y)^2}
                       \right)\nonumber\\
&=&\frac{1}{\pi^4 (x-y)^8}\left(15g_{00}^2-28g_{00}\frac{(x_0-y_0)^2}{(x-y)^2}
                       +16\frac{(x_0-y_0)^4}{(x-y)^4}\right)\,.
\end{eqnarray}
Assuming equal time and setting $x^0=y^0$ we obtain exactly the same result as in the previous calculation in Coulomb gauge.

\subsection{The vacuum contribution:\,Calculation in momentum space}
Again Eq.\,(\ref{WICK}) provides a starting point for the calculation. Insertion of the propagator in covariant gauge yields\\
\newline
$\bra\Theta_{00}(x)\Theta_{00}(y)\ket^{\tiny\mbox{th}}=$
\begin{eqnarray}
&-&\int\!\!\frac{d^4p}{(2\pi)^4}\int\frac{d^4k}{(2\pi)^4}\,
  \left(4p_0^2k_0^2-4p_0k_0p_\mu k^\mu+2p_0^2k_\mu k^\mu
  +2k_0^2p_\mu p^\mu+2p_\mu p^\mu k_\nu k^\nu\right)\,
  \frac{\e^{-ip(x-y)}}{p^2+i\varepsilon}\,
  \frac{\e^{-ik(x-y)}}{k^2+i\varepsilon}\nonumber\\
&+&\frac{g_{00}}{4}
  \int\!\!\frac{d^4p}{(2\pi)^4}\int\frac{d^4k}{(2\pi)^4}\,
  \left(p_\mu p^\mu k_\nu k^\nu+4p_0^2k_\mu k^\mu+4k_0^2p_\mu p^\mu\right)\,
  \frac{\e^{-ip(x-y)}}{p^2+i\varepsilon}\,
  \frac{\e^{-ik(x-y)}}{k^2+i\varepsilon}\nonumber\\
&+&\frac{g_{00}}{4}
  \int\!\!\frac{d^4p}{(2\pi)^4}\int\frac{d^4k}{(2\pi)^4}\,
  \left(p_\mu p^\mu k_\nu k^\nu+4p_0^2k_\mu k^\mu+4k_0^2p_\mu p^\mu\right)\,
  \frac{\e^{-ip(x-y)}}{p^2+i\varepsilon}\,
  \frac{\e^{-ik(x-y)}}{k^2+i\varepsilon}\nonumber\\
&-&\frac{g_{00}^2}{16}
  \int\!\!\frac{d^4p}{(2\pi)^4}\int\frac{d^4k}{(2\pi)^4}\,
  \left(16p_\mu p^\mu k_\nu k^\nu+8p_\mu p^\mu k_\nu k^\nu\right)\,
  \frac{\e^{-ip(x-y)}}{p^2+i\varepsilon}\,
  \frac{\e^{-ik(x-y)}}{k^2+i\varepsilon}\,.
\end{eqnarray}
By virtue of analytic continuation we rotate to euclidean signature and arrive at\\
\newline
$\bra\Theta_{00}(x)\Theta_{00}(y)\ket^{\tiny\mbox{th}}=$
\begin{eqnarray}
&+&\int\!\!\frac{d^4p}{(2\pi)^4}\int\frac{d^4k}{(2\pi)^4}\,
  \left(4p_0^2k_0^2-4p_0k_0p_\mu k_\mu+2p_0^2k_\mu k_\mu
  +2k_0^2p_\mu p_\mu+2p_\mu p_\mu k_\nu k_\nu\right)\,
  \frac{\e^{ip(x-y)}}{p^2}\,
  \frac{\e^{ik(x-y)}}{k^2}\nonumber\\
&-&\frac{(-\delta_{00})}{4}
  \int\!\!\frac{d^4p}{(2\pi)^4}\int\frac{d^4k}{(2\pi)^4}\,
  \left(p_\mu p_\mu k_\nu k_\nu+4p_0^2k_\mu k_\mu+4k_0^2p_\mu p_\mu\right)\,
  \frac{\e^{ip(x-y)}}{p^2}\,
  \frac{\e^{ik(x-y)}}{k^2}\nonumber\\
&-&\frac{(-\delta_{00})}{4}
  \int\!\!\frac{d^4p}{(2\pi)^4}\int\frac{d^4k}{(2\pi)^4}\,
  \left(p_\mu p_\mu k_\nu k_\nu+4p_0^2k_\mu k_\mu+4k_0^2p_\mu p_\mu\right)\,
  \frac{\e^{ip(x-y)}}{p^2}\,
  \frac{\e^{ik(x-y)}}{k^2}\nonumber\\
&+&\frac{(-\delta_{00})^2}{16}
  \int\!\!\frac{d^4p}{(2\pi)^4}\int\frac{d^4k}{(2\pi)^4}\,
  \left(16p_\mu p_\mu k_\nu k_\nu+8p_\mu p_\mu k_\nu k_\nu\right)\,
  \frac{\e^{ip(x-y)}}{p^2}\,
  \frac{\e^{ik(x-y)}}{k^2}\,.
\end{eqnarray}
Finally we obtain
\begin{eqnarray}
\bra\Theta_{00}(x)\Theta_{00}(y)\ket^{\tiny\mbox{th}}&=&
  6\int\!\!\frac{d^4p}{(2\pi)^4}\int\frac{d^4k}{(2\pi)^4}\,
  \left(p_0k_0|\vec{p}||\vec{k}|
  \left(\frac{\vec{p}\vec{k}}{|\vec{p}||\vec{k}|}\right)\right)\,
  \frac{\e^{ip(x-y)}}{p^2}\,
  \frac{\e^{ik(x-y)}}{k^2}\nonumber\\
&+&5\left(\int\!\!\frac{d^4p}{(2\pi)^4}\,p_0^2\,
  \frac{\e^{ip(x-y)}}{p^2}\right)^2\nonumber\\
&+&\int\!\!\frac{d^4p}{(2\pi)^4}\int\frac{d^4k}{(2\pi)^4}\,
  \left(|\vec{p}|^2|\vec{k}|^2
  \left(\frac{\vec{p}\vec{k}}{|\vec{p}||\vec{k}|}\right)^2\right)\,
  \frac{\e^{ip(x-y)}}{p^2}\,
  \frac{\e^{ik(x-y)}}{k^2}\nonumber\\
&+&\frac{1}{2}\left(\int\!\!\frac{d^4p}{(2\pi)^4}\,|\vec{p}|^2\,
  \frac{\e^{ip(x-y)}}{p^2}\right)^2\nonumber\\
&+&4\left(\int\!\!\frac{d^4p}{(2\pi)^4}\,p_0^2\,
  \frac{\e^{ip(x-y)}}{p^2}\right)
  \left(\int\!\!\frac{d^4k}{(2\pi)^4}\,k^2\,
  \frac{\e^{ik(x-y)}}{k^2}\right)\nonumber\\
&+&4\left(\int\!\!\frac{d^4p}{(2\pi)^4}\,k_0^2\,
  \frac{\e^{ip(x-y)}}{p^2}\right)
  \left(\int\!\!\frac{d^4k}{(2\pi)^4}\,p^2\,
  \frac{\e^{ik(x-y)}}{k^2}\right)\,.\nonumber\\
\end{eqnarray}
In the following we proceed exactly in the same manner as in the case of the Coulomb gauge calculation. For simplicity we set $x^0=y^0=0$, rescale to dimensionless variables, introduce hyperspherical coordinates and perform the integrals successively.\\
\newpage
$\bra\Theta_{00}(x)\Theta_{00}(y)\ket^{\tiny\mbox{vac}}$
\begin{eqnarray}
&=&\frac{1}{(2\pi)^8\beta^8}\left(6\cdot 2\pi
  \int d|\tilde{p}|d\psi_p d\theta_p
  \int d|\tilde{k}|d\psi_k d\theta_k
  |\tilde{p}|^3\sin^3\psi_p\cos\psi_p\sin\theta_p
  |\tilde{k}|^3\sin^3\psi_k\cos\psi_k\sin\theta_k\nonumber\right.\\
  &&\hspace{30mm}\cdot\,
  \e^{i|\tilde{p}||\tilde{\vec{z}}|\sin\psi_p\cos\theta_p}
  \e^{i|\tilde{k}||\tilde{\vec{z}}|\sin\psi_k\cos\theta_k}\nonumber\\
  &&\hspace{30mm}\cdot\,
  \int_0^{2\pi}\!d\phi_p
  \left(\sin\theta_p\sin\theta_k\cos\phi_p
         +\cos\theta_p\cos\theta_k\right)\nonumber\\
&&+5\cdot 4\pi^2
  \left(\int d|\tilde{p}|d\psi_p d\theta_p
  |\tilde{p}|^3\sin^2\psi_p\cos^2\psi_p\sin\theta_p
  \e^{i|\tilde{p}||\tilde{\vec{z}}|\sin\psi_p\cos\theta_p}\right)^2\nonumber\\
&&+5\cdot 2\pi
  \int d|\tilde{p}|d\psi_p d\theta_p
  \int d|\tilde{k}|d\psi_k d\theta_k
  |\tilde{p}|^3\sin^4\psi_p\sin\theta_p
  |\tilde{k}|^3\sin^4\psi_k\sin\theta_k\nonumber\\
  &&\hspace{30mm}\cdot\,
  \e^{i|\tilde{p}||\tilde{\vec{z}}|\sin\psi_p\cos\theta_p}
  \e^{i|\tilde{k}||\tilde{\vec{z}}|\sin\psi_k\cos\theta_k}\nonumber\\
  &&\hspace{30mm}\cdot\,
  \int_0^{2\pi}\!d\phi_p
  \left(\sin^2\theta_p\sin^2\theta_k\cos^2\phi_p
         2\sin\theta_p\sin\theta_k\cos\theta_p\cos\theta_k\cos\phi_p
         +\cos^2\theta_p\cos^2\theta_k\right)\nonumber\\
&&+\frac{1}{2}\cdot 4\pi^2
  \left(\int d|\tilde{p}|d\psi_p d\theta_p
  |\tilde{p}|^3\sin^2\psi_p\cos^2\psi_p\sin\theta_p
  \e^{i|\tilde{p}||\tilde{\vec{z}}|\sin\psi_p\cos\theta_p}\right)^2\nonumber\\
&&+4\cdot 4\pi^2
  \left(\int d|\tilde{p}|d\psi_p d\theta_p
  |\tilde{p}|^3\sin^2\psi_p\cos^2\psi_p\sin\theta_p
  \e^{i|\tilde{p}||\tilde{\vec{z}}|\sin\psi_p\cos\theta_p}\right)\nonumber\\
  &&\hspace{30mm}\cdot\,
  \left(\int d|\tilde{k}|d\psi_k d\theta_k
  |\tilde{k}|^3\sin^2\psi_k\sin\theta_k
  \e^{i|\tilde{k}||\tilde{\vec{z}}|\sin\psi_k\cos\theta_k}\right)\nonumber\\
&&+4\cdot 4\pi^2
  \left(\int d|\tilde{p}|d\psi_p d\theta_p
  |\tilde{p}|^3\sin^2\psi_p\sin\theta_p
  \e^{i|\tilde{p}||\tilde{\vec{z}}|\sin\psi_p\cos\theta_p}\right)\nonumber\\
  &&\left.\hspace{30mm}\cdot\,
  \left(\int d|\tilde{k}|d\psi_k d\theta_k
  |\tilde{k}|^3\sin^2\psi_k\cos^2\psi_k\sin\theta_k
  \e^{i|\tilde{k}||\tilde{\vec{z}}|\sin\psi_k\cos\theta_k}\right)\right)
\end{eqnarray}
\begin{eqnarray*}
&=&\frac{1}{(2\pi)^8\beta^8}\left(6\cdot4\pi^2
   \left(\int\limits_0^{\infty}\!\! d|\tilde{p}| \int\limits_0^{\pi}\!\! d\psi \int\limits_{-1}^{+1}\!\! d\cos\theta\,
   |\tilde{p}|^3 \sin^3\psi \cos\psi P_1(\cos\theta) \,\e^{i|\tilde{p}||\tilde{\vec{z}}|\sin\psi\cos\theta}\right)^2
   \right.\nonumber\\
& &\left.+5\cdot4\pi^2
   \left(\int\limits_0^{\infty}\!\! d|\tilde{p}| \int\limits_0^{\pi}\!\! d\psi \int\limits_{-1}^{+1}\!\! d\cos\theta\,
   |\tilde{p}|^3 \sin^2\psi\cos^2\psi P_0(\cos\theta) \,\e^{i|\tilde{p}||\tilde{\vec{z}}|\sin\psi\cos\theta}\right)^2
   \right.\nonumber\\
& &\left.+5\cdot2\pi^2
   \left(\int\limits_0^{\infty}\!\! d|\tilde{p}| \int\limits_0^{\pi}\!\! d\psi \int\limits_{-1}^{+1}\!\! d\cos\theta\,
   |\tilde{p}|^3 \sin^4\psi
   \left(\frac{2}{3}P_0(\cos\theta)-\frac{2}{3}P_2(\cos\theta)\right) \,\e^{i|\tilde{p}||\tilde{\vec{z}}|\sin\psi\cos\theta}\right)^2
   \right.\nonumber\\
& &\left.+5\cdot4\pi^2
   \left(\int\limits_0^{\infty}\!\! d|\tilde{p}| \int\limits_0^{\pi}\!\! d\psi \int\limits_{-1}^{+1}\!\! d\cos\theta\,
   |\tilde{p}|^3 \sin^4\psi
   \left(\frac{1}{3}P_0(\cos\theta)+\frac{2}{3}P_2(\cos\theta)\right) \,\e^{i|\tilde{p}||\tilde{\vec{z}}|\sin\psi\cos\theta}\right)^2
   \right.\nonumber\\
& &\left.+\frac{1}{2}\cdot4\pi^2
   \left(\int\limits_0^{\infty}\!\! d|\tilde{p}| \int\limits_0^{\pi}\!\! d\psi \int\limits_{-1}^{+1}\!\! d\cos\theta\,
   |\tilde{p}|^3 \sin^4\psi P_0(\cos\theta) \,\e^{i|\tilde{p}||\tilde{\vec{z}}|\sin\psi\cos\theta}\right)^2
   \right.\nonumber\\
&&+4\cdot4\pi^2
  \left(\int\limits_0^{\infty}\!\! d|\tilde{p}| \int\limits_0^{\pi}\!\! d\psi_p \int\limits_{-1}^{+1}\!\! d\cos\theta_p\,
  |\tilde{p}|^3 \sin^2\psi_p\cos^2\psi_p P_0(\cos\theta_p) \,\e^{i|\tilde{p}||\tilde{\vec{z}}|\sin\psi_p\cos\theta_p}\right)\nonumber\\
  &&\hspace{20mm}\cdot
  \left(\int\limits_0^{\infty}\!\! d|\tilde{k}| \int\limits_0^{\pi}\!\! d\psi_k \int\limits_{-1}^{+1}\!\! d\cos\theta_k\,
  |\tilde{k}|^3 \sin^2\psi_k P_0(\cos\theta_k) \,\e^{i|\tilde{k}||\tilde{\vec{z}}|\sin\psi_k\cos\theta_k}\right)\nonumber\\
\end{eqnarray*}
\newpage
\begin{eqnarray}
&&+4\cdot4\pi^2
  \left(\int\limits_0^{\infty}\!\! d|\tilde{p}| \int\limits_0^{\pi}\!\! d\psi_p \int\limits_{-1}^{+1}\!\! d\cos\theta_p\,
  |\tilde{p}|^3 \sin^2\psi_p P_0(\cos\theta_p) \,\e^{i|\tilde{p}||\tilde{\vec{z}}|\sin\psi_p\cos\theta_p}\right)\nonumber\\
  &&\left.\hspace{20mm}\cdot
  \left(\int\limits_0^{\infty}\!\! d|\tilde{k}| \int\limits_0^{\pi}\!\! d\psi_k \int\limits_{-1}^{+1}\!\! d\cos\theta_k\,
  |\tilde{k}|^3 \sin^2\psi_k\cos^2\psi_k P_0(\cos\theta_k) \,\e^{i|\tilde{k}||\tilde{\vec{z}}|\sin\psi_k\cos\theta_k}\right)
  \right.\\
&=&
\frac{1}{(2\pi)^8\beta^8}
   \left(48\pi^2
   \underbrace{
   \left(\int\limits_0^{\infty}\!\!d|\tilde{p}| \int\limits_0^{\pi}\!\!d\psi
   |\tilde{p}|^3 \sin^3\psi \cos\psi
   i\,j_1(|\tilde{p}||\tilde{\vec{z}}|\sin\psi)\right)^2}_{=0}
   \right.\nonumber\\
&&+80\pi^2
   \left(\int\limits_0^{\infty}\!\!d|\tilde{p}| \int\limits_0^{\pi}\!\!d\psi
   |\tilde{p}|^3 \sin^2\psi \cos^2\psi
   j_0(|\tilde{p}||\tilde{\vec{z}}|\sin\psi)
   \right)^2\nonumber\\
&&+40\pi^2
   \left(\int\limits_0^{\infty}\!\!d|\tilde{p}| \int\limits_0^{\pi}\!\!d\psi
   |\tilde{p}|^3 \sin^4\psi
   \left(\frac{2}{3}j_0(|\tilde{p}||\tilde{\vec{z}}|\sin\psi)
   +\frac{2}{3}j_2(|\tilde{p}||\tilde{\vec{z}}|\sin\psi)\right)\right)^2
   \nonumber\\
&&+80\pi^2
   \left(\int\limits_0^{\infty}\!\!d|\tilde{p}| \int\limits_0^{\pi}\!\!d\psi
   |\tilde{p}|^3 \sin^4\psi
   \left(\frac{1}{3}j_0(|\tilde{p}||\tilde{\vec{z}}|\sin\psi)
   -\frac{2}{3}j_2(|\tilde{p}||\tilde{\vec{z}}|\sin\psi)\right)\right)^2
   \nonumber\\
&&+\underbrace{
   8\pi^2
   \left(\int\limits_0^{\infty}\!\!d|\tilde{p}| \int\limits_0^{\pi}\!\!d\psi
   |\tilde{p}|^3 \sin^2\psi
   j_0(|\tilde{p}||\tilde{\vec{z}}|\sin\psi)
   \right)^2}_{=0}\\
&&+64\pi^2
   \left(\int\limits_0^{\infty}\!\!d|\tilde{p}|\int\limits_0^{\pi}\!\!d\psi_p
   |\tilde{p}|^3 \sin^2\psi_p \cos^2\psi_p
   j_0(|\tilde{p}||\tilde{\vec{z}}|\sin\psi)\right)
   \underbrace{
   \left(\int\limits_0^{\infty}\!\!d|\tilde{k}|\int\limits_0^{\pi}\!\!d\psi_k
   |\tilde{p}|^3 \sin^2\psi_k
   j_0(|\tilde{k}||\tilde{\vec{z}}|\sin\psi_k)\right)}_{=0}\nonumber\\
&&+\left.64\pi^2
   \underbrace{
   \left(\int\limits_0^{\infty}\!\!d|\tilde{p}|\int\limits_0^{\pi}\!\!d\psi_p
   |\tilde{p}|^3 \sin^2\psi_p
   j_0(|\tilde{p}||\tilde{\vec{z}}|\sin\psi)\right)}_{=0}
   \left(\int\limits_0^{\infty}\!\!d|\tilde{k}|\int\limits_0^{\pi}\!\!d\psi_k
   |\tilde{p}|^3 \sin^2\psi_k \cos^2\psi_k
   j_0(|\tilde{k}||\tilde{\vec{z}}|\sin\psi_k)\right)\right)\nonumber\\
&=&
\frac{1}{(2\pi)^8\beta^8}
   \left(80\pi^2
   \left(\frac{1}{\sqrt{2|\tilde{\vec{z}}}|}
   \int\limits_0^\infty\!\!d|\tilde{p}|\int\limits_0^\pi\!\!d\psi\,
   |\tilde{p}|^{\frac{5}{2}}\sin^{\frac{3}{2}}\psi cos^2\psi
   J_{\frac{1}{2}}(|\tilde{p}||\tilde{\vec{z}}|\sin\psi)\right)^2
   \right.\nonumber\\
&&+\frac{160\pi^2}{9}
   \left(\frac{1}{\sqrt{2|\tilde{\vec{z}}}|}
   \int\limits_0^\infty\!\!d|\tilde{p}|\int\limits_0^\pi\!\!d\psi\,
   |\tilde{p}|^{\frac{5}{2}}\sin^{\frac{7}{2}}\psi
   J_{\frac{1}{2}}(|\tilde{p}||\tilde{\vec{z}}|\sin\psi)\right)^2
   \nonumber\\
&&+\frac{160\pi^2}{9}
   \left(\frac{1}{\sqrt{2|\tilde{\vec{z}}}|}
   \int\limits_0^\infty\!\!d|\tilde{p}|\int\limits_0^\pi\!\!d\psi\,
   |\tilde{p}|^{\frac{5}{2}}\sin^{\frac{7}{2}}\psi
   J_{\frac{5}{2}}(|\tilde{p}||\tilde{\vec{z}}|\sin\psi)\right)^2
   \nonumber\\
&&+\frac{80\pi^2}{9}
   \left(\frac{1}{\sqrt{2|\tilde{\vec{z}}}|}
   \int\limits_0^\infty\!\!d|\tilde{p}|\int\limits_0^\pi\!\!d\psi\,
   |\tilde{p}|^{\frac{5}{2}}\sin^{\frac{7}{2}}\psi
   J_{\frac{1}{2}}(|\tilde{p}||\tilde{\vec{z}}|\sin\psi)\right)^2
   \nonumber\\
&&\left.+\frac{320\pi^2}{9}
   \left(\frac{1}{\sqrt{2|\tilde{\vec{z}}}|}
   \int\limits_0^\infty\!\!d|\tilde{p}|\int\limits_0^\pi\!\!d\psi\,
   |\tilde{p}|^{\frac{5}{2}}\sin^{\frac{7}{2}}\psi
   J_{\frac{5}{2}}(|\tilde{p}||\tilde{\vec{z}}|\sin\psi)\right)^2\right)\,.
\end{eqnarray}
Exploiting the integral formulas which we have proved in the previous Sec.\,(\ref{appendix vacuum2}) yields:
\begin{eqnarray}
\bra\Theta_{00}(x)\Theta_{00}(y)\ket^{\tiny\mbox{vac}}&=&
   \frac{1}{(2\pi^8)\beta^8}
   \left(80\pi^2\left(\frac{2\pi}{|\tilde{\vec{z}}|^4}\right)^2
   +\frac{160\pi^2}{9}\left(\frac{-2\pi}{|\tilde{\vec{z}}|^4}\right)^2
   \nonumber\right.\\
   &&\hspace{10mm}\left.
   +\frac{160\pi^2}{9}\left(\frac{8\pi}{|\tilde{\vec{z}}|^4}\right)^2
   +\frac{80\pi^2}{9}\left(\frac{-2\pi}{|\tilde{\vec{z}}|^4}\right)^2
   +\frac{320\pi^2}{9}\left(\frac{8\pi}{|\tilde{\vec{z}}|^4}\right)^2
   \right)\nonumber\\
&=&\frac{15}{\pi^4|\vec{z}|^8}\,.
\end{eqnarray}
This results is exactly equal to the result obtained in Coulomb gauge. This is another evidence for the gauge invariance of the two-point correlator of energy density.